%% file: thesis.tex
\documentclass[10pt,b5paper,twoside,english]{myphdthesis}
\usepackage{ThALV} % 1st page (frontespizio) style [personal data must be edited in ThALV.sty] 

\usepackage[utf8]{inputenc} % To have accented letters, e.g. option "utf8" works for ubuntu
\DeclareUnicodeCharacter{D7}{$\times$}
\DeclareUnicodeCharacter{B0}{$^\circ$}
\DeclareUnicodeCharacter{B1}{$\pm$}

\usepackage{palatino} % Font family 
\usepackage{babel} % language package 

\usepackage{lettrine}
\usepackage{epigraph} % To have an extra page w/ epigraph (optional)

\makeatletter
\renewcommand{\@dotsep}{10000} %remove dots from toc and lof&lot
\makeatother
\usepackage[nottoc]{tocbibind} % To include in TOC also the bilbiograhy and the index

\usepackage{longtable}
\usepackage{deluxetable}  
\usepackage[table]{xcolor}
\usepackage{arydshln}

\usepackage[figuresright]{rotating} % figures
\usepackage{subfigure}
\usepackage{float}
\usepackage[small,bf]{caption} % Caption style

\usepackage{amsmath,amsfonts,amssymb,amsthm}  
\usepackage{latexsym}

\newcommand{\Epar}{E_{||}}
\newcommand{\Eperp}{E_{\perp}}
\newcommand{\lcdm}{$\Lambda \mathrm{CDM}$}
\newcommand{\be}{\begin{equation}\label}
\newcommand{\ee}{\end{equation}}

\def\deg{\hbox{$^\circ$}}

\def\mean#1{\left< #1 \right>}
\def\abs#1{\vert #1 \vert}

\def\squareb#1{\left [ #1 \right ]}
\def\roundb#1{\left ( #1 \right )}
\def\curlyb#1{\left \{ #1 \right \}}

\usepackage{natbib}
\usepackage{fancyhdr}
\renewcommand{\chaptermark}[1]{\markboth{#1}{}}
\renewcommand{\sectionmark}[1]{\markright{\thesection\ #1}}
\renewcommand{\headrulewidth}{0.2pt}
\renewcommand{\footrulewidth}{0pt}

\fancypagestyle{plain}{
\fancyhf{} % removes everything 
\fancyfoot[C]{\bfseries \thepage} % except the center
\renewcommand{\headrulewidth}{0pt}
\renewcommand{\footrulewidth}{0pt}}
\renewcommand\arraystretch{1.2}

\begin{document}

\frontmatter
\setcounter{page}{1}
\maketitle

% Table of contents
\tableofcontents 
\thispagestyle{plain}

%\thispagestyle{fancy}

% List of figures
\listoffigures %\thispagestyle{fancy}
\renewcommand{\chaptermark}[1]{\markboth{#1}{}}
\renewcommand{\sectionmark}[1]{\markright{\thesection\ #1}}
\fancyhead[LE]{\bf \thepage}
\fancyhead[RO]{\bf \thepage}
\fancyhead[LO]{{\it List of Figures}}
\fancyhead[RE]{\it  List of Figures}
\renewcommand{\headrulewidth}{0.2pt}
\renewcommand{\footrulewidth}{0pt}

% List of tables
\listoftables 
\renewcommand{\chaptermark}[1]{\markboth{#1}{}}
\renewcommand{\sectionmark}[1]{\markright{\thesection\ #1}}
\fancyhead[RE,RO]{\bf \thepage}
\fancyhead[LO]{{\it List of Tables}}
\fancyhead[LE]{\it  List of Tables}
\renewcommand{\headrulewidth}{0.2pt}
\renewcommand{\footrulewidth}{0pt}

% Introduction
\renewcommand{\chaptermark}[1]{\markboth{#1}{}}
\renewcommand{\sectionmark}[1]{\markright{\thesection\ #1}}
\fancyhead[RO]{\bf \thepage}
\fancyhead[LE]{\bf \thepage}
\fancyhead[LO]{{\it Introduction}}
\fancyhead[RE]{\it Thesis overview}
\renewcommand{\headrulewidth}{0.2pt}
\renewcommand{\footrulewidth}{0pt}
\addcontentsline{toc}{chapter}{Introduction}
\input{Intro} \thispagestyle{fancy}

%%%%%%%%%%%%%%%%%%%%%%%%%%%%%%%%%%%%%%%%%%%%%%%%%%%%%%%%%%%%%%
% Mainmatter
%%%%%%%%%%%%%%%%%%%%%%%%%%%%%%%%%%%%%%%%%%%%%%%%%%%%%%%%%%%%%%
\mainmatter

% page style for mainmatter
\renewcommand{\chaptermark}[1]{\markboth{#1}{}}
\renewcommand{\sectionmark}[1]{\markright{\thesection\ #1}}
\fancyhead[RO]{\bf \thepage}
\fancyhead[LE]{\bf \thepage}
\fancyhead[LO]{{\it \leftmark}}
\fancyhead[RE]{\it \rightmark}

\renewcommand{\headrulewidth}{0.2pt}
\renewcommand{\footrulewidth}{0pt}

% Equation numeration for mainmatter
\renewcommand{\theequation}{\arabic{chapter}.\arabic{equation}}

\input{chap1}
\input{chap2}

\vspace{0.2cm}

%Part I 
\input{PartI}\thispagestyle{empty}
\addcontentsline{toc}{part}{\large Part I: Noise characterization of the LSPE/STRIP polarimeters} \thispagestyle{empty}
\input{chap3}

\input{chap4}
\vspace{0.2cm}

% Part II
\input{PartII}\thispagestyle{empty}
\addcontentsline{toc}{part}{\large Part II: The LSPE/STRIP scanning strategy} \thispagestyle{empty}
\input{chap5}
\input{chap6}
\vspace{0.2cm}

% Part III
%% \input{PartIII}\thispagestyle{empty}
%% \addcontentsline{toc}{part}{\large Part III: Measuring the Earth's atmosphere from the ground} \thispagestyle{empty}
%% \input{chap7}
%% \input{chap8}
%% \vspace{0.2cm}

% Conclusions
\input{Concl}

\vspace{0.2cm}

% Appendices
\addcontentsline{toc}{part}{\large Appendices} \thispagestyle{empty}
\input{App}\thispagestyle{empty}

\appendix
\makeatletter
\@addtoreset{equation}{section}  % Equation numeration for appendices
\makeatother
\renewcommand{\theequation}
  {\thesection.\arabic{equation}}

\renewcommand{\theequation}{A.\arabic{equation}}
\input{App1}
\renewcommand{\theequation}{B.\arabic{equation}}
\input{App2}

\renewcommand{\theequation}{\arabic{chapter}.\arabic{equation}}

%%%%%%%%%%%%%%%%%%%%%%%%%%%%%%%%%%%%%%%%%%%%%%%%%%%%%%%%%%%%%%%%%%%%%%%%
% BACKMATTER
%%%%%%%%%%%%%%%%%%%%%%%%%%%%%%%%%%%%%%%%%%%%%%%%%%%%%%%%%%%%%%%%%%%%%%%%

\backmatter

% Bibliography
\pagestyle{fancy}
\fancyhf{}
\renewcommand{\chaptermark}[1]{\markboth{#1}{}}
\renewcommand{\sectionmark}[1]{\markright{\thesection\ #1}}
\fancyhead[LE]{\bf \thepage}
\fancyhead[RO]{\bf \thepage}
\fancyhead[LO]{{\it Bibliography}}
\fancyhead[RE]{\it Bibliography}
\renewcommand{\headrulewidth}{0.2pt}
\renewcommand{\footrulewidth}{0pt}

\bibliography{thesis} 
\bibliographystyle{aa} % style aa.bst

% your references Yourfile.bib
%\input{biblio} \thispagestyle{plain}
%\bibliography{tesi}
%\bibliographystyle{apj} %apj style

%\addcontentsline{toc}{chapter}{Nomenclature} \thispagestyle{plain}
%\printnomenclature % to get the list of symbols in the toc
%\printindex \thispagestyle{plain}   % to get the index in the toc
%\thispagestyle{plain}

% List of publictions
\input{pubs}
\addcontentsline{toc}{chapter}{List of Publications} 
\pagestyle{fancy}
\fancyhf{}
\renewcommand{\chaptermark}[1]{\markboth{#1}{}}
\renewcommand{\sectionmark}[1]{\markright{\thesection\ #1}}
\fancyhead[LE]{\bf \thepage}
\fancyhead[RO]{\bf \thepage}
\fancyhead[LO]{{\it List of Publications}}
\fancyhead[RE]{\it  List of Publications}
\renewcommand{\headrulewidth}{0.2pt}
\renewcommand{\footrulewidth}{0pt}

% Acknowledgments
%% \input{Ackno}  
%% \addcontentsline{toc}{chapter}{Acknowledgments} 
%% \pagestyle{fancy}
%% \fancyhf{}
%% \renewcommand{\chaptermark}[1]{\markboth{#1}{}}
%% \renewcommand{\sectionmark}[1]{\markright{\thesection\ #1}}
%% \fancyhead[LE]{\bf \thepage}
%% \fancyhead[RO]{\bf \thepage}
%% \fancyhead[LO]{{\it Acknowledgments}}
%% \fancyhead[RE]{\it  Acknowledgments}
%% \renewcommand{\headrulewidth}{0.2pt}
%% \renewcommand{\footrulewidth}{0pt}

\end{document}

%% file: Intro.tex
\chapter*{Introduction}
\thispagestyle{plain}

\section*{Motivation}
\addcontentsline{toc}{section}{Motivation}

Who are we and where do we go? These questions interrogate the humankind since the dawn of time and certainly they will never be over. The crave to answer these questions drives man into amazing adventures of knowledge and even bigger challenges. Modern science is one of these. How and when was the Universe born? How and when will it end? Observational cosmology studies the Universe as a whole and tries to find the answer to these dramatic questions. \par
The birth of observational cosmology must be placed in the very recent history. Before the 1920s in fact, the concept of Universe was limited to the Milky Way: external galaxies were called \textit{nebulae} and were supposed to belong to our Galaxy as well as all other objects observed in the sky. In 1915 the astronomer Edwin Hubble identified a Cepheid variable in the Andromeda Galaxy and he established its distance. It was actually greater than the Milky Way radius, so that the Universe had to be broader. In 1929, Hubble also discovered that galaxies are receding from us with radial velocities proportional to their distances. This enabled us to understand that the Universe is expanding: this was the actual beginning of the observational cosmology. \par
A theoretical understanding of how the Universe has formed and evolved was possible in recent times as well. In 1915 in fact, the publication of Einstein's theory of general relativity opened new scenarios where theoretical cosmologists could move. Only seven years later, in 1922, Friedmann proposed a solution to the Einstein field equations describing how the Universe could expand, leading to the concept of the \textit{Big Bang}. The Belgian priest, physicist and astronomer Georges Lemaître arrived to the same conclusion independently in 1927. He explained the linear relation between the velocity and the distance of the receding \textit{nebulae}, discovered by Hubble, through his own model of an expanding Universe \citep{1927ASSB...47...49L}. \par
Nowadays, many more steps towards a better knowledge of the physical laws of our Universe have been done. We live in a very fascinating era indeed, rich of scientific discoveries. I remember, for example, the first detection of the Higgs boson \citep{20121, Chatrchyan:2012xdj} and of the gravitational waves generated by the merging of a binary black hole \citep{PhysRevLett.116.061102} as well as the very recent first-ever direct image of a black hole \citep{Akiyama:2019cqa}. \par  
From the astrophysical point of view, the detection of the gravitational waves by the LIGO and Virgo experiments is certainly the most important one. It confirms, in fact, Einstein's prediction made one hundred years ago, providing further evidence of the robustness of general relativity. It also provide us a brand new ``sense'' that allow us to explore the Universe from a different point of view. \par
Interferometers such as LIGO and Virgo are devoted to detect directly the shortest gravitational waves but, if they exist, primordial gravitational waves must have longer wavelengths that cannot be revealed directly. However, their presence could be inferred by cosmological experiments. In this respect, the \textit{Cosmic Microwave Background} (CMB) experiments could be considered ``detectors'' as well, ``measuring'' gravitational waves on the scale of the whole Universe. \par
Currently, the most accredited cosmological models foresee that the primordial Universe was permeated by a stochastic background of gravitation waves. Their existence must be tracked by the presence of the so-called \textit{B-modes} on the polarization anisotropies pattern of the CMB. \par
The quest for B-modes is one of the major challenges in modern cosmology. Their detection would give us an important evidence in favor of the \textit{inflationary paradigm} and, in general, on the physics of the very early Universe. However, the amplitude of this signal is expected to be very low, at the level of fraction of $\mu\mathrm{K}$. For this reason, its detection requires high sensitivity instruments with tens of thousands of detectors, a rigorous control of systematic effects and a very precise knowledge of the foreground polarized emission produced by our own Galaxy. \par
The ``Large Scale Polarization Explorer'' (LSPE) is a CMB experiment searching for B-modes. It is composed of two instruments: SWIPE, a stratospheric balloon, and STRIP, a ground-based telescope. My thesis has been carried out in the framework of LSPE/STRIP. Within this collaboration, I was part of the simulation and data analysis group. During the three years of my PhD, I contributed to optimize the scanning strategy of LSPE/STRIP and I took part to the unit-level test campaign on the LSPE/STRIP receivers. \par
To be member of such collaboration was for me a honor.               

\section*{Thesis overview}
\addcontentsline{toc}{section}{Thesis overview} 

\subsection*{Abstract} % max 300 words 
Detecting B-mode polarization anisotropies on large angular scales in the Cosmic Microwave Background (CMB) polarization pattern is one of the major challenges in modern observational cosmology since it would give us an important evidence in favor of the inflationary paradigm and would shed light on the physics of the very early Universe. Multi-frequency observations are required to disentangle the very weak CMB signal from diffuse polarized foregrounds originating by radiative processes in our galaxy.\par
The ``Large Scale Polarization Explorer'' (LSPE) is an experiment that aims to constrain the ratio, $r$, between the amplitudes of tensor and scalar modes to $\simeq 0.03$ and to study the polarized emission of the Milky Way. \par
LSPE is composed of two instruments: SWIPE, a stratospheric balloon operating at $140$, $210$ and $240\,\mathrm{GHz}$ that will fly for two weeks in the Northern Hemisphere during the polar night of 2021, and STRIP, a ground-based telescope that will start to take data in early 2021 from the ``Observatorio del Teide'' in Tenerife observing the sky at $43\,\mathrm{GHz}$ (Q-band) and $95\,\mathrm{GHz}$ (W-band).\par
In my thesis, I show the results of the unit-level tests campaign on the STRIP detectors that took place at ``Università degli Studi di Milano Bicocca'' from September 2017 to July 2018 and I present the code I developed and the simulations I performed to study the STRIP scanning strategy. During the unit-level tests, we performed more than $800$ tests on $68$ polarimeters to select the $55$ ($49$ Q-band and $6$ W-band) with the best performance in terms of central frequencies, bandwidths, noise temperatures, white noise levels, slopes of the pink noise spectrum and knee frequencies. The STRIP scanning strategy, instead, is based on spinning the telescope around the azimuth axis with constant elevation in order to overlap the SWIPE coverage, maintaining a sensitivity of $1.6 \; \mu \mathrm{K}$ (on average) per sky pixels of $1\deg$. Individual sources will be periodically observed both for calibration and study purposes. 

\subsection*{Summary}
Detecting \textit{B-mode} polarization anisotropies in the cosmic microwave background (CMB) polarization pattern is one of the major challenges in modern observational cosmology. According to the inflationary paradigm, which predicts an accelerated expansion of the Universe occurred at $t \simeq 10^{-34} \,\mathrm{sec}$ after the Big Bang, these anisotropies are the signature of a stochastic background of tensor perturbations (i.e. gravitational waves) that permeated the primordial Universe. These perturbations concurred with scalar perturbations in producing the matter density fluctuations that we observe today in the Universe. \par
A positive detection of B-modes in the CMB on the degree angular scales would provide us with important evidence in favor of the inflationary paradigm. In particular, it would allow us to constrain the value of $r$, which express the ratio between the amplitudes of tensor and scalar modes, and to shed light on the inflation process and on the physics of the very early Universe. \par
Multi-frequency observations are required to disentangle the very weak CMB signal from diffuse polarized foregrounds originating by radiative processes in our galaxy. At frequencies lower than $100\,\mathrm{GHz}$, the polarized sky emission is dominated by synchrotron radiation from electrons moving in the galactic magnetic field. Above that frequency, thermal emission from the interstellar dust is the major contaminant. \par
The ``Large Scale Polarization Explorer'' (LSPE) is an experiment that aims to constrain the value of $r$ to $\simeq 0.03$ at the $99.7\,\%$ confidence level and to study the polarized emission of the Milky Way. So far, in fact, there are no data of the polarized emission of our own Galaxy at large angular scales and at multiple frequencies from the Earth's Northern Hemisphere, whose sensitivity is higher than the one reached by the Planck satellite.\par
The LSPE experiment is composed of two instruments: SWIPE is a stratospheric balloon operating at $140$, $210$ and $240\,\mathrm{GHz}$ that will fly from the Svalbard Islands (or from Kiruna) for two weeks during the polar night of 2021. STRIP instead is a ground-based telescope that will start to take data in early 2021 from the ``Observatorio del Teide'' in Tenerife. It will observe the sky at $43\,\mathrm{GHz}$ and $95\,\mathrm{GHz}$, even though this channel will be used principally as an atmosphere monitor.\par
In my thesis, I show the results of the unit-level tests on the LSPE/STRIP detectors that have been held at ``Università degli Studi di Milano Bicocca'' from September 2017 to July 2018 and I present the code I developed and the simulations I performed to study the LSPE/STRIP scanning strategy.\par
The detectors used by STRIP are polarimeters that, thanks to the combination of several radio-frequency components, produce an overall response proportional to the four Stokes parameters of the incident radiation field. Furthermore, their electronics is able to reduce the correlated noise, the 1/$f$ component, by many order of magnitudes thanks to the so-called \textit{double demodulation} process. \par
During the unit-level tests campaign of STRIP, we performed more than $800$ tests, at cryogenic temperatures, on $68$ polarimeters to select the $55$ ($49$ Q-band and $6$ W-band) with the best performance. We ran three kinds of tests: a bandpass characterization, a \textit{Y-factor} test to estimate the noise temperature and a long acquisition to measure the noise characteristics. \par
The main output of the tests analysis was a list of the central frequencies, bandwidths, noise temperatures, white noise levels, slopes of the pink noise spectrum and knee frequencies for the whole batch of tested polarimeters. This allowed us to select the $55$ ones with the best noise temperature to deploy on the focal plane. The analysis showed high uncertainties. Possible sources of systematic errors are non-linearity in detector (or ADC) response, uncertainty in ADC offset, non-idealities in the polarimeters or in the set-up, etc. More investigations could not be conducted since fundamental \textit{house-keeping} parameters were not recorded by the acquisition software. \par
In the second part of my thesis, I present the STRIP simulation pipeline, which is called \textit{Stripeline} and is written in the Julia programming language. It is based on several modules that: collect all the information about the focal plane components (horn positions, horn/detector pairings, detector properties, etc.), simulate the scanning strategy, produce realizations of pseudo-instrumental noise and compute output maps from TODs. \par
I have used Stripeline to study the scanning strategy of the STRIP instrument, which is driven by three main goals: to observe the same sky region of SWIPE, to obtain at the same time a good sensitivity per sky pixel and a wide sky coverage, to include specific sources in the field of observation. A good trade-off between these three conditions is obtained by spinning the telescope around the azimuth axis with constant elevation and angular velocity. In this way, each receiver will observe also the same air column collecting the same signal due to the atmosphere. \par
I found that for constant elevation angles between $20\deg$ and $25\deg$ from the zenith, the STRIP coverage overlaps the SWIPE one at the $80\%$ level but we also need to ensure a duty cycle greater than $50\%$ to satisfy the STRIP sensitivity requirement, which is set to $1.6 \; \mu \mathrm{K}$ (on average) per resolution elements of $1\deg$. The combination of the telescope motion with the Earth rotation will guarantee the access to the large angular scales. We will observe periodically the Crab Nebula as well as the Perseus molecular complex. The Crab is one of the best known polarized sources in the sky and it will be observed for calibration purposes. The second one is source of \textit{Anomalous Microwave Emission} (AME) that could be characterized both in intensity and polarization.\par
I found also that, to make the sky coverage more uniform, the elevation of STRIP can be modulated from $5\deg$ to $35\deg$ through a proper \textit{bezier} function, while the telescope is spinning. With respect to the nominal scanning mode at $35\deg$, this modulation allow the instrument to increase the average sensitivity but, at the same time, the instrument duty cycle must be ensured to be about $75\%$ to reach the STRIP sensitivity requirement. Further studies are required to assess the effectiveness of this method. In particular, cross-checks with the most recent atmospheric data from Teide Observatory are required to estimate the time-scales on which the atmospheric brightness temperature varies and to define the modulation period. Besides, a fine characterization of the instrumental I $\to$ Q, U leakage must be performed to evaluate the impact of the modulation on the polarization measurements. Finally, the possibility to use a different class of functions to modulate the elevation or to spin the telescope with a non-constant angular velocity (e.g., which changes as a function of the elevation) must be investigated. \par

\subsection*{Organizational note}

The present thesis consists of two parts, for a total of six chapters. The first two chapters are introductory and explain the contexts in which this thesis fits. Part I includes chapters 3 and 4 and is dedicated to show the results of the unit-level tests on the STRIP detectors. Part II is composed by chapters 5 and 6 and presents the code I developed and the simulations I performed to study the STRIP scanning strategy. %Part III, which includes chapters 7 and 8, connects the two previous parts since it shows how to measure the atmosphere's temperature from Tenerife using the scanning strategy described in part II and the noise properties of the detectors shown in part I.
\newline \newline
A more detailed view of the thesis structure is provided in the following description.

\begin{list}{\leftmargin 15pt \itemsep 0pt \topsep 3pt}
\item{\bf Chapter \ref{Chap:1}: The search for the CMB B-modes.}
  I describe the current cosmological framework and the evidences of the accelerating expansion of the Universe. I show how the expansion can described by Einstein and Friedmann equations, which lead to the concept of an initial singularity: the so-called \textit{Big Bang}. Then, I describe the main problems of this theory and show how the concept of \textit{inflation} can solve them. I present the \textit{Cosmic Microwave Background} (CMB) argument outlining its intensity and polarization features. Finally, I introduce the E and B-modes in the power spectra of the CMB anisotropies and show what precious information a B-modes detection could give us. A summary description about the state-of-the art of the search for B-modes concludes the chapter.
\item{\bf Chapter \ref{Chap:2}: The LSPE experiment.}
  I introduce the ``Large Scale Polarization Explorer'' (LSPE) and describe the two instruments the experiment is made by: SWIPE, the high frequencies instrument, and STRIP, the low frequencies one.
\item{\bf Chapter \ref{Chap:3}: The STRIP detection chain.}
  I explain the mathematical and physical model of the STRIP detection chain. It consists in a sequence of an antenna, the so-called \textit{feedhorn}, an orthomode transducer (OMT) and a polarimeter, which is the proper microwave detector. Going through the model, I show how they are able to detect directly the four Stokes parameters of an incident radiation field. I illustrate how the instrumental 1/$f$ noise is reduced of several orders of magnitude through the \textit{double demodulation} process.
\item{\bf Chapter \ref{Chap:4}: Unit-level tests.}
  I address the functionality and performance tests carried out on the STRIP polarimeters during the unit-level tests campaign at the University of ``Milano Bicocca''. I describe the experimental set-up and discuss the performed tests, both at room and cryogenic temperature. I illustrate in detail the methods, the calculations and the codes used to estimate the bandwidths, the central frequencies and the noise temperatures. Then, I focus on the noise characterization of the detectors. Therefore, I show the distribution of the knee frequency, slope of the pink noise spectrum and white noise level of the analyzed polarimeters.   
\item{\bf Chapter \ref{Chap:5}: Stripeline: the STRIP simulation pipeline.}
  I present the STRIP simulation pipeline. First, I provide a general overview of the three main parts of the code: the instrument database, the pointing generation and the map-making. Then, I focus on the pointing simulation code that is my main contribution. Here, I define a ground (local) reference system and show how to translate it to the sky (absolute) reference system. I mention also the secondary effects that affects the pointing accuracy in the case of ground observations. 
\item{\bf Chapter \ref{Chap:6}: Scanning strategy analysis.}
  I deal with the problem of the STRIP scanning strategy. I initially describe the guidelines driving the analysis. I show how spinning the telescope at constant elevation allow us to maximize the overlap with the sky region observed by SWIPE, to trade-off the sky coverage with the noise per pixel distribution and to include specific sources in the instrument field of view. I present the results of my simulations in terms of coverage, noise and hit count maps as a function of the telescope elevation. I discuss the results including some consideration about the instrument duty cycle. Finally, I explore the possibility to slowly modulate the elevation angle, while the telescope is spinning, in order to make the sky coverage as uniform as possible.
\item{\bf Conclusions.}
  I review the main results of the thesis and give an outlook on possible future developments.
\item{\bf Appendix \ref{App:1}: Math of the polarimeter model.}
  I address the mathematical details of the STRIP polarimeter model, described in Ch. \ref{Chap:3}. Given the incident radiation field, I compute the expected value of the electric fields, along with their amplitudes, at the four detector outputs in terms of the Stokes parameters.     
\item{\bf Appendix \ref{App:2}: Math of the polarimeter model in the unit-level tests configuration.} I repeat the same calculation of Appendix \ref{App:1} in the case of the experimental set-up used during the unit-level tests campaign.
\end{list}

%% file: chap1.tex
\chapter{The search for the CMB B-modes}
\label{Chap:1}
\thispagestyle{plain}

\section{The expanding Universe}
\label{expuni}
The \textit{cosmological principle} asserts that on the largest scales the Universe is homogeneous and isotropic, which means that its physical properties are the same everywhere and there are no special directions. Observations confirm this principle on very large scales \citep[$\gtrsim 100 \;\mathrm{Mpc}$,][]{Lahav2001}. \par
Furthermore, as discovered by Hubble, galaxies are receding from us with an apparent recession velocity increasing linearly with the distance. \par
In this section, I give an outlook to the mathematical models and the physical laws that, starting from an expanding homogeneous and isotropic Universe, lead to concept of the \textit{Big Bang}.

\subsection{The Hubble's law}
\label{hlaw}
In 1929, Hubble noticed that the galaxies have red-shifted\footnote{See Eq. \ref{redshift} for a definition of the cosmological \textit{redshift}.} spectral lines meaning that they are moving far away from us \citep{1929PNAS...15..168H}. He also observed that their velocity is increasing with distance (Fig. \ref{hubblelaw}), according to the law:

\begin{figure}[H]
\centering
\includegraphics[scale=0.2]{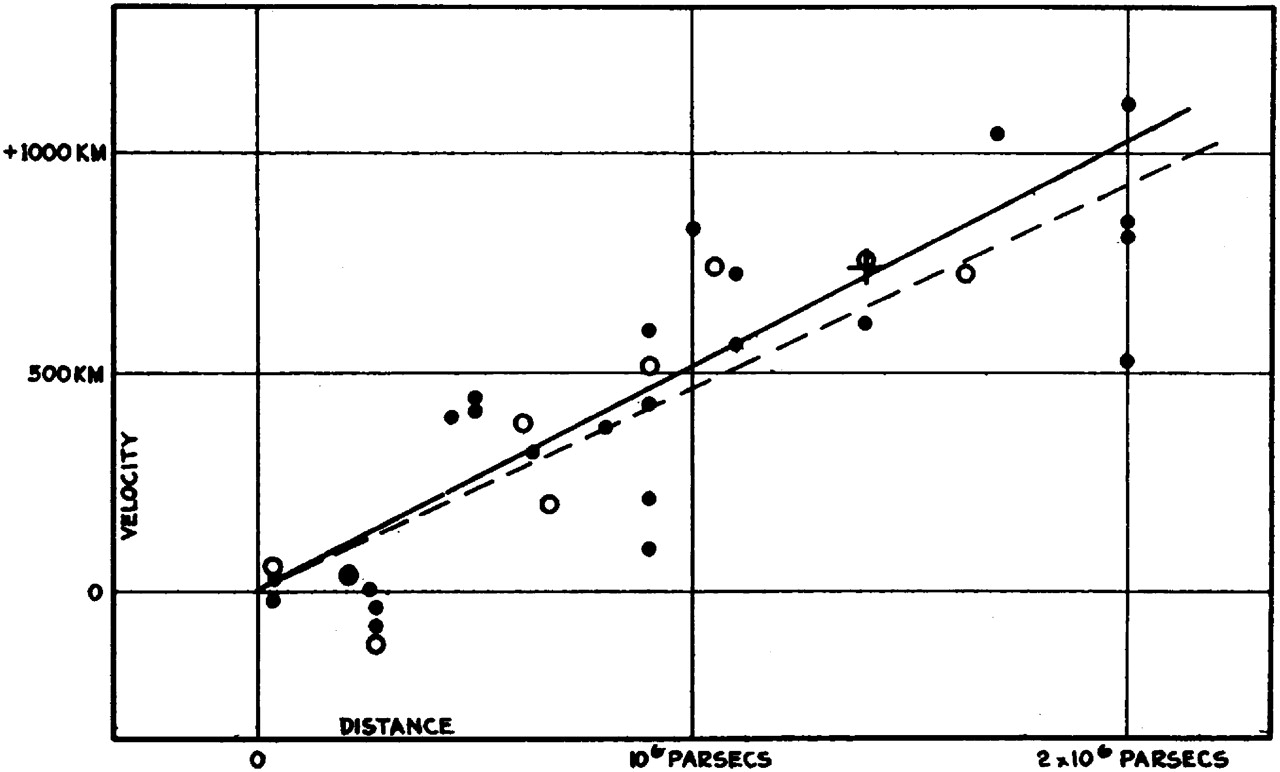}
\caption{Hubble's original diagram showing the radial velocity (units should be $\mathrm{km\,s}^{-1}$) of galaxies as a function of their distance. The solid line is the best-fit to the data (filled points) taking into account the solar motion correction. The dashed line represents the best-fit to the data combined into groups (open points) and neglecting the solar motion. The cross represents the mean velocity of a group of twenty-two galaxies whose distance could not be estimated independently.}\label{hubblelaw} 
\end{figure}

\be{hubble}
\textbf{v} = H_0 \, \textbf{d} \,,
\ee
where $H_0$ is the \textit{Hubble constant}, and its value\footnote{See \citep{2017NatAs...1E.121F} for a discussion about the \textit{$H_0$ tension} or the problem of the relevant discrepancies persistent among local and cosmological measurements of the Hubble constant.} has been recently measured with a $\lesssim 1\%$ precision \citep{Aghanim:2018eyx}:  
\be{planckH0}
H_0 = 67.37 \pm 0.54 \; \mathrm{km \, s^{-1} \, {Mpc}^{-1}} \, .
\ee \par 
This law is exactly what is expected in an expanding Universe. We can introduce a coordinate system that follows the time expansion of the Universe:
\be{coordcomov}
\textbf{r}(t) = a(t) \, \textbf{x} \,,
\ee
where $\textbf{r}(t)$ are called \textit{proper coordinates} and represent the physical distances, $a(t)$ is the \textit{scale factor} that describes the temporal dynamics, and $\textbf{x}$ are the \textit{comoving coordinates} that are independent of the expansion. Conventionally $a(t)$ is chosen so that, at the present time, $t_0$:
\be{acondition}
a(t_0) = 1 \,,
\ee
and, since today the Universe is expanding:
\be{acondition2}
\dot{a}(t_0) > 0 \,.
\ee \par
Taking the derivative of Eq. \ref{coordcomov}, we find Hubble's law:
\be{hubblederiv}
\textbf{v}(\textbf{r}, t) = \dot{\textbf{r}} = \dot{a}\textbf{x} = \frac{\dot{a}(t)}{a(t)}\textbf{r} \,,
\ee
with:
\be{hubbleconstant0}
\frac{\dot{a}(t)}{a(t)} \equiv H(t) \,.
\ee \par
We define the Hubble constant as:
\be{hubbleconstant}
H_0 \equiv H(t_0) = \frac{\dot{a}(t_0)}{a(t_0)} = \dot{a}(t_0) \,,
\ee
where we have used the convention defined by Eq. \ref{acondition}. It is possible to show that this is the only expansion law compatible with the cosmological principle \citep[p. 5]{mukhanov_2005}.\par
If the Universe as a whole expands, the radiation emitted by a far source (whose wavelength is $\lambda_\mathrm{emit}$) will reach us with an increased wavelength ($\lambda_\mathrm{obs}$). Then we can define \textit{z}:
\be{redshift}
1 + z(t) \equiv \frac{\lambda_\mathrm{obs}}{\lambda_\mathrm{emit}} = \frac{1}{a(t)} \,,
\ee 
which, for non-cosmological distances is due to the relative (proper) motion between sources and can be a \textit{redshift} for objects that are receding or a \textit{blueshift} for objects that are approaching; for cosmological distances instead, it is due to the expansion of the Universe so that it is always a \textit{redshift}. \par

\subsection{Einstein's equations}
With the discovery of general relativity, Einstein succeeded in relating the mass of the objects to corresponding metric deformations. This relation is given by the \textit{Einstein field equations}: 
\be{einstein}
G_{\mu \nu} + g_{\mu \nu} \Lambda = \frac{8 \pi G}{c^4} T_{\mu \nu} \,,
\ee
where: $G_{\mu \nu}$ is the \textit{Einstein tensor} that describe the field of curvature of the Universe; $T_{\mu \nu}$ is the \textit{energy-momentum tensor} that defines the density and flux of energy and momentum in space-time; $\Lambda$ is the so-called \textit{cosmological constant}; $g_{\mu\nu}$ is the \textit{metric tensor} that must be determined.\par
The term $\Lambda$ was introduced \textit{ad hoc} by Einstein, because it allows static solutions that Einstein considered the only physically meaningful. When the expansion of the Universe was demonstrated by Hubble, Einstein rejected the $\Lambda$ term and considered it the ``biggest blunder'' in his life. Nowadays, the cosmological constant plays again a central role in modern cosmology to explain the currently observed accelerated expansion (see Sect. \ref{stacos}). \par

\subsection{Friedmann-Lemaître-Robertson-Walker metric}
\label{flrwm}
From special relativity we know that, in four dimensions, the space-time interval is invariant under changes of inertial reference frame:
\be{metric}
d s^2 = \sum_{\mu, \nu} g_{\mu\nu} d x^\mu \, d x^\nu ,
\ee
with $g_{\mu\nu}$:
\be{gmunu}
g_{\mu\nu} = \left( \begin{matrix} -1 & 0 & 0 & 0 \\ 0 & 1 & 0 & 0 \\ 0 & 0 & 1 & 0 \\ 0 & 0 & 0 & 1 \end{matrix} \, \right) \,,
\ee
where $\mu$ and $\nu$ range from $0$ to $4$, with $0$ being the time coordinate while the remaining are the spatial ones.\par
In an expanding Universe the spatial terms in Eq. \ref{gmunu} should be multiplied by the scale factor $a(t)$ since the distance between two points is always proportional to it. Besides, by transforming Eq. \ref{metric} into spherical coordinates and normalizing for the curvature of the Universe it is possible to obtain the so-called \textit{Friedmann-Lemaître-Robertson-Walker} (FLRW) space-time metric:
\be{robertsonwalker}
c^2 \, d {\tau}^2 = c^2 \, d t^2 - a^2(t) \, \squareb{d x^2 + f_k^2(x) \, d{\Omega}^2} \,,
\ee
where $x$ is the comoving distance, $d {\Omega}^2 = d {\theta}^2 + {\sin \theta}^2 \, d{\phi}^2$ is the solid angle element and:
\be{fk}
f_k (x) = \begin{cases} 
  1 / \sqrt k \sin{(\sqrt{k x})}     & \text{if } \; k > 0 \,, \\
  x                                  & \text{if } \; k = 0 \,,\\
  1 / \sqrt{-k} \sinh{(\sqrt{-k x})} & \text{if } \; k < 0 \,.
\end{cases}
\ee
This metric is an exact solution of Einstein field equations. \par
The quantity $k$ is related to the curvature of the space-time (see Eq. \ref{curvature}): $k > 0$ implies a spherical (\textit{closed}) curvature of the Universe; $k = 0$ corresponds to an euclidean (\textit{flat}) geometry of the Universe; $k < 0$ leads to a hyperbolic (\textit{open}) geometry . \par

\subsection{Friedmann's equations}
\label{feq}
By including the FLRW metric (Eq. \ref{robertsonwalker}) in the Einstein's equations (Eq. \ref{einstein}), it is possible to obtain, for a perfect fluid with a given mass density $\rho$ and pressure $p$, two relations describing the evolution of the scale factor $a(t)$. These equations are the so-called \textit{Friedmann's equations}:
\begin{align}\label{friedmann1noL}
{\roundb{\frac{\dot{a}}{a}}}^2 &= \frac{8 \pi G}{3} \rho - \frac{k c^2}{a^2} \,, \\
\frac{\ddot a}{a} &= - \frac{4 \pi G}{3} \roundb{\rho + \frac{3 p}{c^2}} \,. \label{friedmann1noL2}
\end{align} \par
We consider now a three-component Universe made by matter, radiation and cosmological constant, whose associated pressures and densities are expressed respectively by $p_i$ and $\rho_i$, with $i = \curlyb{m, r, \Lambda}$. The total density is the sum of the three densities where the density of the cosmological constant is defined by:
\be{rhotot}
\rho_{\Lambda} \equiv \frac{\Lambda}{8 \pi G} \,.
\ee \par
Given an equation of state for each component, the Friedmann's equations can be solved leading to a model of evolution of the Universe. For a dilute gas, such as the Universe can be approximated, the equation of state can be written as:
\be{eos}
p = w \rho \,, 
\ee
where the value of $w$ depends on the nature of the gas:
\be{w}
w = \begin{cases} 
  0     & \mathrm{non\textrm{-}relativistic} \,,\\
  \frac13 & \mathrm{relativistic} \,,\\
  -1 & \mathrm{vacuum \ energy} \,.
\end{cases}
\ee
Matter and radiation contribute to the equation of state as respectively non-relativistic and relativistic terms. In both cases, as emerges from Eq. \ref{friedmann1noL2}, they give a positive pressure leading to $\ddot a < 0$ that corresponds to a slowing down expansion. In the case instead of the \textit{vacuum energy}, the zero-point energy that exists in space throughout the Universe, as well as in all the cases in which $w < -1/3$, we obtain a negative contribution to the pressure and $\ddot a > 0$ implying an accelerating expansion. Today we assume that this is the role played by the cosmological constant (see Sect. \ref{da}). 
\par
Let us define now, for $k = 0$, a \textit{critical density}:
\be{criticaldensity}
{\rho}_{cr}(t) = \frac{3 \, H^2(t)}{8 \pi G} \,,
\ee
and then the \textit{density parameters} as:
\be{densityparameters}
\Omega_i (t) = \frac{\rho_i (t)}{\rho_{cr} (t)} \,,
\ee
Substituting these parameters in Eq. \ref{friedmann1noL} and considering quantities at the present time ($t = t_0$) we obtain a relation between the curvature of the Universe and the density parameters:
\be{curvature}
k c^2 = H_0^2 \roundb{\Omega_0 - 1} \,,
\ee
where $\Omega_0 = \sum_i \Omega_i (t_0)$. Notice that $\Omega_0 = 1$ implies a flat geometry.\par
Finally, we can express Eq. \ref{friedmann1noL} in terms of the density parameters:
\be{friedmanndensity}
H^2(t) = H_0^2 \squareb{\frac{\Omega_r(t_0)}{a^4(t)} + \frac{\Omega_m(t_0)}{a^3(t)} + \frac{1 - \Omega_0}{a^2(t)} + \Omega_{\Lambda}(t_0)}.
\ee
Eq. \ref{friedmanndensity} shows that for $a \ll 1$ the radiation term is dominant, so that the primordial Universe was dominated by radiation. As the time evolves, the matter contribution starts to be relevant and finally for $a \to \infty$ the cosmological constant term become preeminent.\par
\begin{figure}[H]
\centering
\includegraphics[scale=0.4]{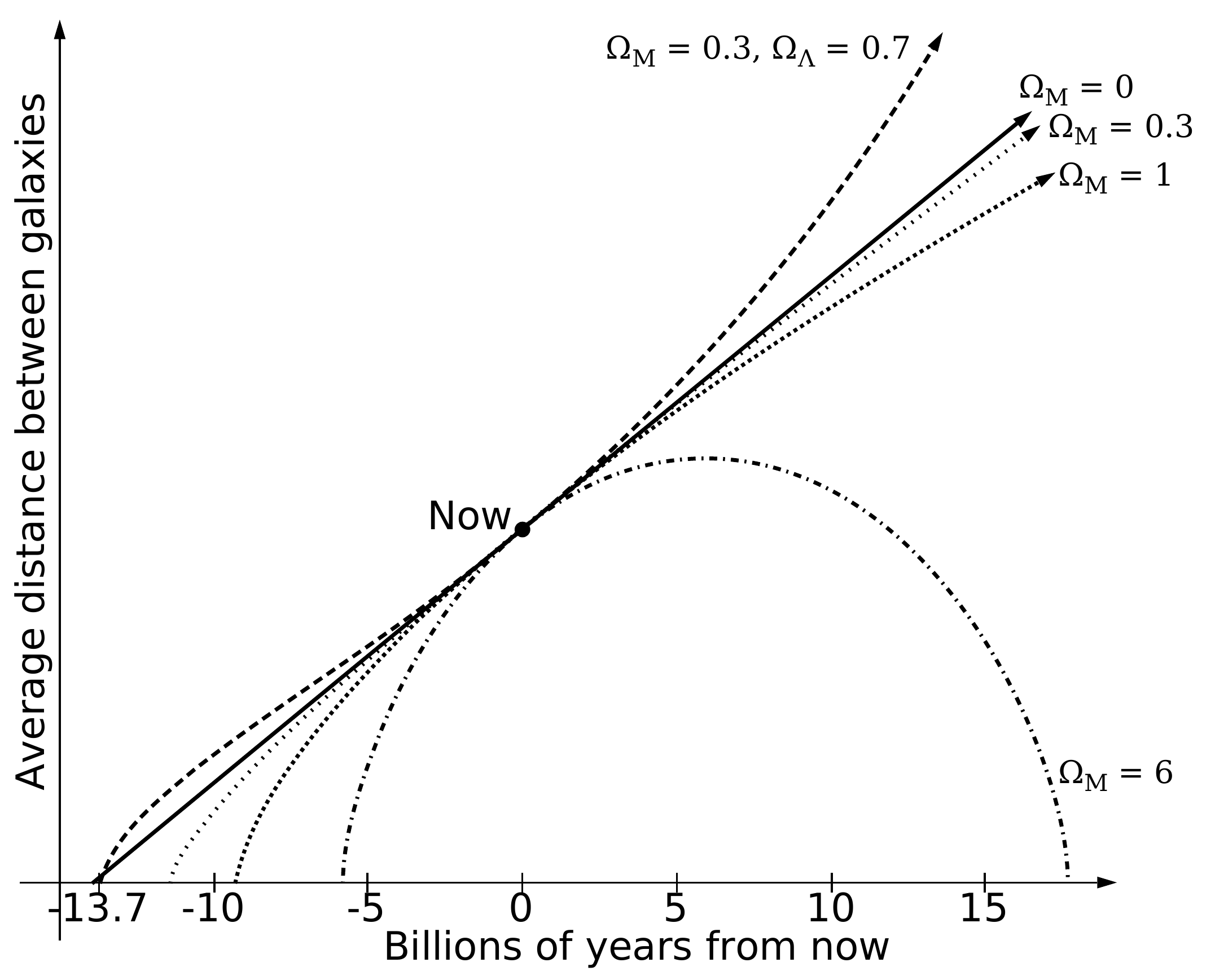}
\caption{The evolution of the scale factor as a function of time. Different scenarios are possible according to the values of the density parameters. A Universe ending in a singularity is possible, in principle, as well as a stationary or an infinitely expanding Universe. Currently, the most accredited model is the latter.}\label{friedmannmodels} 
\end{figure}

Eq. \ref{friedmanndensity} shows also that the evolution of the scale factor depends on the density parameters at the present time. This shows that knowing the current values of the density parameters is key to understand the dynamics of the Universe (Fig. \ref{friedmannmodels}), and this represents one of the main goals of modern observational cosmology. \par

\subsection{Big Bang}
\label{bigbang}
We have seen that expansion implies $\dot a > 0$. Furthermore, in past eras matter and radiation were dominating the evolution of the Universe, so that, as shown previously, pressure was positive and then $\ddot{a} < 0$. This imply that the scale factor $a(t)$ had to go to zero asymptotically. Thus, Friedmann's equations suggest that Universe started to expand and cooling from a hot and high density state, called \textit{Big Bang}. \par
The expansion of the Universe is one of the most important points in favor of the Big Bang. However, there are two others key observations supporting it: the existence and the features of the \textit{cosmic microwave background} radiation (Sect \ref{cmb}) and the excellent agreement between observations and predictions of the abundance of light elements produced by the primordial nucleosynthesis (Sect. \ref{brithe}). \par
The Big Bang is a model that describes the evolution of our Universe, whose birth must be placed $13.801 \;\pm\; 0.024 \; \mathrm{Gyr}$ ago \citep{Aghanim:2018eyx}. In the next section, I briefly focus on the characteristics of this model, which is currently assumed as the standard model of cosmology.\par

\section{The standard cosmological model}
\label{stacos}
The current standard model describing the birth and the evolution of the Universe is called \textit{$\Lambda$ cold dark matter} (\lcdm) model since it is based on two unknown components: the so-called \textit{dark energy}, and \textit{dark matter}. In this section, I deal with these two elements, provide an overview of the cosmological parameters, explore briefly the thermal history of the Universe and, finally, I show the open issues related to this model and how the concept of \textit{inflation} could solve them.

\subsection{Dark energy}
\label{da}
The measurement of the recession velocity and the distance of some high-redshift type Ia supernovae \citep{Perlmutter1999} proves that the expansion rate of the Universe is actually increasing. An accelerated expansion is compatible with several models of flat Universe where the role of ``repulsor'' is played by the cosmological constant. \par
\begin{figure}[H]
\centering
\includegraphics[scale=0.4]{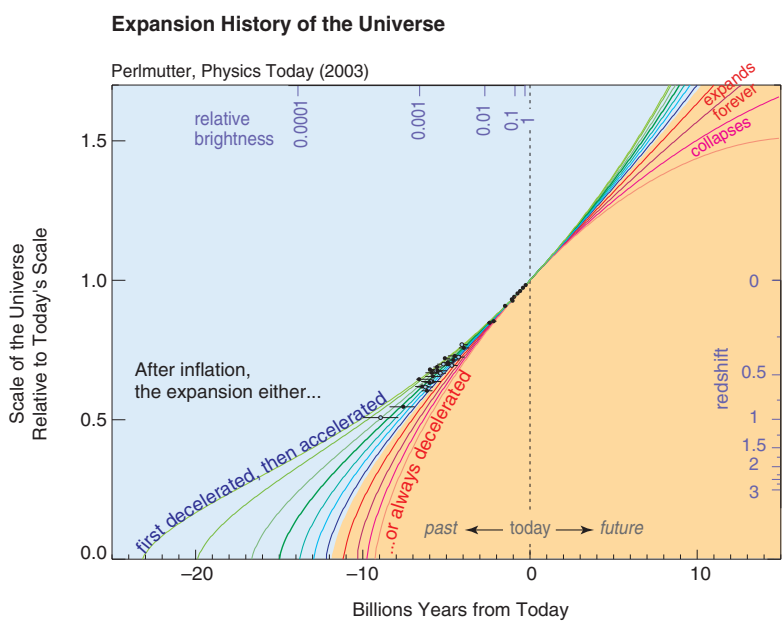}
\caption{Distant supernova measurements (the black data dots). The curves in the blue (yellow) shaded region represent cosmological models in which the dark energy repulsion (matter attraction) overcomes the effect of the matter attraction (dark energy repulsion). The plot shows that we are in a Universe in accelerated expansion. \citep{Perlmutter2003}}\label{perlmutterSN} 
\end{figure}

Fig. \ref{perlmutterSN} shows how the distant supernovae measurements allow to discriminate among several cosmological models. \par
\par
The cosmological constant ($\Lambda$) is responsible of the accelerated expansion of the Universe and it is also called \textit{dark energy}. The physical origin of the dark energy is still unknown. \par
The best measure to date of the abundance of dark energy is \citep{Aghanim:2018eyx}:
\be{lambdadens}
\Omega_\Lambda = 0.6852 \pm 0.0074 \, .
\ee

\subsection{Cold dark matter}
There are many evidences \citep[rotation curves of galaxies, weak lensing measurements, hot gas in clusters, etc. see][]{Freese} that in our Universe a kind of matter exists, along with ordinary matter, which does not interact with electromagnetic radiation. Its physical nature is unknown but it must be non-baryonic and actually it constitutes the bulk of matter: this is the so-called \textit{dark matter}. \par
The total amount of matter in our Universe is \citep{Aghanim:2018eyx}:
\be{mdens}
\Omega_m = 0.3147 \pm 0.0074 \,,
\ee
which is divided in baryonic (ordinary, $\simeq 16\,\%$) and non-baryonic (dark matter, $\simeq 84\,\%$). \par
There are several and very different hypotheses on the nature of the dark matter ranging from weakly interacting particles, such as WIMPs or sterile neutrinos, to primordial black holes that could have formed soon after the Big Bang. So far though, no convincing detections of any of it have been provided. \par
Dark matter is called \textit{cold} because its primeval particles have stopped to scatter with matter when they were non-relativistic, by contrast with \textit{hot} dark matter, whose primeval particles should have been relativistic at that time. \par
The ``temperature'' of the dark matter is very important in the galaxy formation scenario: cold dark matter implies that smallest observable structures (i.e., galaxies) formed first, then cluster and then super-clusters; on the contrary, if dark matter were hot, largest structures should have formed first and than fragmented to the smallest ones. The first scenario (i.e., the \textit{bottom-up} scheme) is consistent with the observed relative ages of galaxies and super-clusters. 

\subsection{Brief thermal history of the Universe}
\label{brithe}
Our current knowledge of physics allows us to understand what happened in the earliest moments of the Universe up to $\simeq 10^{-12}$ seconds after the Big Bang. Before that moment, its energy, density and temperature were extremely high. These conditions are unique in the history of the Universe and are not reproducible in laboratories. As the expansion proceeds, these conditions approach to more reproducible ones, becoming more accessible to our knowledge.\par
In its earliest phases, the Universe dynamics was dominated by radiation so that its temperature and energy decreased according to:
\be{tedecrate}
T(t), E(t) \propto t^{-1/2} \,.
\ee
During this period we can distinguish several eras.\par
From $\sim 10^{-12}$ to $\sim 10^{-6}$ seconds after the Big Bang (\textit{quark epoch}) the Universe was filled with a hot quark-gluon plasma, containing quarks, leptons and their antiparticles. \par
At $\sim 10^{-6}$ seconds (\textit{hadron epoch}) the primordial plasma cooled so that hadrons, including baryons such as proton and neutrons, could form. Initially matter and anti-matter were in thermal equilibrium but, as the temperature continued to fall, the hadrons and the anti-hadrons annihilated producing high-energy photons. A small residue of hadrons remained but no anti-hadrons. Nowadays indeed, anti-matter is essentially not observed in nature. \par
At approximately $\sim 1$ second after the Big Bang, the scattering processes that embedded neutrinos into the matter stopped because their free mean path had increased. In this way, neutrinos decoupled from matter starting to freely propagate through space. \par
Between $\sim 10$ and $\sim 1000$ seconds, the temperature dropped to the point that nuclear fusion was allowed: at this time nuclei of light elements could form. At the end of this process, there were nuclei of Hydrogen ($\sim 75\%$), Helium ($\sim 25\%$) and Lithium (few percents). During this epoch, the plasma was composed of nuclei, electrons, photons and dark matter, and the Universe expansion was still dominated by radiation. \par
After $\sim 47000$ years, the matter contribution started to be dominant. \par
At $\sim 380000$ years after the Big Bang, the temperature of the Universe was cold enough that electrons and nuclei combined to form neutral atoms: this process is called \textit{recombination}. Just before that moment, electrons and photons were in thermal equilibrium through Thomson scattering processes, but when electrons and protons combined, the free main path of the photons suddenly grows, so that they started to propagate freely in the expanding Universe, from the so-called \textit{last scattering surface}. This radiation fills the Universe as a uniform and isotropic background. Because the expansion has stretched its wavelength into the microwave region, this radiation has been defined \textit{cosmic microwave background} (CMB). \par
After recombination, the Universe became transparent for the first time and the only photons were those of the CMB without any other source of light: these are the so-called \textit{dark ages}, from $\sim 380000$ years to $\sim 0.3$ billions of years after the Big Bang. First galaxies and stars (Population I) could form in this era when dense regions collapsed due to gravity. Gravitational attraction among galaxies allowed for cluster and super-cluster to form. \par
Between $\sim 0.3$ and $\sim 1$ billions of years, when stars and galaxies were formed, new high-energetic photons reionized the neutral hydrogen back to plasma of ions, electrons and photons. Even if this time the plasma was much more diffuse, the optical depth of the Universe changed a bit, leaving a signature on the CMB spectrum. As the Universe continued to cool down and expand, reionization gradually ended. \par
The dark energy dominated era started $\sim 9.8$ billions of years after the Big Bang and lasts until present days. 

\subsection{The cosmic microwave background}
\label{cmb}
The CMB started to propagate when matter and radiation combined to form neutral atoms. The initial condition of thermal equilibrium between matter and radiation implies that the CMB photons distribution was a black-body spectrum, given by the Planck's law: 
\be{blackbodyintensity}
B(\nu, T) = \frac{2 h \nu^3}{c^2} \frac{1}{e^{\frac{h \nu}{kT}} - 1} \,.
\ee
The shape of the spectrum remained unchanged during the expansion since both the temperature and the frequency grow proportionally to $1 + z$. \par
\begin{figure}[H]
\centering
\includegraphics[scale=0.45]{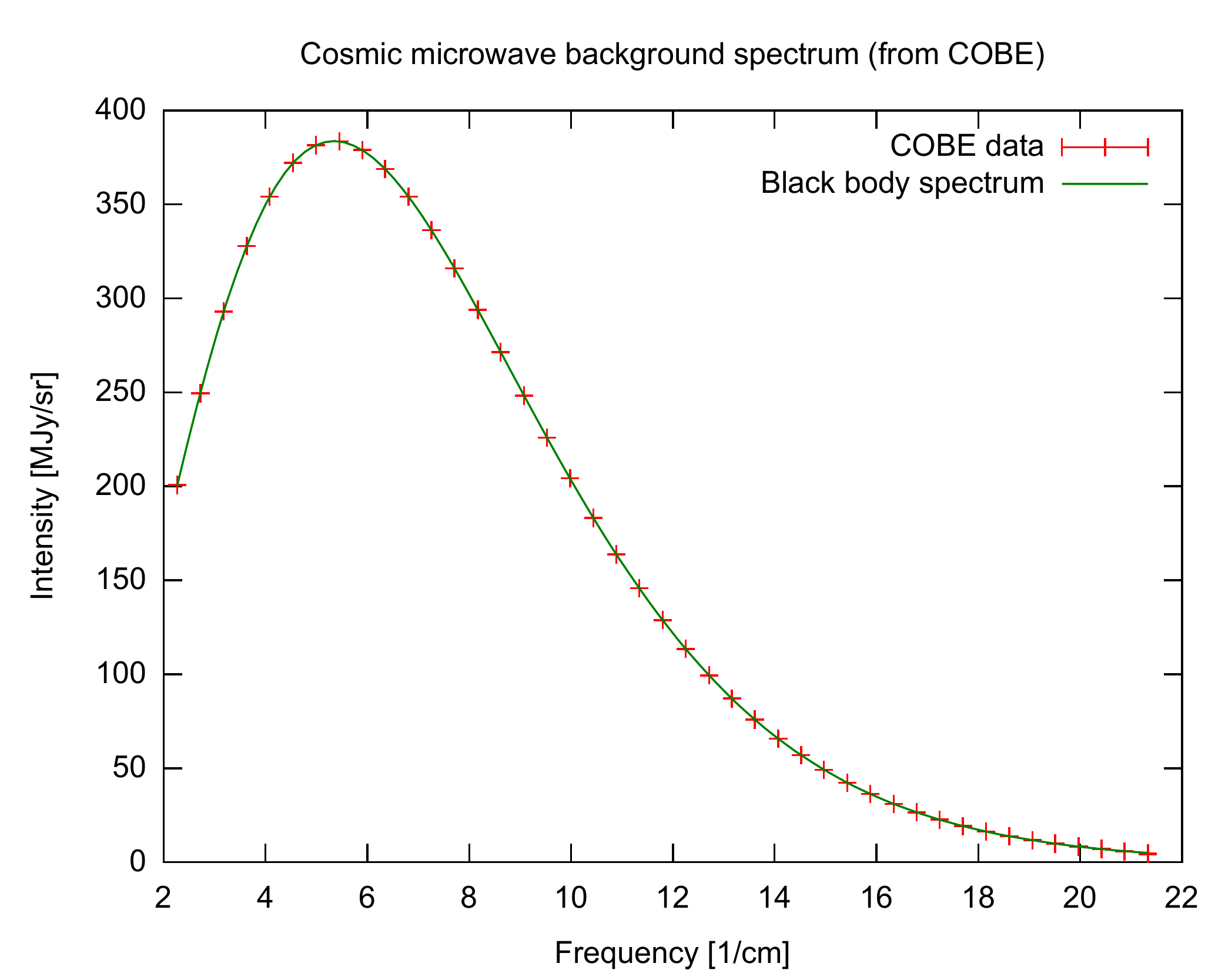}
\caption{CMB spectrum. The green line is the theoretical black body spectrum at the temperature of $\simeq \; 2.7 \;\mathrm{K}$ and the red crosses are the data collected by COBE \citep{Boggess}.}\label{CMBspectrum} 
\end{figure}
\begin{figure}[H]
\centering
\includegraphics[scale=0.7]{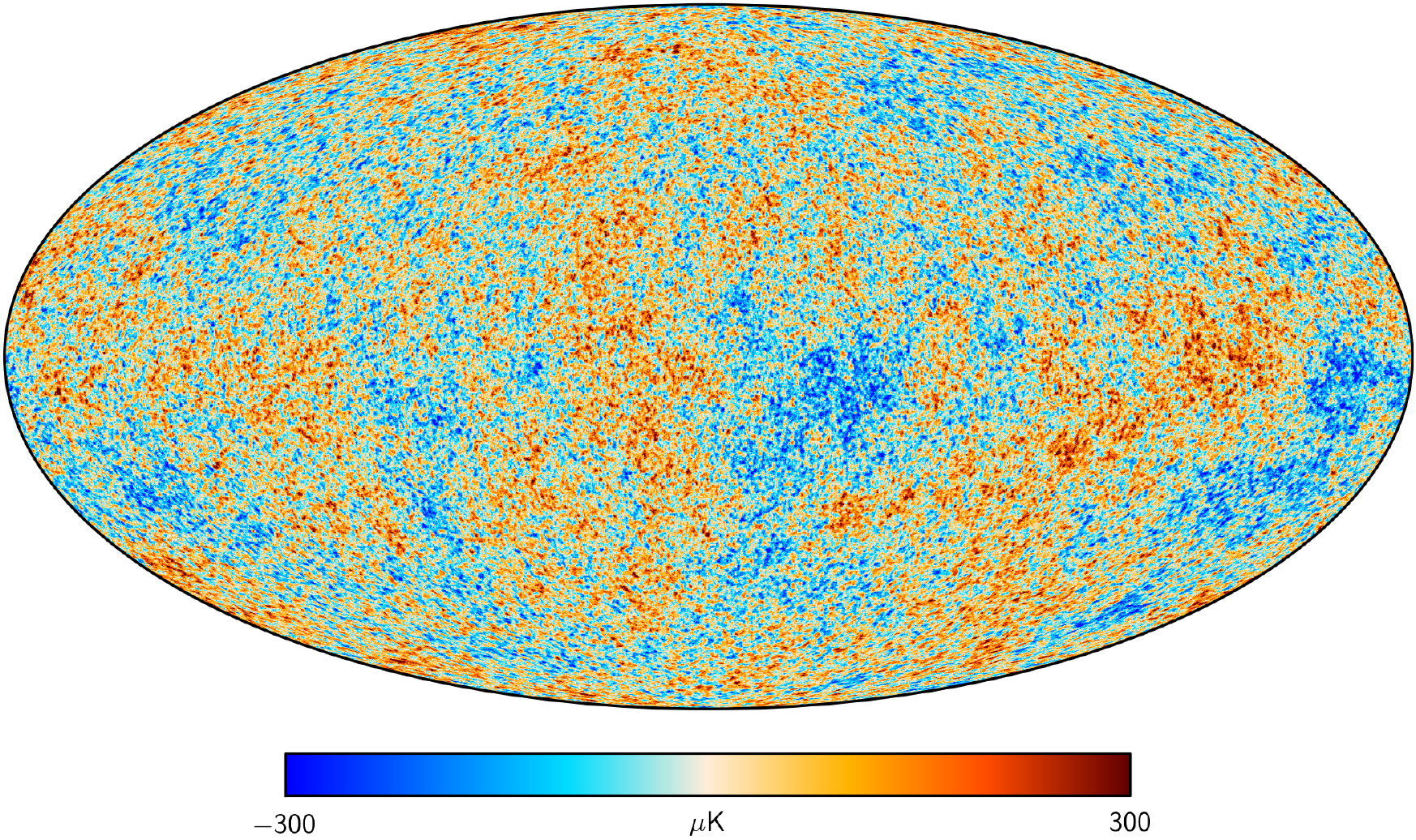}
\caption{Map of the CMB temperature anisotropies measured by Planck \citep{Planck2015-I}. }\label{CMB} 
\end{figure}

The CMB radiation that we observe today is a near-perfect black body radiation with average temperature $T_\mathrm{CMB} = 2.7260 \pm 0.0013 K$ \citep[][see Fig. \ref{CMBspectrum}]{Fixsen}.
Temperature \textit{anisotropies} are of the order of $\Delta T / T \approx 10^{-5}$ \citep[][see Fig. \ref{CMB}]{Smoot} and reflect the density fluctuations in the primordial plasma, which were the seeds of the primeval structures in the Universe. \par
The characteristic angular size of the fluctuations in the CMB is called \textit{acoustic scale} $\Theta_\ast$ and is one of the fundamental cosmological parameters of the \lcdm \ model.  

\subsection{Cosmological parameters}
The \lcdm \ model is based on six fundamental cosmological parameters, which can be constrained by fitting the shape of the \textit{power spectrum} of the CMB anisotropies (Sect. \ref{angpowspe}).\par
These parameters are $\Theta_\mathrm{MC}$, which is an approximation to the acoustic scale $\Theta_\ast$, the barionic matter density $\Omega_b$, the dark matter density $\Omega_c$, the amplitude of the scalar fluctuations $A_s$ (Sect. \ref{infpar}), the scalar spectral index $n_s$ (Sect. \ref{infpar}) and the optical depth of reionization $\tau$. Their values are listed in Table \ref{cospartab}.\par
All the other parameters, such as the Hubble constant $H_0$ or the the dark energy density $\Omega_\Lambda$, can be inferred from these.  
\begin{table}[H]
\centering
\begin{tabular}{lc}
\hline
Parameter & Value \\ \hline \hline 
$\Omega_b h^2$ & $0.02233 \pm 0.00015$ \\
$\Omega_c h^2$ & $0.1198 \pm 0.0012$ \\
$100 \, \Theta_\mathrm{MC}$ & $1.04089 \pm 0.00031$ \\ 
$\tau$ & $0.0540 \pm 0.0074$ \\
$\ln(10^{10}A_s)$ & $3.043 \pm 0.014$ \\
$n_s$ & $0.9652 \pm 0.0042$ \\ \hline
\end{tabular}\caption{The six fundamental cosmological parameters determined by \citet{Aghanim:2018eyx}. The $h$ factor is related to the Hubble constant through: $h = H_0 / (100 \ \mathrm{km \; s^{-1} {Mpc}^{-1}})$.}\label{cospartab}
\end{table}

\subsection{Open issues of the Big Bang cosmology}
Unfortunately, the Big Bang model does not lead to a complete description of the events occurred in the primordial Universe since it leaves unsolved three fundamental questions.
\begin{list}{\leftmargin 15pt \itemsep 0pt \topsep 3pt}
\item{\bf Initial conditions problem.}
  The \lcdm \ model provides no physical mechanism to explain the origin of the primordial fluctuations traced by the CMB anisotropies. In other words, what are the initial condition of the Universe that gave rise to the density fluctuations in the primordial plasma? 
\item{\bf Horizon problem.}
  The distance travelled by light from $t = 0$ to the recombination era ($t_{rec}$) is given by:
\be{distancetraveld}
d = a(t_{rec}) \int_0^{t_{rec}} \frac{c \, dt}{a(t)} \,.
\ee
This distance is a measure of the size of the regions that have been in causal contact until the photons decoupled from the baryonic plasma.\par
Nowadays, this distance corresponds to angular size of $\simeq 1\deg$ in the sky, which means that sky regions separated by more than $\simeq 1\deg$ could never have been in causal contact in the past. Why, instead, do they show the same CMB temperature to a very high degree? 
\item{\bf Flatness problem.}
  Current observations show that the density of the Universe is very close to the critical one:
\be{mdens}
\Omega_k \equiv \Omega_0 - 1 \simeq 0.001  \, . %\pm 0.002 \, .
\ee
Furthermore, from Eq. \ref{curvature} we know that:
\be{densityevo}
\frac{k c^2}{\dot{a}^2} = \Omega(t) - 1 \,. 
\ee
For past eras, $a(t) \propto t^\alpha$ with $\alpha < 1$, so that $\roundb{\Omega(t) - 1} \propto t^\beta$, with $\beta > 0$. This means that, at early times, the density must have been even closer to the critical one. This is possible only for a fine tuning of the initial condition. How could these special conditions occur? 
\end{list}

\subsection{The inflationary paradigm}
\label{infpar}
The \textit{inflationary paradigm}, or simply \textit{inflation}, proposed in the 1980s by Alan Guth, Andrei Linde, Paul Steinhardt and Alexei Starobinsky, solves the issues listed previously. \par
This model assumes that an accelerated expansion occurred between $10^{-33}$ and $10^{-32}$ seconds after the Big Bang. This is possible only if during this period $\ddot{a} > 0$ and thus, from Eq. \ref{friedmann1noL2}, $p < - \frac{1}{3} \rho c^2$. Then, we can imagine that, for a short period, $H = \frac{\dot{a}}{a} = \mathrm{const}$, so that:
\be{inf}
a(t) \propto e^{Ht} .
\ee \par
This kind of expansion ensures that $\ddot{a} > 0$ and, at the same time, it responds to the open problems of the \lcdm \ model. In fact, substituting Eq. \ref{inf} into Eq. \ref{densityevo}, we obtain:
\be{densityevobecomes}
\Omega(t) - 1 = \frac{k c^2}{\dot{a}^2} = \frac{k c^2}{H^2}e^{-2Ht} \underset{t\to\infty}{\longrightarrow} 0. 
\ee
In this way, at the end of inflation the space is flat, to account for the level of flatness observed today. \par
Furthermore, during the exponential expansion, the physical scales grew superluminal, together with the entire Universe, crossing the horizon (i.e. the distance traveled so far by the light). After this short time they return inside the horizon, which expands at the speed of light. In this way, the thermalization could occur, but regions that were causally connected during inflation become disconnected. \par
Moreover, inflation intrinsically introduces a way to explain the CMB anisotropies and the density inhomogeneities. In fact, it is possible to describe inflation by a quantum scalar field called \textit{inflaton}:
\be{inflaton}
\phi(\textbf{x}, t) = \phi^{(0)}(t) + \delta \phi(\textbf{x}, t) \,.
\ee
The homogeneous term $\phi^{(0)}(t)$ drives the background expansion while the perturbative term $\delta \phi(\textbf{x}, t)$ generates fluctuations. \par
During this epoch, pressure and density are related to the homogeneous part of inflaton through a potential term $V$:
\be{densityinfl}
\rho = \frac{1}{2}\roundb{\dot{\phi}^{(0)}}^2 + V \roundb{\phi^{(0)}} \,,
\ee
\be{pressureinfl}
p = \frac{1}{2}\roundb{\dot{\phi}^{(0)}}^2 - V \roundb{\phi^{(0)}} \,.
\ee \par
\begin{figure}[H]
\centering
\includegraphics[scale=0.45]{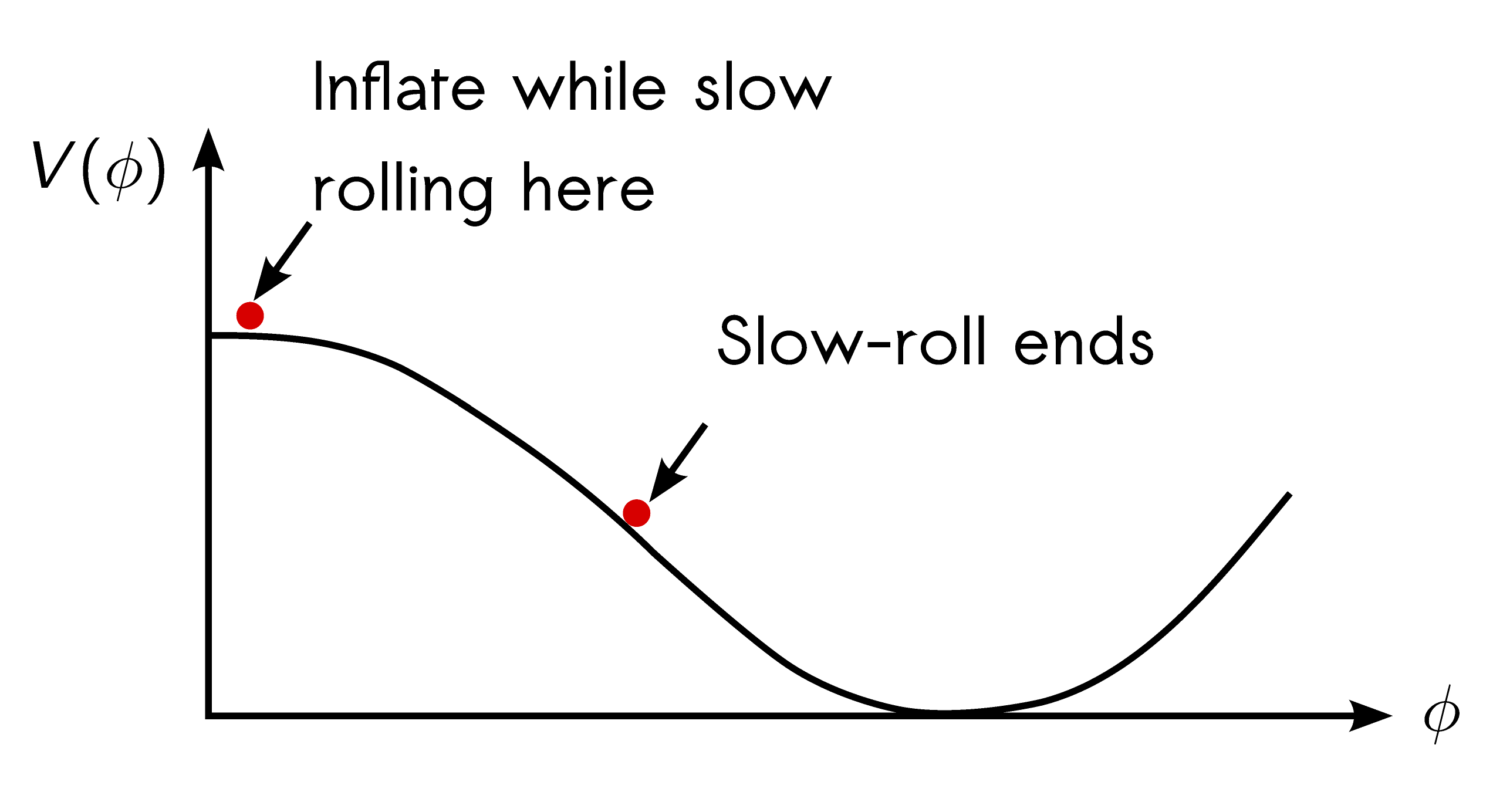}
\caption{A sketch of the slow-roll potential.}\label{slowroll} 
\end{figure}

The inflationary condition of negative pressure can be obtained only if the potential term dominates: 
\be{slowrollcond}
\roundb{\dot{\phi}^{(0)}}^2 \ll V \roundb{\phi^{(0)}} \,.
\ee
This condition can be achieved for different shapes of the potential, which define several models of inflation. Currently, the most reliable model is the so-called \textit{slow-roll} inflation (Fig. \ref{slowroll}). \par
The energy of inflation is related to the potential through the relation:
\be{energyinfl}
\varepsilon_\mathrm{inflation} = \frac{1}{16 \pi G} \roundb{\frac{d V}{d \phi} \frac{1}{V}}^2 \,.
\ee

\par 
The quantum fluctuations $\delta \phi(\textbf{x}, t)$ perturb both the matter distribution, through \textit{scalar} perturbations, and the space-time metric, via \textit{tensor} perturbations.\par
Scalar perturbations couple to the density of matter and radiation and are ultimately responsible for most of the inhomogeneities and anisotropies in the Universe. \par
The tensor fluctuations instead introduce in the Universe a stochastic background of gravitational waves that induces a pattern with null divergence in the polarization of CMB. These signatures, predicted by inflationary models but not detected yet, are called \textit{B-modes} and are considered the \textit{smoking gun} of inflation. \par 
The spectra of these perturbations are given respectively by:
\be{scalarspec}
P_S(k) = A_S^2 k^{n_S-1} \,,
\ee
\be{tensorspec}
P_T(k) = A_T^2 k^{n_T}\,.
\ee
where $n_S$ and $n_T$ are the spectral indices; $n_{S/T} = 1$ corresponds to scale invariant fluctuations. \par 
It is possible to define the \textit{tensor-to-scalar ratio} $r$ as:
\be{tensortoscalar}
r = \frac{A_T^2}{A_S^2} \,,
\ee
which parametrizes the amplitude of the B-mode signature in the CMB polarization anisotropies. This parameter is related to the energy of inflation through:
\be{energyinfltsr}
\varepsilon_\mathrm{inflation} = 1.06 \times 10^{16} GeV \roundb{\frac{r}{0.01}}^{1/4} \,.
\ee \par
Detecting the B-mode signal in the CMB radiation on large angular scales (low multipoles) would thus give us an important evidence in favor of the inflationary paradigm. Moreover, reconstructing the B-mode power spectrum at low multipoles would allow us to constrain the value of $r$ and to shed light on the inflation process and the physics of the very early Universe.

\section{CMB anisotropies}
\label{cmbani}
The matter density fluctuations that we observe today in the Universe can be related to primordial quantum fluctuations in the inflaton field. Today, we see the trace of these fluctuations in the CMB as well as in the matter distribution. \par
The \textit{anisotropies} in the temperature CMB pattern, in particular, are the signature of these primordial perturbations. Precise measurements of such anisotropies allow us to discriminate between different cosmological models.\par 
In this section, I introduce the concept of angular power spectrum of anisotropies and discuss the CMB polarization properties.

\subsection{Angular power spectrum of the CMB temperature anisotropies}\label{angpowspe}
We can write the temperature field as:
\be{temperature}
T\roundb{\hat{p}} = T \squareb{1 + \Theta \roundb{\hat{p}}} \,,
\ee
where $\Theta = \frac{\Delta T}{T}$ are the anisotropies observed in each direction of the sky, $\hat{p}$. 
We now expand the field in terms of spherical harmonics:
\be{expansion}
\Theta \roundb{\hat{p}} = \sum_{\ell=1}^{\infty} \sum_{m=-\ell}^{\ell} a_{\ell m}^T \, Y_{\ell m} \roundb{\hat{p}} .
\ee 
The $Y_{\ell m}$ terms are the Legendre polynomials, which are orthogonal with normalization:
\be{normalization}
\int Y_{\ell m}\roundb{\hat{p}} Y_{\ell'm'}^{\ast}\roundb{\hat{p}} \hspace{2pt} d\Omega = \delta_{\ell \ell'} \delta_{mm'} \,,
\ee
where $d \Omega$ is the infinitesimal solid angle in the direction $\hat{p}$. Therefore, Eq. \ref{expansion} can be inverted by multiplying both sides by $Y_{\ell m}^{\ast} \roundb{\hat{p}}$ and integrating:
\be{alm}
a_{\ell m}^T = \int \Theta \roundb{\hat{p}} Y_{\ell m}^{\ast} \roundb{\hat{p}} \hspace{1pt} d\Omega .
\ee \par
The mean value of all the $a_{\ell m}$ is zero due to the homogeneity but their variance is not. The variance of the $a_{\ell m}$ as function of $\ell$ is called \textit{angular power spectrum} $C_\ell$:
\be{almsquared}
\mean{a_{\ell m}^T} = 0 \; ; \hspace{30pt} \mean{a_{\ell m}^Ta_{\ell'm'}^{T \ast}} = \delta_{\ell \ell'} \delta_{mm'} C_\ell .
\ee \par
For any given $\ell$ the variance of the $a_{\ell m}$ is computed over the set of $2 \ell + 1$ samples in the distribution. Thus, there is a fundamental uncertainty in the knowledge we may get about the $C_\ell$. This uncertainty is called \textit{cosmic variance} and equals to:
\be{cosmicvariance}
\roundb{\frac{\Delta C_\ell}{C_\ell}}_\mathrm{cosmic \ variance} = \sqrt{\frac{2}{2\ell + 1}} \,.
\ee 
This term is dominant at the lowest $\ell$ reflecting the lack of statistical information we have at the largest angular scales since there is only one Universe we can sample from.   \par
\begin{figure}[H]
\centering
\includegraphics[scale=0.7]{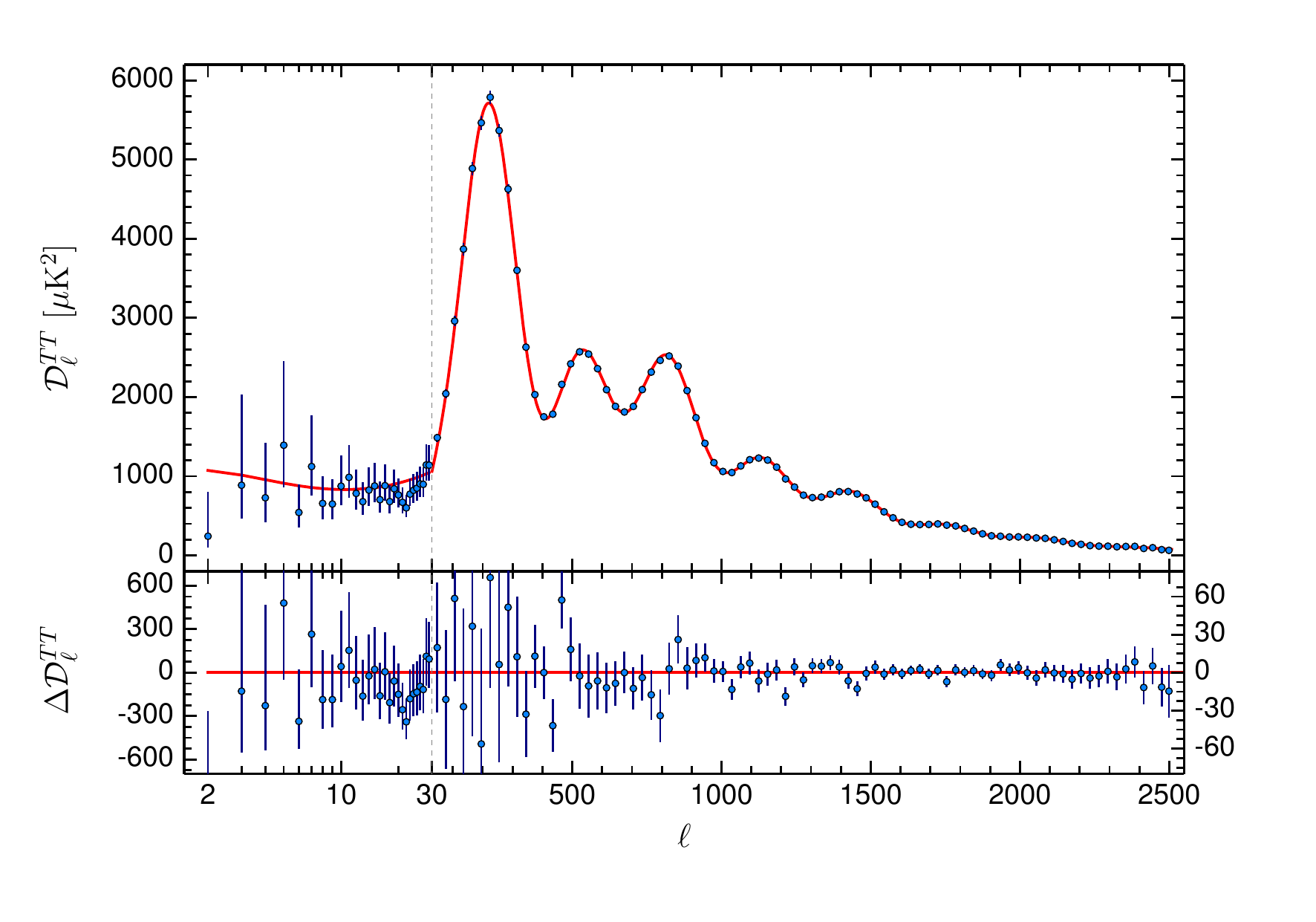}
\caption{The power spectrum of the temperature anisotropies of the CMB \citep{Planck2015-XIII}. The red solid line represents the best \lcdm \ model fit. Error bars show $\pm 1 \sigma$ uncertainties. The lower panel shows the residuals. The spectrum peaks at $\ell \sim 100$ which corresponds to angular scales of $ \Theta \sim 1\deg$.}\label{powerspectrumTT} 
\end{figure}

The most recent estimate of the power spectrum of the temperature anisotropies of the CMB is shown in Fig. \ref{powerspectrumTT}. Sometimes it is preferred to plot $D_\ell$, defined by \ref{dl}, rather than $C_\ell$:
\be{dl}
D_\ell = \frac{\ell (\ell+1)}{2 \pi} \hspace{3pt} C_\ell \,.
\ee
\par
The temperature power spectrum in Fig. \ref{powerspectracomparison} shows eight peaks that correspond to acoustic oscillations in the primordial plasma. The position of the peaks assess the value of the acoustic scale $\Theta_\ast$. The relative heights of the acoustic peaks depends on the barionic matter density, $\Omega_b$: as it increases, the odd peaks become enhanced over the even peaks. The overall amplitude of the peaks depends on the value of the dark matter density parameter $\Omega_c$ while the dark energy density parameter $\Omega_\Lambda$ contributes in shifting the angular location of the peaks left and right. Reionization, whose optical depth is measured by $\tau$, suppresses the heights of the acoustic peaks uniformly. The damping tail is due to the photon diffusion that smooths initial fluctuations from inflation\footnote{It is possible to find animations showing the relation between the cosmological parameters and the CMB power spectrum at: \url{http://background.uchicago.edu/~whu/metaanim.html}.}. 

\subsection{The CMB polarization}\label{cmbpol}
The CMB radiation that we observe today is linearly polarized at the $\simeq 10 \,\%$ level ($\delta P \sim \frac1{10} \, \delta T$). The CMB polarization, detected for the first time by the DASI instrument \citep{Leitch}, is due to Thomson scattering during the recombination era. \par
The relationship between the Thomson scattering cross section and the radiation polarization is \citep{Chandrasekhar}: 
\be{sectho}
\frac{d\sigma_T}{d\Omega} \propto \abs{\hat{\epsilon} \cdot \hat{\epsilon}'} \,,
\ee
where $\hat{\epsilon}' \ (\hat{\epsilon})$ are the incident (scattered) polarization directions. \par
\begin{figure}[H]
\centering
\includegraphics[scale=0.5]{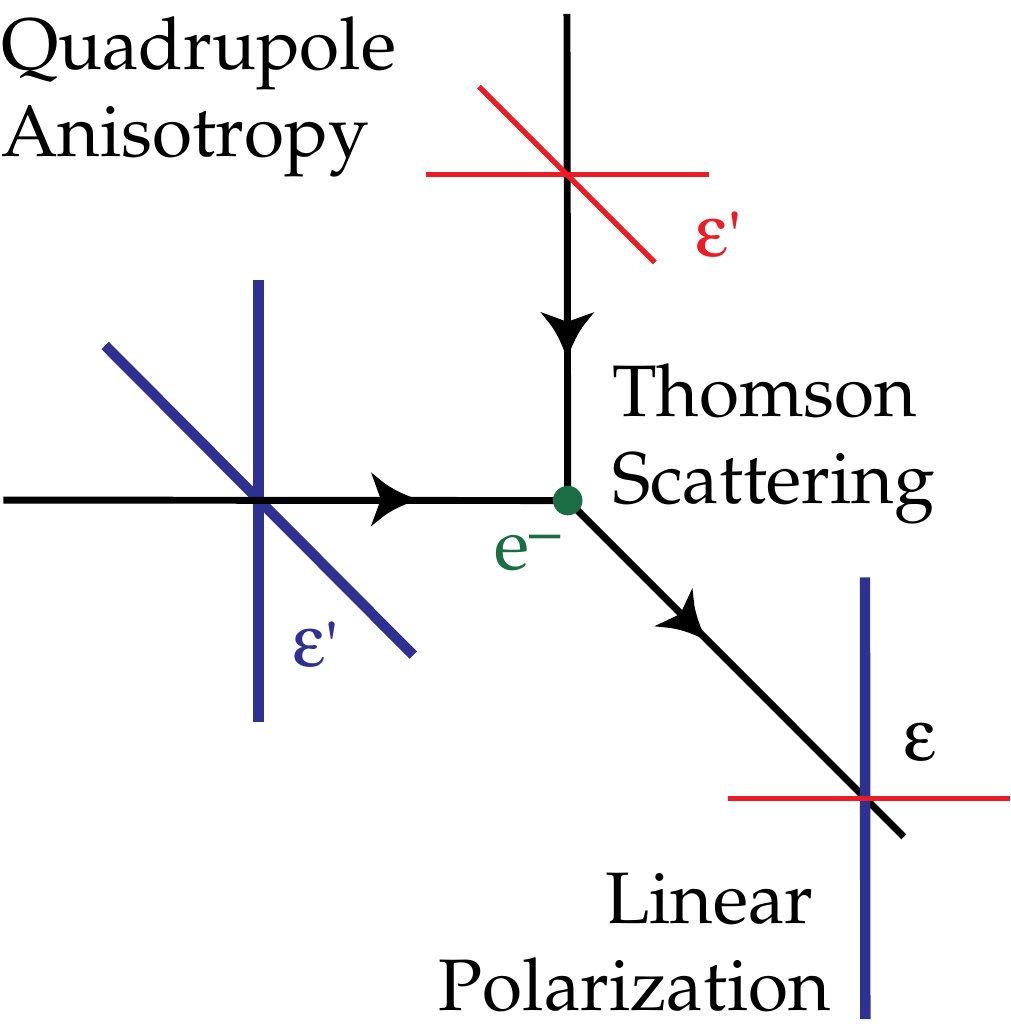}
  \caption{Quadrupole anisotropy in the incident radiation generates linear polarization after Thomson scattering. In this sketch blue (red) lines represent hot (cold) radiation. Picture from \citet{Hu}.}\label{thomsonscatt}
\end{figure}
The incoming radiation field makes the target electron oscillate in the direction of the incident electric field \textbf{E} converting unpolarized radiation into polarized radiation. While the polarization is concordant to the motion of the electron, the scattered radiation intensity peaks in the direction normal to the incident polarization. \par
Thomson scattering alone is not enough to generate a linearly polarized signal but it is necessary a \textit{quadrupole} anisotropy in the temperature of the incident radiation field. In fact, if it were isotropic, the orthogonal polarization states from all incident directions would balance each other and no polarized signal would be generated. In the case of \textit{dipole} anisotropy instead, the intensity peaks at $180\deg$ separation and the radiation would possess only one linear polarization state with the average value that would balance with the average polarization states coming from $\pm 90\deg$. But, if separation among the peaks is $90\deg$, the distribution has a quadrupole pattern and orthogonal contributions will be different, leaving a net linear polarization in the scattered radiation (Fig. \ref{thomsonscatt}). \par
A reversal in sign of the temperature fluctuation corresponds to a $90\deg$ rotation of the polarization \citep[for more detail see][]{Hu}. \par

\subsection{The Stokes parameters}\label{stopar}
The electric field of a plane monochromatic wave with wave vector $\textbf k$ can be decomposed along the two axes transverse to the direction of propagation:
\be{Edecomposition}
\textbf{E} = \operatorname{Re} \squareb{ \Epar(t) \textbf{e}_{||} + \Eperp(t) \textbf{e}_{\perp}} \,,
\ee
where the unit vectors $\roundb{\textbf{e}_{||} , \textbf{e}_{\perp} , \textbf k}$ form a cartesian coordinate system. \par 
A polarized radiation field can be completely described in terms of the four \textit{Stokes parameters}, defined by:
\be{stokes}
\begin{aligned} 
I &= \mean{\abs{\Epar}^2} + \mean{\abs{\Eperp}^2} \,, \\
Q &= \mean{\abs{\Epar}^2} - \mean{\abs{\Eperp}^2} \,, \\
U &= \mean{\Epar \Eperp^{\ast}} + \mean{\Epar^{\ast} \Eperp} = 2 \operatorname{Re} \mean{\Epar \Eperp^{\ast}} \,, \\
V &= i \, \roundb{\mean{\Epar \Eperp^{\ast}} - \mean{\Epar^{\ast} \Eperp}} = -2 \operatorname{Im} \mean{\Epar \Eperp^{\ast}} \,.
\end{aligned}
\ee
The parameter I is the total intensity of radiation. V is related to the circular polarization of the radiation and does not appear in the CMB polarization pattern. Q and U describe the linear polarization.\par
\begin{figure}[H]
\centering
\includegraphics[scale=0.7]{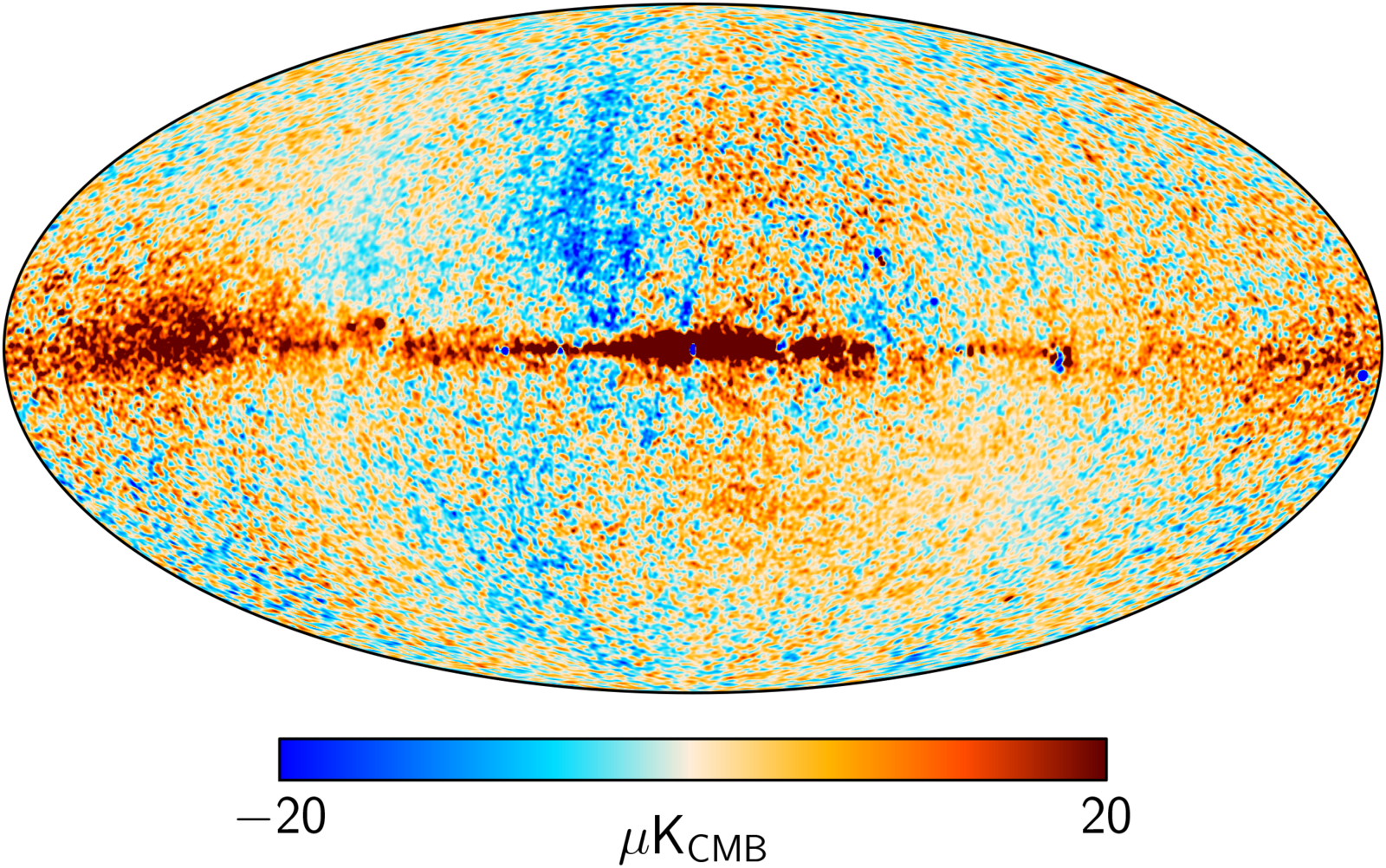}
\includegraphics[scale=0.165]{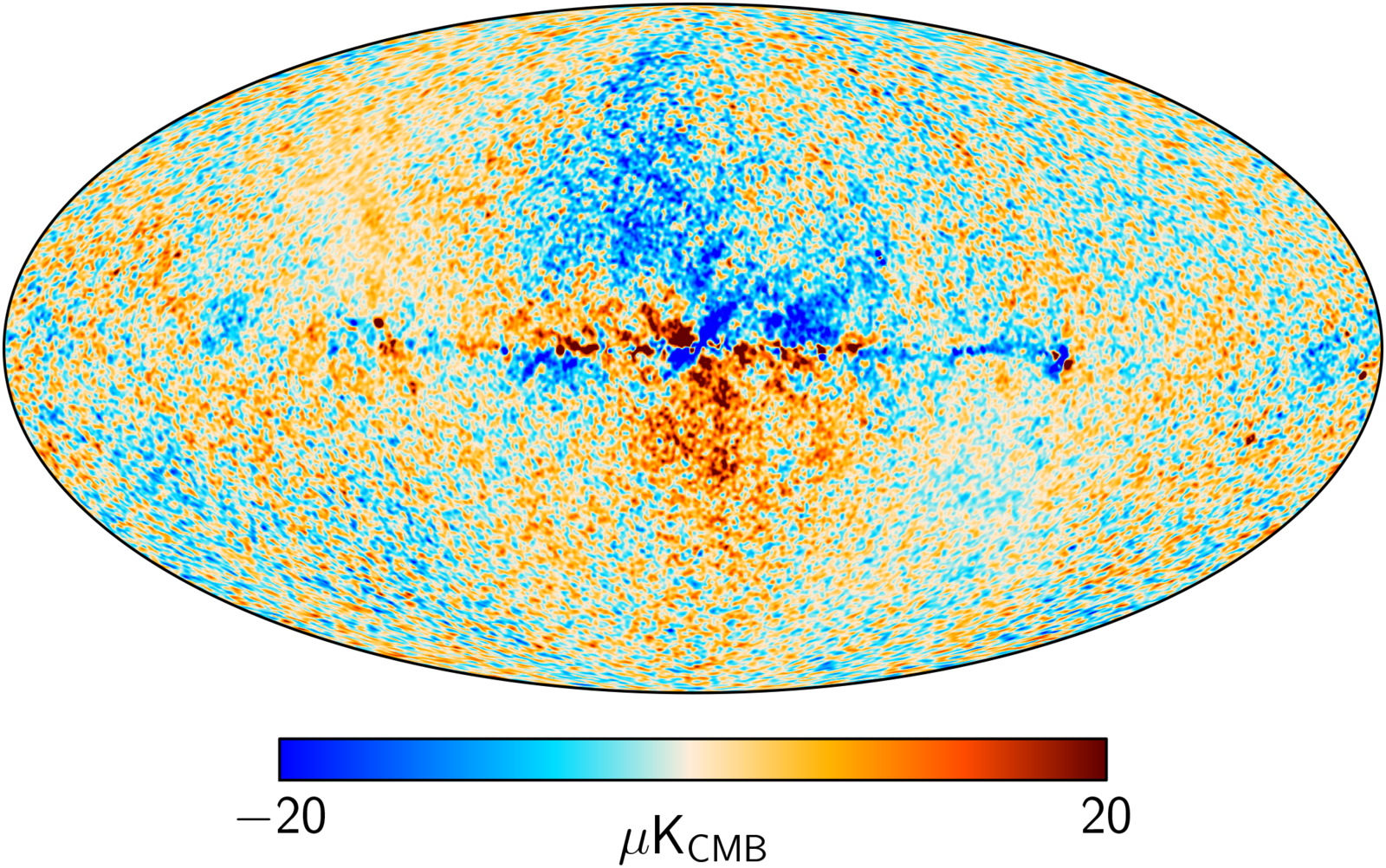}
\caption{Q (top) and U (bottom) maps at $44\,\mathrm{GHz}$ by \citet{Planck2015-I} including the galactic plane emission.}\label{QUmaps}
\end{figure}

Fig. \ref{QUmaps} shows two Q and U maps of the whole sky as measured by Planck.

\subsection{E and B modes}\label{ebmod}
We can construct two quantities from the Stokes Q and U parameters with a definite value of spin\footnote{In this context, the term ``spin'' is referred to the way in which some functions defined on the sphere transform under rotations. These functions are defined in \citet[][Appendix A]{ZaldarriagaPhD}.}:
\be{spin2objects}
\roundb{Q \, \pm \, iU}(\hat{p}) = \sum_{\ell, m}a_{\ell m}^{\pm 2} Y_{\ell m}^{\pm 2}(\hat{p}).
\ee
The coefficients $a_{\ell m}^{\pm 2}$ are generalizations of the ordinary coefficients of the spherical harmonics expansion. Their construction is described in detail in \citet[][p. 18]{ZaldarriagaPhD}. \par
Instead of $a_{\ell m}^{\pm 2}$, it is convenient to introduce a linear combinations of them:
\be{linearcombinations}
a_{\ell m}^E = \frac{a_{\ell m}^2 - a_{\ell m}^{-2}}{2}, \hspace{25pt} a_{\ell m}^B = -i \, \frac{a_{\ell m}^2 + a_{\ell m}^{-2}}{2}\,.
\ee \par
We can now define two quantities in real space:
\be{EandBmodes}
E(\hat{p}) = \sum_{\ell,m} a_{\ell m}^E Y_{\ell m}(\hat{p}), \hspace{25pt} B(\hat{p}) = \sum_{\ell,m} a_{\ell m}^B Y_{\ell m}(\hat{p}) \,;
\ee
they are the so-called \textit{E-mode} and \textit{B-mode}. \par
\begin{figure}[H]
\centering
\includegraphics[scale=0.4]{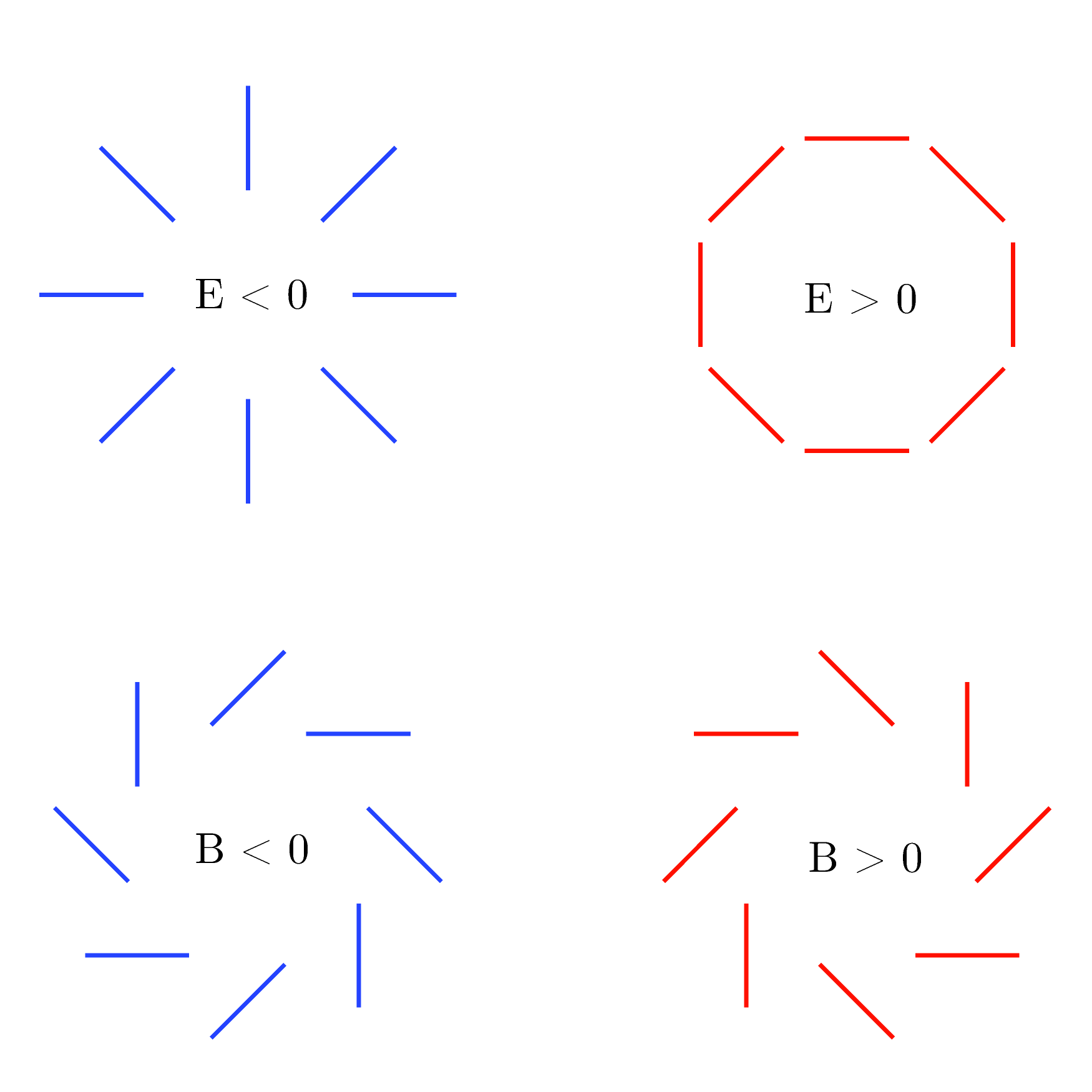}
\caption{E and B-mode characteristic patterns \citep{Krauss}. Note that they behave differently under parity transformation: while the E-patterns remain unchanged, the positive and negative B-patterns get interchanged.}\label{EandBpolarizationpattern}
\end{figure}

The E and B-modes of the CMB completely specify all the statistical proprieties of the linear polarization field. These quantities are both invariant under rotation but behave differently under inversion of the spatial coordinates: E remains unchanged and B changes its sign, in analogy with electric and magnetic fields. Furthermore, E and B-modes present different features in the sky polarization pattern: E-polarization vectors are radial around cold spots and tangential around hot spots in the sky while B-polarization vectors have vorticity around any given point in the sky, as shown in Fig. \ref{EandBpolarizationpattern}.
Given E, B and T (as defined in Eq. \ref{EandBmodes} and \ref{temperature}), it is possible to construct the following angular power spectra:
\be{allpowerspectra}
C_\ell^{XY} = \frac{1}{2\ell+1} \sum_{m=-\ell}^{m=\ell} \mean{a_{\ell m}^{X \ast} a_{\ell m}^Y} , \hspace{20pt} X,Y = \{T,E,B\} \,.
\ee \par
The polarization pattern, described through E and B modes, is a projection of the quadrupole anisotropy, as mentioned in Sect. \ref{cmbpol}. But which are the physical processes capable to generate a quadrupole anisotropy? \par
In terms of the multipole decomposition of the radiation field into spherical harmonics, the quadrupole moments are represented by $\ell = 2$, $m = 0, \, \pm 1, \, \pm 2$ that correspond respectively to scalar, vector and tensor perturbations of the metric. In particular, we know that:
\begin{list}{\leftmargin 15pt \itemsep 0pt \topsep 3pt}
\item Scalar modes are perturbations in the density of the cosmological fluid at the epoch of recombination and lead only to E-mode polarization pattern in the sky.
\item Vector perturbations represent vortical motions of the primordial matter. Nevertheless, the vorticity is damped by the expansion of the Universe so that they do not leave an imprint in the CMB polarization pattern.
\item Tensor modes can be generated only by perturbations of the metric in the primordial Universe. They contribute to the CMB polarization signal producing both E and B-modes.
\end{list}\par

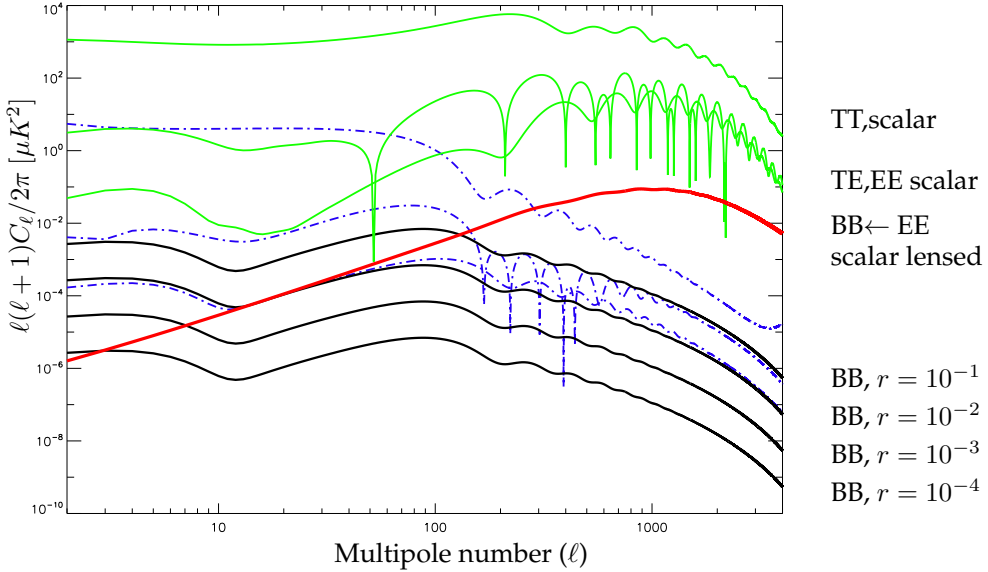
\begin{figure}[H]
\begin{center}
\vskip 1.9cm
{
\setlength{\unitlength}{1cm}
\begin{picture}(14, 6)(0,0)
\put(11,5.7){TT,scalar}
\put(11,4.9){TE,EE scalar}
\put(11,4.3){BB$\leftarrow $ EE}
\put(11,3.9){scalar lensed}
\put(11,2.3){BB, $r=10^{-1}$}
\put(11,1.8){BB, $r=10^{-2}$}
\put(11,1.3){BB, $r=10^{-3}$}
\put(11,0.8){BB, $r=10^{-4}$}
\put(4.5,0){Multipole number ($\ell $)}
\put(0.5,3){\begin{rotate}{90} $\ell (\ell +1)C_{\ell }/2\pi ~[\mu K^2]$ \end{rotate}}
\includegraphics[scale=0.6]{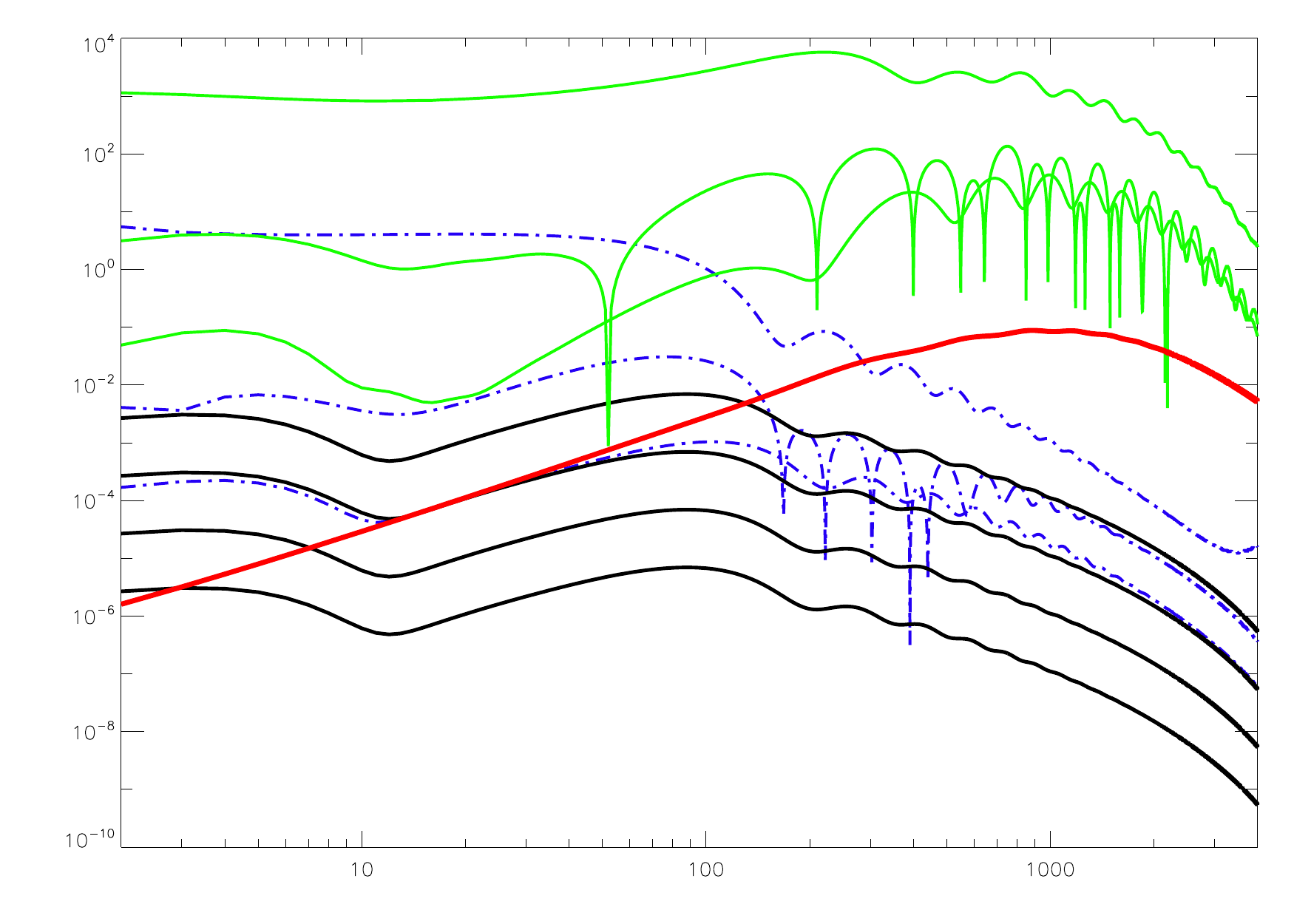}
\end{picture}}
\end{center}
\vskip -0.5cm
\caption{Theoretical power spectra of the CMB generated assuming the parameters from the best-fit model of the WMAP seven-year data ($H_0 = 71.4 \; \mathrm{km \, s^{-1} \, {Mpc}^{-1}}$, $\Omega_b = 0.045$, $\Omega_c = 0.220$, $\Omega_\Lambda = 0.73$, $\tau = 0.086$, and $n_S = 0.969$). The green curves show from top to bottom respectively the TT, TE, and EE power spectra generated by the scalar modes. The red curve indicates the BB scalar component resulting from the gravitational lensing of the EE modes. The three dashed blue curves indicate, from top to bottom, the TT, TE (logarithm of the absolute value), and EE spectra resulting from the tensor mode with $r=0.1$. The solid black curves indicate the BB spectra for the decreasing values of $r = 0.1, 0.01, 0.001$ and $0.0001$ \citep{Core}.} 
\label{powerspectracomparison} 
\end{figure}\par

Although tensor perturbations lead to E-mode as well, the major contribution to them is due to the scalar perturbations. Besides, as shown in Fig. \ref{powerspectracomparison}, at small angular scales ($\ell \gtrsim 100$) E-mode can be turned into B-mode from gravitational lensing produced by massive structures, such as galaxy clusters. \par
From Fig. \ref{powerspectracomparison}, it appears clear how to observe low multipoles, corresponding to degree angular scales or more, is strongly required in order to distinguish between primordial and lensed B-modes.

\section{State-of-the-art and perspectives}
\label{staart}
The CMB radiation was detected accidentally for the first time in 1964, by two American radio astronomers: Arno Penzias and Robert Wilson. Since its discovery, several ground-based and ballon-borne experiments, as well as three space missions, have studied the properties of the CMB. \par
The COBE satellite \citep{Boggess} was the first experiment to measure precisely the CMB spectrum and to detect its spatial anisotropies. Its successors were WMAP \citep{Bennet} and Planck \citep{Planck2015-I}. \par
Polarization anisotropies have been observed for the first time by the ground based experiment DASI \citep{Kovac_2002} and then measured with a better sensitivity in a full sky survey by WMAP and Planck. \par
In the last decade, several ground based experiments have been proposed and deployed, mostly in the Antarctica, as BICEP/Keck Array \citep{BicepKeck2} and SPT \citep{Ruhl_2004}, and in the Atacama Desert (Chile), like Polarbear \citep{PolarBear}, CLASS \citep{EssingerHileman}, QUIET \citep{QUIET_wp} and ACT \citep{Thornton_2016}. Moreover, several balloon experiments as EBEX \citep{Oxley_2004} and SPIDER \citep{Spider} flew recently some years ago. \par
\begin{figure}[H]
\centering
\includegraphics[scale=0.3]{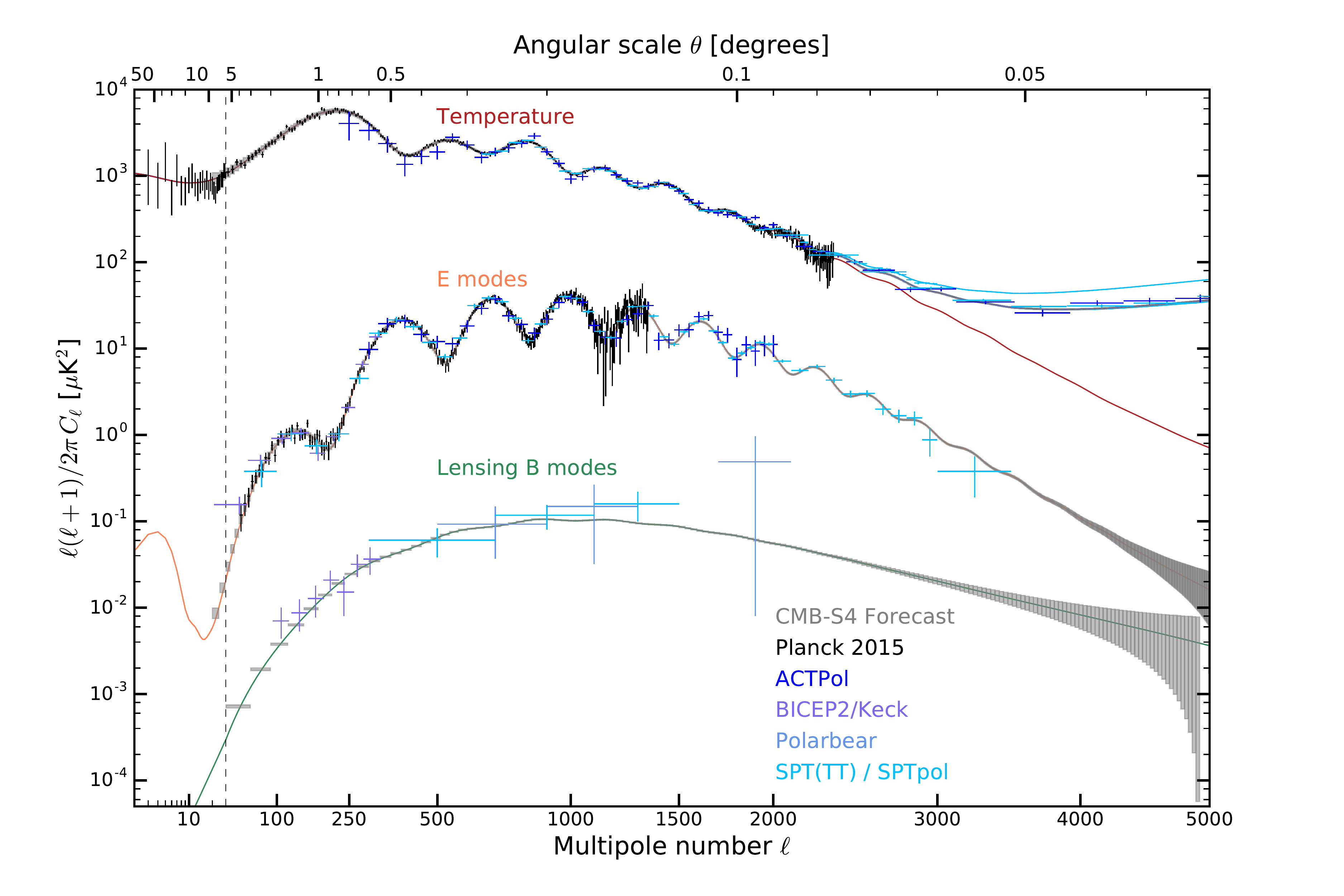}
\caption{Current measurements of the angular power spectrum of the CMB temperature and polarization anisotropies \citep{2016arXiv161002743A}.}\label{presentstatus}
\end{figure}

Fig. \ref{presentstatus} summarizes the status of the E and B modes detection.\par
Nowadays, a number of experiments\footnote{A complete list of operating and planned CMB experiments is available at \url{http://lambda.gsfc.nasa.gov/product/expt/}.} are being designed in order to detect the B-mode signal. In particular, the Simons Observatory \citep{Ade_2019} will observe the sky from Atacama Desert in six frequency bands: $27$, $39$, $93$, $145$, $225$ and $280\,\mathrm{GHz}$. It will combine information from three small-aperture telescopes ($0.5\,\mathrm{m}$, SATs) and one large-aperture telescope ($6\,\mathrm{m}$, LAT), with a total of $60000$ \textit{transition-edge sensors} (TES) bolometers. Furthermore, a new satellite mission, named LiteBIRD \citep{Suzuki2018}, has been recently approved by the ``Japan aerospace exploration agency'' (JAXA). LiteBIRD will observe the CMB through a $400\,\mathrm{mm}$ diameter telescope and $2622$ cryogenic bolometers. It will survey the whole sky with $15$ frequency bands from $40$ to $400 \mathrm{GHz}$.\par
 
\subsection{CMB-S4}
\begin{figure}[H]
\centering
\includegraphics[scale=0.5]{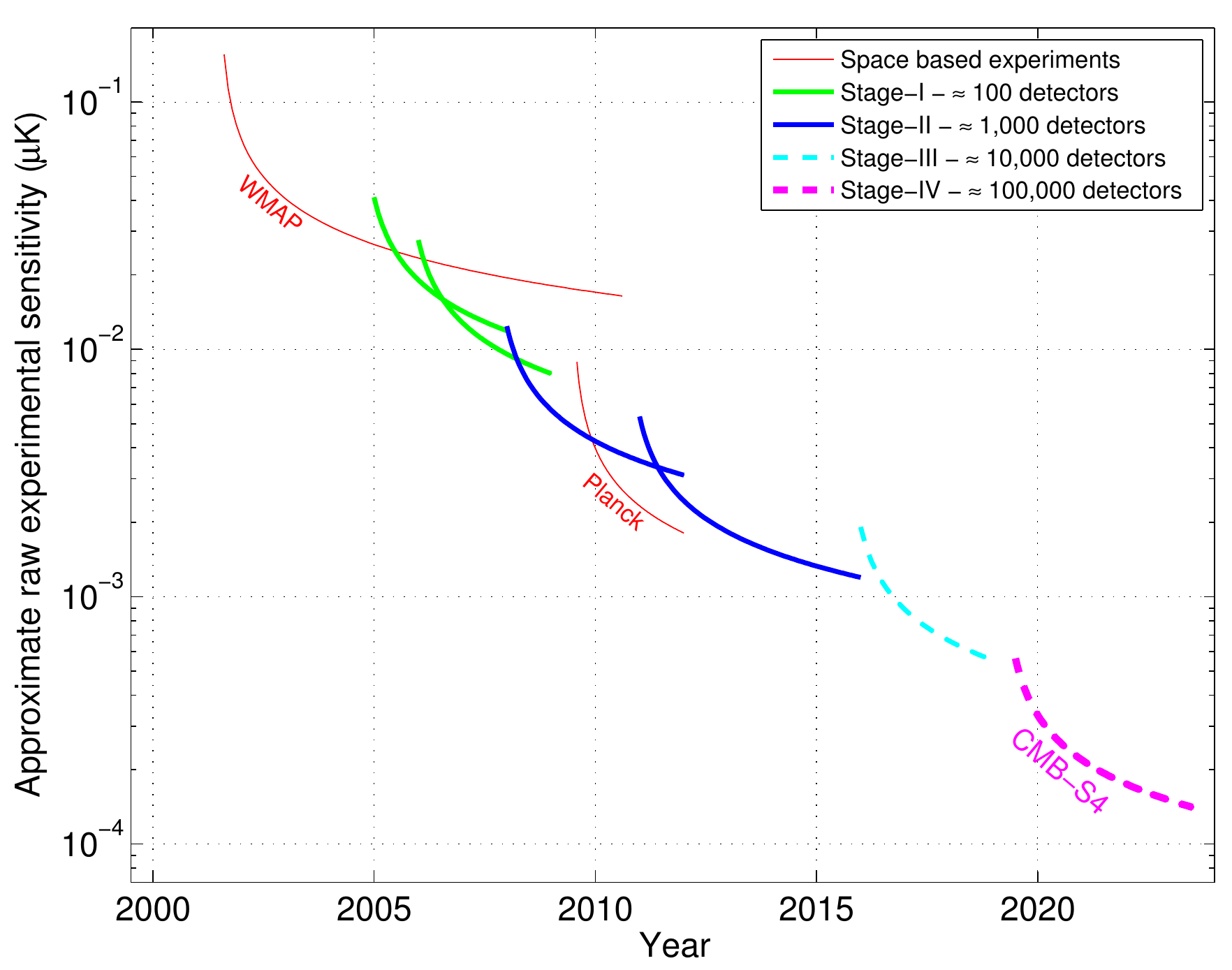}
\caption{Sensitivity of CMB experiments as a function of time. Ground-based CMB experiments are classified into stages \citep{2016arXiv161002743A}.}\label{s4sens}
\end{figure}

As shown in Fig. \ref{s4sens}, in the last two decades the CMB experiments have increased their sensitivities following a scaling law, which depends on the total number of bolometers. To maintain this scaling, more focal plane pixels and more telescopes are required.\par
Ground-based CMB experiments are classified into stages according to the number of detectors from Stage I ($\sim 100$ detectors) to Stage IV ($\sim 100000$ detectors), by steps of a factor $10$ in the number of detectors. Currently, we are at Stage III.\par
The CMB-S4 project \citep{2016arXiv161002743A} aims to combine information from several high-sensitivity telescopes. CMB-S4 science goals require sensitivity of order of $1\,\mu\mathrm{K\,arcmin}$ and order of $500000$ detectors for a four-year survey. To take advantage of the best atmospheric conditions, the South Pole and the Chilean Atacama sites are baselined, with the possibility of adding a northern site to increase the sky coverage.

\par

\par

\subsection{The Large Scale Polarization Explorer}
The ``Large Scale Polarization Explorer'' (LSPE) will observe about the $30\%$ of the Northern Sky at large angular scales and at multiple frequencies. It was initially conceived as a single balloon-borne experiment but, recently, it has been converted into two different telescope: the high frequency instrument, SWIPE ($140$, $220$, $240\,\mathrm{GHz}$), will remain on board of the balloon while the low frequencies instrument, STRIP ($43\,\mathrm{GHz}$), will become a stand-alone ground-based telescope. \par
The combination of the data of the two instruments will provide a map of the sky with unprecedented sensitivity, in the Earth's Northern Hemisphere. The LSPE project, in fact, aims either to detect B-modes or to constrain the value of $r$ to $\simeq 0.03$ at the $99.7\,\%$ confidence level. In any case, its data will be useful to study the polarized emission of the Milky Way in the Northern Sky. Furthermore, the LSPE/STRIP instrument, which will be installed at ``Observatorio del Teide'' in Tenerife, will have a second channel at $95\,\mathrm{GHz}$ that is devoted to study the properties of the Tenerife's atmosphere. This information will be crucial to test the goodness of the Tenerife site, which could be possibly exploited for future CMB experiments from the Northern Hemisphere. 

\subsection{Main issues in detecting B modes}\label{miidbm}
The Planck collaboration has published the best limit to date on tensor modes using CMB temperature data alone: $r < 0.11$ at $95 \,\%$ confidence level. A slightly more stringent upper limit was found by exploiting the latest data from the Keck array and BICEP2 telescopes: with $r < 0.09$ at $95 \,\%$ confidence \citep{BicepKeck2}. \par 
Whether to improve this upper limits or to finally detect B-modes two main issues have to be addressed:
\begin{list}{\leftmargin 15pt \itemsep 0pt \topsep 3pt}
\item{\bf Instrumental sensitivity.} The signal of B-modes is expected to be vary faint calling for sensitivities in the order of few tens of $n\mathrm{K}$ (at $1\deg$ angular resolution) or less. \par To reach such fine sensitivity a large number of detectors as well as a high control of systematic effects are required. While increasing the number of detector is always possible, to remove systematic effects is not. This make crucial to manage with them. \par
  The instrumental noise depends typically on the kind of receiver used to measure the signal. In CMB experiments principally two kinds of detectors are used: the so-called \textit{bolometers} and the \textit{radiometers}. Bolometers are used to measure the power of the incoming radiation field: they are essentially thermometers as they heat proportionally to the power of the accident field. They are non-coherent receivers, used, nowadays, in a range of frequencies from $\sim 40\,\mathrm{GHz}$ to $\gtrsim 100 \; \mathrm{GHz}$. Radiometers are instead coherent detector as they preserve the information on the phase of the electromagnetic signal. They collect the microwave signal and process it through radio-frequency components such as amplifiers, phase switches and hybrids. At the end of the radiometric chain the signal is converted in electric signal through a diode. They are typically used at lower frequencies ($\lesssim 100 \; \mathrm{GHz}$). \par
  Despite the typology of detector, a number of systematic effects are common to both. The most typical one is due to long-term gain fluctuations into amplification of the signal in receivers. This effect is called \textit{1/$f$ noise} or \textit{pink noise} (Sect. \ref{noisechara}) since it introduces a correlated component, typically a power law, in the noise spectra and it differs from the so-called \textit{white noise} which is statistical noise, i.e. a flat component in the noise spectrum. The noise spectrum is then described by three parameters: the white noise level, the \textit{slope} of the power law associated to the pink spectrum and the \textit{knee frequency}, which is the frequency at which the pink and the white noises have the same power.\par
  A family of systematic effects is related to the pick-up of the incoming radiation. They are beam pattern asymmetries that introduce spurious polarizations or secondary lobes that collects radiation from uncontrolled directions of the sky as well as stray-light contamination and spillover.\par
  Other systematic effects could be due to spurious signals in the data-stream such as cosmic rays (typically in satellite experiments), non-linearity in the amplification chain or errors in pointing and calibration. \par
  Several systematic effects result, ultimately, in leakages from a Stokes parameter to another. Typically, I $\to$ Q/U and Q/U $\leftrightarrow$ U/Q leakages are due to hardware characteristics and detector non-idealities.
\item{\bf Foregrounds.} The other main challenge in detection of B-modes is to distinguish between the cosmological signal and the diffuse emissions produced by our own galaxy, the so-called \textit{foregrounds}. \par
  Although the Milky Way emits microwave radiation due to different effects only two of them contribute in polarization signal: the synchrotron emission due to cosmic ray electrons spiralizing around the lines of the galactic magnetic field and the thermal dust emission from large molecules heated by starlight. The polarization degree of such components is not constant over the sky and it results in dominating over the CMB polarized signal (Fig. \ref{foreg}) even far from the galactic plane.\par
\begin{figure}[H]
\centering
\includegraphics[scale=0.2]{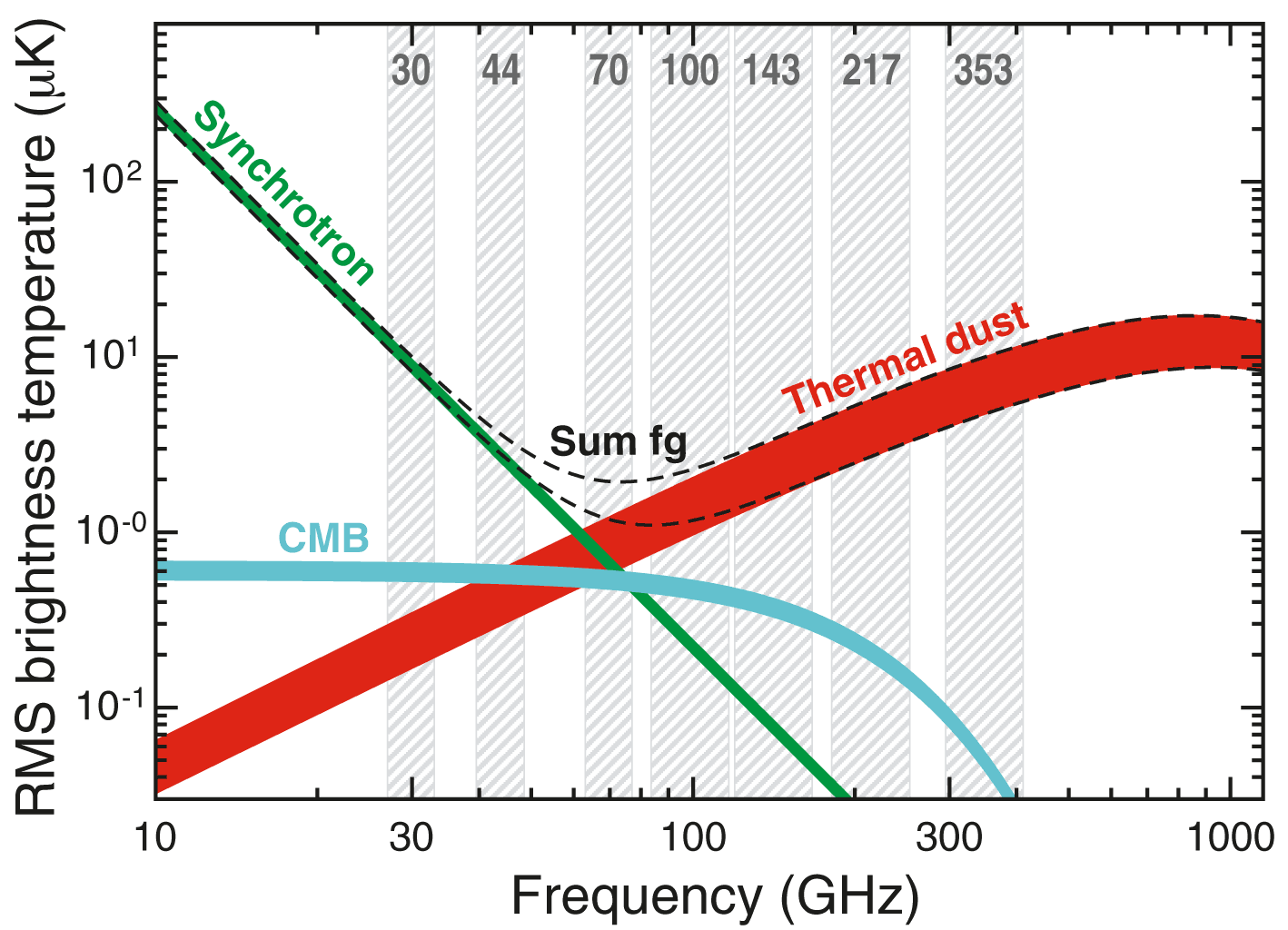}
\caption{Brightness temperature RMS as a function of frequency and astrophysical components for polarization emission. The rms is calculated on maps at angular resolution of $40 \,\mathrm{arcmin}$ on sky fraction between $73\%$ and $93\%$, corresponding to the lower and upper edges of each line. \citep{Planck2015-X}.}\label{foreg}
\end{figure} 
Assuming that the galactic magnetic field is uniform the first effect results dominant for low microwaves frequencies ($\lesssim 70 \; \mathrm{GHz}$) and it could be modeled, in terms of brightness temperature, with a power law:
\be{syncro}
T(\nu) \propto \nu^{\beta_S} \,,
\ee
with
\be{betas}
\beta_S = -\frac{p-3}{2} \,,
\ee
where $p$ is the spectral index of the energy distribution of the electrons propagating in the magnetic field.\par
The thermal dust emission from interstellar dust grains such as graphites and silicates dominates for higher microwave frequencies ($\gtrsim 100 \; \mathrm{GHz}$). Its spectral intensity and then its brightness temperature can be modeled by the following law:
\be{thedast}
T(\nu) \propto \nu^{\beta_d} \, B_\nu(T_d) \,,
\ee
where $B_\nu$ is the Planck spectrum and $T_d$ is the physical temperature of the grains. The polarization mechanism depends on the fact that grains could have a non-spherical shape with the major axis tending to align perpendicular to the local magnetic field. So that, the alignment degree depends on the size distribution of the grains.\par
The principle of the component separation is based on the fact that CMB and foreground emissions behave differently with the frequency dependence, as shown in Fig. \ref{foreg}. Furthermore, they are supposed to be uncorrelated, so that it is theoretically possible to perform parametric fits. For this reason, measurements of the sky at multiple frequencies are crucial to disentangle the weak CMB signal from the foreground emission.  
\end{list}

\subsection{A novel approach to CMB experiments: the case of QUBIC}
A special mention deserves the ``Q \& U Bolometric Interferometer for Cosmology'' \citep[QUBIC,][]{Battistelli} since it is currently the only experiment searching for B-modes based on interferometry. \par
\begin{figure}[H]
\begin{centering}
\includegraphics[scale=0.43]{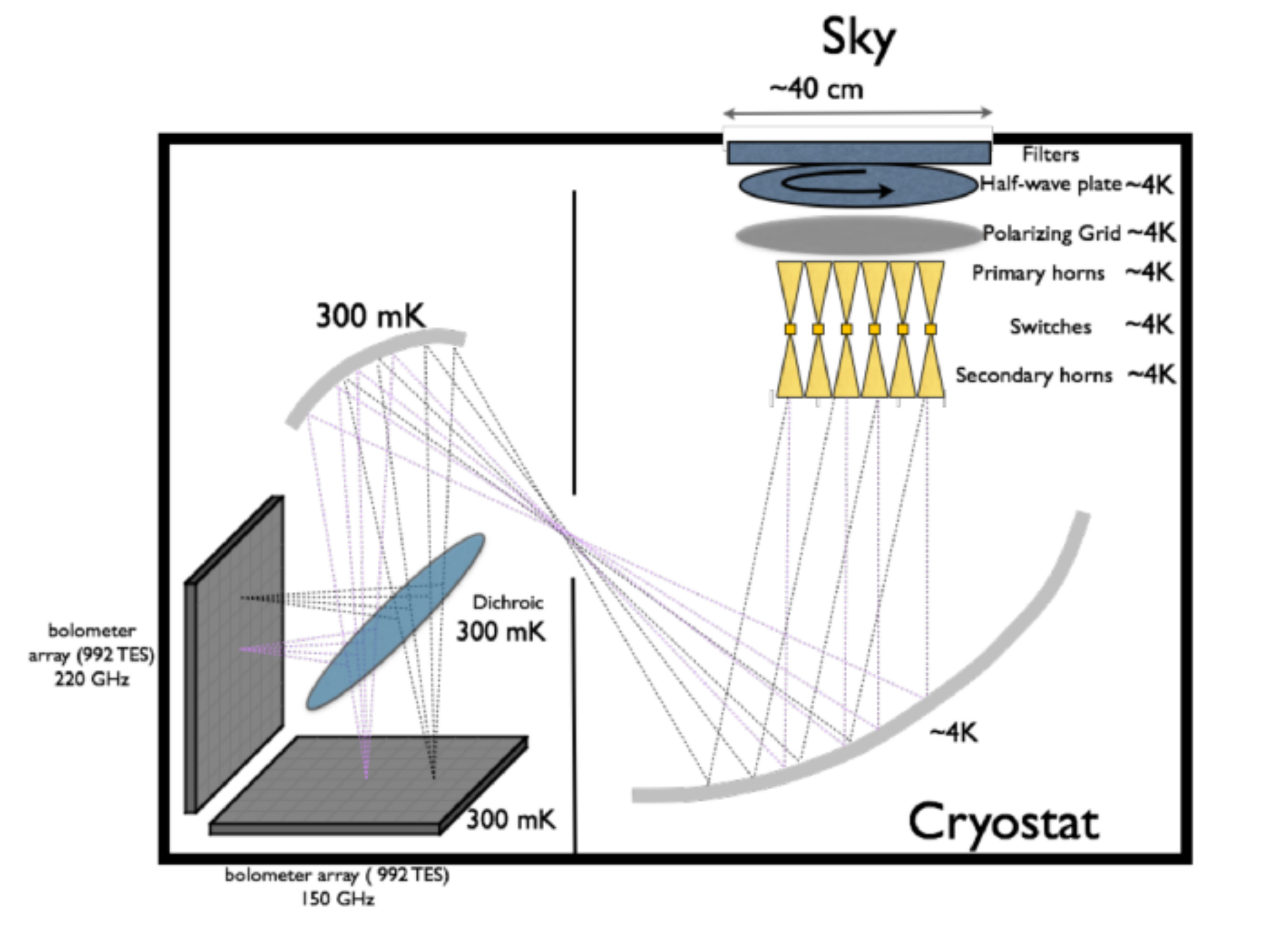}
\caption{Sketch of the QUBIC concept.}\label{design} 
\end{centering}
\end{figure}

QUBIC observes the sky through an array of $400$ back-to-back corrugated antennas (the so-called \textit{horns}) operating at the frequencies of $150$ and $220\,\mathrm{GHz}$ with $25\,\%$ bandwith. They are placed behind the window of a cryostat and act as diffractive pupils. The electric field coming from a given sky direction experiences phase differences due to the distance between the input horns. The back horns re-emit the electric field preserving this phase difference inside the cryostat. An image showing the working principle of QUBIC is given in Fig. \ref{design}. \par
The signal is collected by a dual-reflector telescope that focus rays launched at a given angle from the re-emitting horn array to a single point on the focal plane. In this way, equivalent baselines will produce identical fringe patterns. This principle is at the base of the so called \textit{self-calibration} technique that allows to reduce dramatically the impact of systematic effects. \par
After the telescope, a dichroic filter selects the two frequency bands which propagate in two orthogonal directions. The interference fringe patterns arising from all pairs of horns are formed on the two focal plane of the optical combiner. On each focal plane, is placed a $992$-element bolometer array cooled to $\sim 300 \hspace{2pt} \mathrm{mK}$. \par
The polarization of the incoming field is modulated using a \textit{half-wave plate} (HWP) located before the horns. In order to reduce the possibility of leakages of I into Q and U a polarizing grid placed after the half-wave plate completely rejects one polarization direction.\par
The self-calibration technique exploits the fact that equivalent baselines should measure the same quantity, in the absence of systematic effects. Using the switches, it is possible to modulate on/off a single pair of horns while leaving all the others open (or closed) in order to access the visibility measured by this pair of horns alone. By repeating this process with a subset of all available baselines, equivalent and different, it is possible to construct a system of equations whose unknowns are the systematic effects parameters for each channel. For a large enough array of primary horns, the system is over-constrained and can be solved. So that, the accuracy of this procedure depends only by the time spent on self-calibration. \par
A technological demonstrator of QUBIC is currently under test. The full instrument is expected to be installed in 2021 at the Alto Chorillo site in Argentina and it is expected to constrain the tensor-to-scalar ratio to $r \simeq 0.01$ in two years of data taking at the $90\,\%$ confidence level. \par 

%% file: chap2.tex
\chapter{The LSPE experiment}
\label{Chap:2}
\thispagestyle{plain}

The ``Large Scale Polarization Explorer'' \citep[LSPE,][]{Lspe} project aims to constrain the tensor-to-scalar ratio to $r \simeq 0.03$ at the $99.7\,\%$ confidence level and, more widely, to study the polarized emission of the Milky Way. So far, in fact, there are no data of the polarized emission of our own Galaxy at large angular scales and at multiple frequencies from the Earth's Northern Hemisphere whose sensitivity is better than Planck's. Observing in the Northern Hemisphere confers great importance to LSPE since today most of the CMB polarization experiments observe mostly the Southern Sky. \par
The LSPE project is composed of two experiments: SWIPE and STRIP.\par
SWIPE \citep{SWIPE_wp} is a stratospheric balloon that will fly from the Svalbard Islands (or from Kiruna) for two weeks during the polar night of 2021. It will observe the sky at three different frequencies: $140$, $210$ and $240\,\mathrm{GHz}$ adopting two arrays of $163$ \textit{transition-edge sensors} (TES bolometers).\par
STRIP \citep{STRIP_wp} is a ground-based telescope that will operate for two years from the ``Observatorio del Teide'' in Tenerife, starting in 2021. Its focal plane is composed of an array of $49$ coherent polarimeters operating at $43\,\mathrm{GHz}$ plus $6$ elements at $95\,\mathrm{GHz}$, which will be exploited as an atmospheric monitor. \par
STRIP and SWIPE will observe approximately the same sky covering about the $30\%$ of the Northern Sky (Fig. \ref{ss}). \par
The LSPE project is founded by: ``Agenzia Spaziale Italiana'' (ASI) and ``Istituto Nazionale di Fisica Nucleare'' (INFN). 

\section{SWIPE: the high frequency instrument}
\begin{figure}[H]
\centering
\includegraphics[scale=0.2]{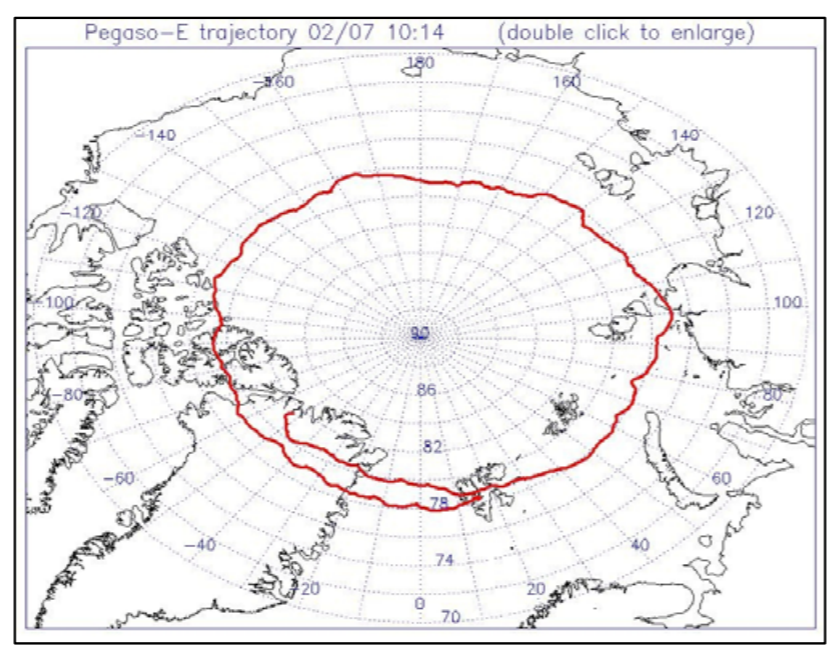}
\includegraphics[scale=0.23]{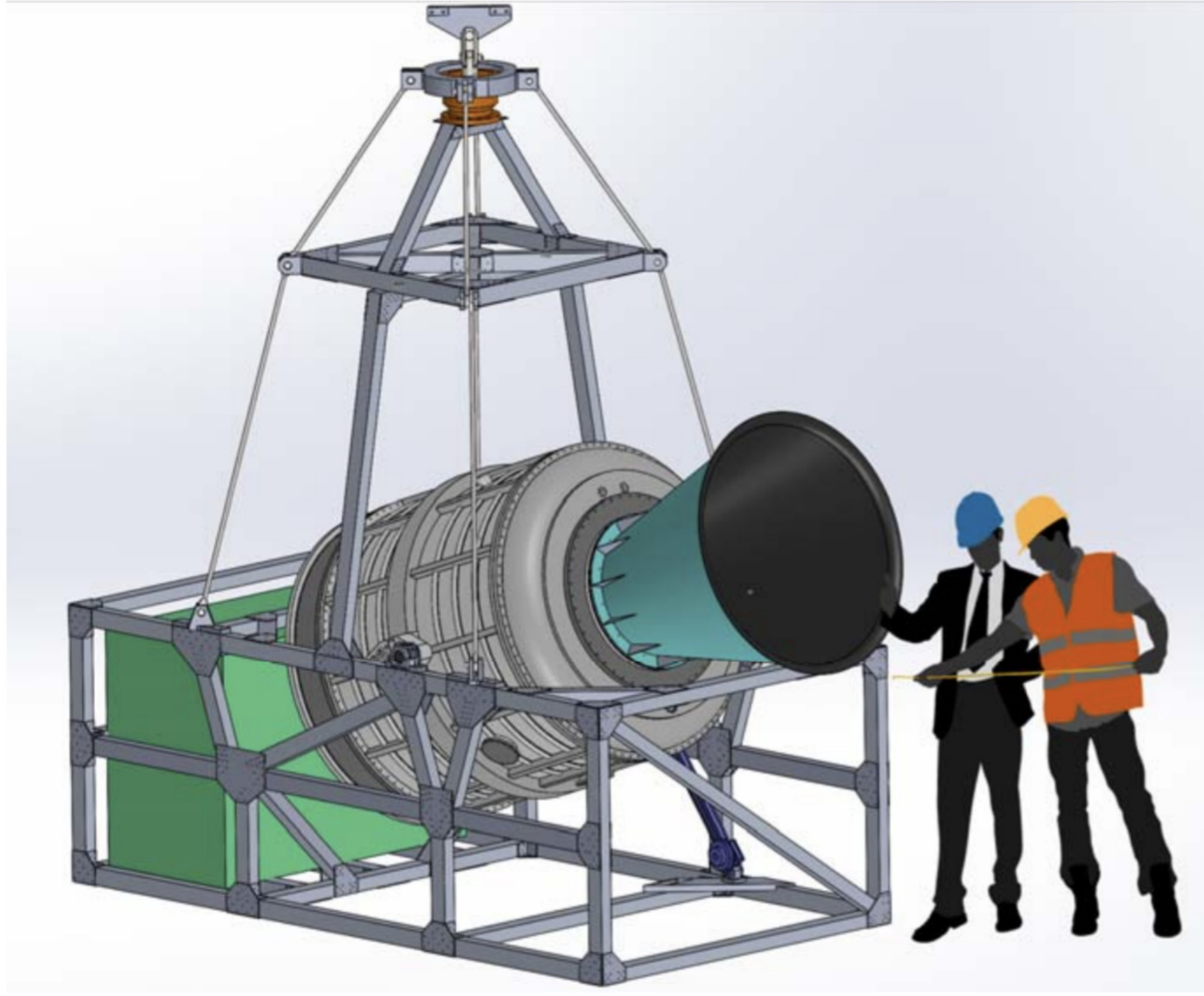}
\caption{\textit{Left}: Path of one of the test flights performed by ASI to study the stratospheric circulation near the North Pole. The SWIPE trajectory might be similar to this. \textit{Right}: Sketch of the SWIPE payload.}\label{path}
\end{figure}

SWIPE is an instrument that will fly on a stratospheric balloon for a long duration flight during the arctic winter of 2021 (Fig. \ref{path}) leaving either from the Svalbard islands (Norway) or from Kiruna (Sweden). It will fly at an altitude of about $35\,\mathrm{km}$ to avoid atmospheric emission principally due to water vapor. The flight is supposed to last fifteen days in order to reach the requested sensitivity on large angular scales. \par
The entire telescope is cooled down to $0.3\,\mathrm{K}$ by a cryostat in order to reduce the radiative background as much as possible. The cryostat is mounted in a frame, the so-called \textit{gondola}, which allows for azimuth scans and spin (Fig. \ref{path}).\par
The polarization of the signal coming from the sky is modulated through a \textit{half-wave plate} (HWP) and focused on the two focal planes by mean of a small ($\simeq 490\,\mathrm{mm}$) refractor telescope. The signal is split by a large wire-grid polarizer and then collected by two curved focal plane, each populated with $163$ multi-moded feedhorns (Fig. \ref{focalplane}). \par
Frequencies are distributed in vertical bands, rather than in the conventional concentric distribution because, given the SWIPE scanning strategy, all frequency must cover the same elevation range. Furthermore, the multi-moded nature of the SWIPE horns imply that the STRIP coverage must extend slightly in order to cover the larger sidelobes of SWIPE. The mask used for component separation takes into account the different angular resolutions. \par
Each horn transfers the radiation to a TES bolometer. \par
 \begin{figure}[H]
\centering
\includegraphics[scale=0.25]{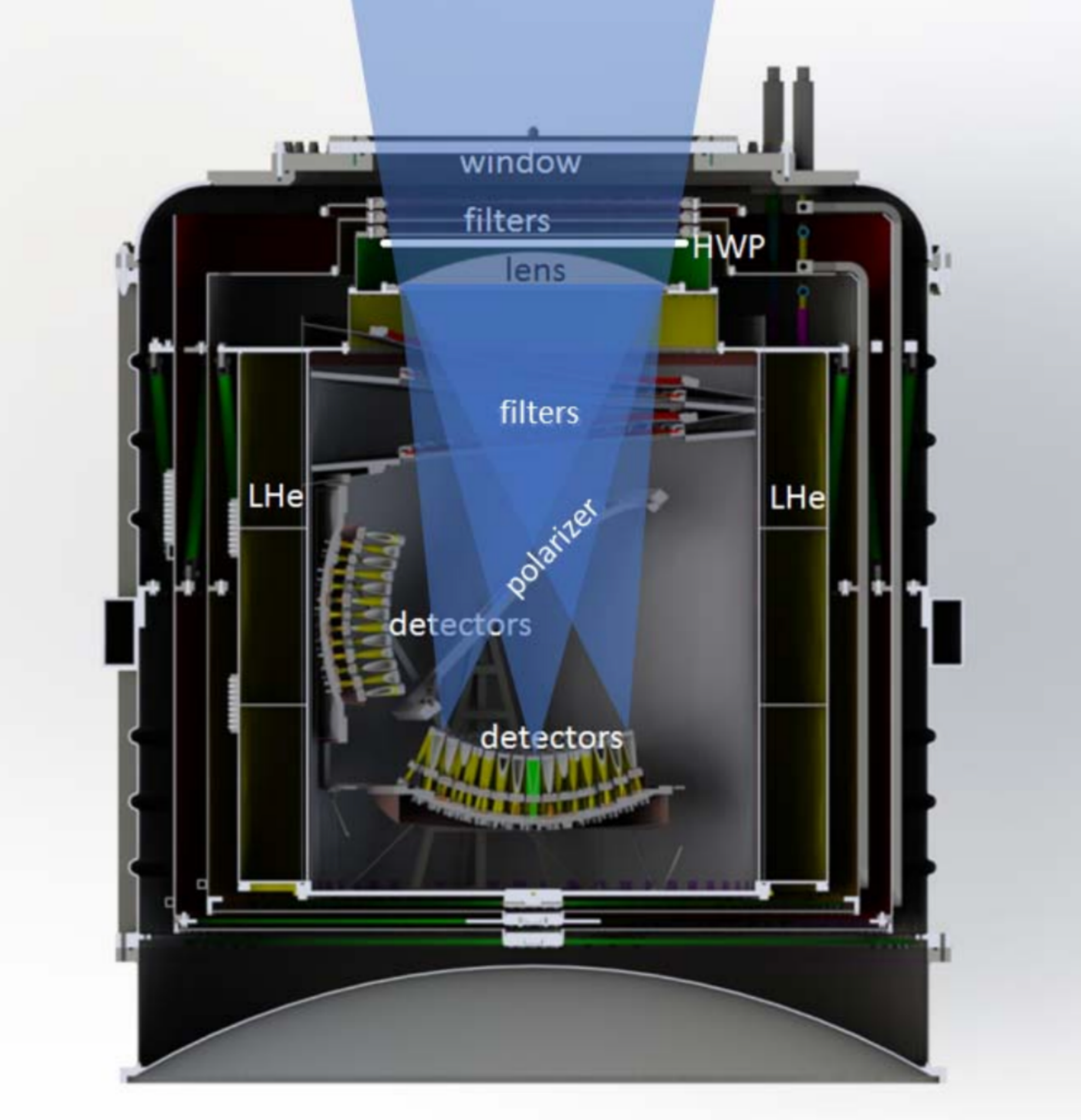}
\includegraphics[scale=0.45]{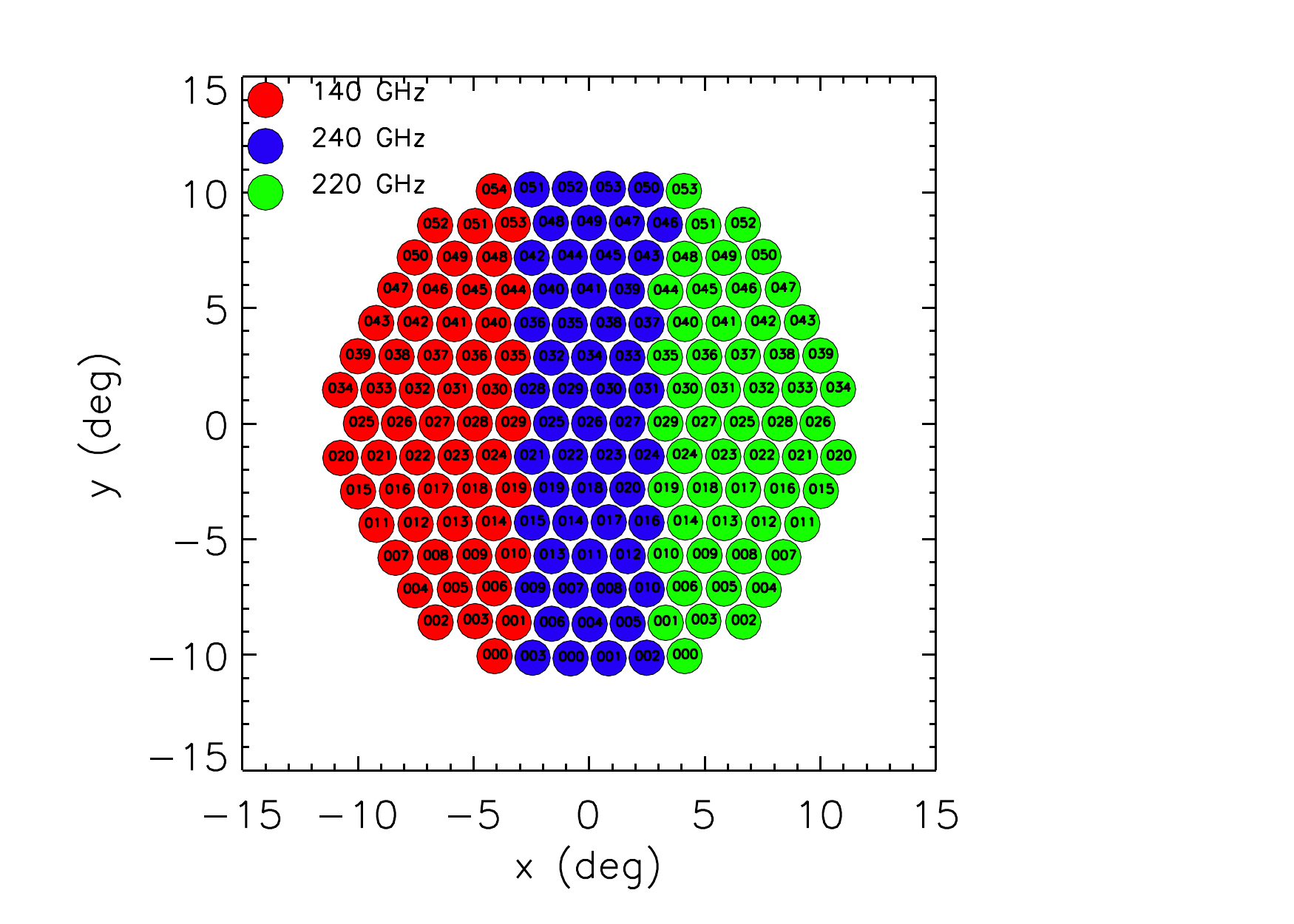}
\caption{Sketch of the SWIPE optical system (left) and of a focal plane (right).}\label{focalplane}
\end{figure}
SWIPE will measure the sky signal in three frequency bands: $140$, $210$ and $240 \, \mathrm{GHz}$. The \textit{noise equivalent temperature} (NET) for each band is reported in Table \ref{net}. \par
\begin{table}[H]
\centering
\begin{tabular}{|c|c|c|c|}
  \hline
$\mathrm{\mathbf{Channel}}$ [$\mathrm{GHz}$] & $\mathbf{140}$ & $\mathbf{210}$ & $\mathbf{240}$ \\  \hline 
$\mathrm{\mathbf{NET}}$ [$\mu\mathrm{K}_\mathrm{CMB}\,\sqrt{s}$] & $12.7$ & $15.7$ & $30.9$ \\ \hline
\end{tabular}\caption{NET for the SWIPE channels.}\label{net}
\end{table}

\section{STRIP: the low frequency instrument}\label{STRIPlfi}
STRIP is a ground-based telescope that will operate from ``Observatorio del Teide'' in Tenerife, starting from early 2021. \par
\begin{figure}[H]
\centering
\includegraphics[scale=0.35]{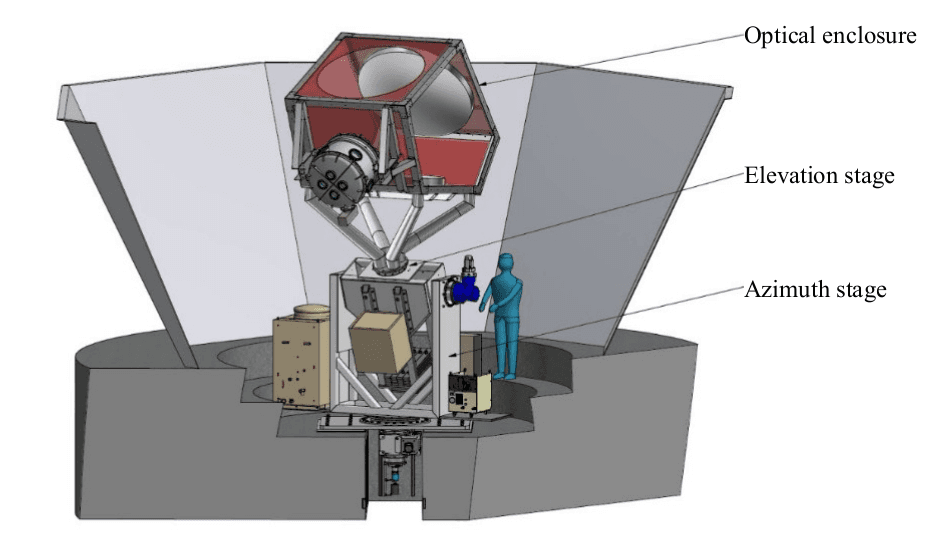}
\caption{Sketch of the STRIP telescope and of the external shields that protect it from the ground radiation.}\label{telescope}
\end{figure}

STRIP will use the same telescope that was originally designed and built for the CLOVER experiment \citep{Clover_wp}: a $1.5\,\mathrm{m}$ \textit{Crossed-Dragone} \citep{dragone} telescope with an angular resolution of $20\,\mathrm{arcmin}$. The telescope can rotate around the elevation and the azimuth axes (Fig. \ref{telescope}).\par
The radiation coming from the sky is focused onto the window of a cryostat that contains the focal plane, which is cooled down to $20\,\mathrm{K}$. \par
The focal plane is composed of forty-nine coherent polarimeters\footnote{Nineteen polarimeters have been already used by the QUIET experiment \citep{QUIET_wp}.} operating at $43\,\mathrm{GHz}$ (Q-band) plus a second frequency channel with six elements at $95\,\mathrm{GHz}$ (W-band), each with about $18\%$ bandwidth.\par
The Q-band channel is the one properly devoted to astrophysical measurements, even if it is able to measure the atmospheric signal as well. On the contrary, the $95\,\mathrm{GHz}$ channel alone is not able to detect the astrophysical signal, due to its poor sensitivity (Table \ref{deltaTT}). So that, this channel will be used to monitor and study the atmospheric emission in situ, both in intensity and in polarization. Besides, its data will be used for cross-checking purposes as well as to assess the feasibility of future W-band CMB experiments from Tenerife. At present, the presence of this channel will not affect the scanning strategy.\par
\begin{figure}[H]
\centering
\includegraphics[scale=0.5]{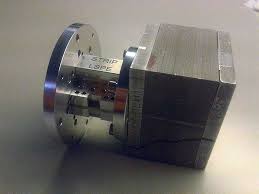}
\includegraphics[scale=0.7]{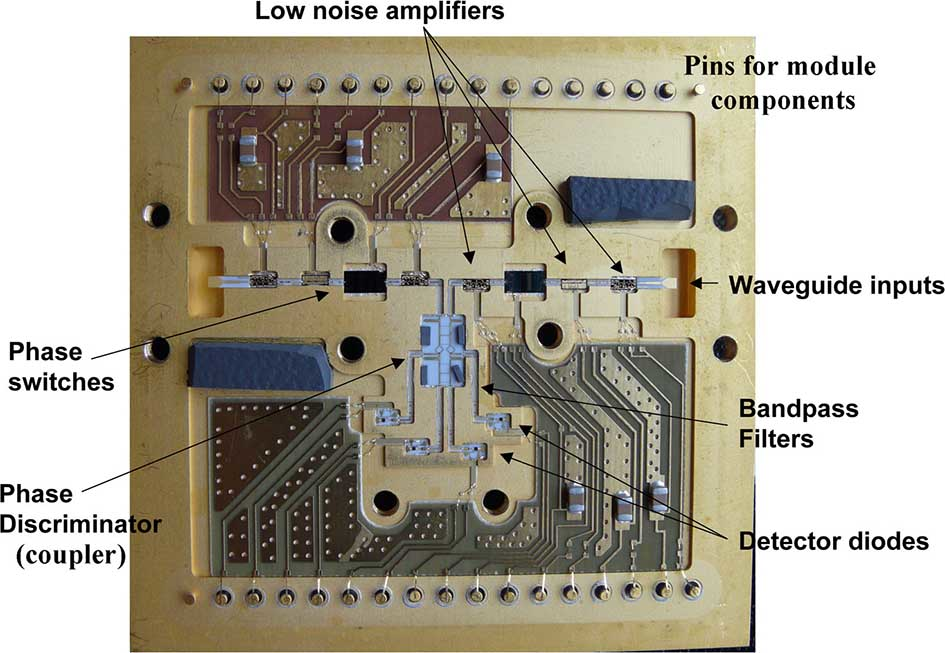}
\caption{\textit{Left}: Polarizer and OMT chain assembled. \textit{Right}: Picture of the polarimeter radiometric chain comprehending RF amplifiers, phase switches, hybrid couplers, bandpass filters and diodes.}\label{polarimeter}
\end{figure}

 
Each polarimeter is coupled to a corrugated horn through a chain (Fig. \ref{polarimeter}) composed by a \textit{polarizer} and an \textit{orthomode transducer} (OMT) that converts linear polarization into left- and right-circularly polarized components. \par
The Q-band radiometric chains are arranged in seven modules, each one composed of seven antennas. The entire focal plane results in a honeycomb structure (Fig. \ref{focalplane}). \par
\begin{figure}[H]
\centering
\includegraphics[scale=0.25]{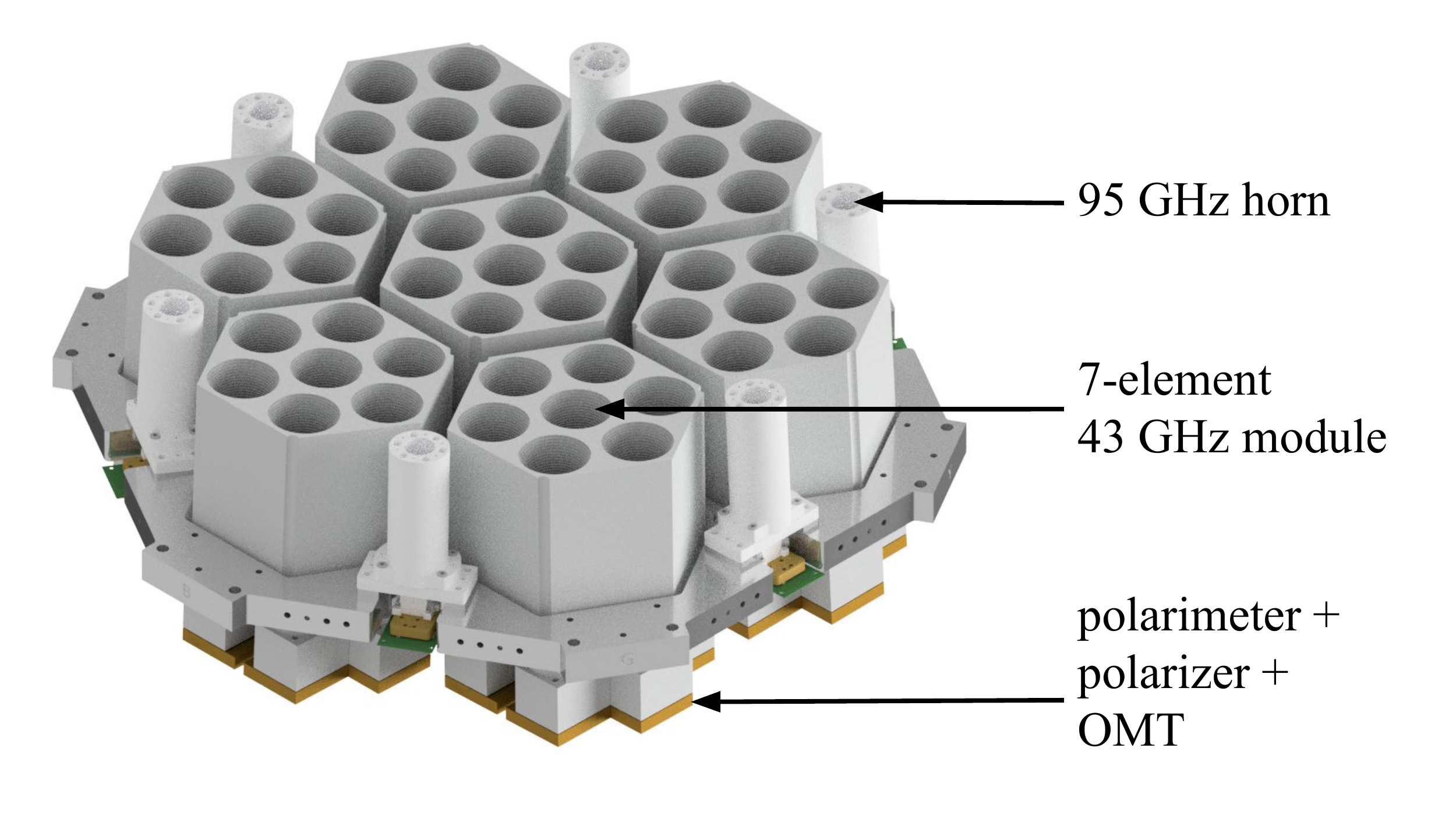}
\includegraphics[scale=0.2]{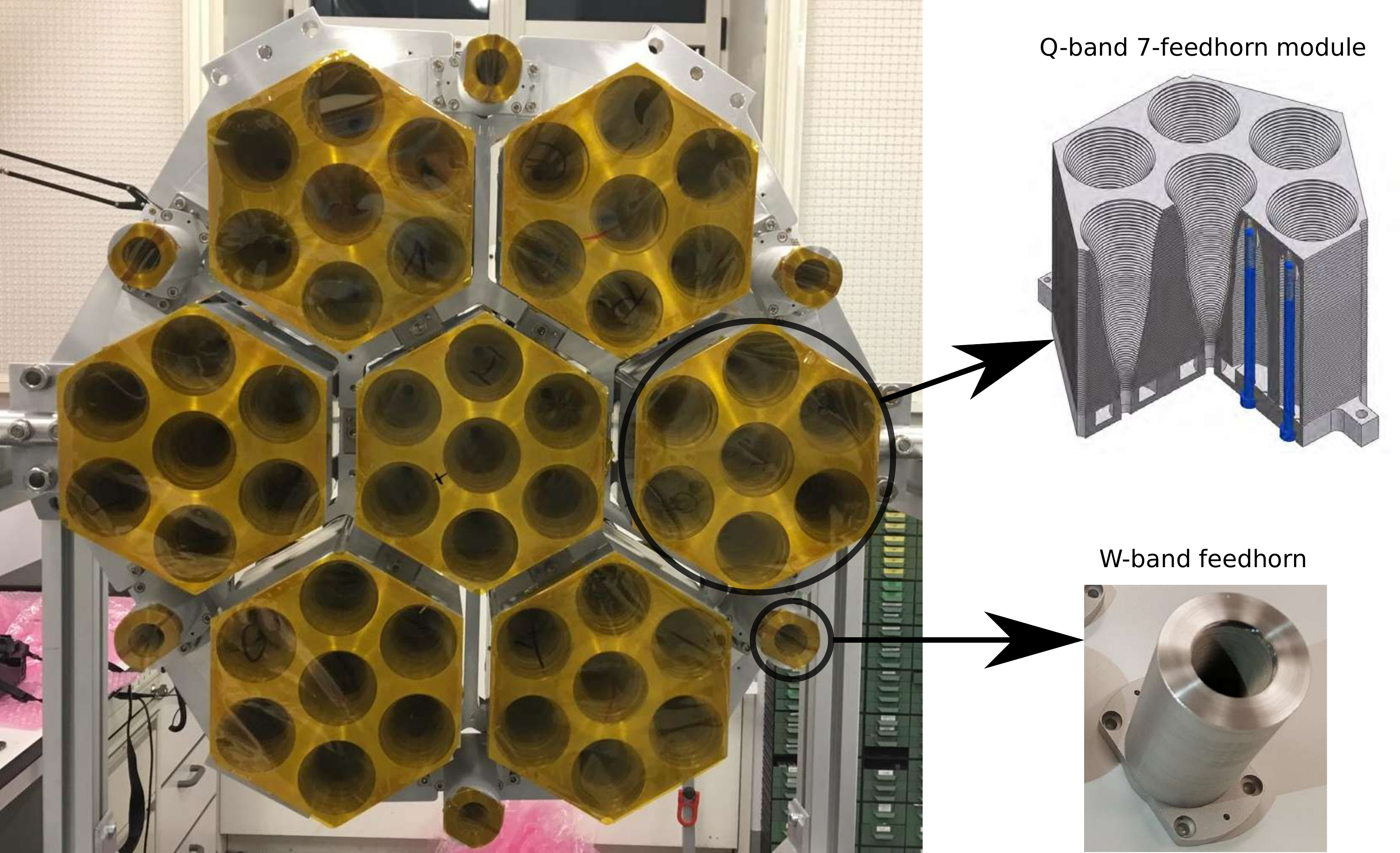}
\caption{\textit{Left}: Sketch of the focal plane with the $49$ feedhorns at $43\,\mathrm{GHz}$, the $6$ horns at $95\,\mathrm{GHz}$ and the other components of the detection chains. \textit{Right}: a picture of the integrated focal plane where the Q- and W-bands feedhorns are highlighted.}\label{focalplane}
\end{figure}
The radiometric chain of the STRIP detectors, Fig. \ref{polarimeter}, allows us to measure directly the four Stokes parameters of the incident radiation field. This is possible by means of radio-frequency (RF) components that combine and/or shift the phase of the signal. A more detailed explanation is provided in Ch. \ref{Chap:3}.
The STRIP sensitivity is measured in terms of the standard deviation of the white noise component of the Q and U Stokes parameters measured in thermodynamic temperature. Its sensitivity requirements are reported in Table \ref{deltaT}.\par
\begin{table}[H]
\centering
\begin{tabular}{|c|c|c|}
  \hline
  $\mathrm{\mathbf{Channel}}$ [$\mathrm{GHz}$] & $\mathbf{43}$ & $\mathbf{95}$  \\  \hline 
  $\mathbf{\Delta} \mathrm{\mathbf{Q}}_\mathrm{\mathbf{rms}}$ [$\mu\mathrm{K}_\mathrm{CMB}\,\sqrt{s}$] & $541.5$ & $1099.5$ \\ \hline
\end{tabular}\caption{Sensitivity requirements for the STRIP channels.}\label{deltaT}
\end{table}\par
In Table \ref{deltaTT}, I report the same requirements of Table \ref{deltaT} expressed in brightness temperature per resolution elements of $1\deg$.
\begin{table}[H]
\centering
\begin{tabular}{|c|c|c|}
  \hline
  $\mathrm{\mathbf{Channel}}$ [$\mathrm{GHz}$] & $\mathbf{43}$ & $\mathbf{95}$  \\  \hline 
  $\mathbf{\Delta} \mathrm{\mathbf{Q}}_\mathrm{\mathbf{rms}}$ [$\mu\mathrm{K}\,\mathrm{deg}$] & $1.6$ & $10.54$ \\ \hline
\end{tabular}\caption{Sensitivity requirements for the STRIP channels.}\label{deltaTT}
\end{table}\par

\section{LSPE as a whole}\label{STRIPasaw}
\begin{figure}[H]
\centering
\includegraphics[scale=0.43]{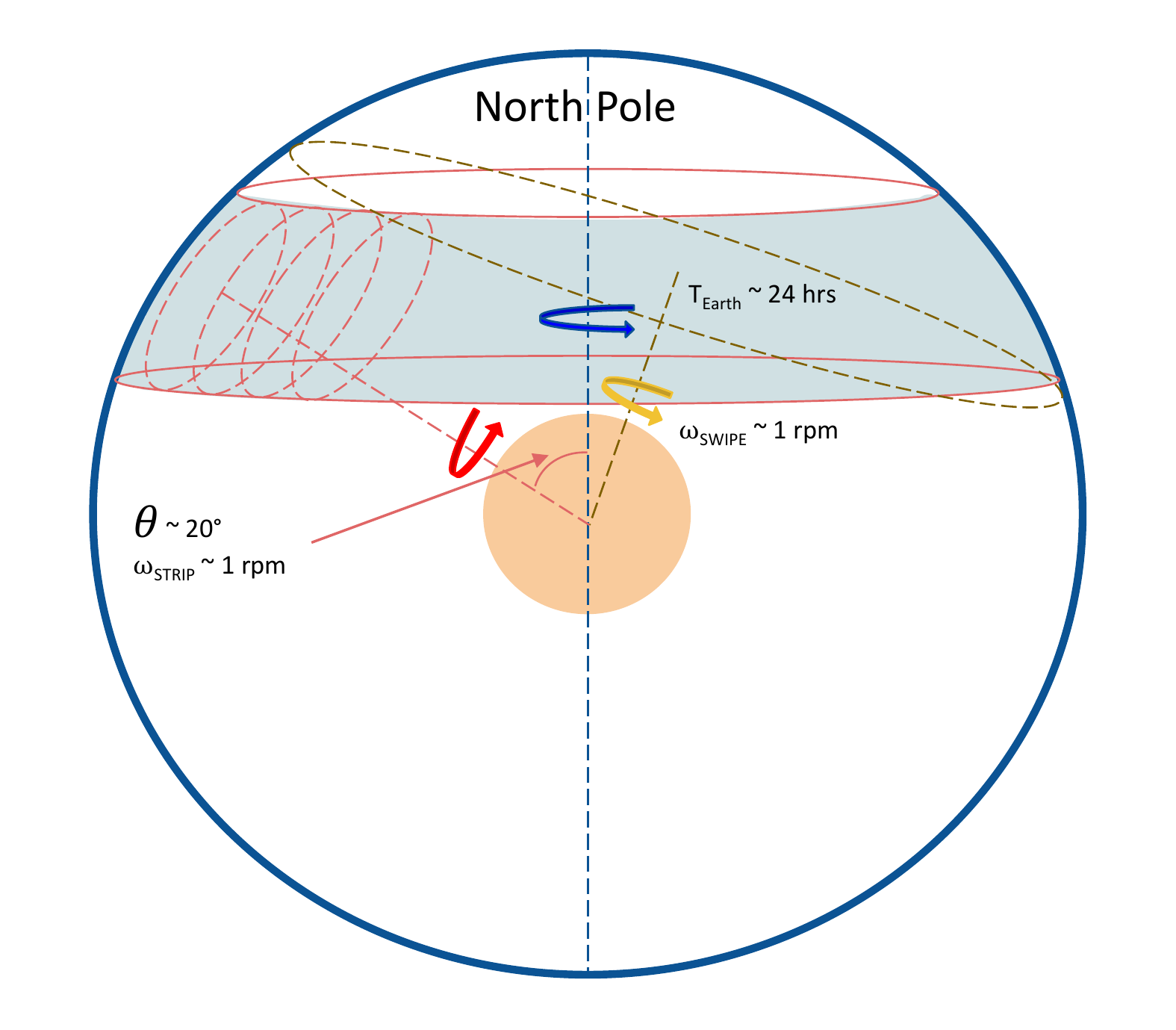}
\caption{Scheme of SWIPE and STRIP scanning strategies: they will observe approximately the same portion of the Northern Sky. STRIP will observe a sky-band whose amplitude depends on its elevation angle, thanks to the combination of its motion and the Earth’s rotation.}\label{ss}
\end{figure}

SWIPE and STRIP will observe approximately the same sky region (Fig. \ref{ss}) thanks to the fact that the STRIP telescope can spin around the azimuth axis at constant elevation, describing circles on the sky whose radius depends on the elevation itself. The daily motion of the Earth changes the sky above the telescope, obtaining to cover a wide-band of the sky. This region matches the SWIPE patch, adjusting the elevation angle (Ch. \ref{Chap:6}). This method allows LSPE to observe a sky fraction of $\simeq 30\%$.\par
\begin{figure}[H]
\centering
\includegraphics[scale=0.6]{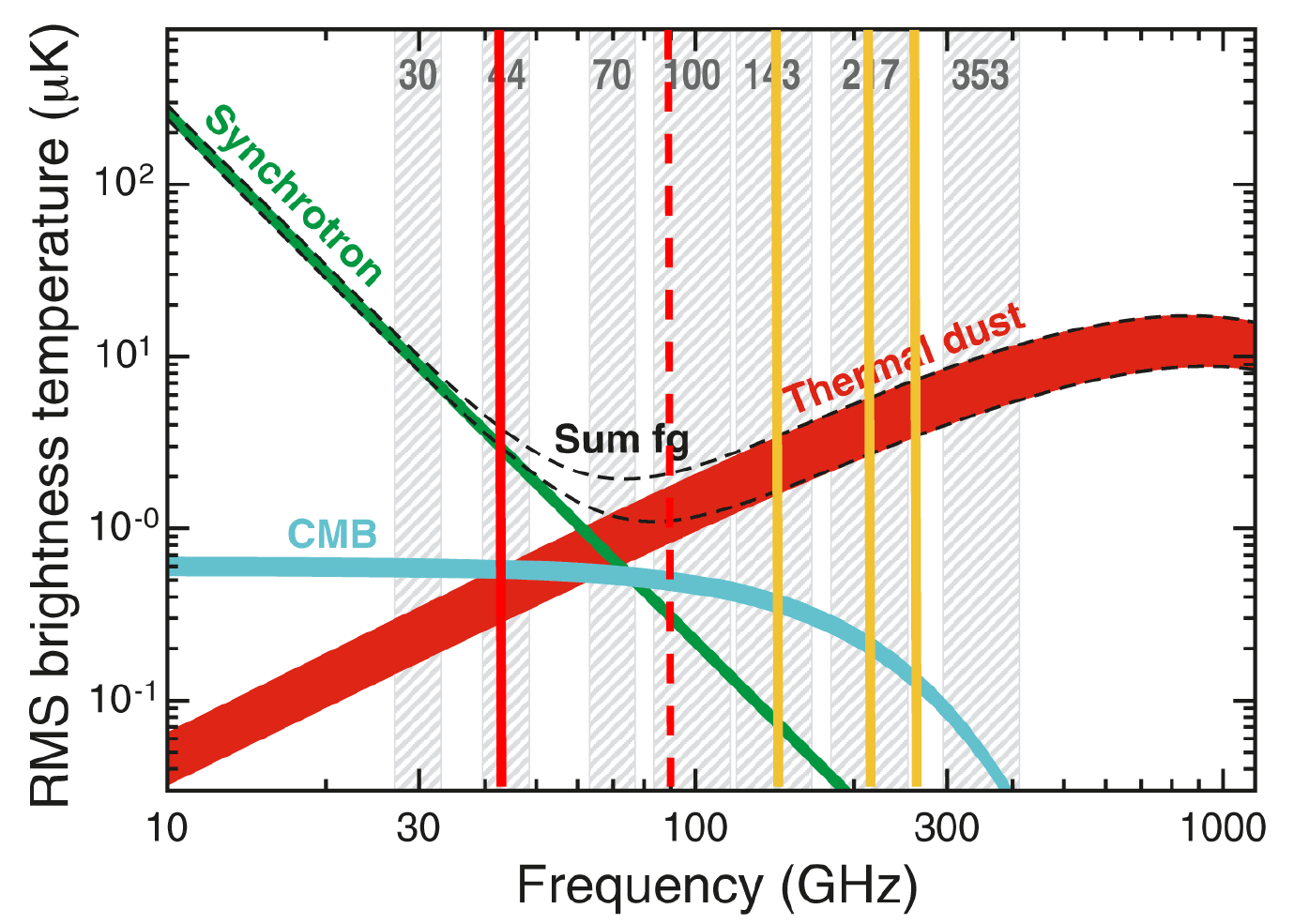}
\caption{SWIPE (yellow) and STRIP (red) frequency channels (left). The dashed line represents the 95 GHz channel of STRIP, which will be exploited as a monitor of the atmosphere.}\label{swipestrip}
\end{figure}

LSPE exploits its four cosmological channels to properly separate astrophysical emissions. While STRIP focuses on the synchrotron emission, SWIPE will measure the interstellar dust with its high frequency bands (Fig. \ref{swipestrip}). \par
The LSPE sensitivity goal is set in order to improve by a factor $\simeq 5$ the sensitivity reached by Planck \citep[Table \ref{deltaTplanck}, ][]{piacentini:inpress}. In the case of STRIP, this goal corresponds to reach the values of sensitivity per resolution element of $1\deg$ reported in Table \ref{deltaTgoal}. 
\begin{table}[H]
\centering
\begin{tabular}{|c|c|c|c|}
  \hline
$\mathrm{\mathbf{Channel}}$ [$\mathrm{GHz}$] & $\mathbf{30}$ & $\mathbf{44}$ & $\mathbf{70}$  \\  \hline 
$\mathbf{\Delta} \mathrm{\mathbf{Q}}_\mathrm{\mathbf{rms}}$ [$\mu\mathrm{K}\,\mathrm{deg}$] & $4.90$ & $5.92$ & $5.65$ \\ \hline
\end{tabular}\caption{Sensitivity per resolution element of $1\deg$ for the LFI channels of Planck.}\label{deltaTplanck}
\end{table}

\begin{table}[H]
\centering
\begin{tabular}{|c|c|c|}
  \hline
$\mathrm{\mathbf{Channel}}$ [$\mathrm{GHz}$] & $\mathbf{43}$ & $\mathbf{95}$  \\  \hline 
$\mathbf{\Delta} \mathrm{\mathbf{Q}}_\mathrm{\mathbf{rms}}$ [$\mu\mathrm{K}\,\mathrm{deg}$] & $1.2$ & $8.2$ \\ \hline
\end{tabular}\caption{LSPE sensitivity goal in Q and W bands.}\label{deltaTgoal}
\end{table}

%% file: PartI.tex
\part*{Part I \vspace{0.5cm}\\Noise characterization of the LSPE/STRIP polarimeters}

%% file: chap3.tex
\chapter{The STRIP detection chain}
\label{Chap:3}
\thispagestyle{plain}
In this chapter, I illustrate the working principle of the STRIP detectors by going through the details of the mathematical model and the electronic processes that drive the acquisition of the sky signal. \par
The STRIP detectors are coherent polarimeters based on \textit{high-electron-mobility transistors} (HEMTs), which are a type of \textit{low noise amplifiers} (LNAs). There are $49$ detectors working at $43\,\mathrm{GHz}$ and $6$ working at $95\,\mathrm{GHz}$, a number of which have been inherited by the QUIET experiment \citep{QUIET_wp}. \par
The working principle and the radio-frequency (RF) components that constitute the detectors are the same as in QUIET: the electromagnetic signal coming from the sky is collected by a corrugated antenna, the so-called \textit{feedhorn} or simply \textit{horn}, and then it propagates inside the radiometric chain until it is converted into an electric signal by a diode. Thanks to the combination of several RF components, the overall response of the detectors is proportional to the four Stokes parameters of the incident radiation field. Furthermore, the electronics of the polarimeter is able to reduce the correlated noise, the 1/$f$ component, by many order of magnitudes thanks to the \textit{double demodulation} process. \par
Even if the STRIP detection chain is the same as those of QUIET, the two experiments are different because the latter was installed at the Atacama Desert observing very small patches of the sky \citep[$\lesssim 1\%$, ][]{B2011} in the Southern Hemisphere at $43$ and $95\,\mathrm{GHz}$. On the contrary, LSPE will observe the Northern Sky, and, in combination with SWIPE, will allow us to obtain a more accurate component-separated CMB map.\par
The following description of the STRIP polarimeter model is mostly qualitative: a more quantitative approach is provided by Appendix \ref{App:1}.\par

\section{Radiometric chain}\label{radiometrichain}

The radiometric chain of STRIP is made by the sequence of RF components each one devoted to a very precise purpose. \par
The first element of the chain is the feedhorn that act as interface between the free space and the waveguide. The horn corrugations (Fig. \ref{focalplane}) in fact, optimize the coupling of the signal from the free space to the waveguide propagation. \par
The second and third elements are the \textit{polarizer} and the \textit{orthomode transducer} (OMT). They divide the incoming signal in two circularly polarized components, the left and the right ones. These two signals are the input for the polarimeter, which is the last element of the chain. \par
\begin{figure}[H]
\centering
\includegraphics[scale=0.57]{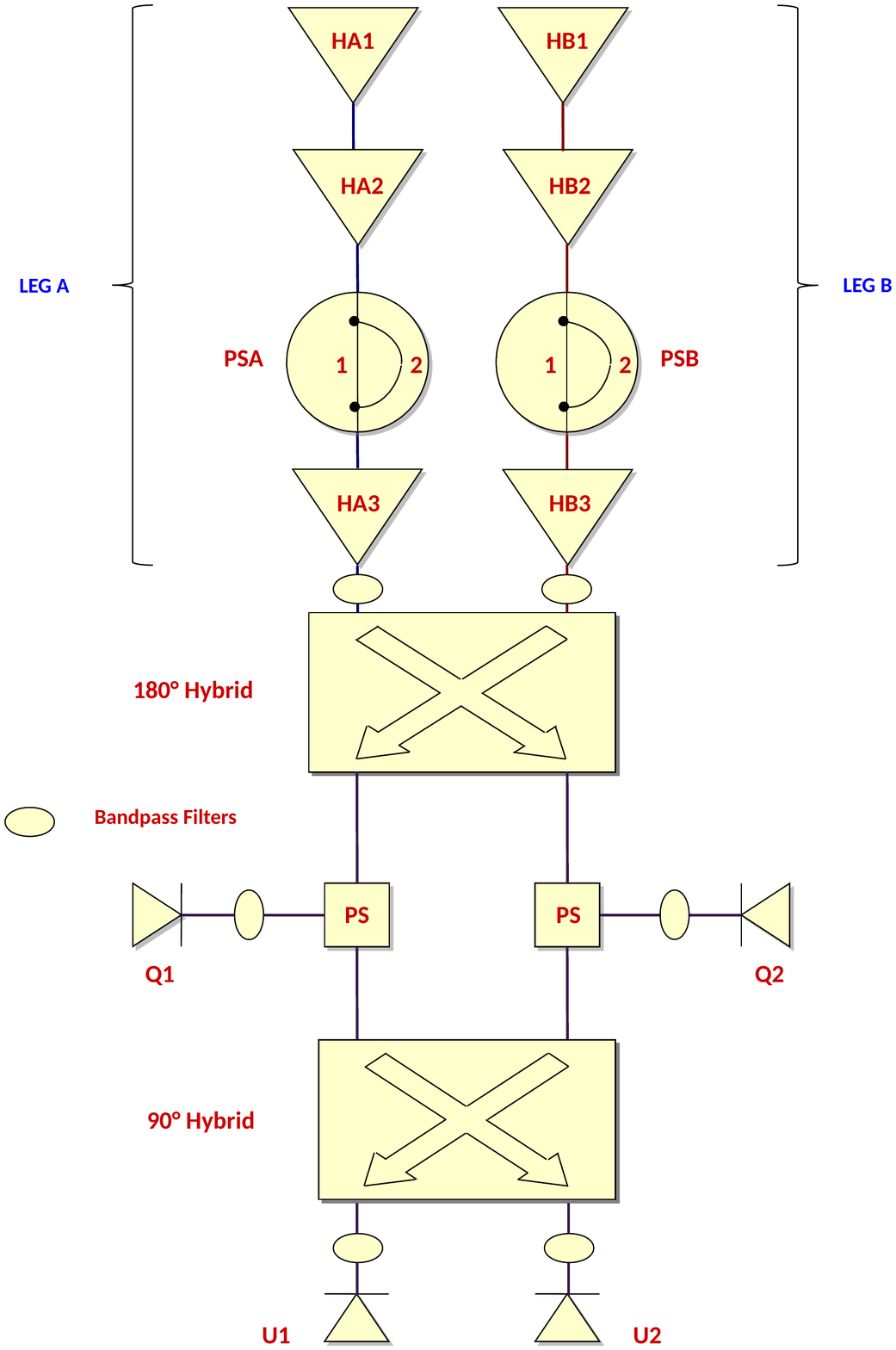}
\caption{Sketch of the RF components that form a STRIP polarimeter. The official naming convention is reported: $\mathrm{HA1}$, $\mathrm{HA2}$, $\mathrm{HA3}$ and $\mathrm{HB1}$, $\mathrm{HB2}$, $\mathrm{HB3}$ are the amplifiers; $\mathrm{PSA}$ and $\mathrm{PSB}$ are the phase switches with their two internal paths $1$ and $2$; the two $\mathrm{PS}$ are the power splitters; $\mathrm{Q1}$, $\mathrm{Q2}$, $\mathrm{U1}$ and $\mathrm{U2}$ are the four diodes.}\label{radchain}
\end{figure}

The polarimeter has the duty to detect the signal. At this purpose, it is composed in turn by a sequence of LNAs, \textit{phase switches}, \textit{bandpass filters}, a \textit{$180\deg$ hybrid}, \textit{power splitters}, a \textit{$90\deg$ hybrid} and \textit{diodes} (Fig. \ref{radchain} and \ref{polarimeter}). \par

\subsection{Polarimeter components}
The radiometric components that constitute the polarimeters are shown in Fig. \ref{radchain}.\par
At its entrance, each polarimeter is made by two waveguides in which the circularly polarized signals propagate. In these two legs (or arms), the electromagnetic field is amplified and switched through a chain of three amplifiers and one phase switch. \par
The phase switch drives the signal towards one of two possible paths whose optical lengths differ by a phase angle of $\pi$. If both ways are simultaneously closed the signal cannot obviously propagate. On the contrary, if both ways are opened at the same time, the two output signals interfere and cancel each other out. A scheme of the possible scenarios are reported in Table \ref{onephaseshift}.\par
As this process happens in both legs of the polarimeter, also the two signals in the two legs could be each other phase shifted or not, according to the values reported in Table \ref{twophaseshift}. \par

\begin{table}[H]
\centering
\begin{tabular}{|c|c|c|} \hline
 $\mathrm{PATH} \, 1$ &  $\mathrm{PATH} \, 2$ & $\phi$ \\ \hline
 0 &  0 &  - \\ \hdashline[0.5pt/1pt]
 0 &  1 &  $\pi$ \\ \hdashline[0.5pt/1pt]
 1 &  0 &  0 \\ \hdashline[0.5pt/1pt]
 1 &  1 &  - \\ \hline
\end{tabular}\caption{Possible phase configuration ($\phi$) of the signal at the exit of each phase switch. The values $0$ and $1$ corresponds respectively to the path being open or closed.}\label{onephaseshift}
\end{table}

\begin{table}[H]
\centering
\begin{tabular}{|c|c|c|c|c|c|c|} \hline
$\mathrm{PSA} \, 1$ & $\mathrm{PSA} \, 2$ & $\mathrm{PSB} \, 1$ & $\mathrm{PSB} \, 2$ & $\phi_A$ & $\phi_B$  & $\Delta\phi$ \\ \hline
0 & 0 & 0 & 0 & - & - & -                       \\ \cdashline{1-4}[0.5pt/1pt] \cdashline{5-7}[2pt/2pt]
0 & 0 & 1 & 0 & - & 0 & -                       \\ \cdashline{1-4}[0.5pt/1pt] \cdashline{5-7}[2pt/2pt]
0 & 0 & 1 & 1 & - & - & -                       \\ \cdashline{1-4}[0.5pt/1pt] \cdashline{5-7}[2pt/2pt] 
0 & 0 & 0 & 1 & - & $\pi$ & -                   \\ \hdashline[2pt/3pt]
0 & 1 & 0 & 0 & $\pi$ & - & -                   \\ \cdashline{1-4}[0.5pt/1pt] \cdashline{5-7}[2pt/2pt]\rowcolor{yellow!25}
0 & 1 & 1 & 0 & $\pi$ & 0 & $\pi$               \\ \cdashline{1-4}[0.5pt/1pt] \cdashline{5-7}[2pt/2pt]
0 & 1 & 1 & 1 & $\pi$ & - & -                   \\ \cdashline{1-4}[0.5pt/1pt] \cdashline{5-7}[2pt/2pt]\rowcolor{yellow!25}
0 & 1 & 0 & 1 & $\pi$ & $\pi$ & 0               \\ \hdashline[2pt/3pt]
1 & 1 & 0 & 0 & - & - & -                       \\ \cdashline{1-4}[0.5pt/1pt] \cdashline{5-7}[2pt/2pt]
1 & 1 & 1 & 0 & - & 0 & -                       \\ \cdashline{1-4}[0.5pt/1pt] \cdashline{5-7}[2pt/2pt]
1 & 1 & 1 & 1 & - & - & -                       \\ \cdashline{1-4}[0.5pt/1pt] \cdashline{5-7}[2pt/2pt]
1 & 1 & 0 & 1 & - & $\pi$ & -                   \\ \hdashline[2pt/3pt]
1 & 0 & 0 & 0 & 0 & - & -                       \\ \cdashline{1-4}[0.5pt/1pt] \cdashline{5-7}[2pt/2pt]\rowcolor{yellow!25}
1 & 0 & 1 & 0 & 0 & 0 & 0                       \\ \cdashline{1-4}[0.5pt/1pt] \cdashline{5-7}[2pt/2pt]
1 & 0 & 1 & 1 & 0 & - & -                       \\ \cdashline{1-4}[0.5pt/1pt] \cdashline{5-7}[2pt/2pt]\rowcolor{yellow!25}
1 & 0 & 0 & 1 & 0 & $\pi$ & $-\pi$              \\ \hline
\end{tabular}\caption{Final configuration of the phase difference between the signal on the two legs ($\Delta\phi$). The values $0$ and $1$ corresponds respectively to the path being open or closed. In both phase switches ($\mathrm{PSA}$ and $\mathrm{PSB}$), the optical lengths of path $1$ and path $2$ differ by a phase angle of $\pi$. Only the four highlighted configurations output a non-null signal.}\label{twophaseshift}
\end{table}
Table \ref{twophaseshift} shows that only four configurations of the phase switches allow the signal to propagate and only three phase configurations are allowed: $\Delta\phi = 0,\;\pm\pi$. Besides, as it will be shown in the next section, the states $\Delta\phi = \pm\pi$ are equivalent. \par
The state of the phase switches can be either fixed in time or continuously modulated. The second option is the one used during the nominal acquisition mode since it allows to reduce drastically the 1/$f$ noise, by exploiting the \textit{double demodulation} process.\par
Two other fundamental RF components of the polarimeter are the \textit{hybrids} (or \textit{couplers}) and the \textit{diodes}. The hybrids combine the signals of the two legs and provide two outputs. If $A$ and $B$ are the two input signals, the output signals are given by: 
\be{hybridoutput}
\begin{aligned}
& A e^{i(\pi-\theta)} + B \,, \\
& A + e^{i(\pi-\theta)}B \,.
\end{aligned}
\ee
where $\theta$ is the phase of hybrid. \par
The four diodes convert the RF signal into electric signal proportionally to the power $P$ of the incident field, i.e., the mean value of the square module of the electric field $E$:
\be{diode}
\Delta V = \alpha P + \beta = \gamma \mean{\abs E ^2} + \delta \,.
\ee \par
\textit{Power splitters} and \textit{bandpass filters} are passive components, which respectively split the signal in two signals of equal amplitude and apply a low-pass filter to obtain the desired bandpass. \par

\subsection{Electronics and software}
The STRIP electronic boards control $330$ LNAs, $110$ phase switches and $220$ detector diodes. The overall design consists of seven identical board units; each one biases seven Q-band and one W-band polarimeters and acquires their output. Each unit transfers the data to the CPU unit via Ethernet LAN and biases the RF components of the polarimetric modules, the \textit{successive approximation register analog-to-digital converters} (SAR ADCs), the \textit{micro-controller} and the \textit{field-programmable gate array} (FPGA) board for data acquisition and handling. The data flow from the ADCs merges into the FPGA that pre-processes the information taking into account the phase switches state. \par
The data are locally stored in a \textit{secure digital} (SD) card and transmitted by the micro-controller to a CPU board, for redundancy. The pre-amplification section is designed to maximize the gain while minimizing the noise and ensuring optimal dynamic range. All the biases are acquired and stored as \textit{house-keeping} (HK).\par
The HK software is used both to supply the desired bias parameters to the diodes, to the LNAs and the phase switches and to read the outputs of the detectors, writing them in a text file. \par

\section{Mathematical model}\label{mamo}
We can understand how the output signals are proportional to the Stokes parameters by going deeper into the mathematical details of the polarimeter model. \par
\begin{figure}[H]
\centering
\includegraphics[scale=0.6]{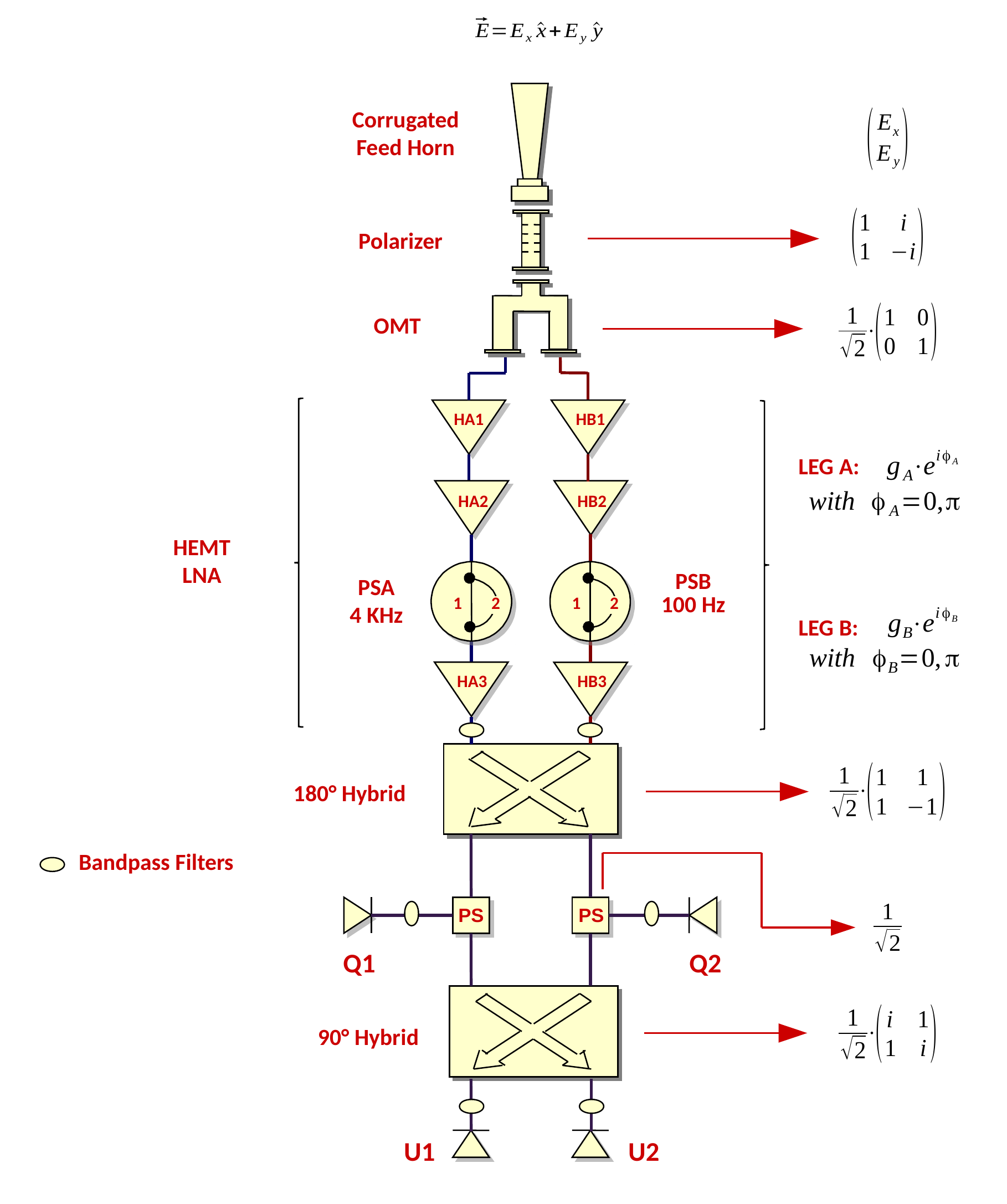}
\caption{Scheme of the STRIP radiometric chain and the mathematical operator associated to each RF component.}\label{radchain2}
\end{figure}

As schematized in Fig. \ref{radchain2}, each radiometric component can be associated to a mathematical operator ($M$) applied to the incident electromagnetic field ($x_\mathrm{input}$), so that the output signal from the given RF component ($y_\mathrm{output}$) will be: $y_\mathrm{output} = M x_\mathrm{input}$.\par
In particular, in the ideal case:

\begin{list}{\leftmargin 15pt \itemsep 0pt \topsep 3pt}
\item{\bf Feedhorn.}
  It can be considered just as the identity matrix:
  \be{fh}
  I_F = \begin{pmatrix} 1 & 0 \\ 0 & 1 \end{pmatrix} \,.
  \ee
  
\item{\bf Polarizer and OMT.}
  They split the signal into two circularly polarized components:
  \be{pOMT}
  O = \frac{1}{\sqrt2}\begin{pmatrix} 1 & i \\ 1 & -i \end{pmatrix} \,.
  \ee
  
\item{\bf HEMT and Phase Switches.}
  Each leg is made by a sequence of two HEMTs amplifiers, a phase switch and a third HEMT element. The overall effect onto the signal is given by:
  \be{legs}
  \begin{aligned}
    \mathrm{LEG}_A &= g_A \cdot e^{i\phi_A} \,, \\
    \mathrm{LEG}_B &= g_B \cdot e^{i\phi_B} \,,
  \end{aligned}
  \ee
  with $\phi_{A/B} = 0,\pi$, as shown in Table \ref{onephaseshift}, and with $g_{A/B}$ that are the cumulative gains of the two legs, given by the product of the three amplifier gains in each leg. So that:
  \be{legsnoise}
  L_N = \begin{pmatrix} \mathrm{LEG}_A & 0 \\ 0 & \mathrm{LEG}_B \end{pmatrix} \,.
  \ee 
  At this level, it is crucial to take into account the noise $\mathbf{N} = (N_A,\; N_B)$ that the amplifiers add to the signal $\mathbf{S}$. This happens because the thermal emission of every RF component located before the amplifiers is amplified itself. This introduces a spurious and uncorrelated signal that results in a white noise component. 
  
\item{\bf $\mathbf{180\deg}$ hybrid.}
  It combines the two signals according to Eq. \ref{hybridoutput}, with $\theta = 180\deg$:
  \be{H180}
  H_{180} = \frac{1}{\sqrt2}\begin{pmatrix} 1 & 1 \\ 1 & -1 \end{pmatrix} \,.
  \ee
  
\item{\bf Power splitter.}
  It is a signal splitter:
  \be{ps}
  P_S = \frac{1}{\sqrt2}\begin{pmatrix} 1 & 0 \\ 0 & 1 \end{pmatrix} \,.
  \ee
  
\item{\bf $\mathbf{90\deg}$ hybrid.}
  It combines the two signals according to Eq. \ref{hybridoutput}, with $\theta = 90\deg$:
  \be{H90}
  H_{90} = \frac{1}{\sqrt2}\begin{pmatrix} i & 1 \\ 1 & i \end{pmatrix} \,.
  \ee
  
\item{\bf Diodes.}
  They convert the RF signal into electric signal, according to Eq. \ref{diode}.
\end{list}

Given an incident electric field $\mathbf{E} = (E_x,\;E_y)$, it is possible to compute the signal that hits the four diodes ($\mathrm{Q1},\,\mathrm{Q2},\,\mathrm{U1},\,\mathrm{U2}$) by applying the previous list of operators and adding the noise introduced by the amplifiers:
\begin{align}\label{signalQ}
  \mathbf{Q} &= P_S \cdot H_{180} \cdot L_N \cdot \bigl[(O \cdot I_F \cdot \mathbf{E}) + \mathbf{N}\bigr] \,, \\
  \mathbf{U} &= H_{90} \cdot P_S \cdot H_{180} \cdot L_N \cdot \bigl[(O \cdot I_F \cdot \mathbf{E}) + \mathbf{N}\bigr] \label{signalU} \,,
\end{align}
where $\mathbf{Q} = (Q1,\; Q2)$ and $\mathbf{U} = (U1,\; U2)$.\par
By doing the math (Appendix \ref{App:1}) and using the definition of the Stokes parameters (Eq. \ref{stokes}), the amplitude of the electric fields at the four diodes are given by:
\begin{align}\label{efields}
  E_{Q1} &= \frac{g_A e^{i\phi_A} + g_B e^{i\phi_B}}{2\sqrt2} E_x + \frac{g_A e^{i\phi_A} - g_B e^{i\phi_B}}{2\sqrt2} i E_y + \frac{g_A e^{i\phi_A}}{2} N_A + \frac{g_B e^{i\phi_B}}{2} N_B  \,, \\
  E_{Q2} &= \frac{g_A e^{i\phi_A} - g_B e^{i\phi_B}}{2\sqrt2} E_x + \frac{g_A e^{i\phi_A} + g_B e^{i\phi_B}}{2\sqrt2} i E_y + \frac{g_A e^{i\phi_A}}{2} N_A - \frac{g_B e^{i\phi_B}}{2} N_B  \,, \\
  E_{U1} &= \frac{g_A e^{i\phi_A} (i + 1) + g_B e^{i\phi_B} (i - 1)}{4} E_x + \frac{g_A e^{i\phi_A}(i - 1) + g_B e^{i\phi_B} (i + 1)}{4} E_y + \\
  &\phantom{{}=1} + \frac{(i + 1) g_A e^{i\phi_A}}{2\sqrt2} N_A + \frac{(i - 1) g_B e^{i\phi_B}}{2\sqrt2} N_B  \notag \,, \\ 
  E_{U2} &= \frac{g_A e^{i\phi_A} (i + 1) - g_B e^{i\phi_B} (i - 1)}{4} E_x + \frac{g_A e^{i\phi_A}(i - 1) - g_B e^{i\phi_B} (i + 1)}{4} E_y + \label{efieldslast} \\
  &\phantom{{}=1} + \frac{(i + 1) g_A e^{i\phi_A}}{2\sqrt2} N_A - \frac{(i - 1) g_B e^{i\phi_B}}{2\sqrt2} N_B  \notag \,. 
\end{align} 
So that, the power measured by the four ADCs is:
\begin{align}\label{epower}
  \mean{\abs{E_{Q1}} ^2} &= \frac14 \bigl[g^2 (I + N) + g_A g_B \cos(\Delta\phi) Q - g_A g_B \sin(\Delta\phi) U - \frac{g_A^2 - g_B^2}{2} V \bigr] \,, \\
  \mean{\abs{E_{Q2}} ^2} &= \frac14 \bigl[g^2 (I + N) - g_A g_B \cos(\Delta\phi) Q + g_A g_B \sin(\Delta\phi) U - \frac{g_A^2 - g_B^2}{2} V \bigr] \,, \\
  \mean{\abs{E_{U1}} ^2} &= \frac14 \bigl[g^2 (I + N) + g_A g_B \sin(\Delta\phi) Q + g_A g_B \cos(\Delta\phi) U - \frac{g_A^2 - g_B^2}{2} V \bigr] \,, \\
  \mean{\abs{E_{U2}} ^2} &= \frac14 \bigl[g^2 (I + N) - g_A g_B \sin(\Delta\phi) Q - g_A g_B \cos(\Delta\phi) U - \frac{g_A^2 - g_B^2}{2} V \bigr] \,, \label{epowerlast0}
\end{align}
with $\Delta\phi = 0,\pm\pi$, as shown in Table \ref{twophaseshift}, and where I have defined:
\be{gsquare}
g^2 = \frac{g_A^2 + g_B^2}{2} \,,
\ee
and:
\be{Ndef}
N = \frac{g_A^2\mean{\abs{N_A} ^2} + g_B^2\mean{\abs{N_B} ^2}}{g^2} \,.
\ee \par
With this notation, the polarimeter has only one noise term $\mathrm{N}$, which is common to the four outputs, and only one bandwidth and one central frequency, which are defined respectively by:
\begin{align}\label{bdwdef}
  \beta &= \frac{\bigl(\int g_A(\nu) g_B(\nu) d\nu \bigr)^2}{\int g^4(\nu)d\nu} \,, \\
  \nu_c &= \frac{\int g_A(\nu) g_B(\nu) \,\nu\, d\nu}{\int g^2(\nu) d\nu} \,. \label{nucdef}
\end{align}\par

By substituting the values of $\Delta\phi$ reported in Table \ref{twophaseshift} in Eqs. from \ref{epower} to \ref{epowerlast0}, it is possible to simplify the notation:
\begin{align}\label{epowersimply1}
  \mean{\abs{E_{Q1}} ^2} &= \frac14 \bigl[g^2 (I + N) \pm g_A g_B Q - \frac{g_A^2 - g_B^2}{2} V \bigr] \,, \\
  \mean{\abs{E_{Q2}} ^2} &= \frac14 \bigl[g^2 (I + N) \mp g_A g_B Q - \frac{g_A^2 - g_B^2}{2} V \bigr] \,, \\
  \mean{\abs{E_{U1}} ^2} &= \frac14 \bigl[g^2 (I + N) \pm g_A g_B U - \frac{g_A^2 - g_B^2}{2} V \bigr] \,, \\
  \mean{\abs{E_{U2}} ^2} &= \frac14 \bigl[g^2 (I + N) \mp g_A g_B U - \frac{g_A^2 - g_B^2}{2} V \bigr] \,, 
\end{align}
where the upper signs refer to the case $\Delta\phi = 0$, while the lower ones refer to $\Delta\phi = \pm \pi$.\par

If the two legs of the polarimeter have the same gain, $g_A = g_B = g$, the previous equations further simplify:
\begin{align}\label{epowersimply2}
  \mean{\abs{E_{Q1}} ^2} &= \frac{g^2}{4} \bigl[I + N \pm Q \bigr] \,, \\
  \mean{\abs{E_{Q2}} ^2} &= \frac{g^2}{4} \bigl[I + N \mp Q \bigr] \,, \\
  \mean{\abs{E_{U1}} ^2} &= \frac{g^2}{4} \bigl[I + N \pm U \bigr] \,, \\
  \mean{\abs{E_{U2}} ^2} &= \frac{g^2}{4} \bigl[I + N \mp U \bigr] \,. \label{epowersimply2latest}
\end{align} \par

%The complete set of calculations leading to Eqs. from \ref{efields} to \ref{epowerlast0} is reported in Appendix \ref{App:1}.

\section{Double demodulation}
In the nominal acquisition mode, the phase switches on the two legs switch continuously between the phase states $0$ and $\pi$. This modulation happens with different frequencies: a faster modulation ($2 \,\mathrm{KHz}$) on one leg is used in data analysis to reduce 1/$f$ noise, while a slower modulation ($50 \,\mathrm{Hz}$) on the other leg helps to reduce the leakage of $I$ into $Q$, due to gain imbalance. \par
During the unit-level tests at ``Università degli Studi di Milano Bicocca'', the two frequencies have been reduced respectively to $1 \,\mathrm{KHz}$ and $25 \,\mathrm{Hz}$. In any case, the sampling frequency of the data corresponds always to the slower modulation frequency. \par
\begin{figure}[H]
\centering
\includegraphics[scale=0.3]{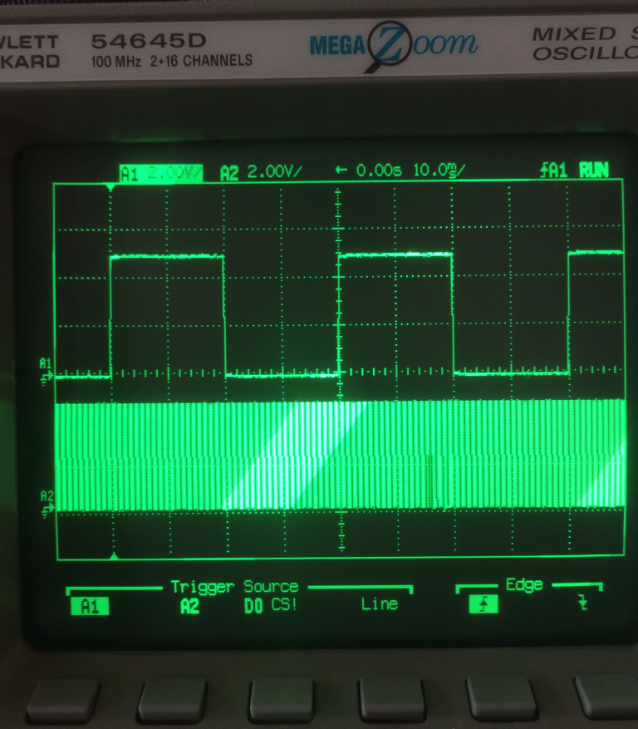}
\caption{Voltages applied to the phase switches as a function of the time. On the x-axis, each side of the square corresponds to $10\,\mathrm{ms}$. One phase switch is modulated at $25 \,\mathrm{Hz}$ (upper curve) while the other at $1 \,\mathrm{KHz}$ (lower curve).}\label{switching}
\end{figure}

Fig. \ref{switching} shows a picture of the voltages applied to the two phase switches as a function of the time.\par
\par
\begin{figure}[H]
\centering
\includegraphics[scale=0.5]{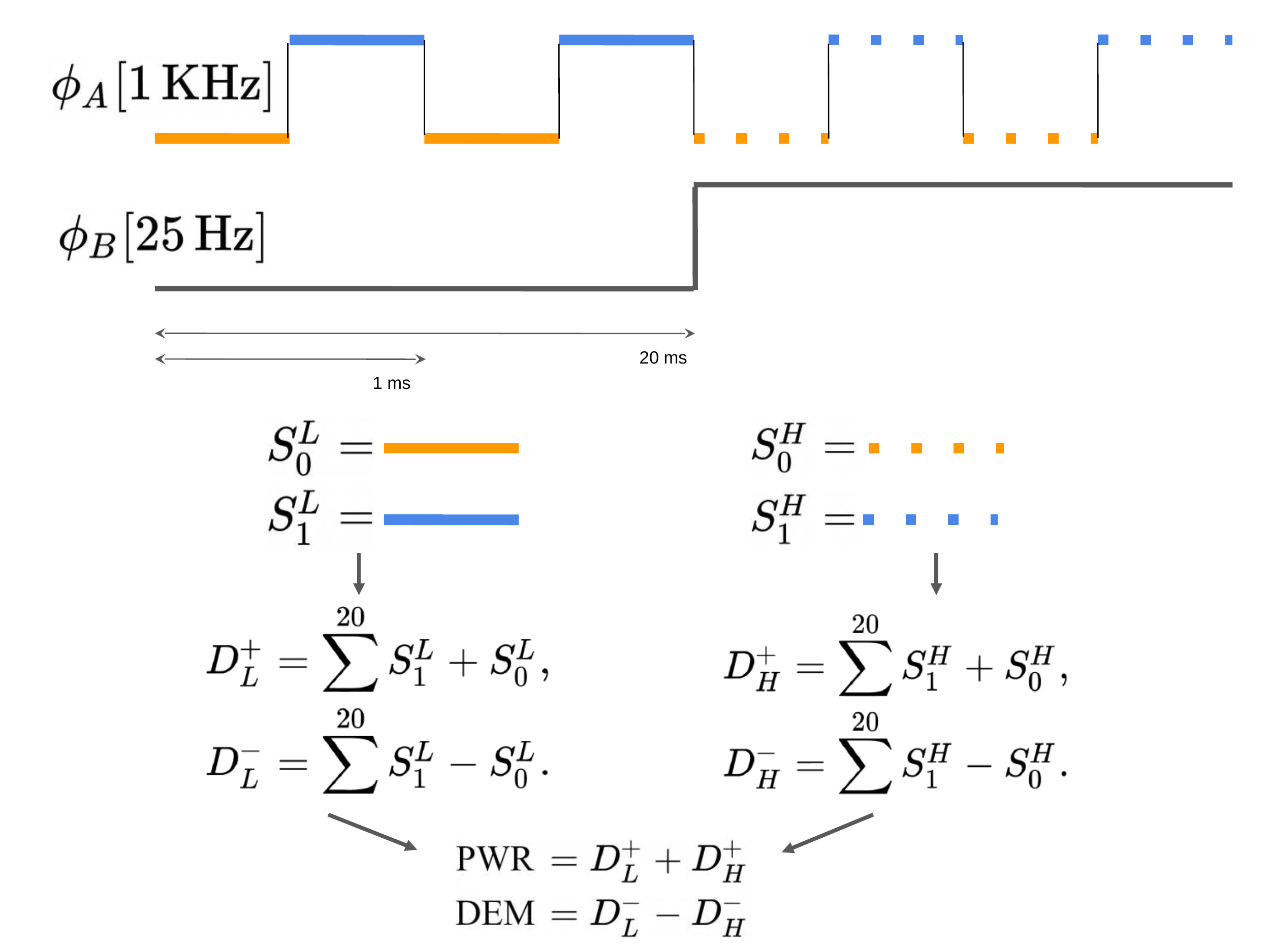}
\caption{Scheme of the double demodulation principle.}\label{doubdem}
\end{figure}\par
Fig. \ref{doubdem} shows the operating scheme of the \textit{double demodulation} principle. \par
As shown in the previous section, the output signal $S$ at the four diodes depends on the state of the phase switches. Let us assign the subscript $0$/$1$ to the two possible states corresponding to the faster modulation and the superscript $L$/$H$ to the two corresponding to the slower modulation. Therefore, the four states of the output signal are named: $S_0^L$, $S_1^L$, $S_0^H$, $S_1^H$.\par
Let us consider now the periods of $40\,\mathrm{ms}$ and $1\,\mathrm{ms}$ for the slow and the fast modulations, respectively.\par
A half-period of the slower modulation contains $20$ periods of the faster modulation. Two consecutive samples correspond to the half-periods of the faster modulation. The signal is processed by the electronics in two ways, at the same time: on one hand, two consecutive samples are summed; on the other, they are subtracted (first \textit{demodulation}). This operations are performed over the $20$ repetitions contained in the first half-period of the slower modulation and the results are summed\footnote{The sum over the $20$ samples should be intended as an estimator of the average for the quantity inside the summation. The dividing factor of $20$ in fact, can be neglected since, at the end of the process, the output signals must be converted from \textit{analog-to-digital units} (ADU) to physical units (V).}. So that, at this level, the signals are given by:
\begin{align}
D_L^+ = \sum_{i=1}^{20} S_{1,i}^L + S_{0,i}^L \,, \label{dl+}\\
D_L^- = \sum_{i=1}^{20} S_{1,i}^L - S_{0,i}^L \,. \label{dl-}
\end{align}\par

The same process is repeated also during the second half-period of the slower modulation:
\begin{align}
D_H^+ = \sum_{i=1}^{20} S_{1,i}^H + S_{0,i}^H \,, \label{dl+}\\
D_H^- = \sum_{i=1}^{20} S_{1,i}^H - S_{0,i}^H \,. \label{dl-}
\end{align} \par

\begin{figure}[H]
\centering
\includegraphics[scale=0.6]{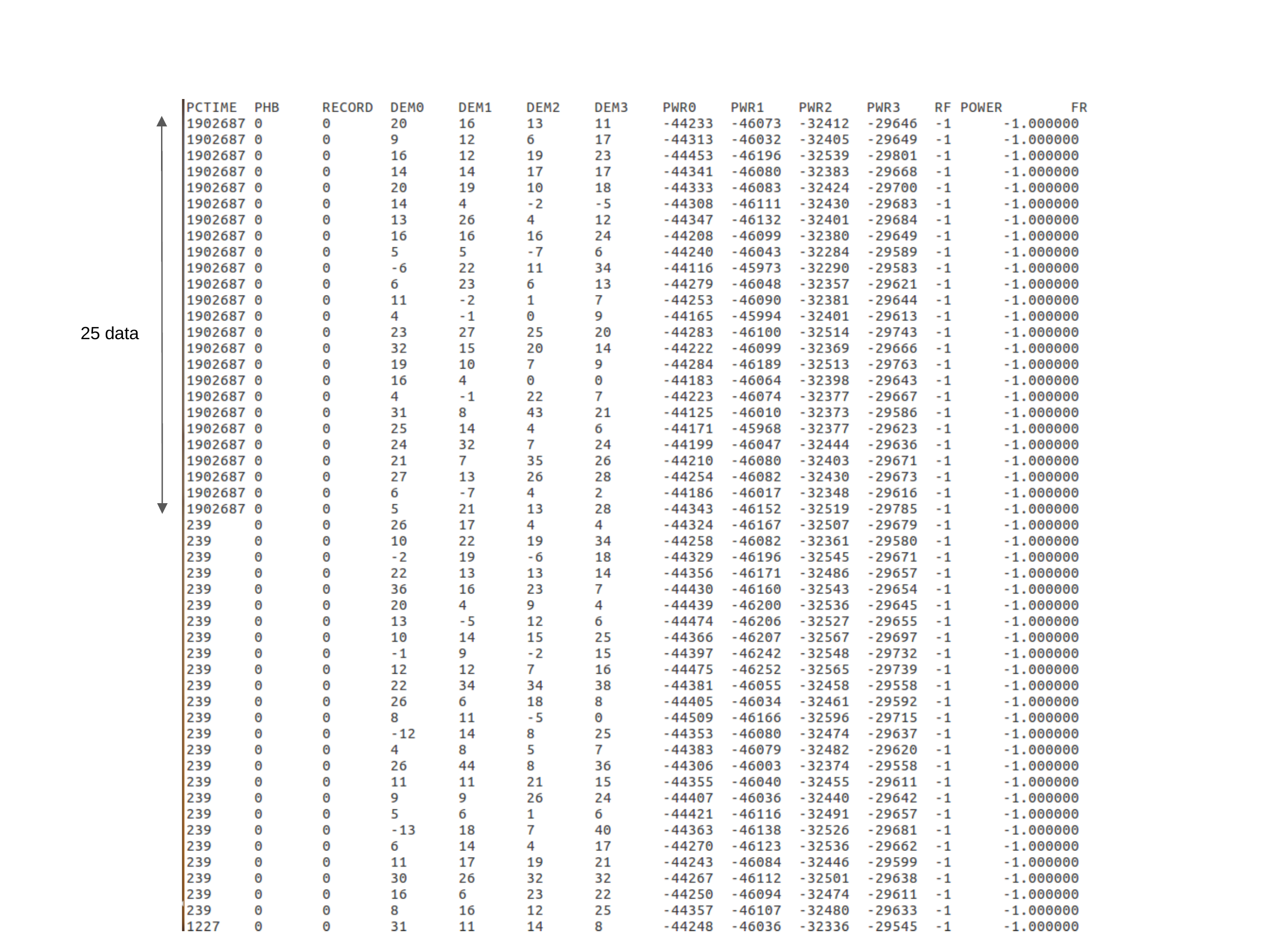}
\caption{Data output format. The first entry is the time: the time label changes every $25$ samples according to the sampling frequency of $25 \,\mathrm{Hz}$. The entries ``DEM'' and ``PWR'' represents respectively the double demodulated and the total power signals. The indices $0,\,1,\,2\,,3$ correspond respectively to the detectors $\mathrm{Q1},\,\mathrm{U1},\,\mathrm{U2},\,\mathrm{Q2}$.}\label{dataformat}
\end{figure}

The final output signals are then the sum $D_L^+ + D_H^+$ and the difference $D_L^- - D_H^-$ (second \textit{demodulation}) corresponding respectively to the \textit{total power} and to the \textit{double demodulated} signals. In this way, at each time there are eight outputs, two for each detector (Fig. \ref{dataformat}). 
\par

The double demodulation process allow us to measure $I$, $Q$ and $U$ in the most direct way. To show that, let us consider the case of detector $\mathrm{Q1}$ in gain balancing condition (Eq. \ref{epowersimply2}) and let us compute by hand the value of the signal as processed by the electronics. \par
In the total power case, the signal results proportional to the total intensity of the incident radiation field:
\begin{align}\label{q1pwr}
Q1^{\mathrm{PWR}} &= \frac{g^2}{4} \bigl[ (I + N \pm Q)_1^L + (I + N \mp Q)_0^L + (I + N \pm Q)_1^H + (I + N \mp Q)_0^H \bigr] =  \notag \\
 &\phantom{}= g^2 (I + N / 2)  \,.
\end{align}\par
In the case of the double demodulation, the signal is proportional to $Q$ only:
\begin{align}\label{q1dem}
Q1^{\mathrm{DEM}} &= \frac{g^2}{4} \bigl\{ (I + N \pm Q)_1^L - (I + N \mp Q)_0^L - \bigl[(I + N \mp Q)_1^H - (I + N \pm Q)_0^H \bigr] \bigr\} = \notag \\
 &\phantom{}= \pm g^2 (Q + N / 2) \,.
\end{align}
The noise is treated as an uncertainty and is added in quadrature. \par
In the same way, it is possible to compute the output signals for the other detectors. They are reported in Table \ref{alloutput}. In the case of gain unbalance between the two legs, the result does not change significantly\footnote{Less than a factor proportional to $V$ that must be added to the total power output.}. 
\begin{table}[H]
\centering
\begin{tabular}{|c|c|c|} \hline
  & $\mathrm{PWR}$ & $\mathrm{DEM}$ \\ \hline
 $\mathrm{Q1}$ &  $g^2 (I + N / 2)$ &  $\pm g^2 (Q + N / 2)$ \\ \hline
 $\mathrm{Q2}$ &  $g^2 (I + N / 2)$ &  $\mp g^2 (Q + N / 2)$ \\ \hline
 $\mathrm{U1}$ &  $g^2 (I + N / 2)$ &  $\pm g^2 (U + N / 2)$ \\ \hline
 $\mathrm{U2}$ &  $g^2 (I + N / 2)$ &  $\mp g^2 (U + N / 2)$ \\ \hline
\end{tabular}\caption{List of the possible outputs of the four detectors.}\label{alloutput}
\end{table}\par
The four demodulated outputs provide an estimation of the Stokes parameters of the incident radiation field. Once the diodes are calibrated and the gain $g$ is known, the output signals can be further combined in order to reduce the uncertainty on the measure of $I$, $Q$ and $U$:
\begin{align}
I &= \frac{Q1^\mathrm{PWR} + Q2^\mathrm{PWR} + U1^\mathrm{PWR} + U2^\mathrm{PWR}}{4} \,, \label{I}\\ %+ \frac{N}{4} 
Q &= \frac{\bigl| Q1^\mathrm{DEM} - Q2^\mathrm{DEM} \bigr| }{2}  \,, \label{Q} \\ %+ \frac{\sqrt{2}N}{4}
U &= \frac{\bigl| U1^\mathrm{DEM} - U2^\mathrm{DEM} \bigr| }{2}  \,. \label{U}  %+ \frac{\sqrt{2}N}{4}
\end{align} \par
Typically, the output power by the LNAs is dominated by the 1/$f$ noise for frequencies less than a few hundred $\mathrm{Hz}$ for Q-band amplifiers (and up to $1 \; \mathrm{KHz}$ for W-band amplifiers). By demodulating at $2 \; \mathrm{KHz}$, drifts in the total power level at longer timescales are differenced away. Residual 1/$f$ noise comes from gain fluctuations multiplying small offsets in the demodulated signal due to I $\to$ Q leakage, which is strongly reduced by the second demodulation.

%% file: chap4.tex
\chapter{Unit-level tests}
\label{Chap:4}
\thispagestyle{plain}

In this chapter, I report the main results of the unit-level tests campaign that took place at ``Università degli Studi di Milano Bicocca'' from September 2017 to July 2018.\par
During the campaign, $68$ polarimeters have been tested in order to select the $55$ ($49$ Q-band and $6$ W-band) with the best performance; these polarimeters have been integrated in the focal plane in November 2018. More than $800$ tests\footnote{All the test reports are stored in a web database whose address is \url{https://striptest.fisica.unimi.it/unittests/}.} have been performed by a team made by the STRIP instrument scientist and four students (including me). \par
We tested each polarimeter twice. At first, we performed functionality tests at room temperature. Then, we placed the polarimeter in a cryogenic chamber and cooled it down to $20\,\mathrm{K}$. Here, we tested again its functionality and finally we measured its performance. \par
Using a curve tracer, we verified the functionality of each electronic component in each polarimeter subsystem. As for the performance measurements, we ran three tests: a \textit{bandpass characterization}, a \textit{Y-factor test} to estimate the noise temperature and a \textit{long acquisition} to measure the noise characteristics. \par   
I do not report the results of the functionality tests here, but I focus on the performance tests. In particular, I was mostly involved on the data analysis of the noise properties and I will describe it in greater detail.\par
Further tests will be performed in Bologna during the system-level tests campaign that will be carried out in $2020$, when the instrument will be integrated for the first time. Consistency checks with the results from unit-level tests will be executed, the instrument biases will be optimized and the instrument will be tested again, as a whole. \par

\section{Experimental set-up}
We numbered the STRIP polarimeters progressively from $1$ to $84$: units $1 \div 70$ are Q-band and $71 \div 84$ are W-band. We tested all the polarimeters in a cryogenic chamber exploiting the experimental set-up described in the following section. \par
The detector bias supply and the data acquisition were managed by a \textit{house-keeping} (HK) software developed for the unit-level tests campaign. We used this software to perform run-time checks but it did not record most of the HK that could have been useful during the data analysis, such as biases and temperatures.\par

\subsection{Set-up scheme and math}\label{setup}
The experimental set-up was based on the idea to inject controllable signals into the two entrances of the polarimeter. These signals originated from the thermal radiation of two loads and from a radio-frequency (RF) generator. The thermal loads temperatures could be set and managed by the test operator. \par
\begin{figure}[H]
\centering
\includegraphics[scale=0.7]{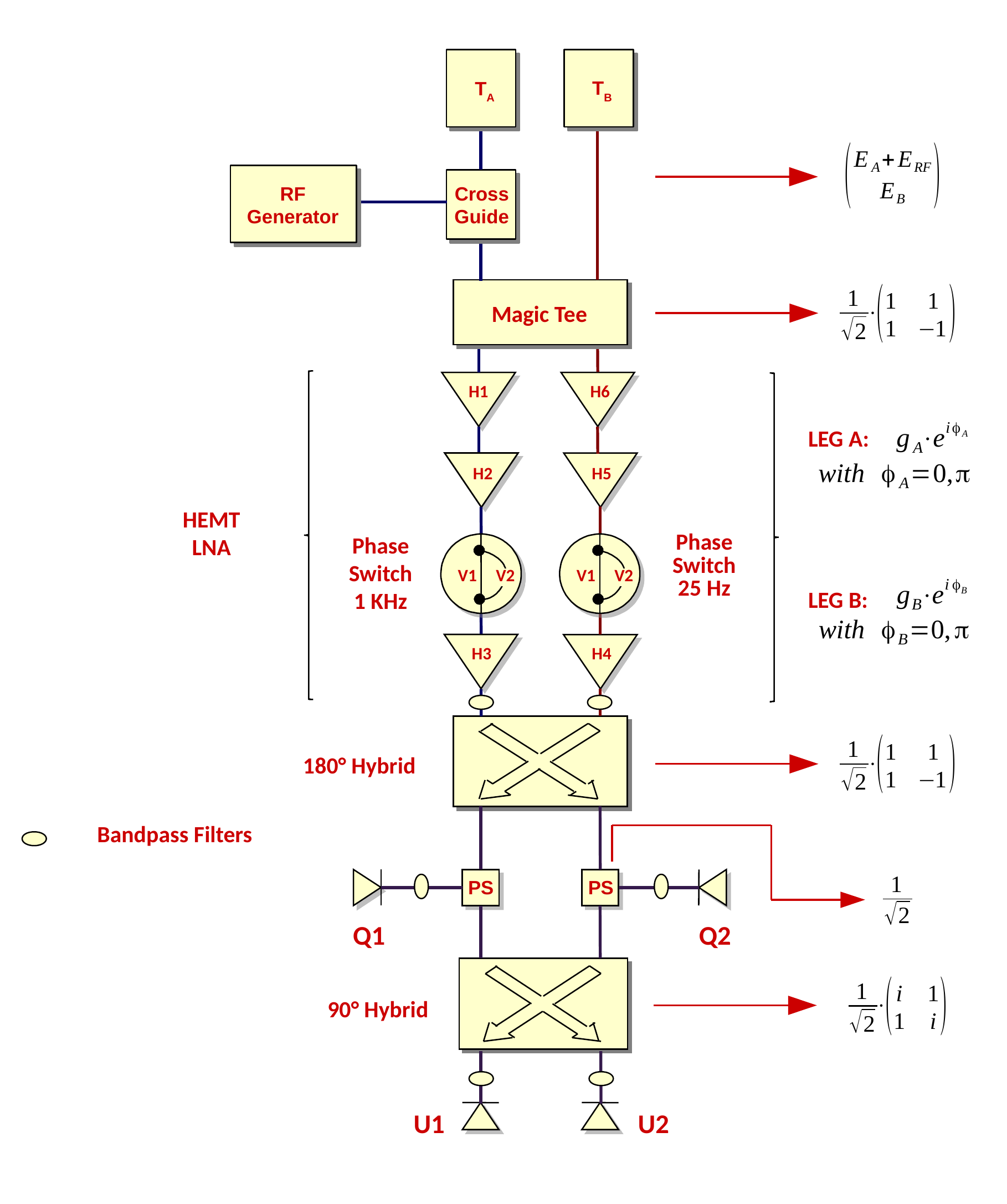}
\caption{Scheme of the STRIP radiometric chain and the mathematical operator associated to each RF component in the unit-level tests configuration.}\label{expsteup}
\end{figure}
\begin{figure}[H]
\centering
\includegraphics[scale=0.135]{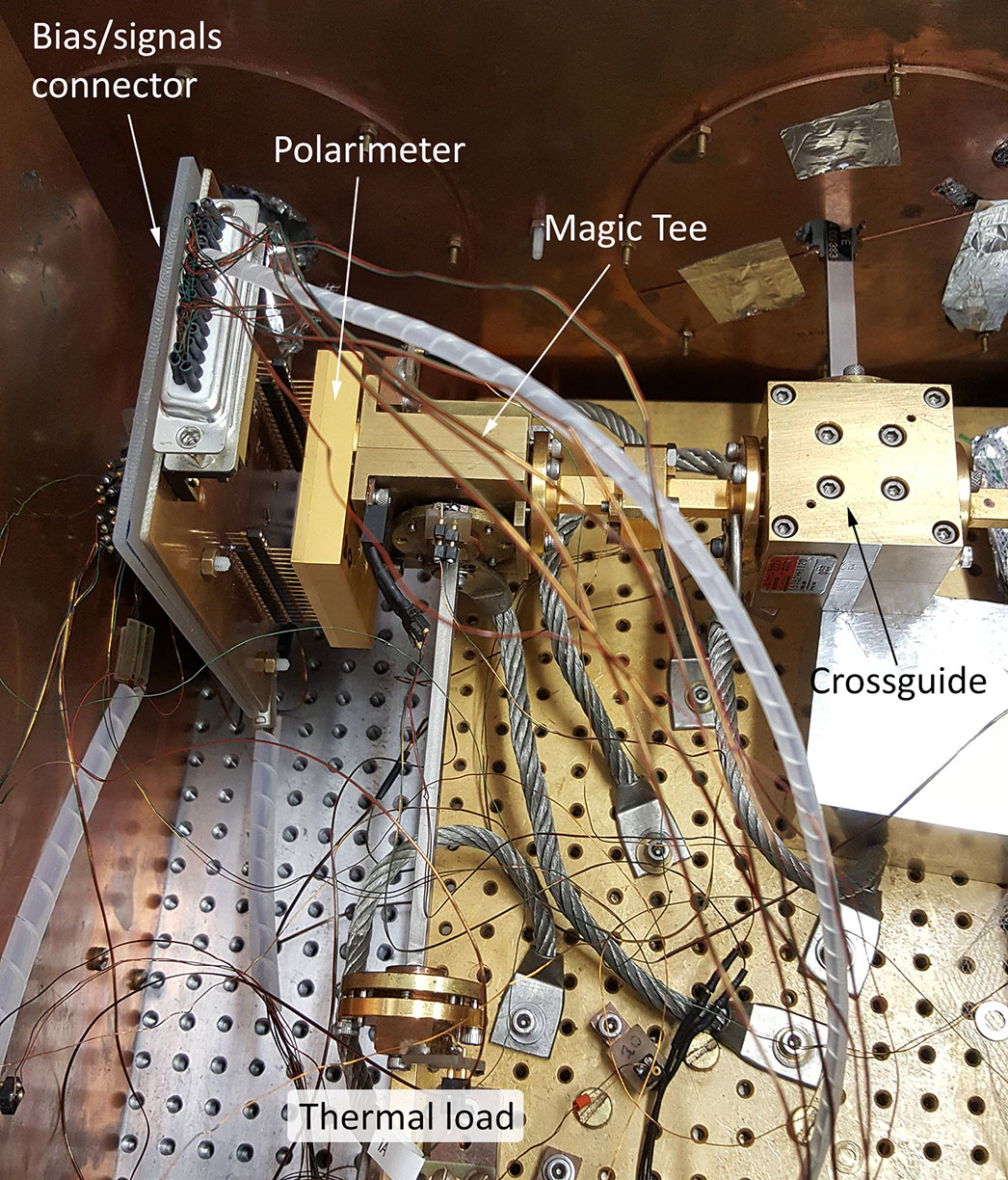}
\includegraphics[scale=0.166]{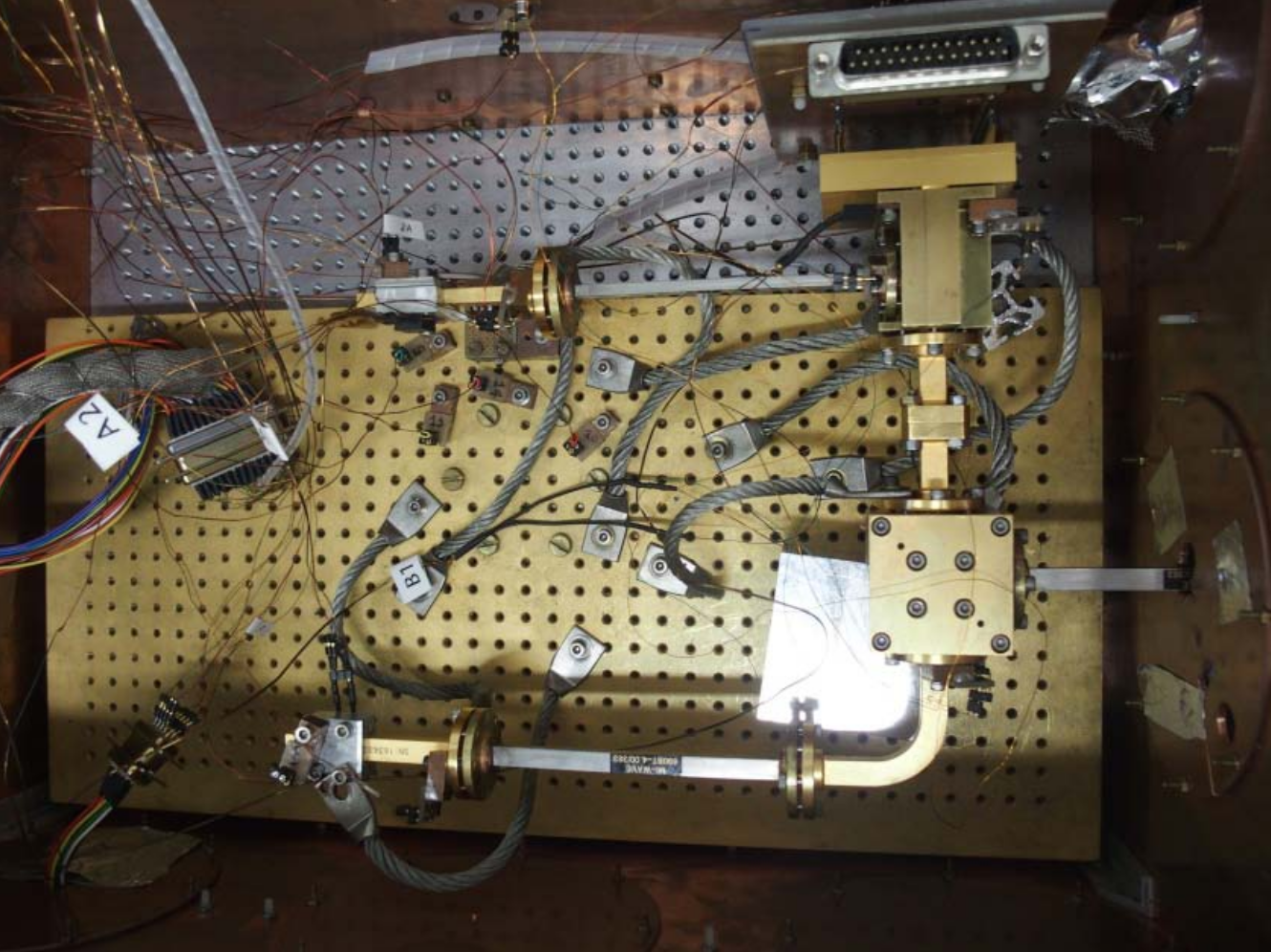}
\caption{Pictures of the experimental set-up. \textit{Right}: View from the top.}\label{picturesetup}
\end{figure}

The radiometric chain used during the test was different with respect to the one illustrated in Sect. \ref{radiometrichain}. In fact, the sequence of horn, polarizer and OMT was replaced by a \textit{magic tee} and two waveguides connected to the thermal loads. The RF signal produced by the generator could be added to the thermal signal of one of the loads exploiting a \textit{cross-guide}. I show, in Fig. \ref{expsteup}, a scheme of the experimental set-up while, in Fig. \ref{picturesetup}, I show some pictures of the real set-up.\par
\par
\par

The signals produced by the two loads ($E_A$ and $E_B$) are coupled by mean of the magic tee that acts as a $180\deg$ hybrid (Eq. \ref{H180}). An additional RF signal $E_\mathrm{RF}$ can be added to $E_A$ through the cross-guide:
\be{ea'}
E_A' = E_A + E_\mathrm{RF} \,.
\ee \
Note that $E_A$, $E_B$ and $E_\mathrm{RF}$ signals are uncorrelated among them since they originate from different sources.\par
Let us call $T$ the mathematical operator associated to the magic tee:
\be{magictee}
T = \frac{1}{\sqrt2}\begin{pmatrix} 1 & 1 \\ 1 & -1 \end{pmatrix} \,.
\ee \par
According to the mathematical model reported in Sect. \ref{mamo}, the expected signal at the four detector diodes is:
\begin{align}\label{signalQbicocca}
  \mathbf{Q} &= P_S \cdot H_{180} \cdot L_N \cdot \bigl[T \cdot \mathbf{E} + \mathbf{N}\bigr] \,, \\
  \mathbf{U} &= H_{90} \cdot P_S \cdot H_{180} \cdot L_N \cdot \bigl[T \cdot \mathbf{E} + \mathbf{N}\bigr] \label{signalUbicocca} \,,
\end{align}
where $\mathbf{E} = (E_A',\,E_B)$ and $\mathbf{Q} = (Q1,\; Q2)$; $\mathbf{U} = (U1,\; U2)$.\par
By working out the math (Appendix \ref{App:2}) and using the definition of the Stokes parameters (Eq. \ref{stokes}), it is possible to show that the power of the electric field at the four diodes is:
\begin{align}\label{epowerbicocca}
  \mean{\abs{E_{Q1}} ^2} = \frac14 \bigl[&g^2 \bigl(\mean{E_A^2} + \mean{E_\mathrm{RF}^2} + \mean{E_B^2} + N \bigr) + \notag \\
    \phantom{} &g_A g_B \cos(\Delta\phi) \bigl(\mean{E_A^2} + \mean{E_\mathrm{RF}^2} - \mean{E_B^2} \bigr) \bigr] \,, \\
  \mean{\abs{E_{Q2}} ^2} = \frac14 \bigl[&g^2 \bigl(\mean{E_A^2} + \mean{E_\mathrm{RF}^2} + \mean{E_B^2} + N \bigr) - \notag \\
    \phantom{} &g_A g_B \cos(\Delta\phi) \bigl(\mean{E_A^2} + \mean{E_\mathrm{RF}^2} - \mean{E_B^2} \bigr) \bigr] \,, \\
  \mean{\abs{E_{U1}} ^2} = \frac14 \bigl[&g^2 \bigl(\mean{E_A^2} + \mean{E_\mathrm{RF}^2} + \mean{E_B^2} + N \bigr) + \notag \\
    \phantom{} &g_A g_B \sin(\Delta\phi) \bigl(\mean{E_A^2} + \mean{E_\mathrm{RF}^2} - \mean{E_B^2} \bigr) \bigr] \,, \\
  \mean{\abs{E_{U2}} ^2} = \frac14 \bigl[&g^2 \bigl(\mean{E_A^2} + \mean{E_\mathrm{RF}^2} + \mean{E_B^2} + N \bigr) - \notag \\
    \phantom{} &g_A g_B \sin(\Delta\phi) \bigl(\mean{E_A^2} + \mean{E_\mathrm{RF}^2} - \mean{E_B^2} \bigr)  \bigr] \,, \label{epowerlastbicocca}
\end{align}
were $\Delta\phi = 0,\pm\pi$ (Table \ref{twophaseshift}) and:
\be{gsquare}
g^2 = \frac{g_A^2 + g_B^2}{2} \,,
\ee
\be{Ndef}
N = \frac{g_A^2\mean{\abs{N_A} ^2} + g_B^2\mean{\abs{N_B} ^2}}{g^2} \,.
\ee \par
The terms proportional to $\sin(\Delta\phi)$ disappear since $\Delta\phi = 0,\pm\pi$. Furthermore, during the tests we have always tried to achieve the gain balance condition, $g_A = g_B = g$. Under these assumptions (and with $\Delta\phi = 0$), the measured signals are: 
\begin{align}\label{epowerbicocca}
  \mean{\abs{E_{Q1}} ^2} &= \frac{g^2}{4} \bigl(2\mean{E_A^2} + 2\mean{E_\mathrm{RF}^2} + N \bigr) \,, \\
  \mean{\abs{E_{Q2}} ^2} &= \frac{g^2}{4} \bigl(2\mean{E_B^2} + N \bigr) \,, \label{blind} \\
  \mean{\abs{E_{U1}} ^2} &= \frac{g^2}{4} \bigl(\mean{E_A^2} + \mean{E_\mathrm{RF}^2} + \mean{E_B^2} + N \bigr) \,, \\
  \mean{\abs{E_{U2}} ^2} &= \frac{g^2}{4} \bigl(\mean{E_A^2} + \mean{E_\mathrm{RF}^2} + \mean{E_B^2} + N \bigr) \,. \label{epowerlast}
\end{align}
When $\abs{\Delta\phi} = \pi$, signals $Q1$ and $Q2$ are swapped. \par
Table \ref{twophaseshift} shows that the phase switches configurations $0101$ and $1010$ are equivalent since they both lead to $\Delta\phi = 0$ between the signals in the two legs. In the same way, $0110$ and $1001$ are equivalent because $\abs{\Delta\phi} = \pi$.\par %A few tests to check empirically these equivalences have been performed, with good results. \par
I report, in Table \ref{alloutputbicocca}, the total power and the demodulated outputs of the four detectors when the double demodulation is active.
\begin{table}[H]
  \centering
\begin{tabular}{|c|c|c|} \hline
  & $\mathrm{PWR}$ & $\mathrm{DEM}$ \\ \hline
 $\mathrm{Q1}$ &  $g^2 \bigl(\mean{E_A^2} + \mean{E_\mathrm{RF}^2} + \mean{E_B^2} + N/2 \bigr)$ &  $\pm g^2 \bigl(\mean{E_A^2} + \mean{E_\mathrm{RF}^2} - \mean{E_B^2} + N/2 \bigr)$ \\ \hline
 $\mathrm{Q2}$ &  $g^2 \bigl(\mean{E_A^2} + \mean{E_\mathrm{RF}^2} + \mean{E_B^2} + N/2 \bigr)$ &  $\mp g^2 \bigl(\mean{E_A^2} + \mean{E_\mathrm{RF}^2} - \mean{E_B^2} + N/2 \bigr)$ \\ \hline
 $\mathrm{U1}$ &  $g^2 \bigl(\mean{E_A^2} + \mean{E_\mathrm{RF}^2} + \mean{E_B^2} + N/2 \bigr)$ &  $g^2N/2$ \\ \hline
 $\mathrm{U2}$ &  $g^2 \bigl(\mean{E_A^2} + \mean{E_\mathrm{RF}^2} + \mean{E_B^2} + N/2 \bigr)$ &  $g^2N/2$ \\ \hline
\end{tabular}\caption{List of the possible outputs of the four detectors in the unit-level tests configuration.}\label{alloutputbicocca}
\end{table}\par

\subsection{Hardware and software}
\begin{figure}[h]
\centering
\includegraphics[scale=0.2]{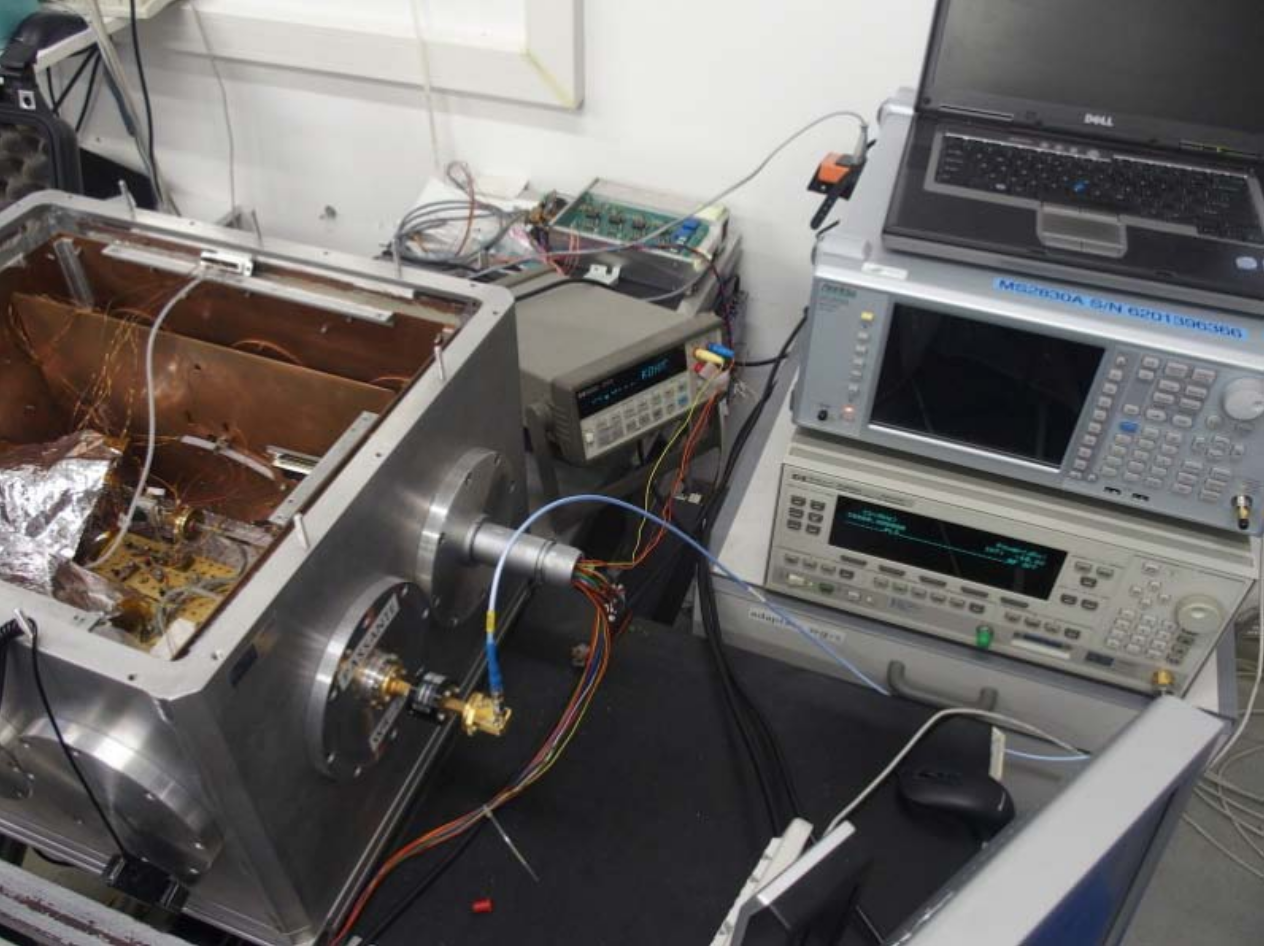}
\caption{Picture of the cryogenic chamber. The RF generator is visible on the right.}\label{cryocha}
\end{figure}

The $550\times450\times250\,\mathrm{mm}$ vacuum chamber (Fig. \ref{cryocha}) containing the experimental set-up is made by aluminum. Vacuum is obtained using an \textit{oil-free Varian\textsuperscript{TM} system}, coupled with a scroll primary pump and a turbo pump. The minimum pressure available is $10^{-6}\,\mathrm{mbar}$. \par
\par
\begin{figure}[h]
\centering
\includegraphics[scale=0.176]{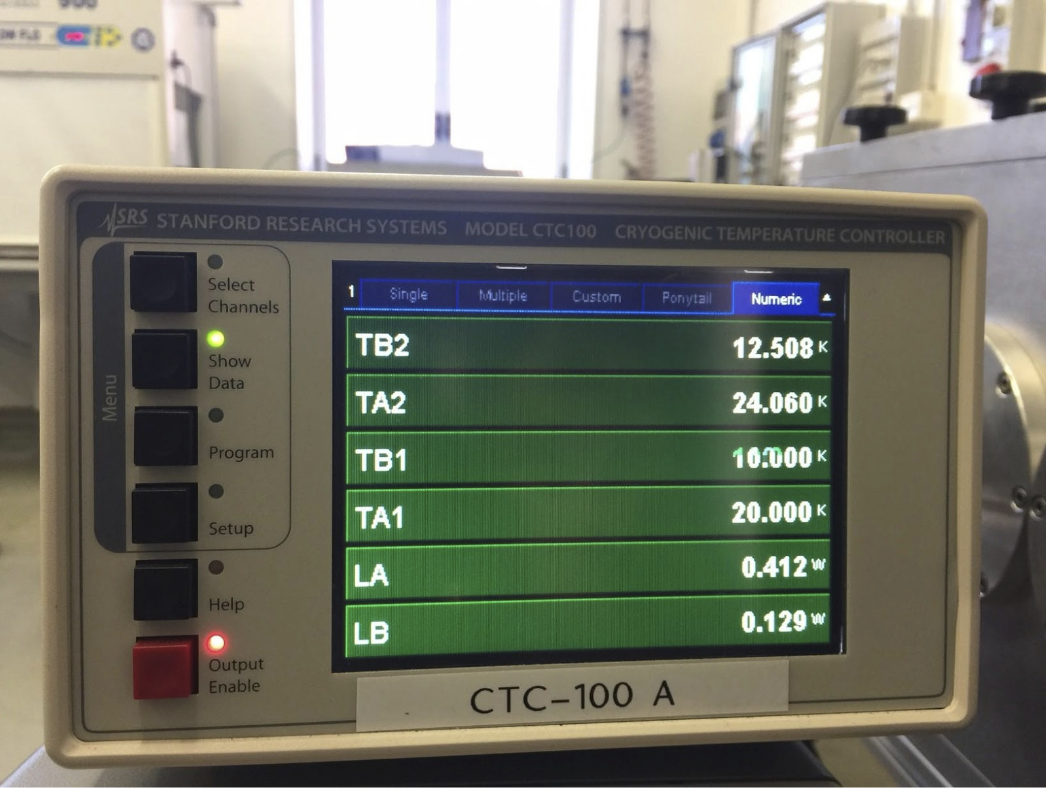}
\includegraphics[scale=0.18]{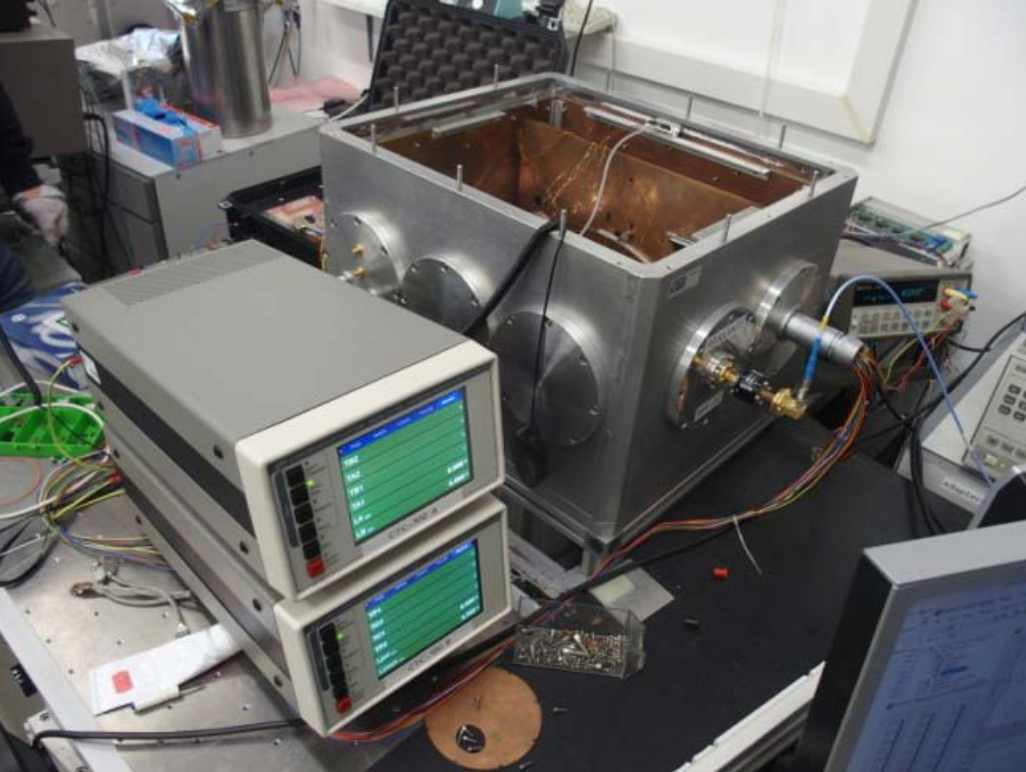}
\caption{Pictures of the temperature controller. The vacuum chamber is visible on the right picture.}\label{temcon}
\end{figure}

A water-cooled two-stage cryocooler with a cooling power of $1\,\mathrm{W}$ at $4.2\,\mathrm{K}$ allow the vacuum chamber to reach cryogenic temperatures. A cryogenic temperature controller (Fig. \ref{temcon}) allow us to set temperatures between $4\,\mathrm{K}$ and $300\,\mathrm{K}$. The controller monitors the temperature and uses one or more heaters to stabilize it within $\pm 0.05\,\mathrm{K}$, thanks to the large heat capacity of the copper cold plate. Cooling time is of the order of $8\,\mathrm{hours}$ while the warm-up procedure lasts about $6\,\mathrm{hours}$. \par
Temperature sensors are spread all over the chamber and cross-calibrated by means of a calibrated Rhodium-Iron sensor. \par
\par
To achive a better thermalization of the polarimeter, the complete assembly is covered with multiple layers of Mylar sheets in order to ensure a higher radiative insulation. \par
The RF generator (Fig. \ref{cryocha}) operates in a wide frequency range from $10\,\mathrm{MHz}$ to $50\,\mathrm{GHz}$. It could be remote controlled by the HK software to set manually the output power, the frequency band, the frequency step and the sweeping speed.\par
\begin{figure}[H]
\centering
\includegraphics[scale=0.4]{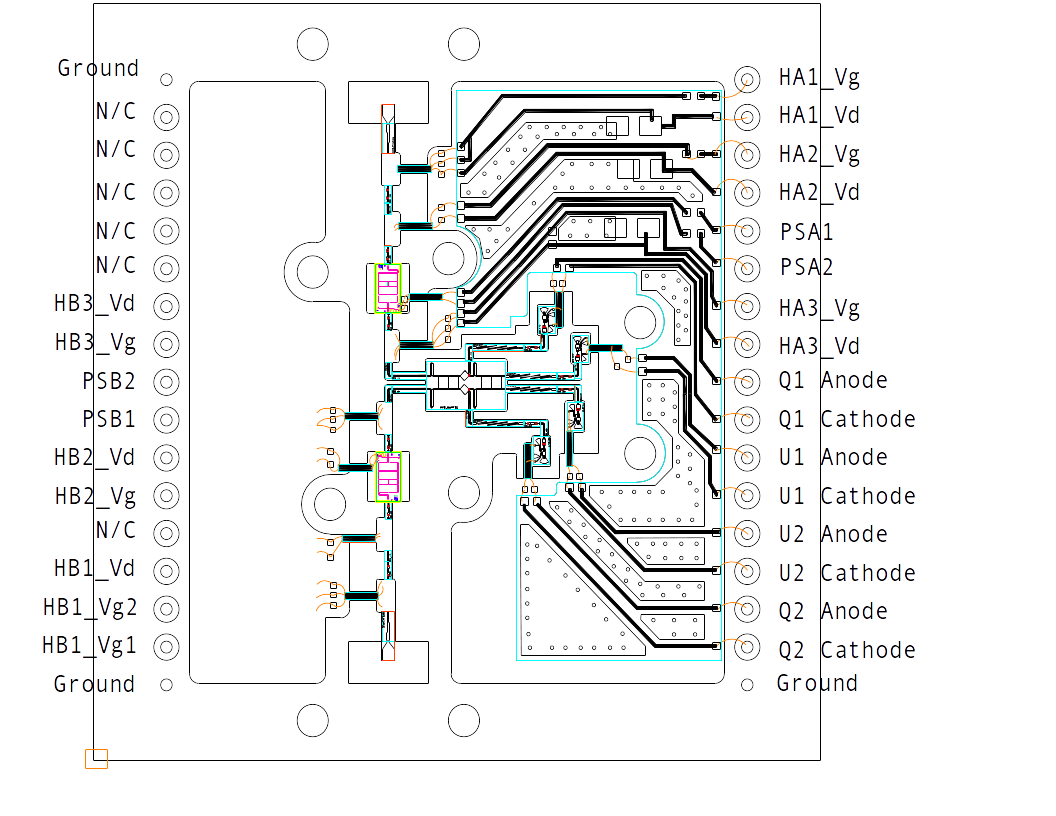}
\caption{Scheme of the electronic board of a STRIP polarimeter used during the tests.}\label{elename}
\end{figure}

Each polarimeter is powered and controlled by an electronic board expressly developed for the tests. It provides gain and drain voltages to the HEMTs amplifiers, anode and cathode voltages to the diodes and the proper biases to the phase swicthes (Fig. \ref{elename}). 
\par
\begin{figure}[H]
\centering
\includegraphics[scale=0.18]{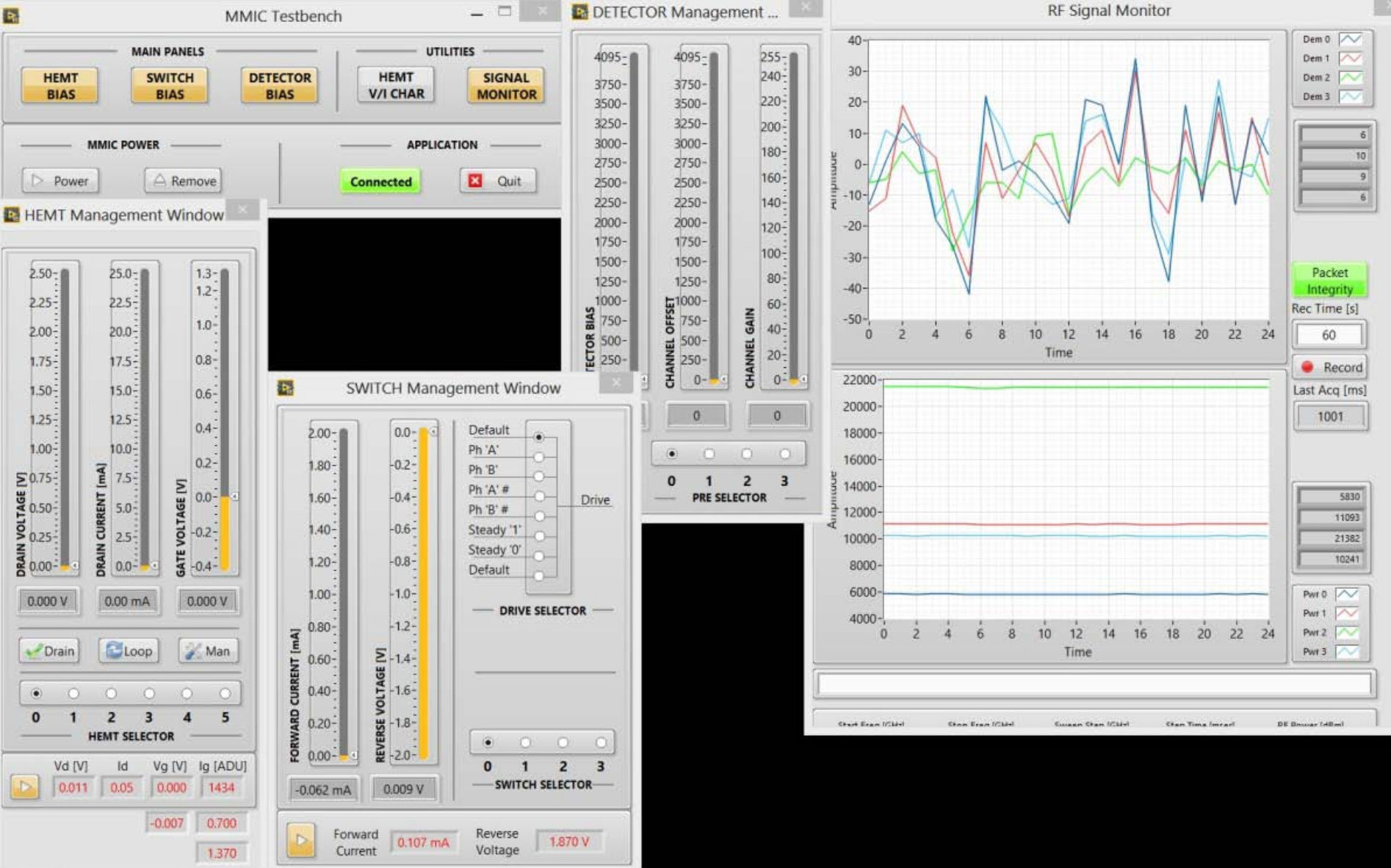}
\caption{Screenshot of the graphic interface of the HK software.}\label{hksoft}
\end{figure}

The HK software has three purposes: to supply the desired bias parameters to the detectors, to control the RF generator and to read the detector outputs in order to acquire and save the data. The code has been developed in LabVIEW and provides graphic interface (Fig. \ref{hksoft}). 
\par

\section{Bandwidth and central frequency calculation}\label{bp}
During the unit-level tests, we characterized the frequency response of each polarimeter. To perform the bandwidth characterization, we used the RF generator to inject into the receiver a tone whose frequency swept the range $38\div50\,\mathrm{GHz}$ for Q-band\footnote{The range was enlarged to $36\div50\,\mathrm{GHz}$ for the last $20$ polarimeters tested.} and $80\div110\,\mathrm{GHz}$ for W-band, with steps of $0.1\,\mathrm{GHz}$, while we kept the two thermal loads at $20\,\mathrm{K}$.\par
We measured the frequency responses maintaining a fixed phase switch configuration. Given a phase switch configuration, one of the four detectors was ``blind'', namely not sensitive to the variation of the injected signal (see Eq. \ref{blind}). Therefore, in order to get information on all the detectors, we repeated the frequency response test twice for each polarimeter, one time for each independent phase switch configuration (e.g., $0101$ and $0110$). \par
\begin{figure}[H]
\centering
\includegraphics[scale=0.3]{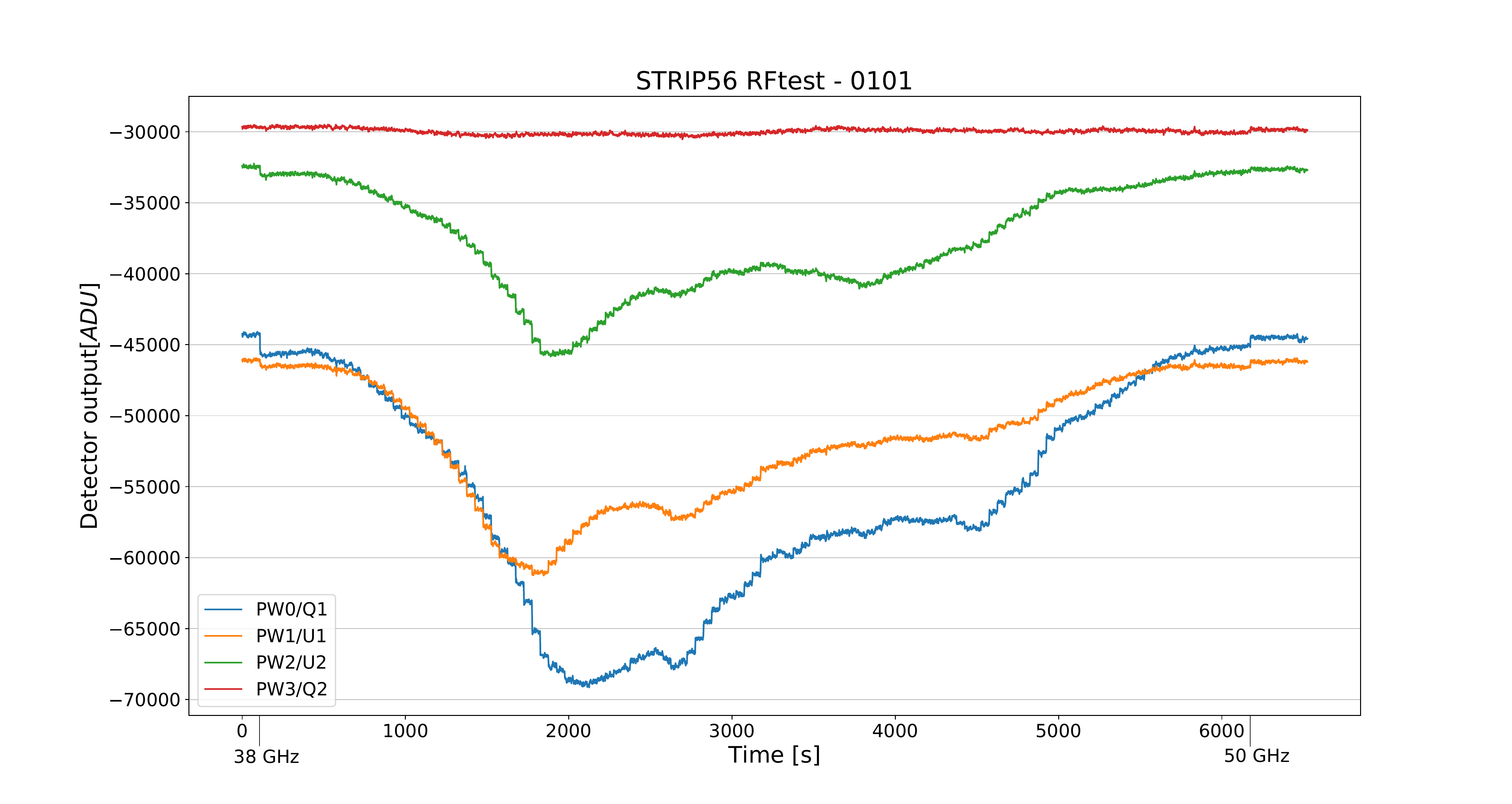}
\caption{\label{fig:band_rowdata}Raw output of the bandpass test for STRIP56 performed with the phase switch configuration $0101$. Detector $\mathrm{Q2}$ is blind, as expected.}
\label{STRIP56_rowdata}
\end{figure}\par
The four detectors produce very different output powers (Fig. \ref{STRIP56_rowdata}). This is mostly due to the presence of an electronic offset, set by hand by the operator in order to avoid ADC saturation. To correctly compute the bandwidth and the central frequency, we must estimate and subtract the offset from data.\par
Fig. \ref{STRIP56_rowdata} also shows the small ``steps'' that appear at the beginning of the curves. They were due to switching on/off of the generator: the polarimeter resulted sensitive at frequencies even lower than $38\,\mathrm{GHz}$. After discovering this behavior (for the last $\sim 20$  polarimeters tested) the swept range was extended to $36\,\mathrm{GHz}$ and these lower steps disappeared. Unfortunately, it was not possible to test higher frequencies since $50\,\mathrm{GHz}$ was the highest frequency allowed by the generator.\par
\begin{figure}[h]
\centering
\includegraphics[scale=0.23]{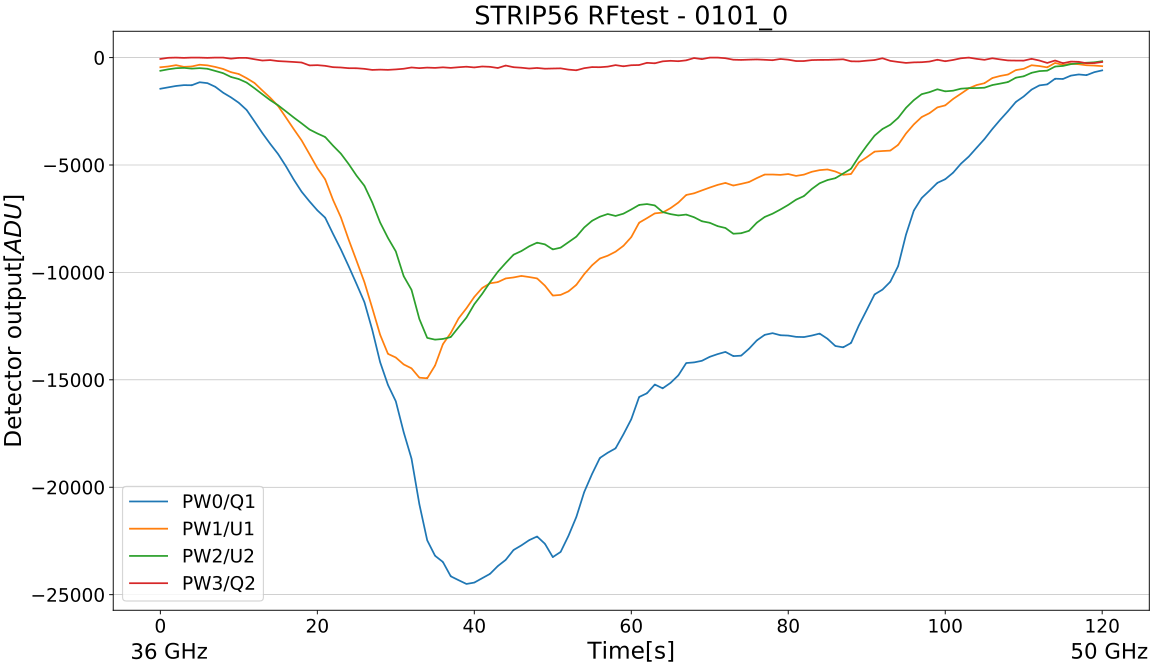}
\includegraphics[scale=0.23]{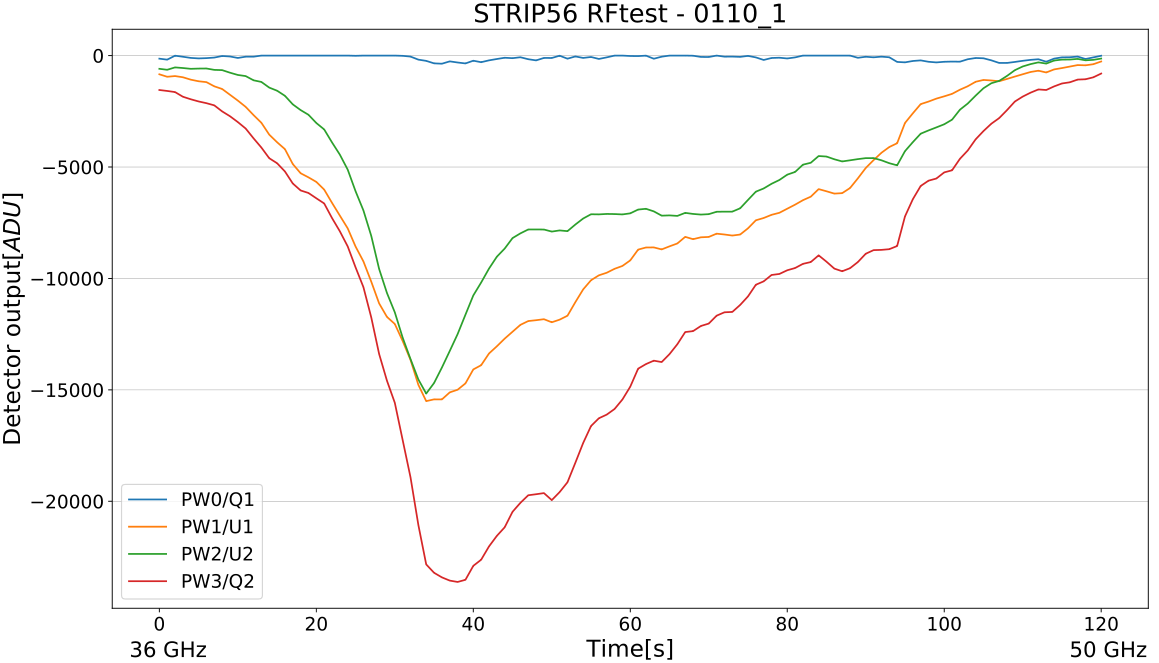}
\caption{Output of the bandpass test for the STRIP56 polarimeter, after the subtraction of the electronic offset. \textit{Up}: test performed with phase switch configuration: $0101$. \textit{Down}: test performed with phase switch configuration: $0110$.}
\label{STRIP56no_offset}
\end{figure}

We estimated the electronic offset for the four detectors by taking the mean value of the output before switching on and after switching off the generator, and by performing a linear fit between these two points. In this way, we automatically take into account possible drifts of the electronics during the test period. Fig. \ref{STRIP56no_offset} shows the output signal of the four diodes of the STRIP56 polarimeter after the offset removal. 

We used the frequency response measurements to calculate bandwidth and central frequency, defined by Eqs.\ref{bdwdef} and \ref{nucdef} where $g_A(\nu)g_B(\nu) = g^2(\nu)$ is the detector response. In our case, the frequency steps are discrete and constant ($\Delta\nu = 0.1,\,\mathrm{GHz}$) thus, we simplify the previous equations as follows:

\be{BWsimpl}
\beta = \frac{[\sum g^2(\nu)]^{2} \Delta\nu}{\sum g^4(\nu)} \,,
\ee
\be{centfreqsimpl}
\nu_c = \frac{\sum g^2(\nu)\nu}{\sum g^2(\nu)} \,.
\ee\par

\begin{figure}[H]
\centering
\includegraphics[scale=0.32]{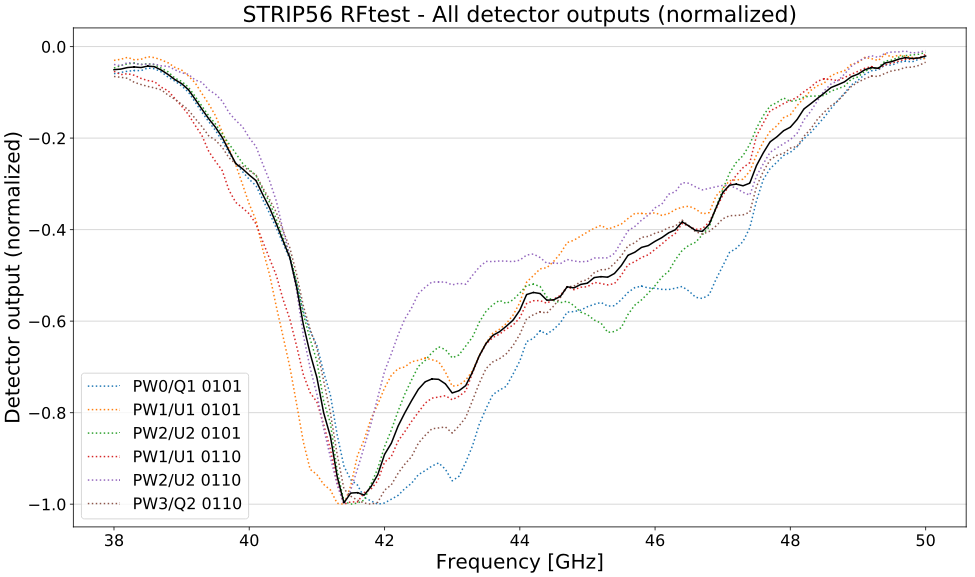}
\caption{All detector outputs from both frequency response tests (phase switch configuration $0101$ and $0110$) for STRIP56 normalized to the maximum. The best estimate of the bandwidth is shown in black.}
\label{STRIP56_alldet}
\end{figure}\par
We computed bandwidths and central frequencies for each detector, using both tests. We put together the six bands estimates ($3$ non-blind detectors, $2$ tests) to obtain the best estimate of the polarimeter bandwidth and central frequency. To do this we normalized each bandshape in the range $0\div1$ and we took the median of the six output values at each frequency as the best estimate of the polarimeter bandwidth. Fig. \ref{STRIP56_alldet} reports all the normalized detector outputs for STRIP56 together with the best bandwidth, shown in black.\par
\begin{figure}[H]
\centering
\includegraphics[scale=0.4]{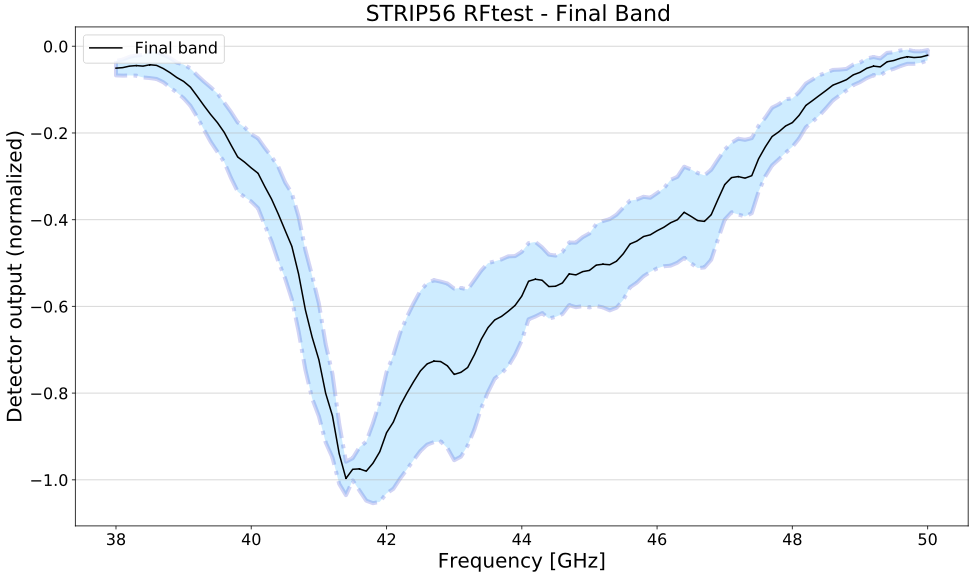}
\caption{The best estimate of the band for the STRIP56 polarimeter, together with its error bar.}
\label{STRIP56_finalband}
\end{figure}\par
Fig. \ref{STRIP56_finalband} shows the best bandwidth for STRIP56 and its error bar. We estimated the latter as the difference between the $97.5$ and the $2.5$ percentile of the distribution of the six output values at each frequency, which corresponds to a $95\%$ confidence level.\par
Finally, our best estimate of the central frequency is the central frequency of the best bandwidth along with its uncertainty, corresponding to the difference between the $97.5$ and the $2.5$ percentile of the distributions.

\subsubsection{Results}
\begin{figure}[H]
\centering
\includegraphics[scale=0.45]{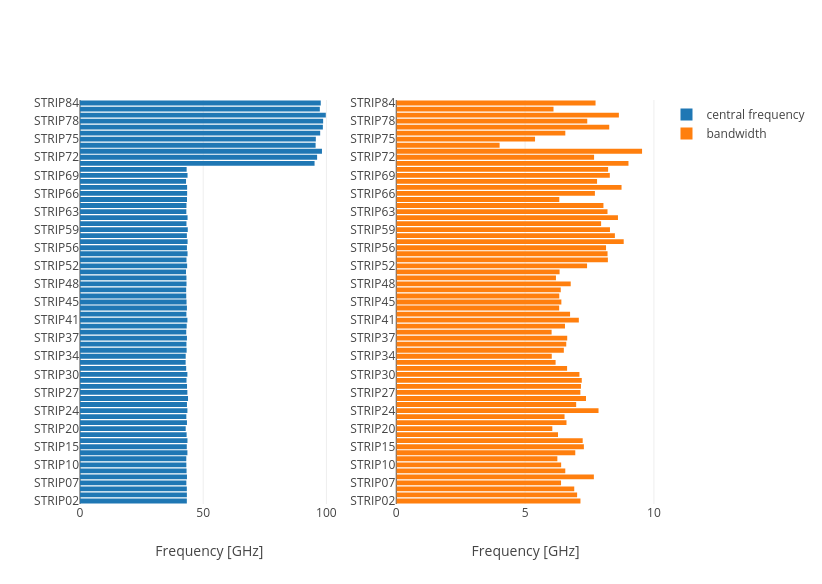}
\caption{Central frequencies (blue) and bandwidths (orange) distributions for all the STRIP polarimeters tested. Polarimeters from STRIP71 to STRIP84 are W-band.}
\label{bwcfdistrib}
\end{figure}

Fig. \ref{bwcfdistrib} shows the distributions of bandwidths and central frequencies for all the STRIP polarimeters. We report the averages values for the $49$ Q-band and $6$ W-band polarimeters integrated in the STRIP focal plane in Table \ref{avalbc}. \par

We found a clear difference in the bandwidth between the Q-band QUIET polarimeters and the remaining, in fact, the former have a larger bandwidth ($\beta\sim 8\,\mathrm{GHz}$) compared to the others ($\beta \sim 6.5\div 7.0\,\mathrm{GHz}$). This difference is probably due to the fact that they have been assembled at different times, using different batches of components.\par
All the polarimeters show a remarkable agreement in the central frequency. \par

\begin{table}[H]
  \centering
\begin{tabular}{|c|c|c|} \hline
  & $\beta$ [$\mathrm{GHz}$] & $\nu_c$ [$\mathrm{GHz}$] \\ \hline
 Q-band &  $7.3 \pm 0.8$ &  $43.4 \pm 0.2$ \\ \hline
 W-band &  $7.9 \pm 1.9$ &  $97.5 \pm 1.8$ \\ \hline
\end{tabular}\caption{Average values of bandwidths and central frequencies for the polarimeters integrated in the focal plane.}\label{avalbc}
\end{table}\par

\section{Noise temperature estimation}\label{tnoise}
The polarimeter noise temperature quantifies the noise associated to the thermal emission of all the radiometric components of the polarimeter itself. Its value is dominated by the passive components located before the amplifiers and from the noise of the first amplification stage. We can estimate the noise temperature using the so-called \textit{Y-factor test}. \par
\begin{figure}[H]
\centering
\includegraphics[scale=0.5]{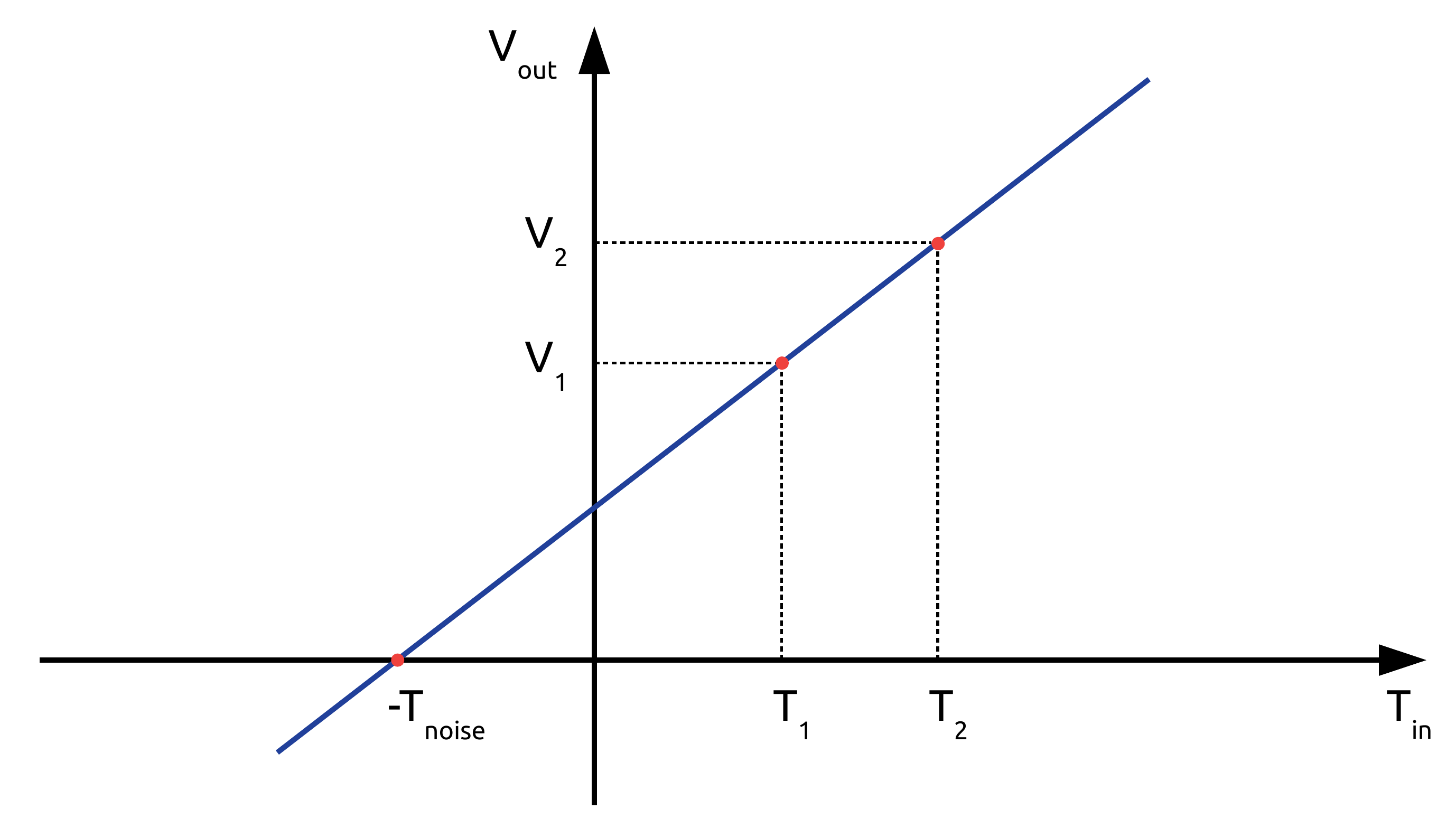}
\caption{Operating principle of the Y-factor test.}\label{yfactor_}
\end{figure}

The Y-factor test is based on the principle that if no source is observed the detected signal should be zero. Since though the radiometric chain has its own physical temperature, it emits as a blackbody at a certain brightness (noise) temperature and an offset will be present always in the output power, corresponding to this temperature (Fig. \ref{yfactor_}). In this way, by changing the temperature of the source between two values, $T_1$ and $T_2$, it is possible to estimate the noise temperature $T_n$ by means of a simple linear fit: 
\be{yfact}
T_n = \frac{T_1 + Y T_2}{Y - 1} \,,
\ee
where $Y = V_1/V_2$ is the ratio between the average voltages, $V_1$ and $V_2$, measured at the two temperature steps. At the same time, the Y-factor test provides also an estimate of the detector gain $g$.\par
 \par
\begin{figure}[H]
\centering
\includegraphics[scale=0.3]{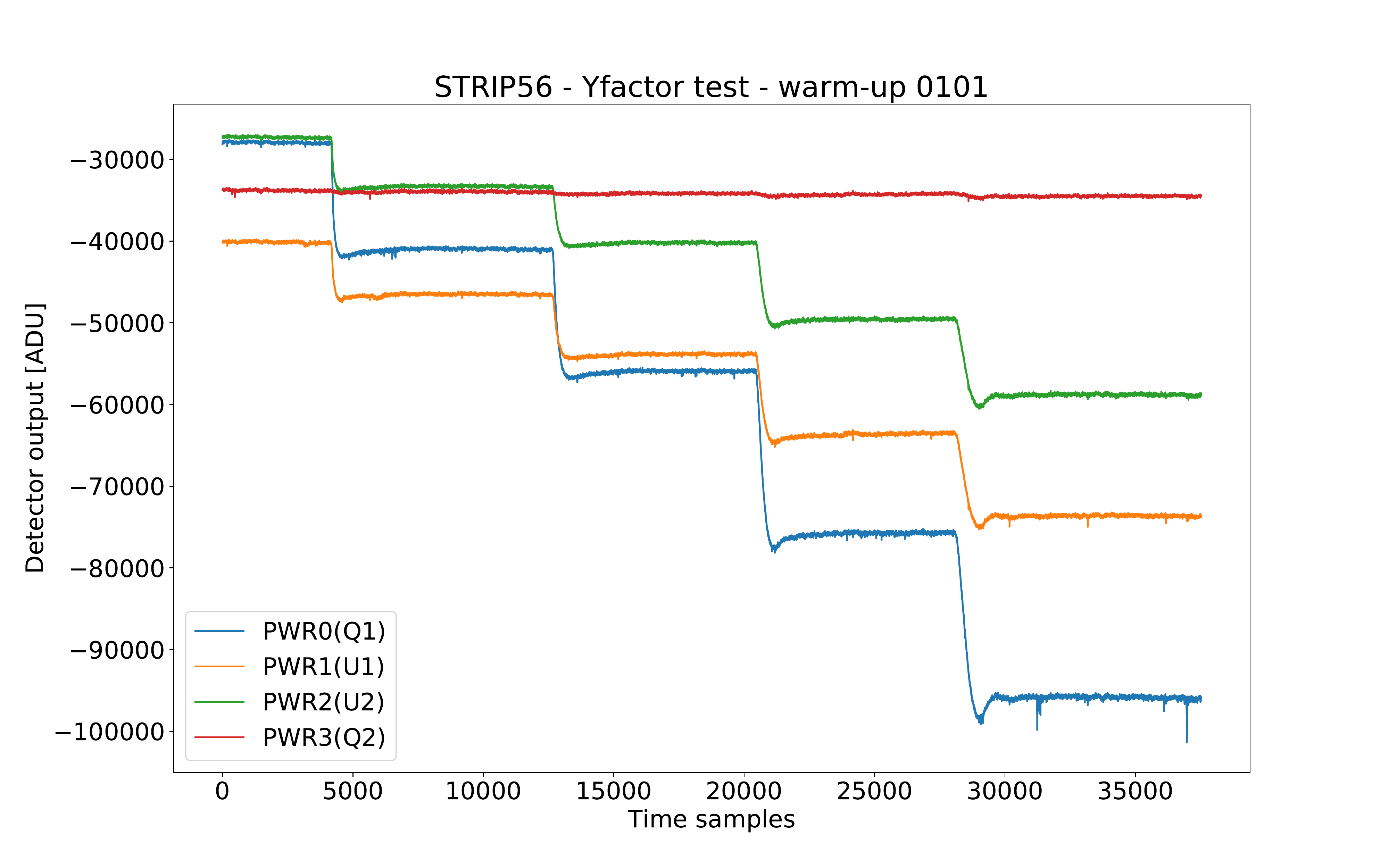}
\includegraphics[scale=0.3]{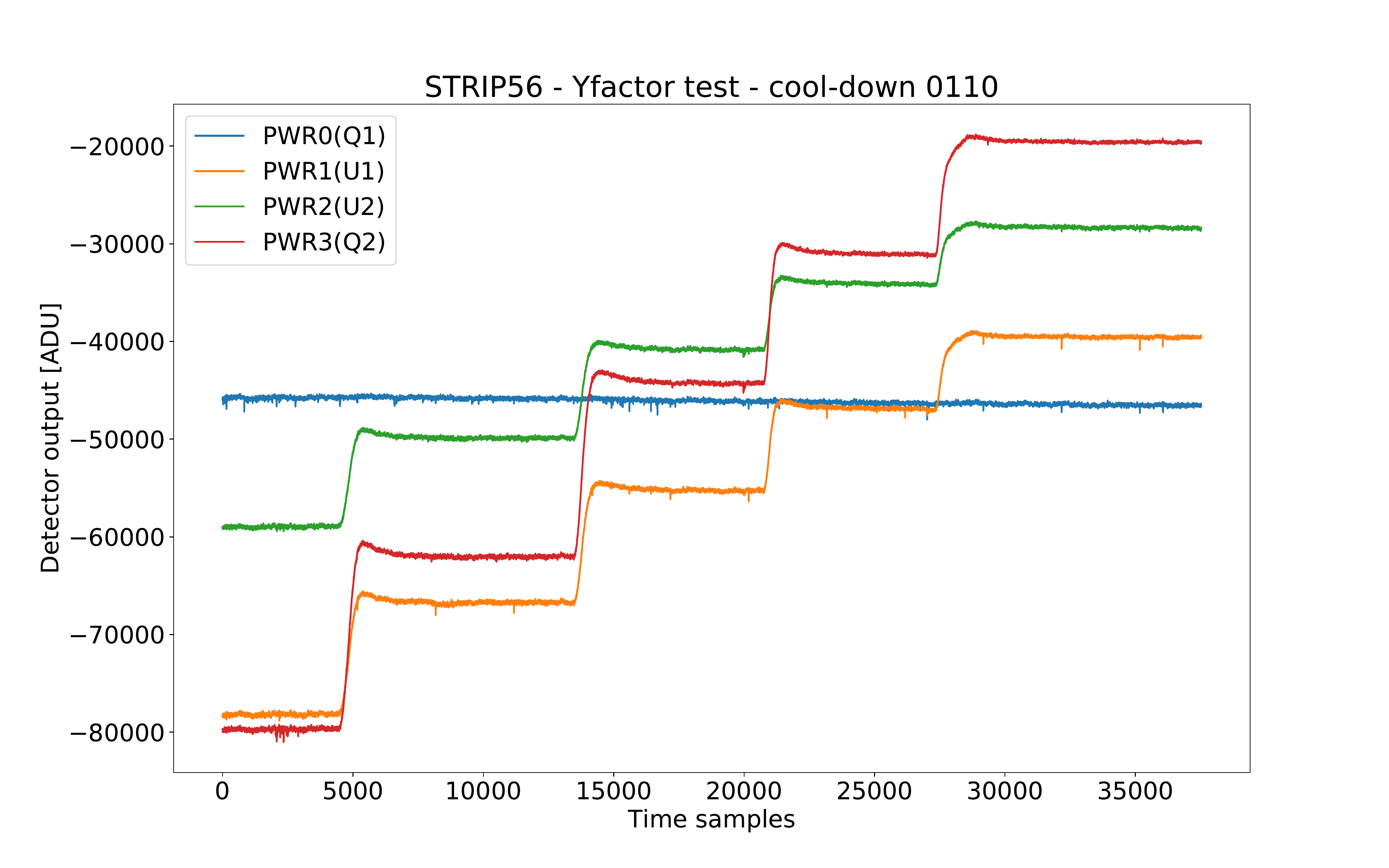}
\caption{The Y-factor test for the STRIP56 polarimeter. \textit{Up}: test performed by warming-up one thermal load and with the phase switch configuration $0101$. \textit{Down}: test performed by cooling-down one thermal load with the phase switch configuration $0110$. The apparent trend inversion is due to the negative response of the ADC.}
\label{yfactorbase}
\end{figure}

Tho run this test, we changed the temperature of one of the two thermal loads from $10\,\mathrm{K}$ to $50\,\mathrm{K}$ by steps of $10\,\mathrm{K}$ and we measured the response of the polarimeter (Fig. \ref{yfactorbase}). In this way, there are many possible temperature pairs ($10$ if there are $5$ steps) and all of them should produce the same estimate for $T_n$ using Eq. \ref{yfact}. 
\par
\begin{figure}[H]
\centering
\includegraphics[scale=0.45]{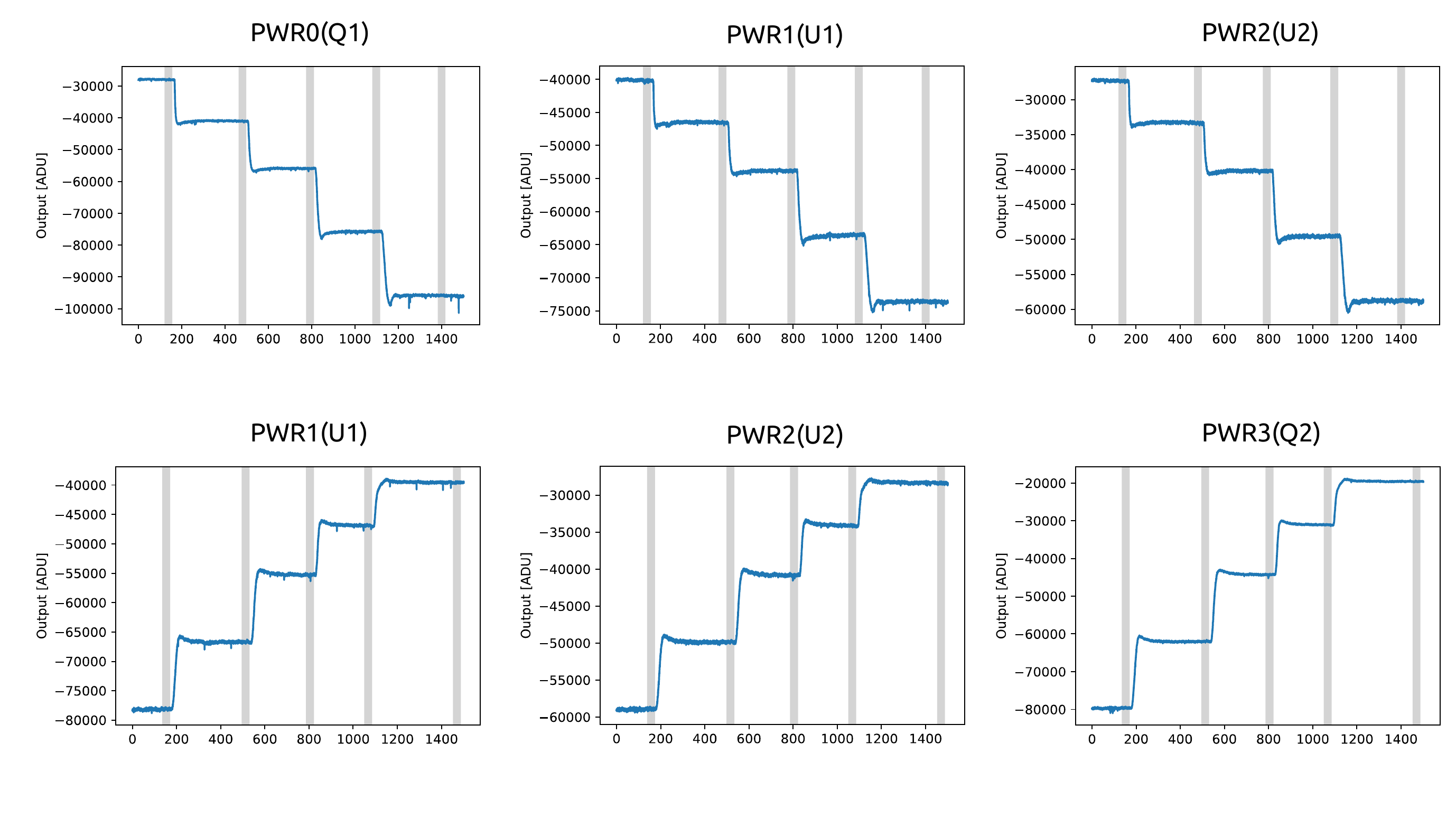}
\caption{The Y-factor test for the STRIP56 polarimeter. Regions where the signal is stable enough to run an analysis of the noise temperature are shown in gray. \textit{Up}: test performed by warming-up one thermal load and with the phase switch configuration $0101$. \textit{Down}: test performed by cooling-down one thermal load with the phase switch configuration $0110$.}
\label{yfregions}
\end{figure}

For each detector and temperature step, we computed the average output response when the thermal load reaches the thermal equilibrium at the new temperature, corresponding to constant regions in the plots. In the case of STRIP56, I show these regions in gray (Fig. \ref{yfregions}).
\par

For almost all the polarimeters, we made two noise temperature tests (Fig. \ref{yfactorbase}): the first in a certain phase switch configuration (e.g., $0101$) when the input load was warmed up and the second with the opposite phase switch configuration (e.g., $0101$) when it was cooled down again. This means that for every polarimeter we produced a large number of noise temperature estimates ($120$ if every test is performed with $5$ temperature steps). \par
As final noise temperature and relative uncertainty we taken the median value of $T_n$ and the $5\mathrm{th}$ and $95\mathrm{th}$ percentiles to fix the upper and lower error bars. We chose to use the median instead of the mean, as the data are not normally distributed.\par
        
\subsection{Results}
\begin{figure}[H]
 \centering
 \includegraphics*[viewport=0 200 700 910, scale=0.55]{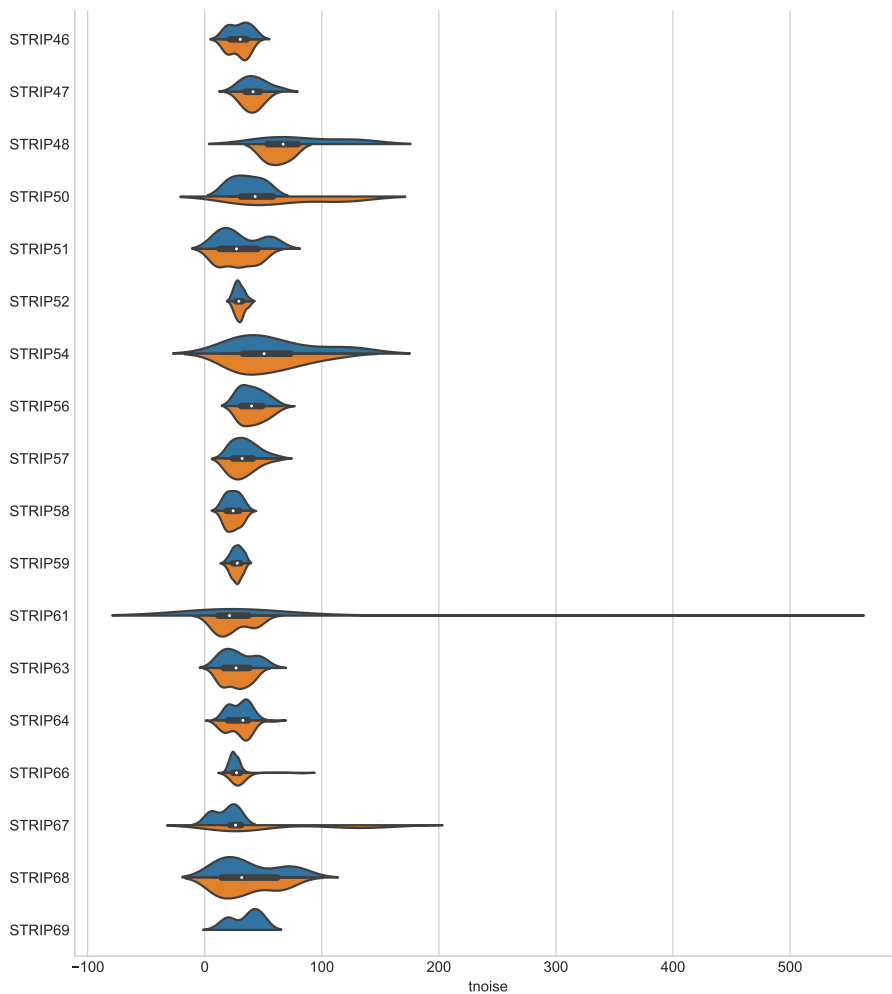}
 \caption{Noise temperature distributions for $54$ Q-band STRIP polarimeters. Blue (orange) distributions refer to the phase switch configuration $0101$ ($0110$). The white dots represent the medians while the black bars are the intervals between the $1\mathrm{st}$ and $4\mathrm{th}$ quartile.}
 \label{violins}
\end{figure}

I show, in Fig. \ref{violins}, the noise temperature distributions for $54$ Q-band polarimeters through violin plots. For each polarimeter, the white dot represents the median value of the blue and orange distributions, referring respectively to the phase switch configuration $0101$ and $0110$. The black error bar is the interval between the $1\mathrm{st}$ and $4\mathrm{th}$ quartile. Violin plots for W-band are similar and are not reported. \par
The analysis showed high uncertainties on the noise temperature measurements. Possible sources of systematic errors could be non-linearity in detector (or ADC) response, uncertainty in ADC offset, imperfect knowledge of $T_1$ and $T_2$ in Eq. \ref{yfact} due to imperfect balance between the two legs, non-idealities in the polarimeters or in the set-up, etc. Unfortunately, more investigations could not be conducted since fundamental HK parameters were not recorded by the software. New measurements of the detector noise temperature will be carried out in early $2020$ during the system-level tests in Bologna. \par
\begin{figure}[H]
 \centering
  \includegraphics*[viewport=0 905 700 1950, scale=0.55]{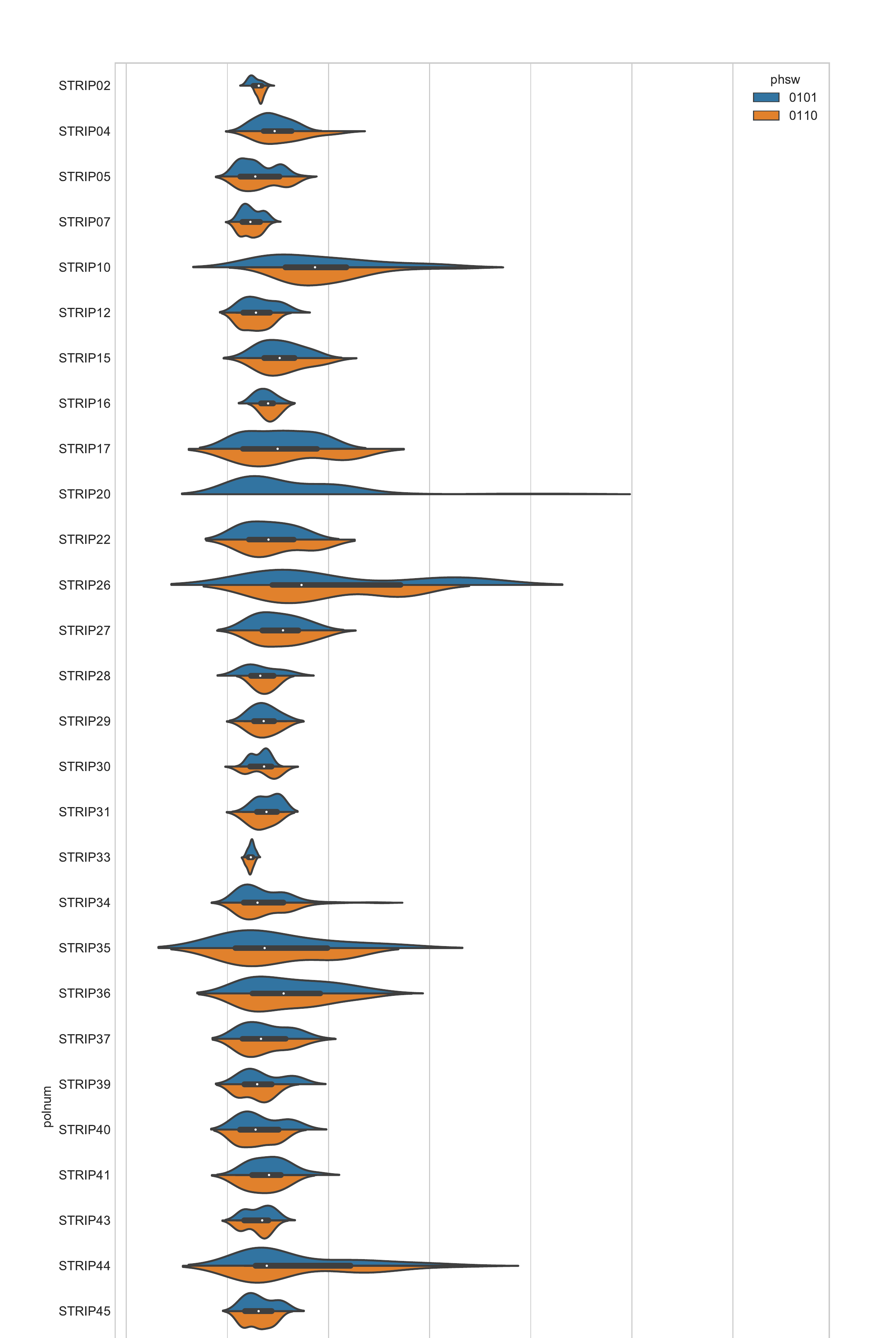}
\end{figure}
I report, in Table \ref{avaltn}, the median value of the noise temperature for the $49$ Q-band and $6$ W-band polarimeters that make up the STRIP focal plane. The uncertainty is given by the median absolute deviation (MAD):
\be{MAD}
\mathrm{MAD} = \mathrm{median}(\abs{x_i - \mathrm{median}(x_i)}) \,.
\ee\par
\begin{table}[H]
  \centering
\begin{tabular}{|c|c|c|} \hline
  & Q-band & W-band \\ \hline
 $T_n$ [$\mathrm{K}$] &  $33.6 \pm 5.8$ & $105.0 \pm 17.0$ \\ \hline
\end{tabular}\caption{Median values of the noise temperatures for the polarimeters of the focal plane.}\label{avaltn}
\end{table}\par

\section{Noise characterization}\label{noisechara}
An ideal receiver would be characterized by a pure \textit{white noise} spectrum, i.e., a flat spectrum with the same power at every frequency:
\be{wn}
P(\nu) = \sigma^2 \,,
\ee
where $P$ is the power of the Fourier transform of the timestream data, which would be normally distributed with standard deviation $\sigma$.\par
Real receivers have always a low-frequency correlated component of the spectrum: this is the so-called \textit{1/$f$ noise} (or \textit{pink spectrum}). Its power is inversely proportional to the frequency:
\be{wn}
P(\nu) = \sigma^2 + \frac{k}{\nu^\alpha} = \sigma^2\Bigl[1 + \Bigl(\frac{\nu_\mathrm{knee}}{\nu}\Bigr)^\alpha\Bigr] \,.
\ee
where $0 \lesssim \alpha \lesssim 2$ is the \textit{slope} of the pink spectrum and the \textit{knee frequency}, $\nu_\mathrm{knee}$, is the frequency at which the power of the white and the pink spectra are equal:
\be{kneefreq}
P_\mathrm{WN} = P_{1 / \mathrm{f}} = P(\nu_\mathrm{knee}) = 2 \sigma^2 \,.
\ee
In general, the knee frequency should be as low as possible to be sure that small time samples are free from correlated noise. \par
The white noise level (WNL) is the spectral density corresponding to $\sigma$, and is related to it by the following relation:
\be{sigmaWNL}
\sigma^2 \bigl[\mathrm{K}^2\bigr] = \frac12 \cdot \nu_\mathrm{sampling} \cdot \mathrm{WNL} \Biggl[\frac{\mathrm{K}^2}{\mathrm{Hz}}\Biggr] \,,
\ee
where $\nu_\mathrm{sampling}$ is the sampling frequency.\par

\subsection{Data analysis}
We performed the \textit{stability test} to estimate $\mathrm{WNL}$, $\alpha$ and $\nu_\mathrm{knee}$ in the output of each polarimeter. \par
The polarimeter under test observed the two thermal loads at the temperature of $\sim 20\,\mathrm{K}$, for a time ranging from $1$ to $60\,\mathrm{hours}$ with the phase switches set on the switching mode. We injected no signal from the RF generator. \par
Data have been acquired with a sampling rate of $25\,\mathrm{Hz}$. Temperatures of the chamber and of the thermal loads have not been recorded, so that any information about temperature fluctuations during the long acquistion were not available.\par
When possible\footnote{All the polarimeters analyzed except STRIP25.}, we calibrated raw data by subtracting to the total power outputs the detector offset (recorded by hand at the beginning of the test) and then dividing by the detector gains (obtained through the noise temperature analysis):
\be{calibration}
\mathrm{data_i} \, [\mathrm{K}] = \frac{\mathrm{raw\,data_i} - \mathrm{offset_i} }{\mathrm{gain_i}} \, \Bigl[\frac{\mathrm{ADU}}{\mathrm{ADU}/\mathrm{K}}\Bigr] \,,
\ee
where the index $i$ refers to the samples. \par
\begin{figure}[H]
\centering
\includegraphics[scale=0.25]{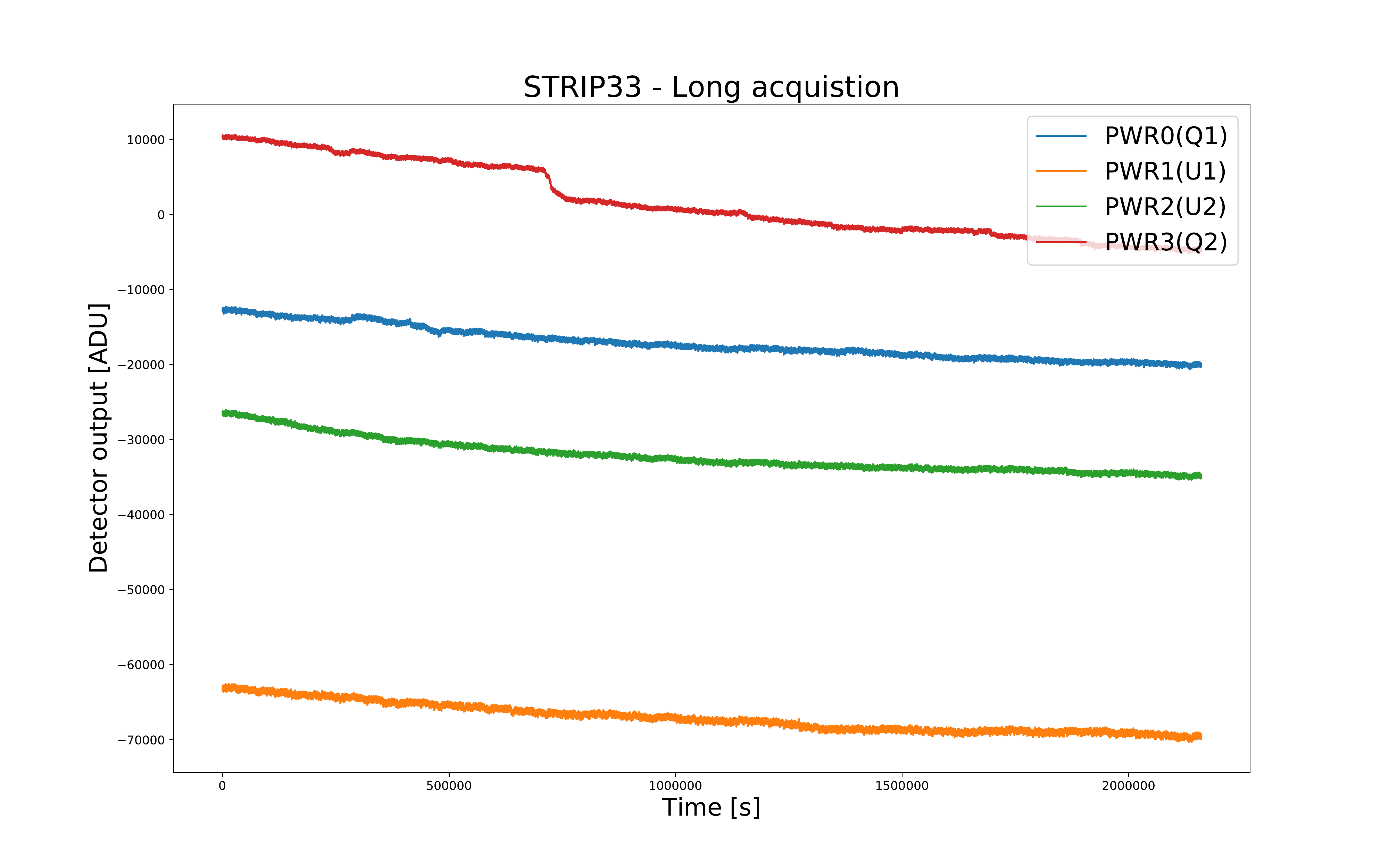}
\includegraphics[scale=0.25]{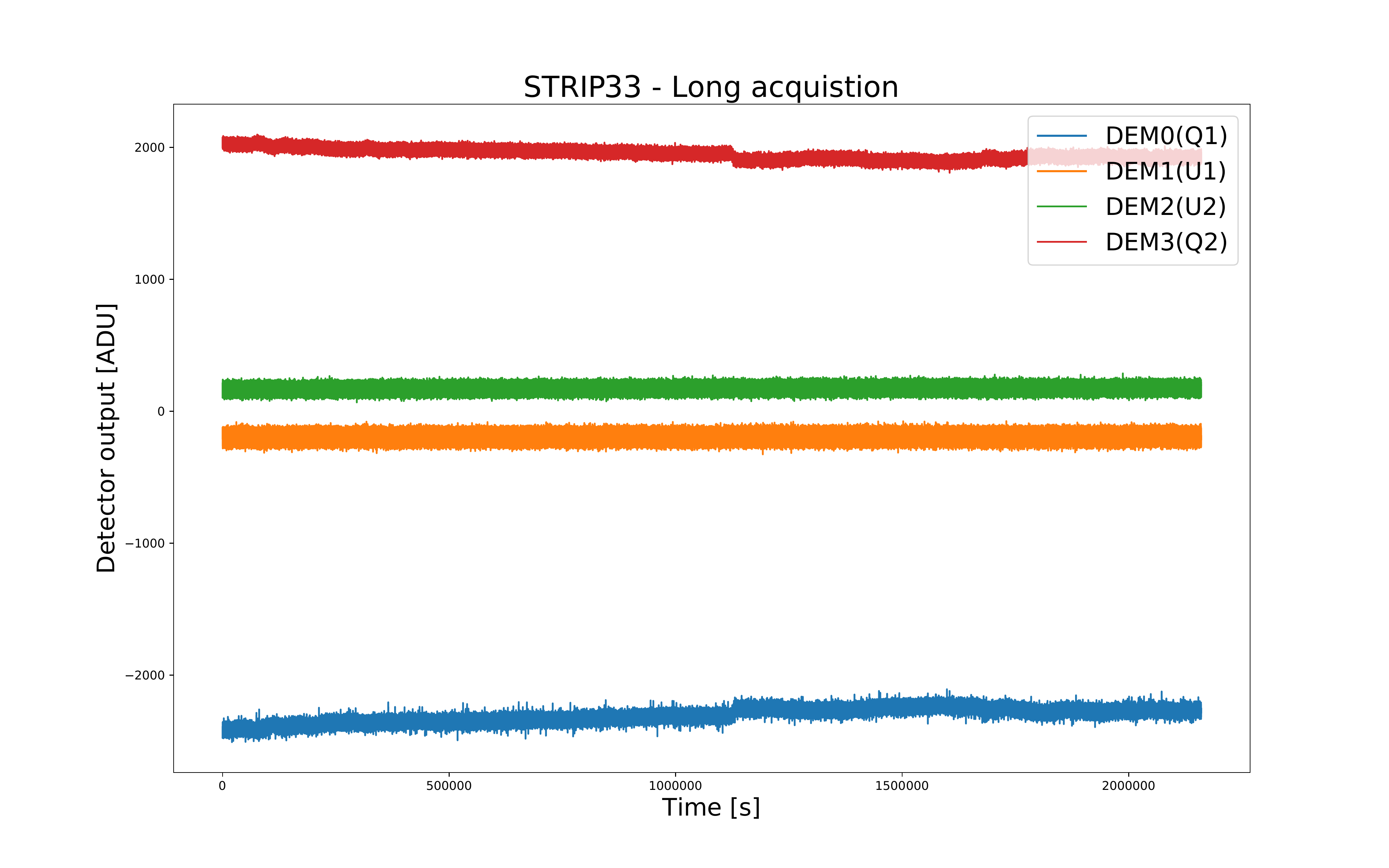}
\caption{The raw output of the long acquisition test for the STRIP33 polarimeter. \textit{Up}: the total power outputs. \textit{Down}: the demodulated outputs.}
\label{stability}
\end{figure}

Fig. \ref{stability} shows an example of the raw output for the long acquisition test of STRIP33.
\par

I estimated\footnote{All the codes developed for this and the previous analysis are available at: \url{https://github.com/lspestrip/striptun}.} the power spectral densities (PSDs) of the eight detector outputs and of their proper combinations to get $I$, $Q$, $U$ (see Eqs.\ref{I}, \ref{Q} and \ref{U}) using Welch's method \citep[pp.$652$-$662$]{Press:2007:NRE:1403886}. I divide the initial data stream into a number of chunks, which I detrend by subtracting a linear fit of the data to remove long thermal drifts. I estimate the PSD for each chunk and then I compute the average PSD of these PSDs. In this way, lowest frequencies of the spectrum are lost but, in return, it is possible to obtain a more precise estimation of the WNL and of the slope, $\alpha$, of the pink spectrum. The longer the test duration the better, because one has to trade-off these two features of the analysis.\par
\begin{figure}[H]
\centering
\includegraphics[scale=0.45]{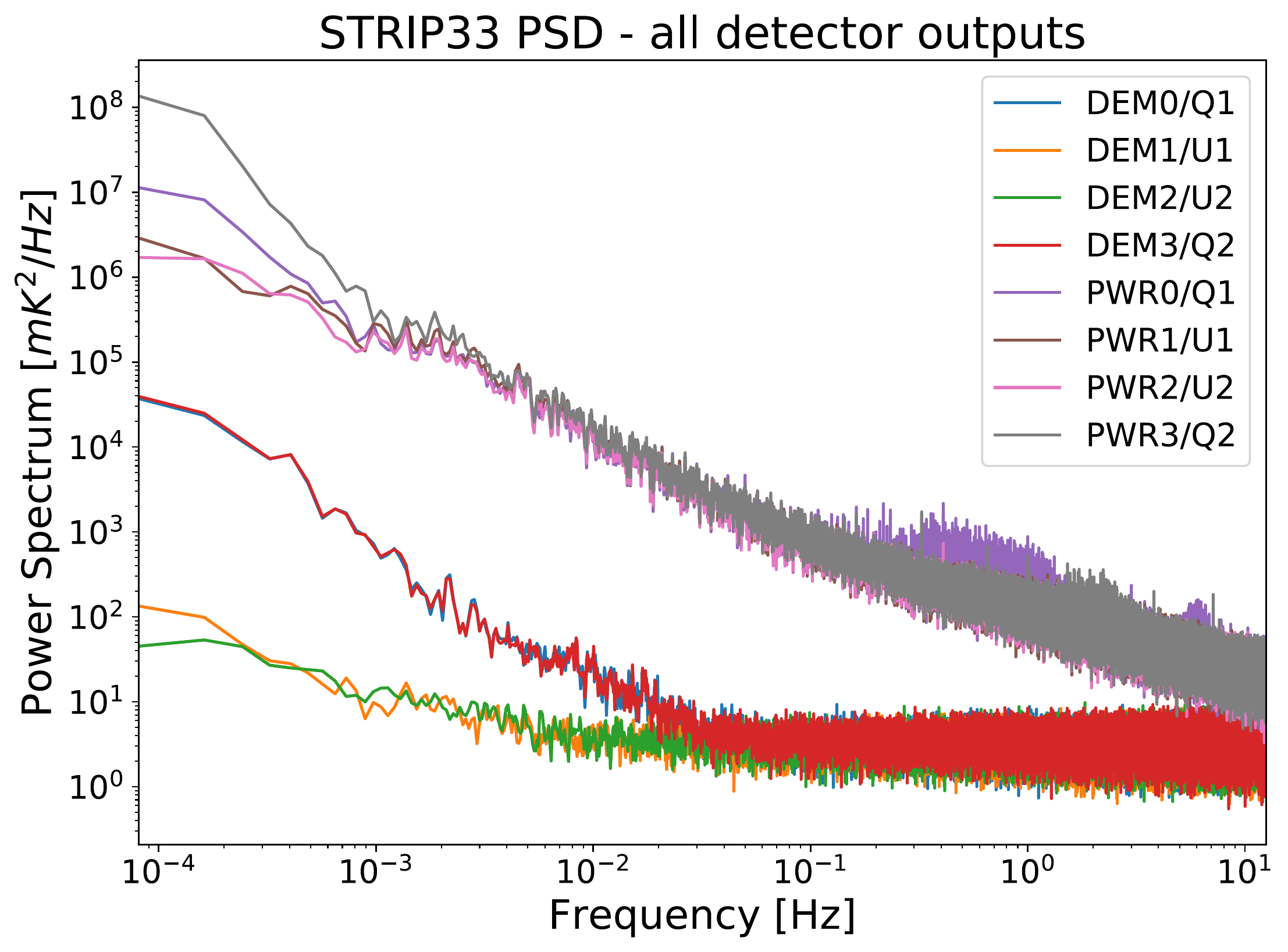}
\caption{The PSDs of all detector outputs of the long acquisition test for STRIP33.}
\label{allspectra}
\end{figure}

Fig. \ref{allspectra} shows the PSDs for all the eight outputs of the four detectors of STRIP33. In this case, the test lasted $24\,\mathrm{hours}$ and I have divided the original data samples into $7$ chunks of equal length. In this case, the double demodulation process has reduced the level of the 1/$f$ noise for the demodulated outputs of $Q$ detectors of a factor $\simeq 10^3$ with respect to the total power outputs and even more for the $U$ ones. \par

\par
\begin{figure}[H]
\centering
\includegraphics[scale=0.45]{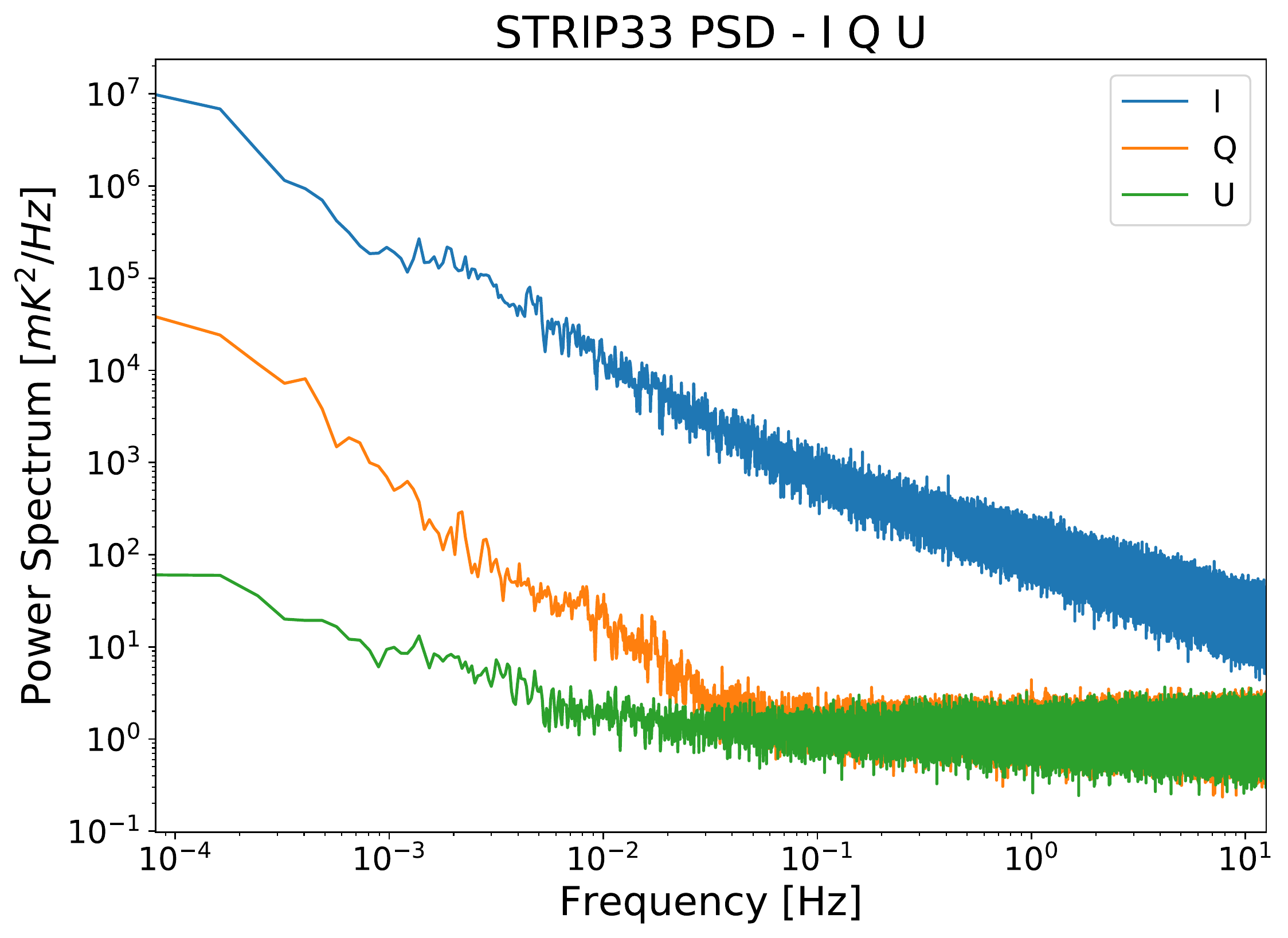}
\caption{The PSDs of the detector outputs combined to obtain $I$, $Q$, $U$.}
\label{stokespectra}
\end{figure}

Fig. \ref{stokespectra} shows the PSDs of the signals combined according to Eqs.\ref{I}, \ref{Q} and \ref{U}.\par
\par

I have estimated the WNL by calculating the median value of the high-frequency part of the spectrum, starting from a proper frequency $\nu_\mathrm{low}$ up to the highest frequencies, and I have computed the slope of the 1/$f$ noise through a linear fit of the low-frequency part (up to some frequency $\nu_\mathrm{high}$). $\nu_\mathrm{low}$ and $\nu_\mathrm{high}$ were set by hand during the analysis. I have estimated the knee frequency as the $x$ coordinate of the intersection between the linear fit and the constant median value.\par
Another possibility would have been to fit the spectra with a three parameters model (WNL, $\nu_\mathrm{knee}$, $\alpha$) but this method resulted ineffective since many spectra presented irregular shapes, affecting the fit results. \par
\begin{figure}[H]
\centering
\includegraphics[scale=0.29]{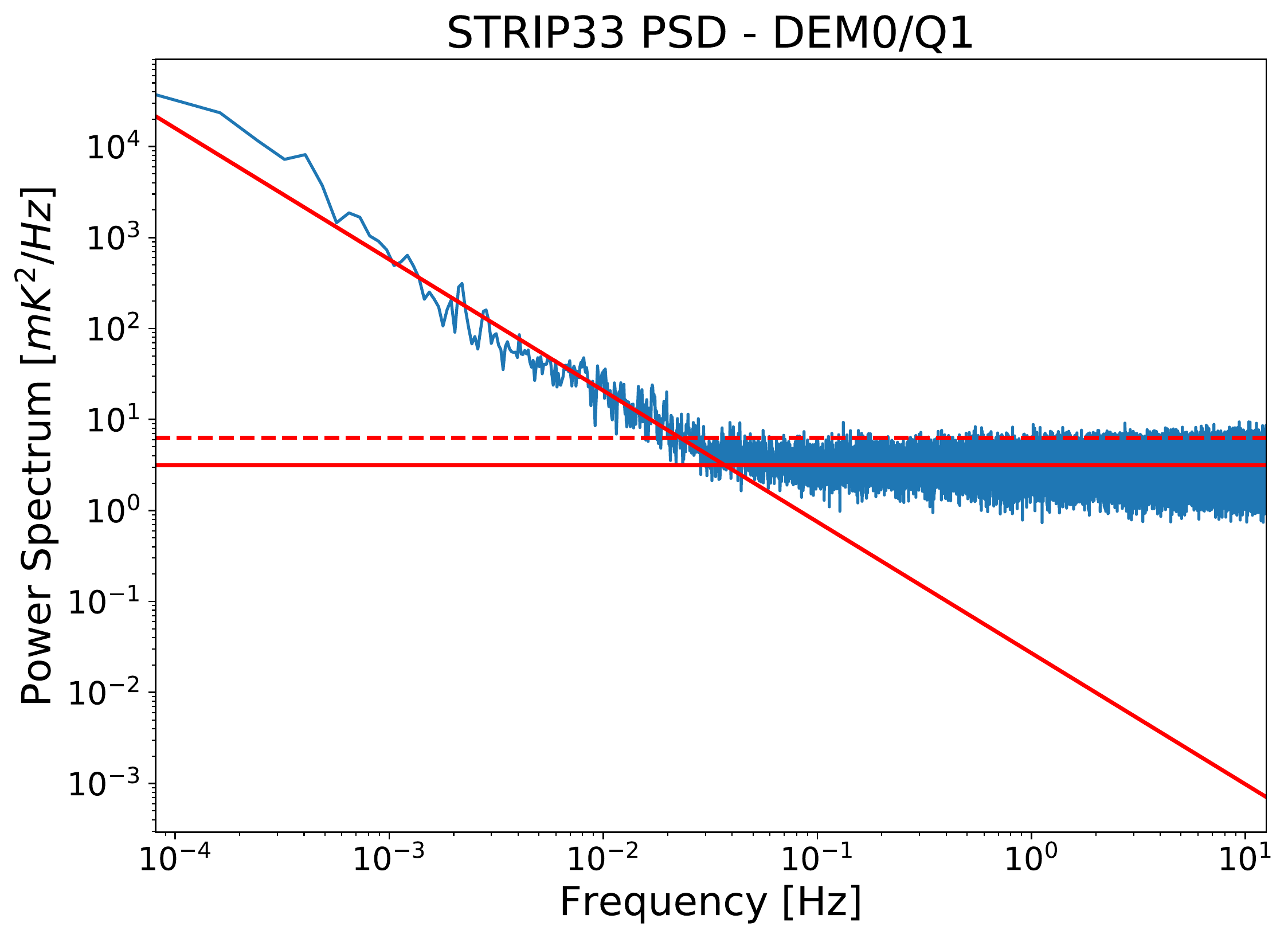}
\includegraphics[scale=0.29]{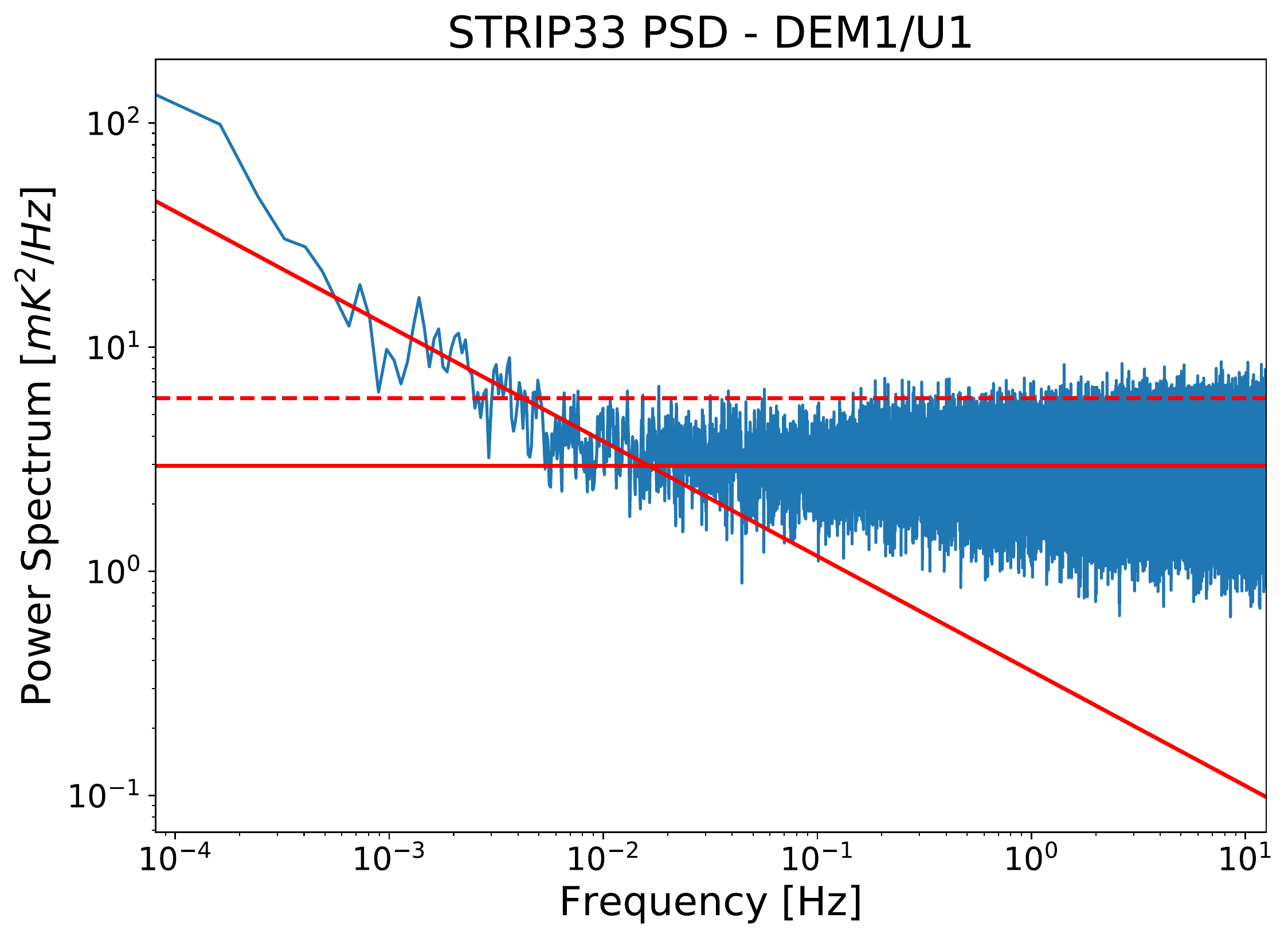}
\includegraphics[scale=0.29]{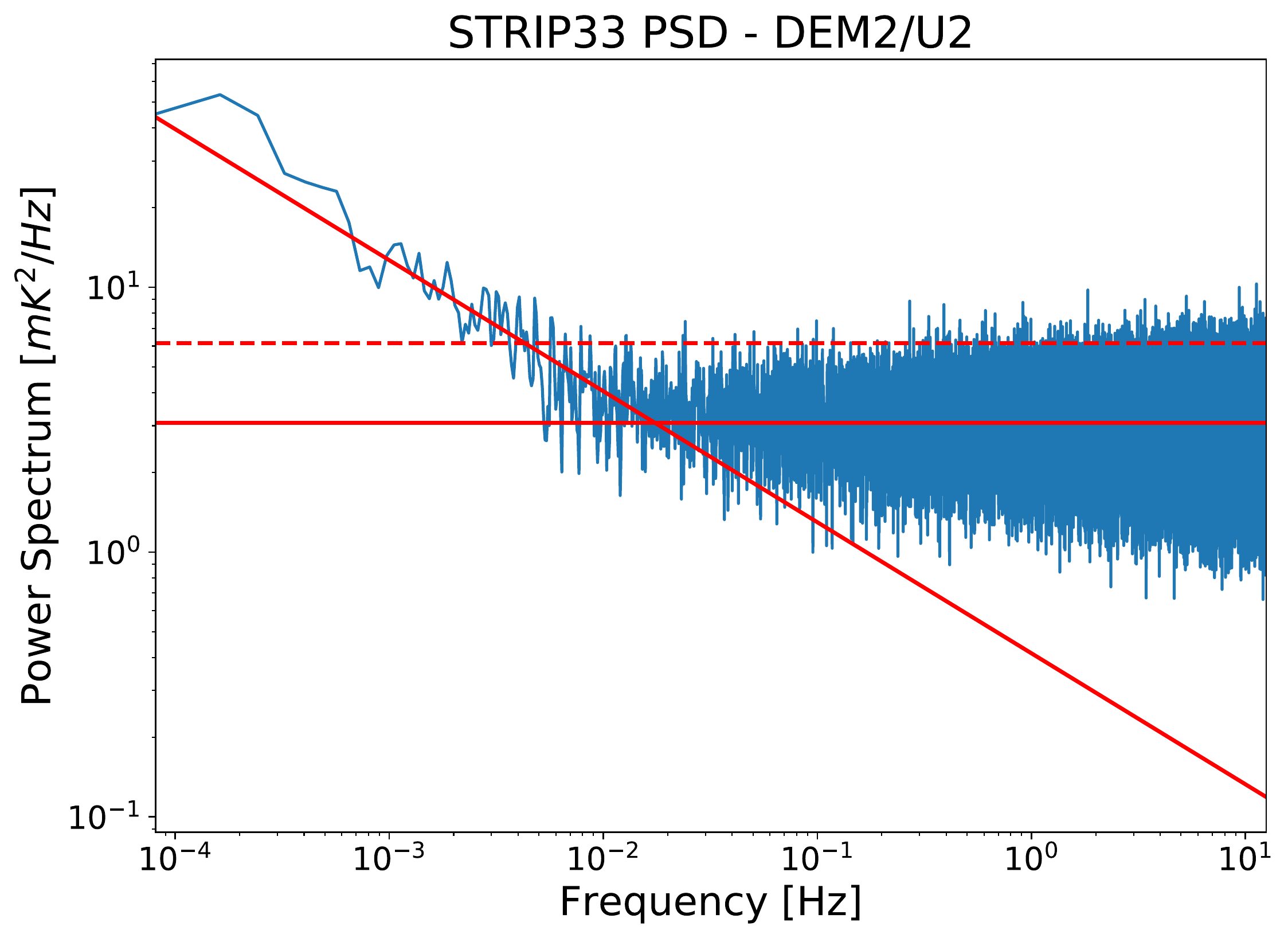}
\includegraphics[scale=0.29]{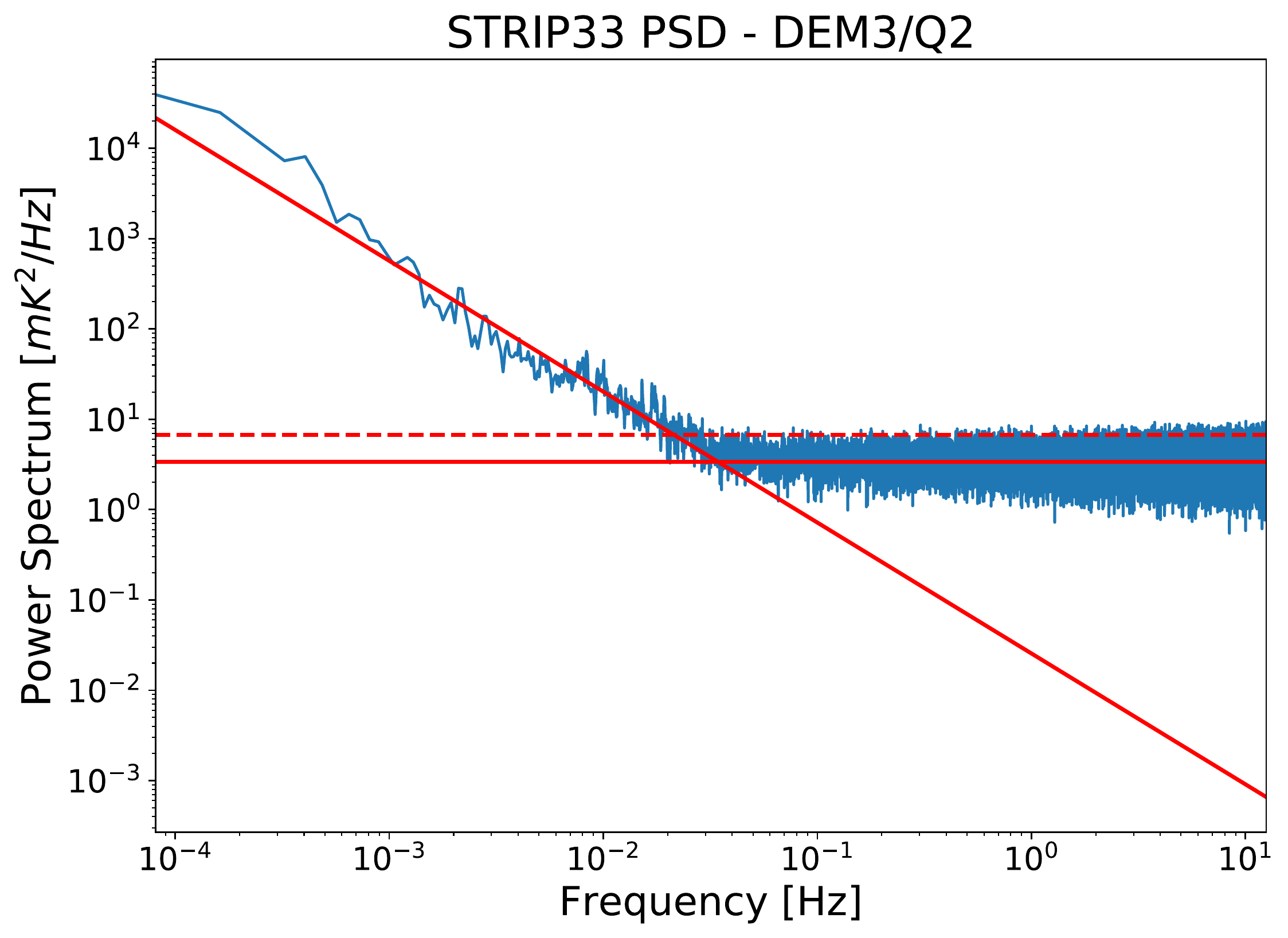}
\includegraphics[scale=0.29]{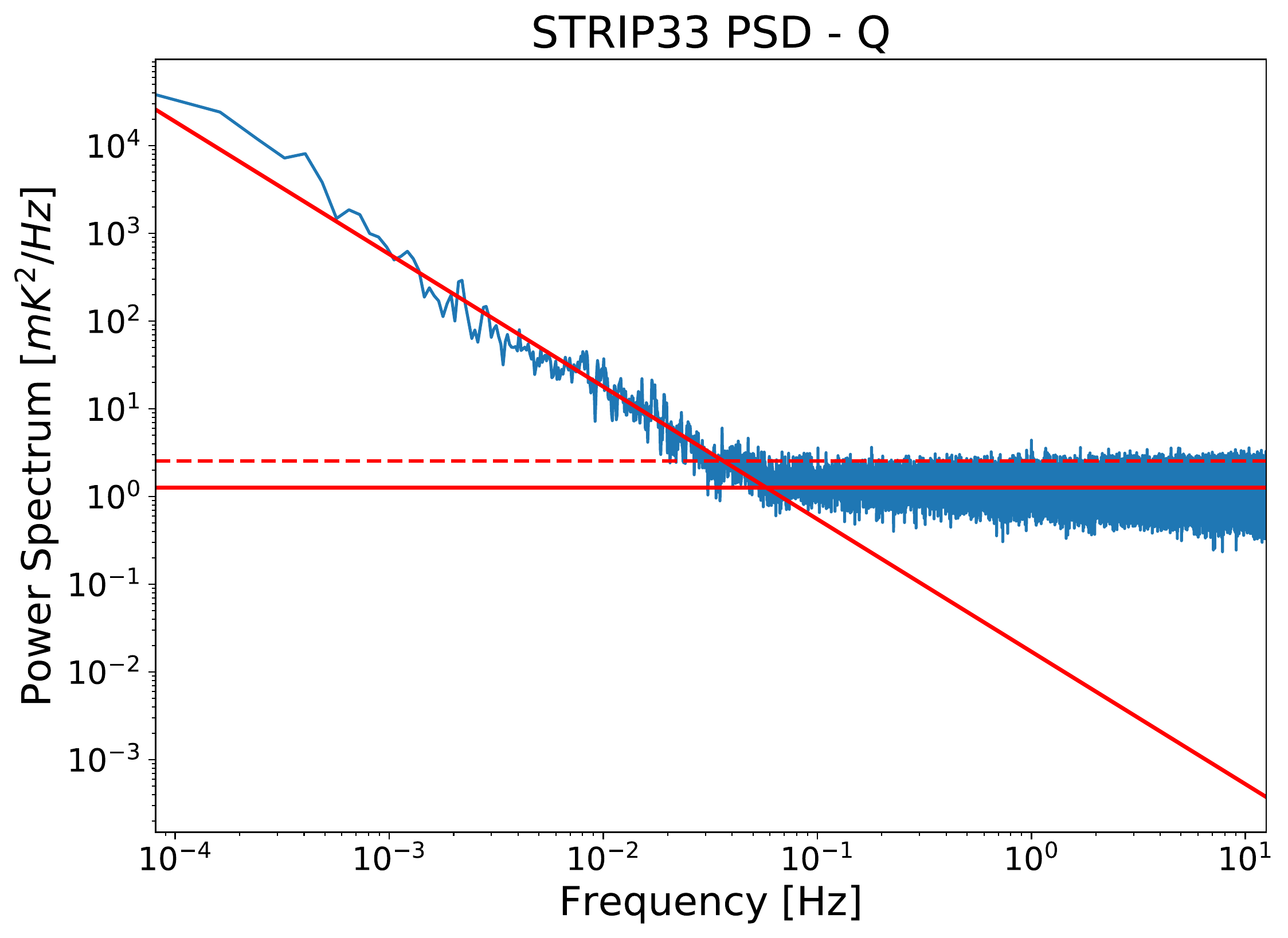}
\includegraphics[scale=0.29]{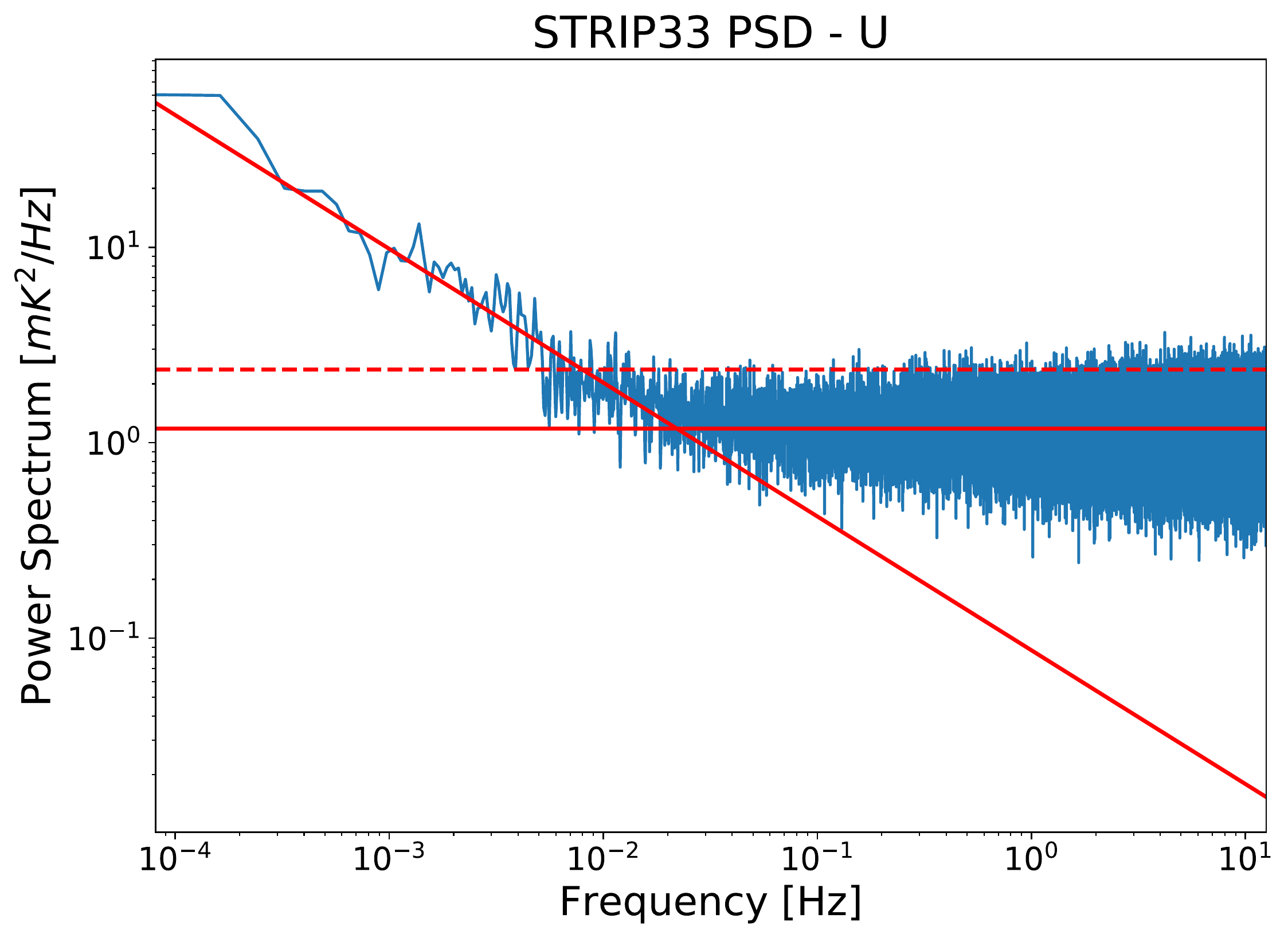}
\caption{The best fit of the noise spectra for the demodulated outputs and for the $Q$ and $U$ combinations for the long acquistion test of STRIP33. The dashed line corresponds to $2\,\sigma^2$.}
\label{demQUspectrum}
\end{figure}

Fig. \ref{demQUspectrum} shows the spectra, along with their best fits, for the double demodulated outputs and for the $Q$ and $U$ combinations in the case of STRIP33. In this case, the frequencies $\nu_\mathrm{low}$ and $\nu_\mathrm{high}$ were set respectively to $1\,\mathrm{Hz}$ and $0.02\,\mathrm{Hz}$.\par
\par
The uncertainty on the median value of the spectrum is given by the mean absolute deviation:
\be{sigmam}
  \centering \sigma_m = \frac1N \sum_{i=1}^{N} \abs{x_i - m} \,,
\ee
where $\mathrm{N}$ is the number of samples and $m$ is the median. \par
If $a \pm \sigma_a$ and $b \pm \sigma_b$ are the best fit coefficients of the pink spectrum, we have that $-a \pm \sigma_a$ is the slope $\alpha$ along with its uncertainty. \par
The uncertainty on the value of the knee frequency is given, according to the simple law of propagation of uncertainty, by:
\be{sigmanu}
  \centering \Delta \nu_\mathrm{knee} = \frac{\nu_\mathrm{knee}}{\abs{a}} \sqrt{ \Bigl (\frac{m - b}{a} \Bigr)^2 \cdot \sigma_a^2  + \sigma_b^2 + \sigma_m^2 } \,.
  \ee\par
\begin{figure}[H]
\centering
\includegraphics[scale=0.4]{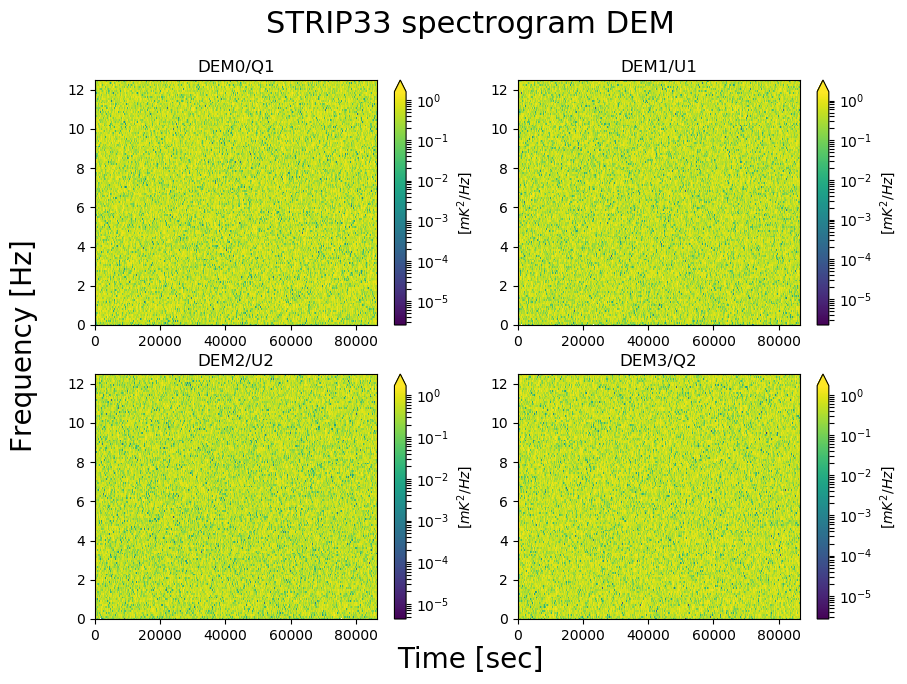}
\caption{The spectrogram of the demodulated outputs for the long acquistion test of STRIP33.}
\label{spgrams}
\end{figure}

The last element of the analysis is the spectrogram that represents the PSD as a function of time. It is useful to study the temporal evolution of possible spikes of the spectrum. In the case of STRIP33, no spikes are present (Fig. \ref{spgrams}).

\subsection{Results}
The noise analysis has been performed over a sub-sample of the whole batch: only $42$ Q-band polarimeters and $10$ W-band ones out of $70$ and $14$. The reason is that long acquisition tests started to be systematically run only a few months after the beginning of the campaign.\par
\begin{figure}[H]
\centering
\includegraphics[scale=0.33]{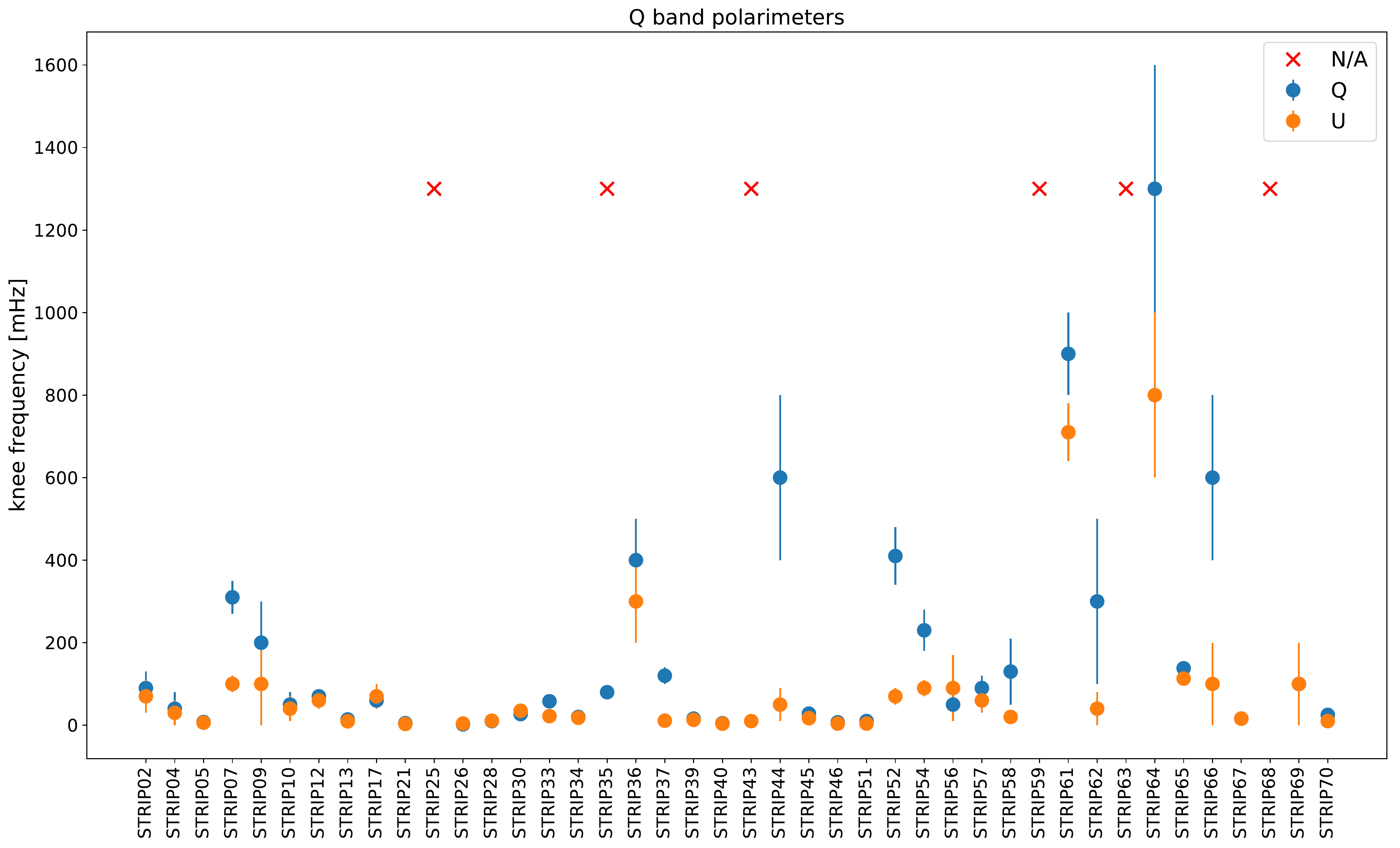}
\includegraphics[scale=0.33]{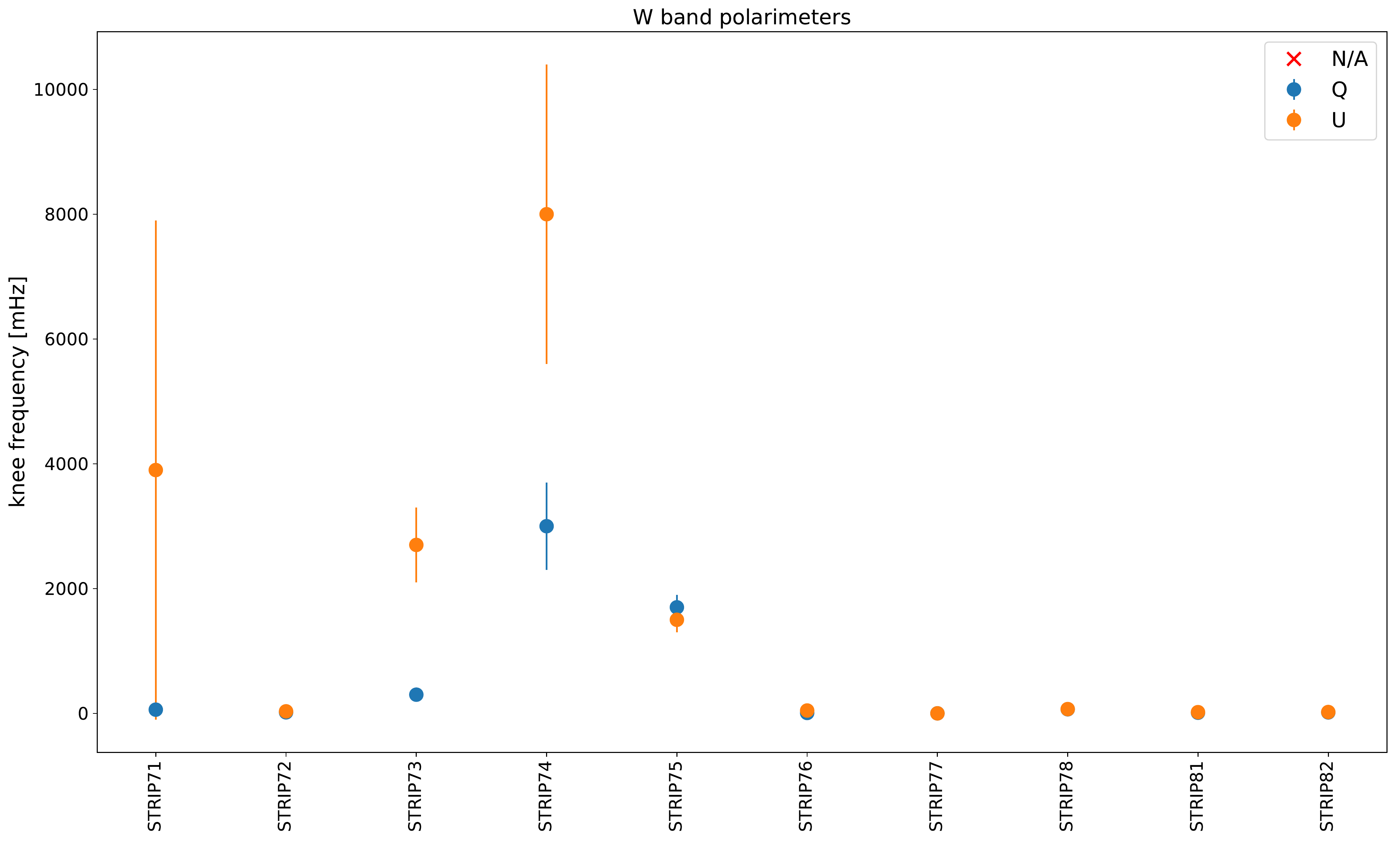}
\caption{Measured values of the knee frequency for the Q-band (up) and W-band (down). Blue and orange markers correspond respectively to the $Q$ and $U$ combinations. Red crosses represent outlier values for more than $3\sigma$.}
\label{kneefdist}
\end{figure}
\begin{figure}[H]
\centering
\includegraphics[scale=0.33]{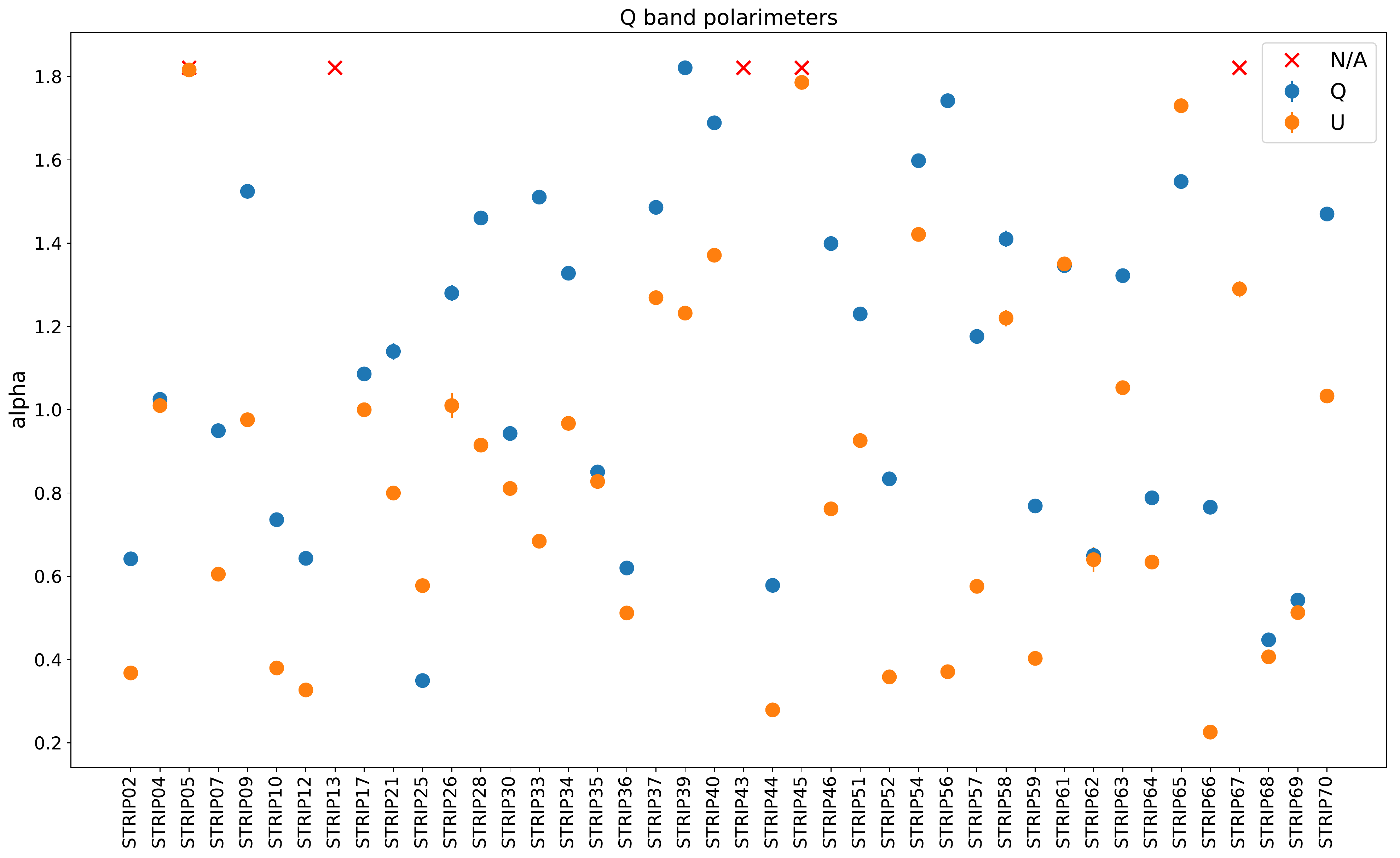}
\includegraphics[scale=0.33]{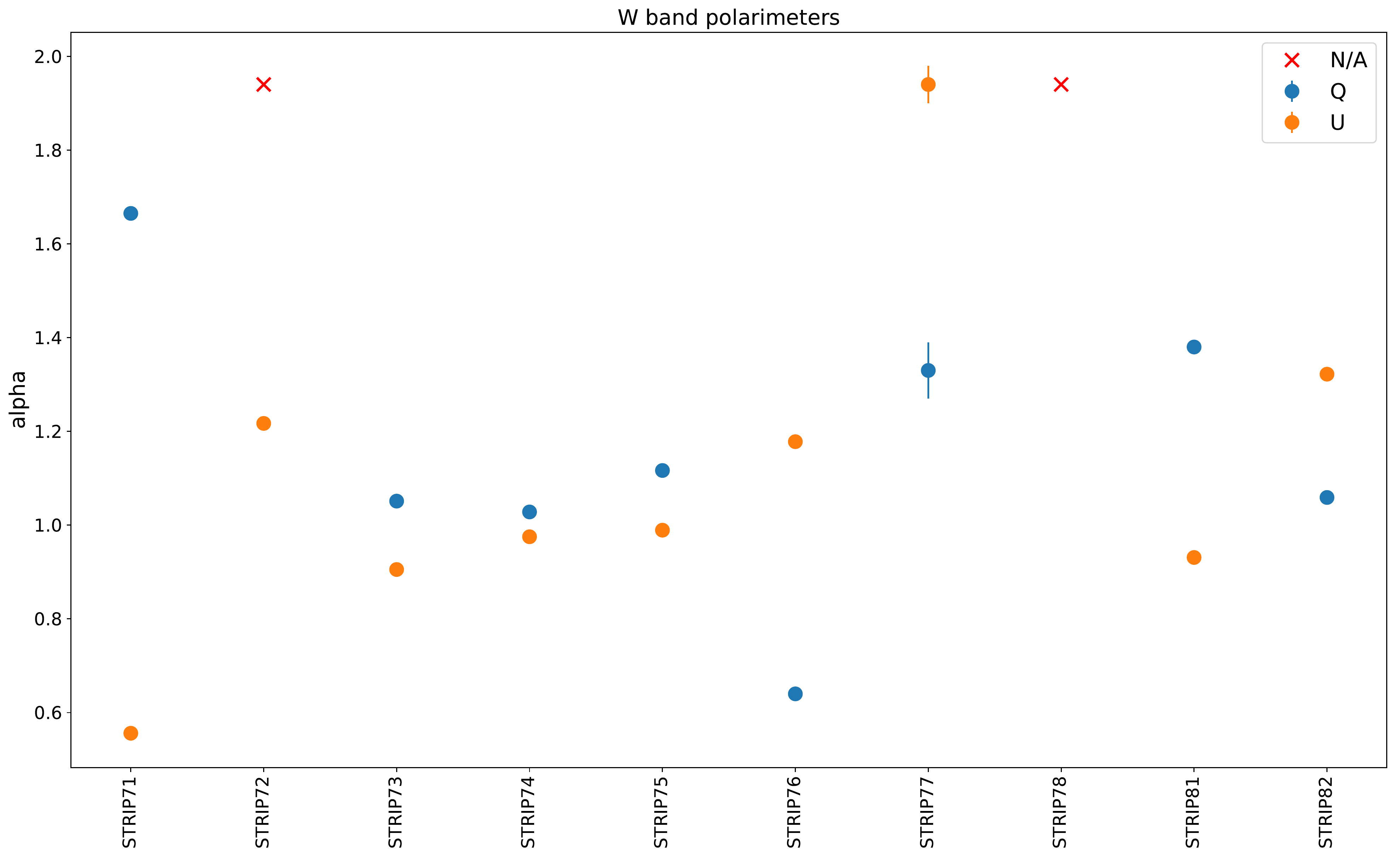}
\caption{Measured values of the slope of the pink spectrum for the Q-band (up) and W-band (down). Blue and orange markers correspond respectively to the $Q$ and $U$ combinations. Red crosses represent outlier values for more than $3\sigma$.}
\label{alphadist}
\end{figure}
\begin{figure}[H]
\centering
\includegraphics[scale=0.33]{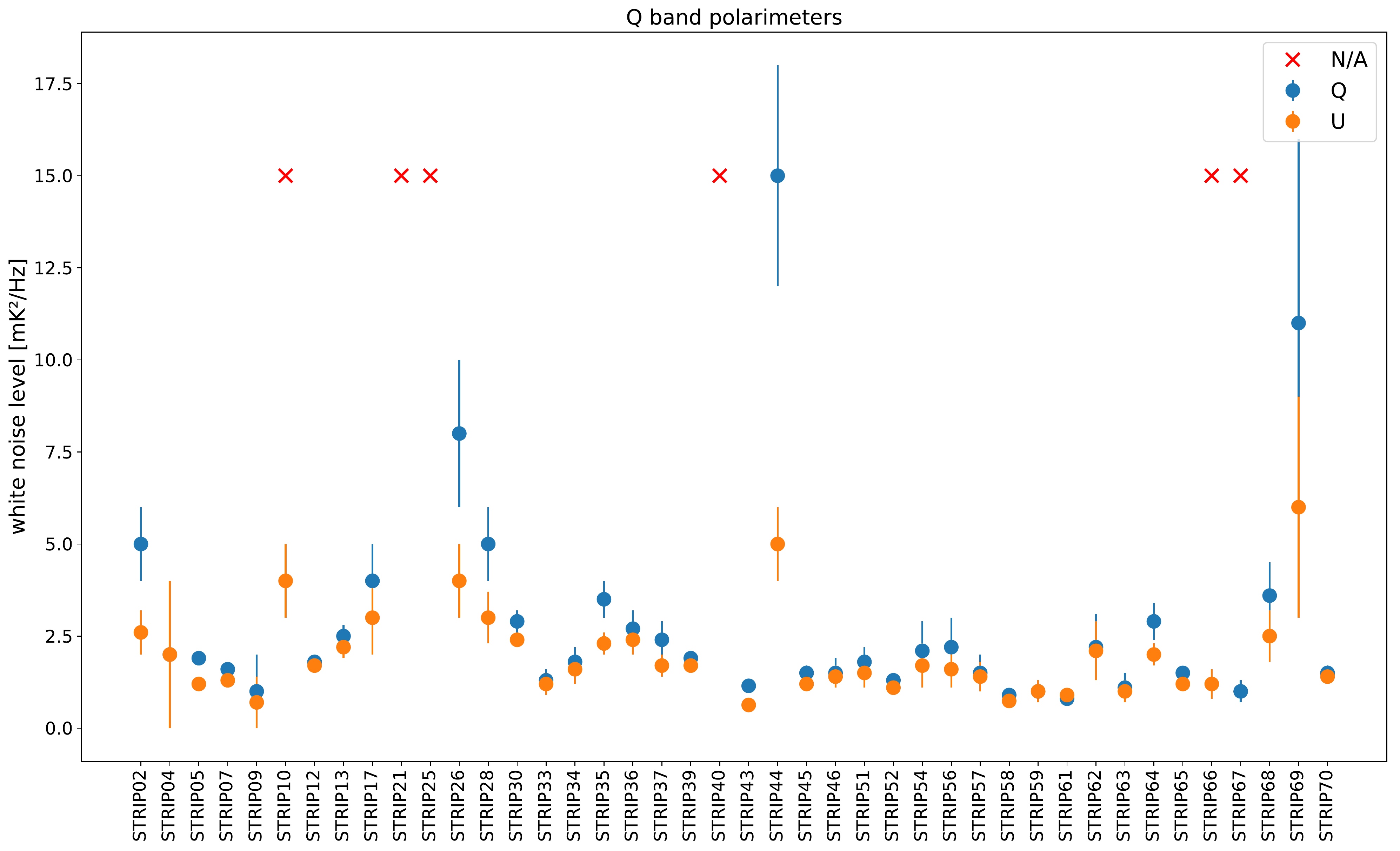}
\includegraphics[scale=0.33]{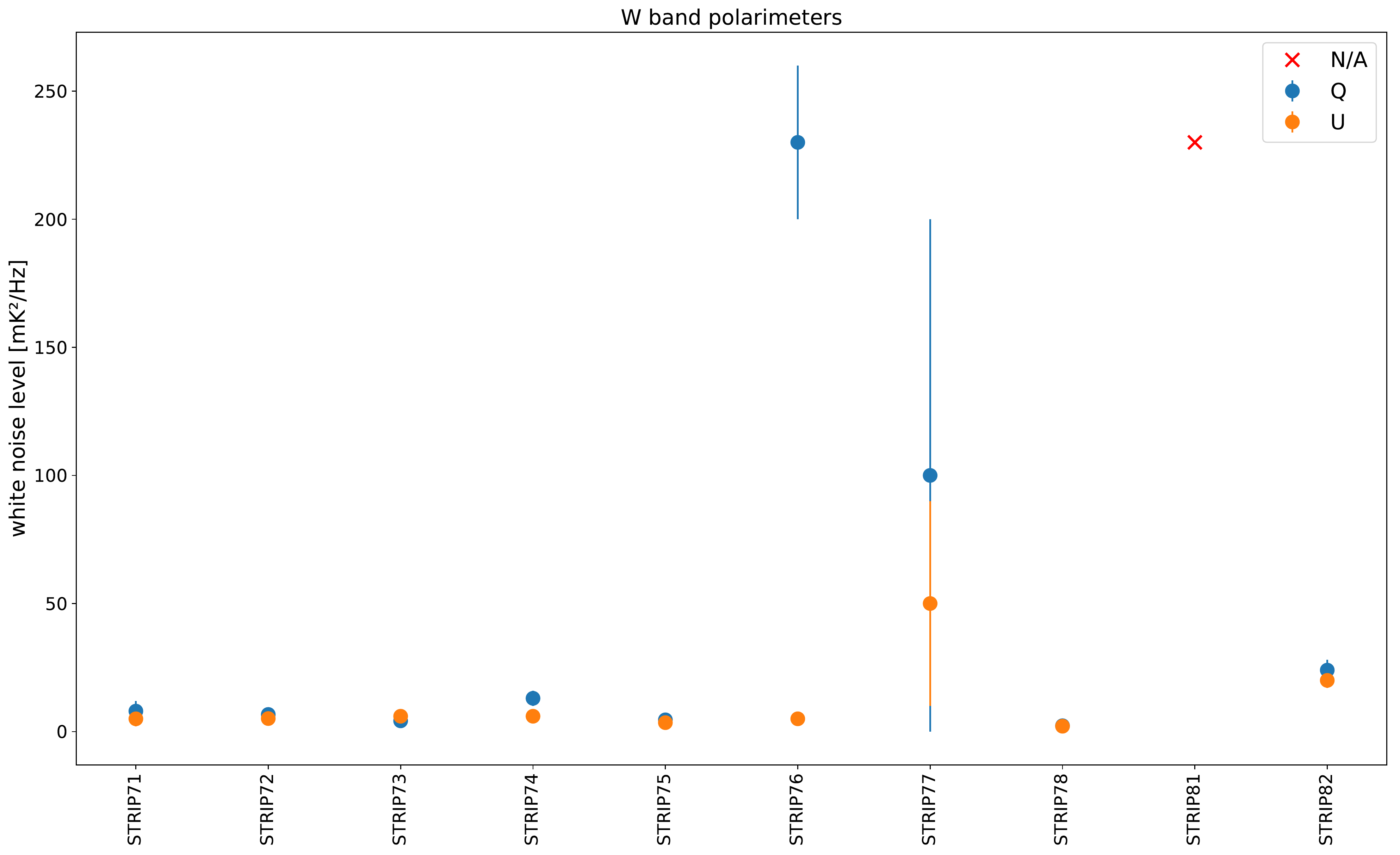}
\caption{Measured values of the white noise level for the Q-band (up) and W-band (down). Blue and orange markers correspond respectively to the $Q$ and $U$ combinations. Red crosses represent outlier values for more than $3\sigma$.}
\label{wnldist}
\end{figure}

I show,  in Figs. \ref{kneefdist}, \ref{alphadist} and \ref{wnldist} respectively, the knee frequency, slope and WNL distributions in terms of the $Q$ and $U$ combinations of the signal. The results of this analysis show non-Gaussian distributions with several outliers and very large error bars.\par
I report, in Table \ref{allparnuqu}, the median values for the $Q$ and $U$ combinations, for all the analyzed Q- and W-band polarimeters. Upper and lower error bars correspond respectively to $5\mathrm{th}$ and $95\mathrm{th}$ percentiles. To estimate these values, I rejected data that differ more than $3\sigma$ from the mean value (assuming a Gaussian distribution). \par
\begin{figure}[H]
\centering
\includegraphics[scale=0.29]{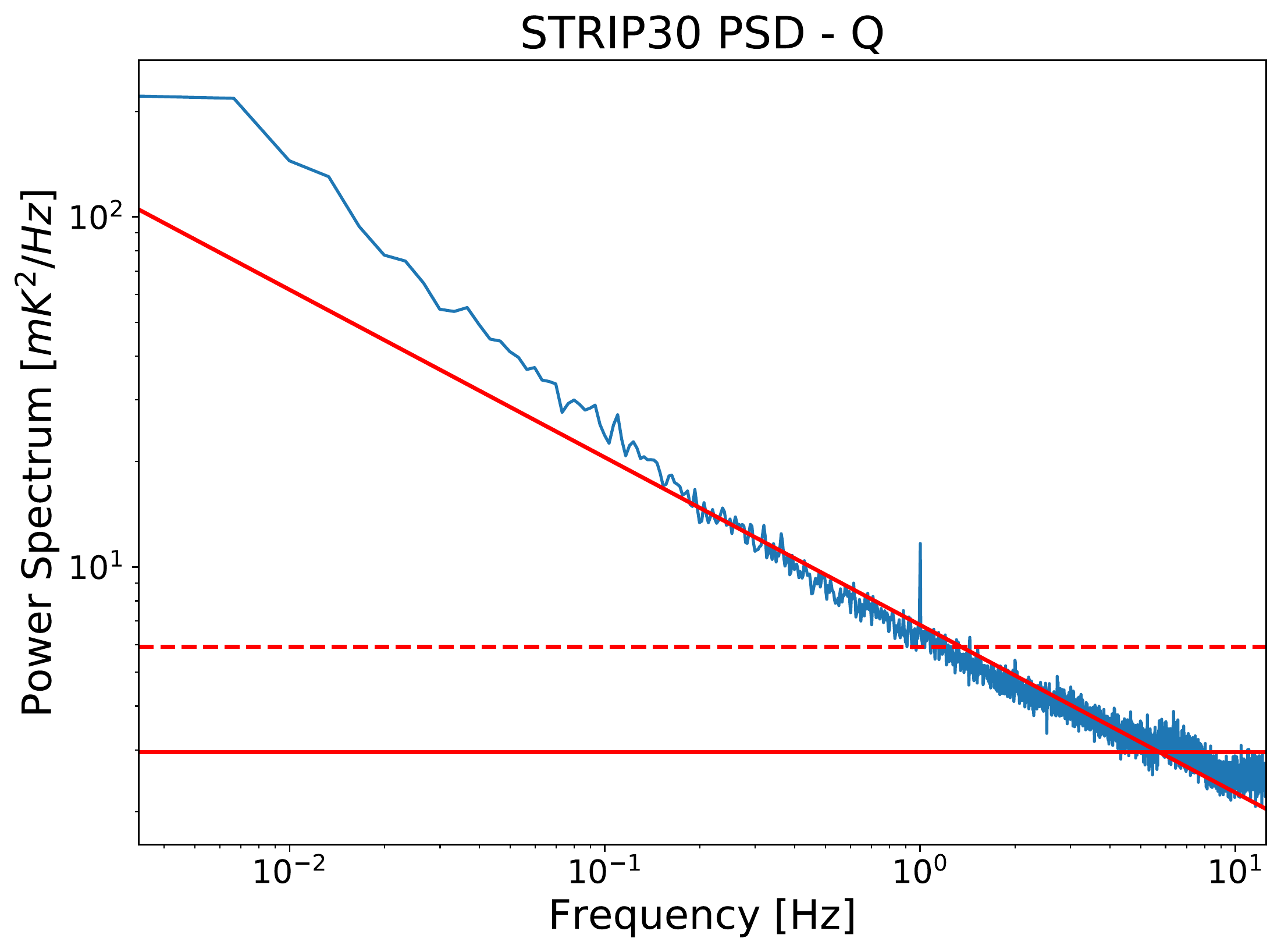}
\includegraphics[scale=0.29]{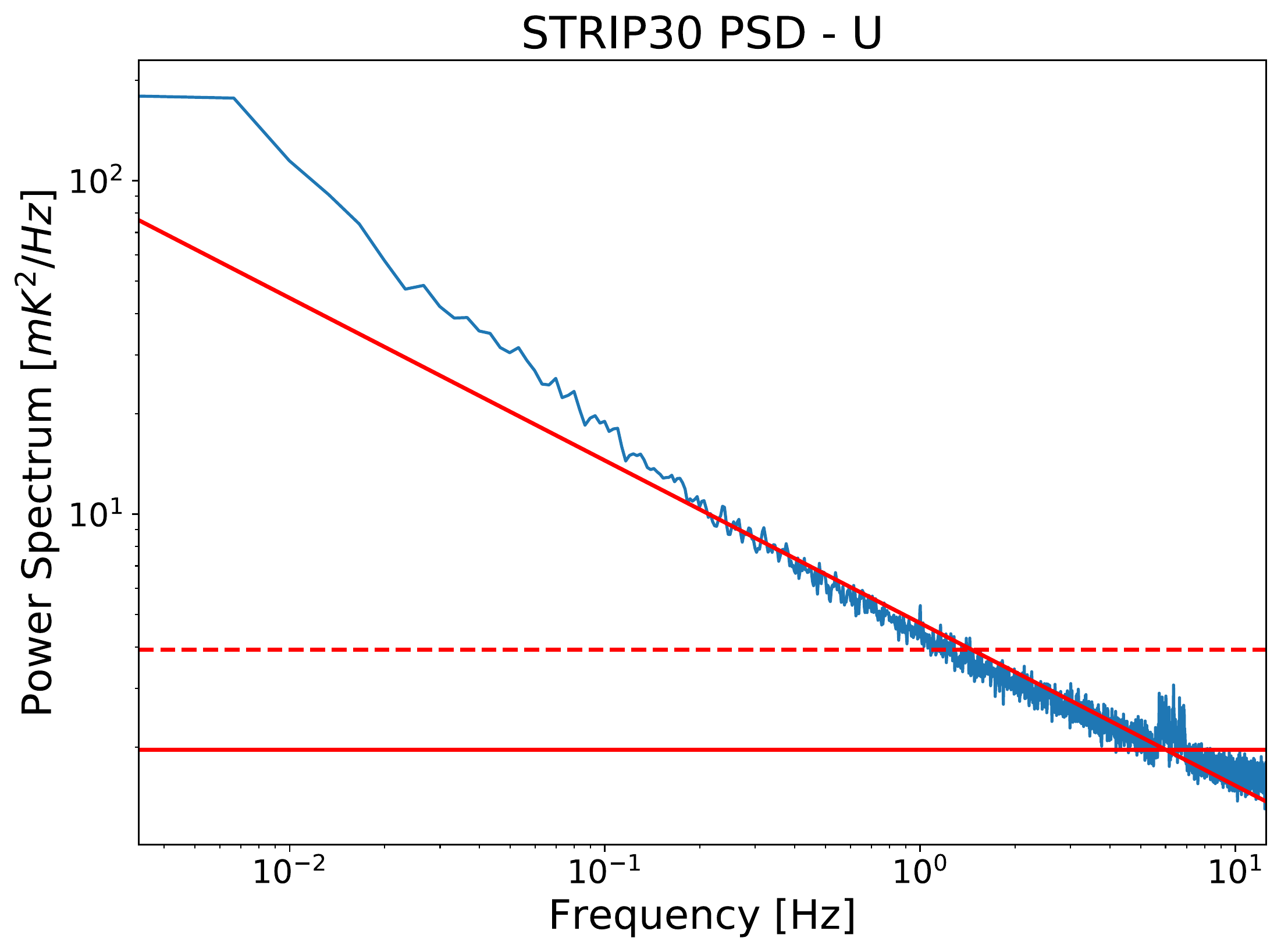}
\includegraphics[scale=0.29]{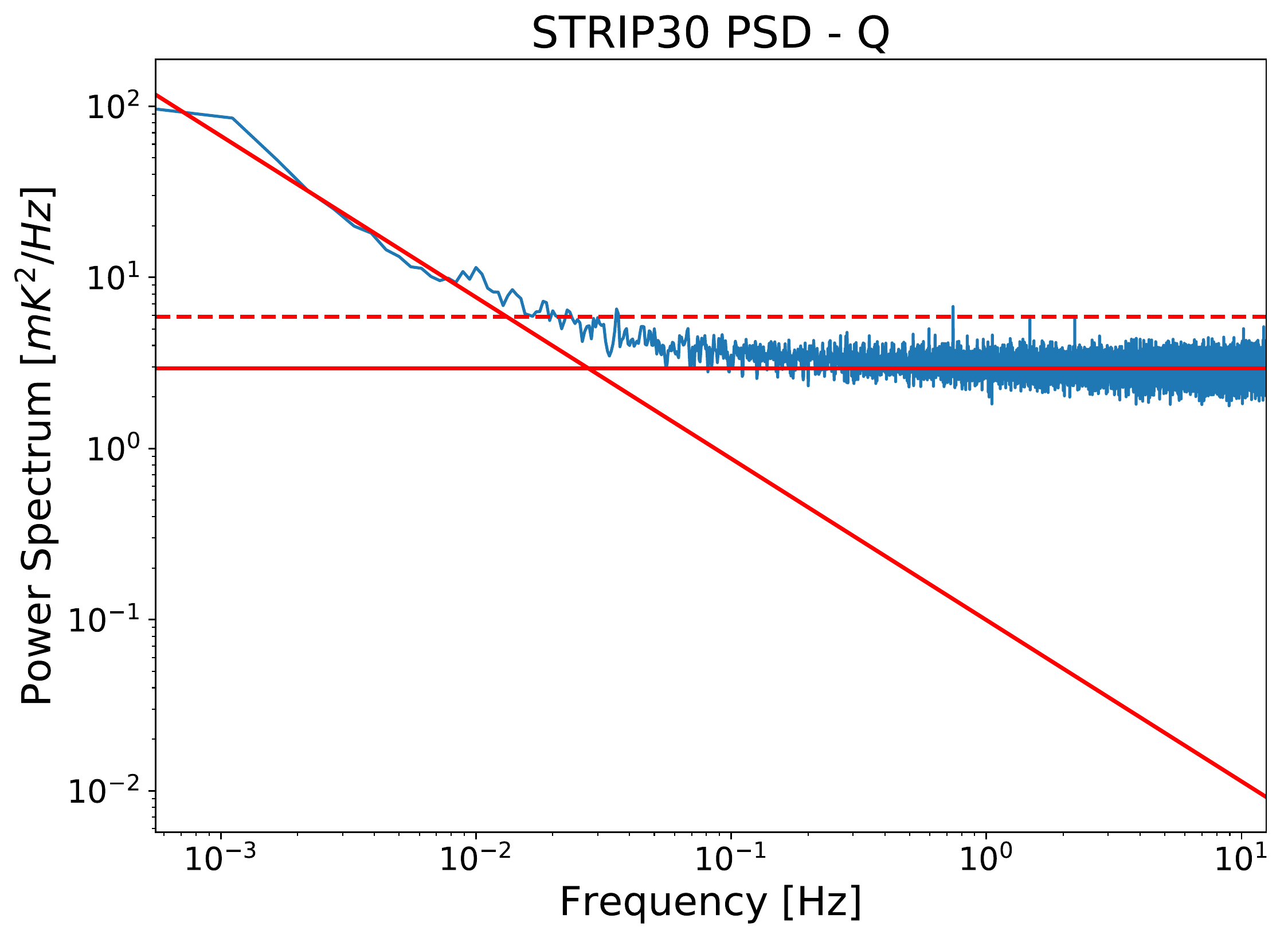}
\includegraphics[scale=0.29]{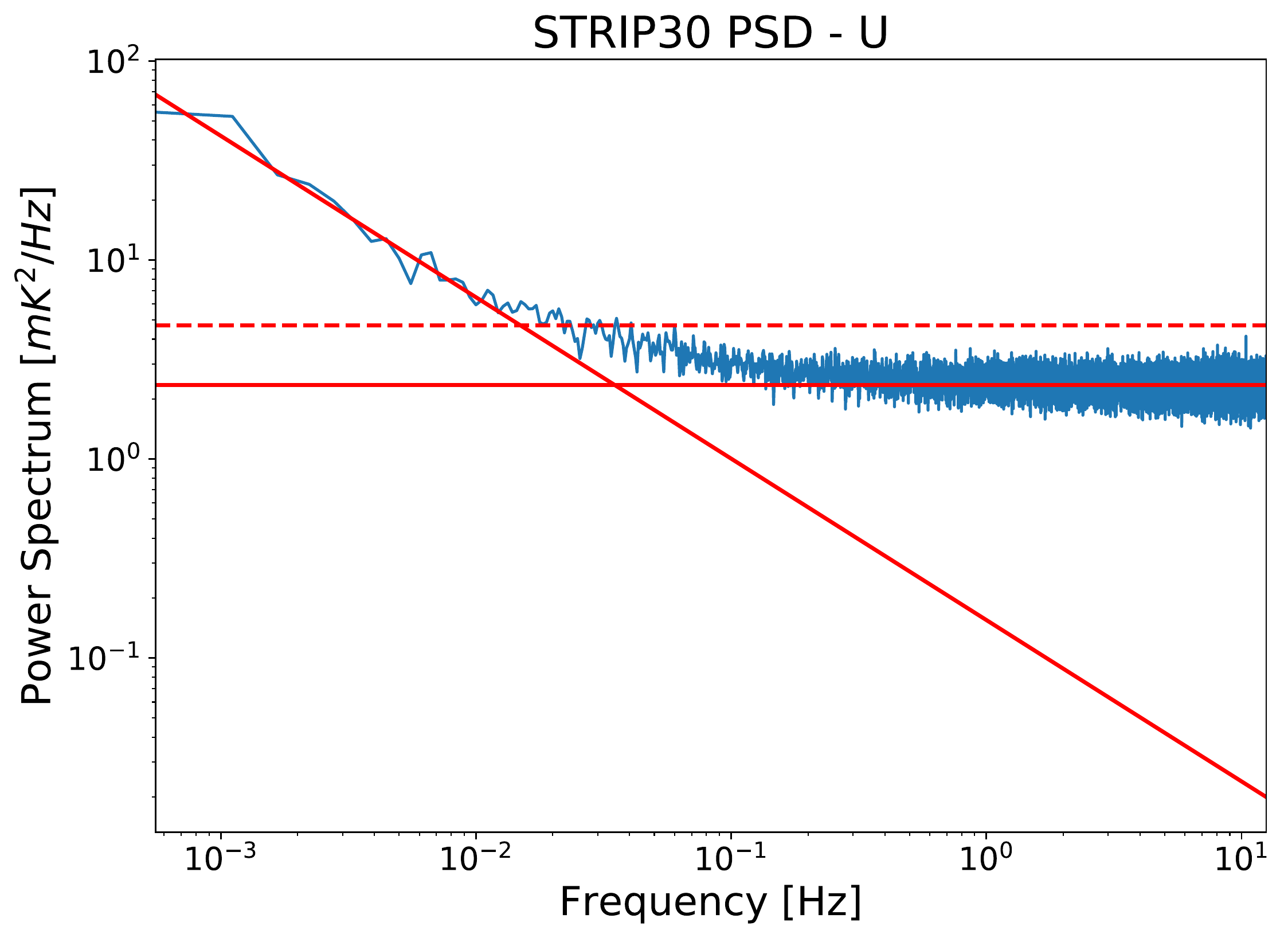}
\caption{Noise spectra of the $Q$ and $U$ combinations for STRIP30, before (upper panels) and after (lower panels) that the UPS was installed. The 1/$f$ noise is drastically reduced.}\label{repeted30}
\end{figure}

The presence of outliers and of some very large error bars is probably due to instabilities of the power generator at the time of the tests. An uninterruptible power supply (UPS) has been installed in the last month of the campaign. We repeated some tests and the results improved (Fig. \ref{repeted30}). It was not possible to repeat all the tests for lack of time. However, new measurements of the detector noise properties will be carried out during the system-level tests in Bologna. 

\begin{table}[H]
  \centering
  \begin{tabular}{|c|c|c|c|c|c|c|} \hline
 & \multicolumn{2}{|c|}{$\nu_\mathrm{knee}$ [$\mathrm{mHz}$]} & \multicolumn{2}{|c|}{$\alpha$} & \multicolumn{2}{|c|}{$\mathrm{WNL}$ [$\mathrm{mK^2 / Hz}$]} \\ \hline
  & $\mathrm{Q}$ & $\mathrm{U}$ & $\mathrm{Q}$ & $\mathrm{U}$ & $\mathrm{Q}$ & $\mathrm{U}$ \\ \hline
 Q-band & $60.0^{+5.0}_{-660}$ & $35.0^{+3.9}_{-382}$ & $1.14^{+0.52}_{-1.69}$ & $0.82^{+0.32}_{-1.73}$ & $1.9^{+1.0}_{-8.6}$ & $1.65^{+0.73}_{-4.15}$ \\ \hline
 W-band & $38.5^{+3.7}_{-2415}$ &  $57.5^{+9.3}_{-6155}$ & $1.08^{+0.77}_{-1.56}$ & $0.99^{+0.70}_{-1.69}$ & $8.1^{+3.1}_{-178}$ & $5.1^{+2.7}_{-38}$ \\ \hline
\end{tabular}\caption{Median values of the knee frequency, slope and WNL distributions. Error bars correspond to $5\mathrm{th}$ and $95\mathrm{th}$ percentiles.}\label{allparnuqu}
\end{table}\par
I report, in Table \ref{allparnuqufp}, the median values for the $Q$ and $U$ combinations, for the Q- and W-band polarimeters installed in the focal plane (whose mesurements exist). Uncertainties are given by MAD (Eq. \ref{MAD}). 

\begin{table}[H]
  \centering
  \begin{tabular}{|c|c|c|c|c|c|c|} \hline
 & \multicolumn{2}{|c|}{$\nu_\mathrm{knee}$ [$\mathrm{mHz}$]} & \multicolumn{2}{|c|}{$\alpha$} & \multicolumn{2}{|c|}{$\mathrm{WNL}$ [$\mathrm{mK^2 / Hz}$]} \\ \hline
  & $\mathrm{Q}$ & $\mathrm{U}$ & $\mathrm{Q}$ & $\mathrm{U}$ & $\mathrm{Q}$ & $\mathrm{U}$ \\ \hline
 Q-band & $75 \pm 59$ & $45 \pm 34$ & $1.3 \pm 0.4$ & $0.9 \pm 0.3$ & $1.8 \pm 0.5$ & $1.5 \pm 0.4$ \\ \hline
 W-band & $63 \pm 5$ &  $78 \pm 76$ & $1.2 \pm 0.2$ & $1.0 \pm 0.2$ & $6.1 \pm 2.4$ & $4.0 \pm 1.7$ \\ \hline
\end{tabular}\caption{Median values of the knee frequency, slope and WNL distributions for the polarimeters installed in the focal plane.}\label{allparnuqufp}
\end{table}\par

\section{Focal plane deployment}
A summary of the parameters derived from the unit-level tests are reported in Table \ref{allparbicocca}. As for the noise properties, I report the mean values of $\alpha$, $\nu_\mathrm{knee}$ and $\mathrm{WNL}$ between the $Q$ and $U$ combinations. \par
The results of STRIP unit-level tests on the Q-band revealed that only $52$ out of $68$ units were not damaged and usable on the STRIP instrument. As there are only $49$ Q-band horns in the focal plane, a selection of them has been done. Since the long acquistion tests were incomplete it has been decided to selected the $49$ polarimeters with the best noise temperatures. \par
\begin{figure}[H]
\centering
\includegraphics[scale=0.15]{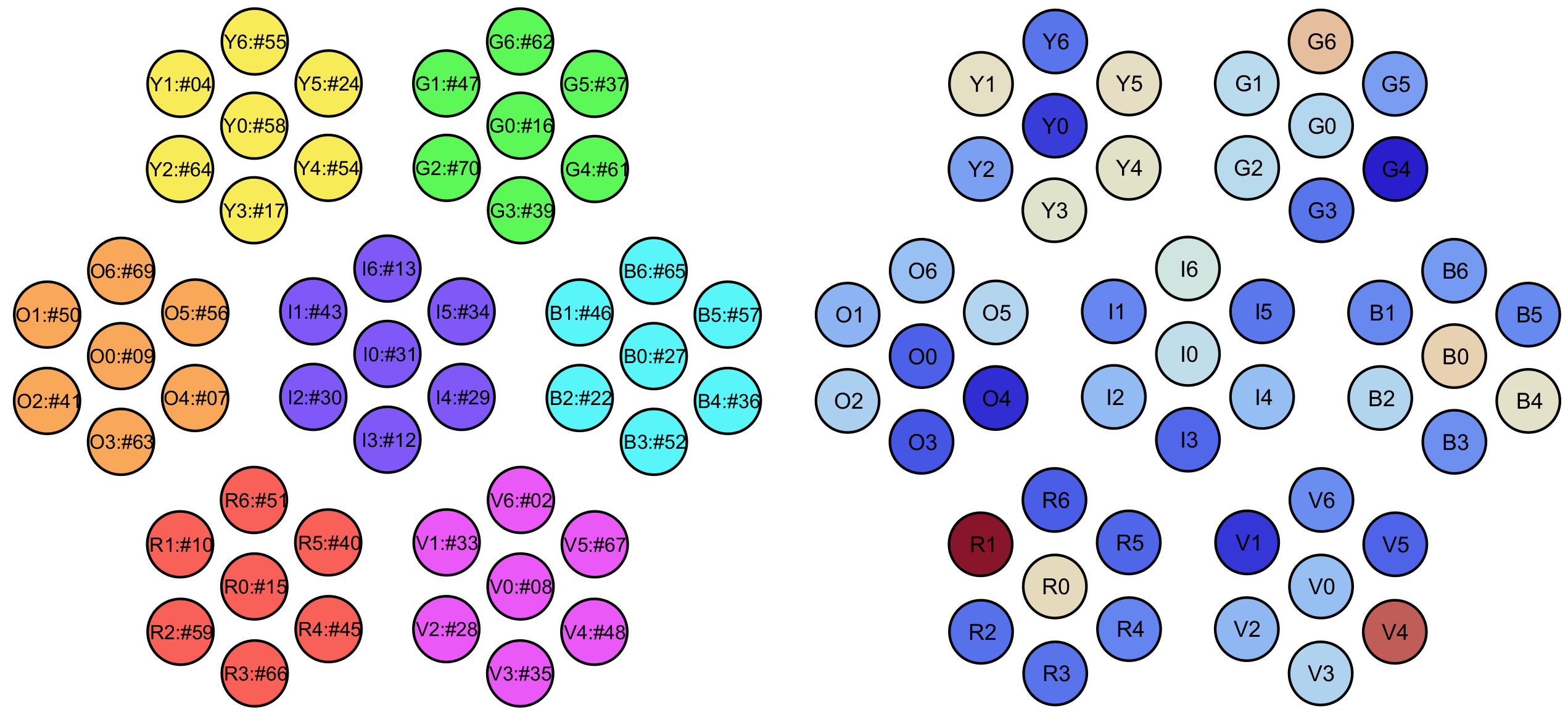}
\caption{\textit{Left}: Pairing between horns and Q-band polarimeters in the focal plane. \textit{Right}: The color scale shows the noise temperature of the polarimeters associated to each horn. Red and blue corresponds to higher and lower temperatures, respectively.}
\label{polpairins}
\end{figure}

As for the pairing between horn and polarimeter, since the orientation of the focal plane with respect to the telescope was not fixed yet, it has been decided to use random pairings. Fig. \ref{polpairins} shows the final focal plane deployment for the Q-band detectors. \par
The same criteria have been used to deploy on the focal plane the $10$ W-band polarimeters, out of $14$. \par

\vspace{10pt}
\par

\begin{footnotesize}
\begin{longtable}{c c c c c c c}
\hline
\hline
& $T_\mathrm{noise}$ & $\beta$ & $\nu_c$ & $\mathrm{WNL}$ & $\nu_\mathrm{knee}$ & $\alpha$ \\
& [$\mathrm{K}$] & [$\mathrm{GHz}$] & [$\mathrm{GHz}$] & [$\mathrm{mK}\times\sqrt{s}$] & [$\mathrm{mHz}$] & \\
\hline \\
\endhead
\input{strip_top-level_performance.tex} \\
\hline \caption{List of all the synthetic parameters derived from the unit-level tests for each polarimeter. No data are reported for damaged polarimeters.}\label{allparbicocca}
\end{longtable}
\end{footnotesize}

\section{A brief summary}
In this chapter, I reported the main results of the unit-level tests campaign during which more than $800$ tests on $68$ polarimeters have been performed in order to select the $55$ ($49$ Q-band and $6$ W-band) with the best performance. For each polarimeter, we ran three tests in a cryogenic chamber cooled down to $20\,\mathrm{K}$: a bandpass characterization, a Y-factor test to estimate the noise temperature and a long acquisition to measure the noise characteristics. \par
As for the bandpass characterization (Sect. \ref{bp}), a clear dichotomy between the Q-band polarimeters inherited by QUIET and the other ones appeared in terms of bandwidth as the former showed a larger one ($\beta \sim 8\,\mathrm{GHz}$) with respect to the others ($\beta \sim 6.5\div 7.0\,\mathrm{GHz}$). All the polarimeters presented a remarkable agreement in the central frequency. I found that, the average values of bandwidth and central frequency for the $49$ Q-band and $6$ W-band polarimeters implemented in the STRIP focal plane are: $\beta^Q = 7.3 \pm 0.8 \,\mathrm{GHz}$, $\nu_c^Q = 43.4 \pm 0.2 \,\mathrm{GHz}$, $\beta^W = 7.9 \pm 1.9 \,\mathrm{GHz}$ and $\nu_c^W = 97.5 \pm 1.8 \,\mathrm{GHz}$. \par
We estimated the noise temperature of the detectors using the so-called Y-factor test (Sect. \ref{tnoise}). To run the test, we changed the temperature of one of the two thermal loads installed on the cryogenic chamber from $10\,\mathrm{K}$ to $50\,\mathrm{K}$ by steps of $10\,\mathrm{K}$ and we measured the response of the polarimeter. For each polarimeter, we computed the noise temperature for each temperature pairs ($10$ if there were $5$ steps) and then we evaluated the median value. Then, I found the median value of the noise temperature for the $49$ Q-band and $6$ W-band polarimeters that make up the STRIP focal plane: $T_\mathrm{noise}^Q = 33.6 \pm 5.8 \,\mathrm{K}$; $T_\mathrm{noise}^W = 105.0 \pm 17.0 \,\mathrm{K}$. The analysis showed high uncertainties. Possible sources of systematic errors could be non-linearity in detector (or ADC) response, uncertainty in ADC offset, imperfect balance between the two legs, non-idealities in the polarimeters or in the set-up, etc. More investigations could not be conducted since fundamental HK parameters were not recorded by the software. \par
We performed the noise analysis (Sect. \ref{noisechara}) over a sub-sample of the whole batch of polarimeters: only $42$ Q-band and $10$ W-band ones out of $70$ and $14$. The analysis consisted in measuring the polarimeter response to the signals emitted by the two thermal loads at the temperature of $\sim 20\,\mathrm{K}$, for a long time, ranging from $1\,\mathrm{hour}$ to $60\,\mathrm{hour}$, with the phase switches set on the switching mode. No signal was injected by the RF generator. Data have been acquired with a sampling rate of $25\,\mathrm{Hz}$. The results of this analysis showed non-Gaussian distributions with several outliers and very large error bars, which were probably due to instabilities of the power generator. The median values of knee frequency, slope of the 1/$f$ spectrum and WNL distributions in terms of the $Q$ and $U$ combinations of the detector outputs, for the Q-band polarimeters installed in the focal plane are: $\nu_\mathrm{knee}^{Q} = 75 \pm 59\,\mathrm{mHz}$, $\nu_\mathrm{knee}^{U} = 45 \pm 34\,\mathrm{mHz}$, $\alpha^{Q} = 1.3 \pm 0.4$, $\alpha^{U} = 0.9 \pm 0.3$, $\mathrm{WNL}^{Q} = 1.8 \pm 0.5\,\mathrm{mK^2 / Hz}$, $\mathrm{WNL}^{U} = 1.5 \pm 0.4\,\mathrm{mK^2 / Hz}$; while for the W-band ones are: $\nu_\mathrm{knee}^{Q} = 63 \pm 5\,\mathrm{mHz}$, $\nu_\mathrm{knee}^{U} = 78 \pm 76\,\mathrm{mHz}$, $\alpha^{Q} = 1.2 \pm 0.2$, $\alpha^{U} = 1.0 \pm 0.2$, $\mathrm{WNL}^{Q} = 6.1 \pm 2.4\,\mathrm{mK^2 / Hz}$, $\mathrm{WNL}^{U} = 4.0 \pm 1.7\,\mathrm{mK^2 / Hz}$. \par

%% file: strip_top-level_performance.tex
STRIP01& --& --& --& --& --& --\\
STRIP02& $31.5^{+4.7}_{-7.2}$& $7.15\pm0.38$& $43.36\pm0.03$& $1.4\pm0.8$& $80.0\pm10.0$& $0.5\pm0.1$\\
STRIP03& --& --& --& --& --& --\\
STRIP04& $51.9^{+15.9}_{-18.6}$& $7.02\pm0.06$& $43.38\pm0.05$& $1.0\pm0.1$& $35.0\pm5.0$& $1.0\pm0.1$\\
STRIP05& $28.1^{+5.9}_{-7.1}$& $6.91\pm0.01$& $43.35\pm0.14$& $0.9\pm0.4$& $6.9\pm0.8$& $2.0\pm0.2$\\
STRIP06& --& --& --& --& --& --\\
STRIP07& $22.7^{+6.3}_{-4.4}$& $6.40\pm0.15$& $43.16\pm0.01$& $0.9\pm0.3$& $205.0\pm105.0$& $0.8\pm0.2$\\
STRIP08& $37.0^{+8.8}_{-11.9}$& $7.65\pm0.31$& $43.57\pm0.07$& --& --& --\\
STRIP09& $27.7^{+119.9}_{-92.7}$& $6.56\pm0.15$& $43.25\pm0.02$& $0.7\pm0.3$& $150.0\pm50.0$& $1.3\pm0.3$\\
STRIP10& $76.9^{+38.2}_{-109.8}$& $6.11\pm0.02$& $43.19\pm0.02$& $1.7\pm0.9$& $45.0\pm5.0$& $0.6\pm0.2$\\
STRIP11& --& --& --& --& --& --\\
STRIP12& $28.2^{+5.7}_{-8.0}$& $6.60\pm0.01$& $43.15\pm0.01$& $0.9\pm0.2$& $65.0\pm5.0$& $0.5\pm0.2$\\
STRIP13& $45.8^{+8.6}_{-6.1}$& $6.95\pm0.12$& $43.59\pm0.07$& $1.1\pm0.3$& $11.7\pm2.3$& $2.3\pm0.2$\\
STRIP14& --& --& --& --& --& --\\
STRIP15& $53.1^{+19.3}_{-11.6}$& $7.02\pm0.06$& $43.41\pm0.01$& --& --& --\\
STRIP16& $40.8^{+12.7}_{-7.9}$& $7.23\pm0.10$& $43.61\pm0.07$& --& --& --\\
STRIP17& $50.1^{+9.8}_{-17.5}$& $6.28\pm0.01$& $43.27\pm0.01$& $1.3\pm0.5$& $65.0\pm5.0$& $1.0\pm0.1$\\
STRIP18& --& --& --& --& --& --\\
STRIP19& --& --& --& --& --& --\\
STRIP20& --& $6.06\pm0.04$& $42.90\pm0.03$& --& --& --\\
STRIP21& $109.7^{+40.8}_{-71.8}$& $6.61\pm0.08$& $43.41\pm0.08$& $6.3\pm3.2$& $4.0\pm1.0$& $1.0\pm0.2$\\
STRIP22& $40.6^{+8.3}_{-6.3}$& $6.53\pm0.56$& $43.15\pm0.00$& --& --& --\\
STRIP23& --& --& --& --& --& --\\
STRIP24& $52.1^{+23.1}_{-25.0}$& $7.85\pm0.24$& $43.59\pm0.07$& --& --& --\\
STRIP25& --& $6.98\pm0.00$& $43.41\pm0.02$& --& $(5.3\pm0.8)\times 10^3$& $0.5\pm0.1$\\
STRIP26& $107.5^{+75.6}_{-127.8}$& $7.36\pm0.03$& $43.84\pm0.04$& $1.7\pm1.0$& $3.0\pm1.0$& $1.1\pm0.1$\\
STRIP27& $55.0^{+19.4}_{-14.9}$& $7.15\pm0.12$& $43.57\pm0.06$& --& --& --\\
STRIP28& $36.0^{+13.3}_{-16.1}$& $7.17\pm0.14$& $43.43\pm0.01$& $1.4\pm0.7$& $10.5\pm0.5$& $1.2\pm0.3$\\
STRIP29& $36.7^{+11.2}_{-11.1}$& $7.16\pm0.09$& $43.36\pm0.08$& --& --& --\\
STRIP30& $36.4^{+8.5}_{-8.9}$& $7.38\pm0.16$& $43.74\pm0.18$& $1.2\pm0.4$& $31.0\pm4.0$& $0.9\pm0.1$\\
STRIP31& $42.7^{+14.4}_{-12.0}$& $6.63\pm0.06$& $43.09\pm0.00$& --& --& --\\
STRIP32& --& --& --& --& --& --\\
STRIP33& $23.6^{+5.5}_{-4.2}$& $6.17\pm0.10$& $42.88\pm0.02$& $0.8\pm0.2$& $40.0\pm18.0$& $1.1\pm0.4$\\
STRIP34& $29.6^{+15.6}_{-65.2}$& $6.03\pm0.02$& $42.90\pm0.03$& $0.9\pm0.2$& $19.0\pm1.0$& $1.1\pm0.2$\\
STRIP35& $40.2^{+16.0}_{-17.2}$& $6.50\pm0.01$& $43.27\pm0.05$& $1.2\pm0.5$& $80.0\pm0.1$& $0.8\pm0.1$\\
STRIP36& $51.1^{+16.4}_{-29.8}$& $6.60\pm0.02$& $43.19\pm0.07$& $1.1\pm0.3$& $350.0\pm50.0$& $0.6\pm0.1$\\
STRIP37& $33.2^{+6.8}_{-8.7}$& $6.63\pm0.09$& $43.40\pm0.10$& $1.0\pm0.4$& $65.5\pm54.5$& $1.4\pm0.1$\\
STRIP38& --& --& --& --& --& --\\
STRIP39& $29.4^{+6.6}_{-8.6}$& $6.03\pm0.16$& $43.11\pm0.05$& $0.9\pm0.2$& $14.5\pm1.5$& $1.5\pm0.3$\\
STRIP40& $28.1^{+6.2}_{-5.7}$& $6.83\pm0.01$& $43.40\pm0.02$& $0.9\pm0.4$& $4.2\pm0.5$& $1.5\pm0.2$\\
STRIP41& $39.5^{+13.0}_{-21.4}$& $7.08\pm0.05$& $43.59\pm0.04$& --& --& --\\
STRIP42& --& --& --& --& --& --\\
STRIP43& $31.0^{+7.7}_{-10.7}$& $6.75\pm0.10$& $43.18\pm0.02$& $0.7\pm0.4$& $15.0\pm5.0$& $2.2\pm0.1$\\
STRIP44& $178.1^{+154.4}_{-3371.0}$& $6.32\pm0.19$& $43.39\pm0.11$& $2.2\pm1.6$& $325.0\pm275.0$& $0.4\pm0.1$\\
STRIP45& $30.8^{+8.1}_{-4.1}$& $6.41\pm0.06$& $43.24\pm0.05$& $0.8\pm0.3$& $22.5\pm5.5$& $2.0\pm0.2$\\
STRIP46& $31.1^{+7.6}_{-4.7}$& $6.33\pm0.05$& $43.12\pm0.05$& $0.9\pm0.2$& $5.5\pm1.5$& $1.1\pm0.3$\\
STRIP47& $41.5^{+11.4}_{-17.7}$& $6.38\pm0.20$& $43.09\pm0.07$& --& --& --\\
STRIP48& $68.8^{+18.4}_{-57.3}$& $6.77\pm0.18$& $43.22\pm0.01$& --& --& --\\
STRIP49& --& --& --& --& --& --\\
STRIP50& $35.6^{+4.3}_{-1609.2}$& $6.20\pm0.04$& $43.20\pm0.02$& --& --& --\\
STRIP51& $27.3^{+7.9}_{-4.5}$& $6.34\pm0.03$& $43.01\pm0.01$& $0.9\pm0.3$& $7.0\pm3.0$& $1.1\pm0.2$\\
STRIP52& $31.9^{+5.9}_{-6.2}$& $7.40\pm1.25$& $43.48\pm0.26$& $0.8\pm0.2$& $240.0\pm170.0$& $0.6\pm0.2$\\
STRIP53& --& --& --& --& --& --\\
STRIP54& $50.8^{+21.3}_{-15.9}$& $8.18\pm0.06$& $43.29\pm0.05$& $1.0\pm0.3$& $160.0\pm70.0$& $1.5\pm0.1$\\
STRIP55& $29.5^{+2.1}_{-13.1}$& $8.20\pm0.04$& $43.64\pm0.04$& --& --& --\\
STRIP56& $40.7^{+12.6}_{-18.3}$& $8.14\pm0.02$& $43.39\pm0.05$& $1.0\pm0.4$& $70.0\pm20.0$& $1.1\pm0.7$\\
STRIP57& $31.3^{+8.5}_{-20.7}$& $8.83\pm0.09$& $43.67\pm0.05$& $0.9\pm0.2$& $75.0\pm15.0$& $0.9\pm0.3$\\
STRIP58& $24.3^{+6.1}_{-4.4}$& $8.48\pm0.40$& $43.37\pm0.02$& $0.6\pm0.2$& $75.0\pm55.0$& $1.3\pm0.1$\\
STRIP59& $29.1^{+6.8}_{-5.6}$& $8.29\pm0.12$& $43.73\pm0.04$& $0.7\pm0.1$& $110.0\pm60.0$& $0.6\pm0.2$\\
STRIP60& --& --& --& --& --& --\\
STRIP61& $21.2^{+5.4}_{-4.1}$& $8.16\pm0.24$& $43.25\pm0.01$& $0.7\pm0.2$& $805.0\pm95.0$& $1.3\pm0.1$\\
STRIP62& $58.1^{+23.7}_{-35.6}$& $8.60\pm0.19$& $43.66\pm0.03$& $1.0\pm0.2$& $170.0\pm130.0$& $0.6\pm0.1$\\
STRIP63& $26.8^{+6.5}_{-4.9}$& $8.20\pm0.07$& $43.15\pm0.03$& $0.7\pm0.2$& $70.0\pm20.0$& $1.2\pm0.1$\\
STRIP64& $33.6^{+9.8}_{-5.5}$& $8.25\pm0.52$& $43.35\pm0.11$& $1.1\pm0.5$& $(1.1\pm0.2)\times 10^3$& $0.7\pm0.1$\\
STRIP65& $32.9^{+7.2}_{-4.5}$& $6.33\pm0.12$& $43.42\pm0.03$& $0.8\pm0.3$& $125.5\pm12.5$& $1.6\pm0.1$\\
STRIP66& $29.3^{+6.9}_{-33.1}$& $7.71\pm0.23$& $43.48\pm0.06$& $0.7\pm0.2$& $350.0\pm250.0$& $0.5\pm0.3$\\
STRIP67& $27.9^{+7.9}_{-54.6}$& $8.74\pm0.01$& $43.47\pm0.03$& $0.7\pm0.2$& $16.0\pm0.1$& $1.7\pm0.4$\\
STRIP68& $31.6^{+8.1}_{-6.2}$& $7.79\pm0.03$& $43.03\pm0.03$& $1.2\pm0.5$& $(5.5\pm0.2)\times 10^3$& $0.4\pm0.1$\\
STRIP69& $37.2^{+7.5}_{-4.7}$& $8.29\pm0.31$& $43.64\pm0.01$& $2.1\pm1.1$& $100.0\pm0.1$& $0.5\pm0.1$\\
STRIP70& $41.3^{+7.0}_{-5.6}$& $8.40\pm0.05$& $43.35\pm0.00$& $0.9\pm0.2$& $17.5\pm7.5$& $1.3\pm0.2$\\
STRIP71& $68.1^{+22.8}_{-20.1}$& $9.01\pm0.27$& $95.19\pm0.22$& $1.8\pm0.9$& $(2.0\pm1.9)\times 10^3$& $1.1\pm0.6$\\
STRIP72& $91.7^{+32.6}_{-30.5}$& $7.68\pm0.65$& $96.23\pm0.16$& $1.7\pm0.6$& $25.0\pm8.0$& $1.9\pm0.6$\\
STRIP73& $114.6^{+27.5}_{-20.5}$& $10.34\pm1.42$& $98.75\pm0.57$& $1.6\pm0.7$& $(1.5\pm1.2)\times 10^3$& $1.0\pm0.1$\\
STRIP74& $108.4^{+64.7}_{-25.4}$& $4.01\pm0.36$& $95.61\pm0.34$& $2.2\pm1.3$& $(5.5\pm2.5)\times 10^3$& $1.0\pm0.1$\\
STRIP75& $80.2^{+38.6}_{-17.2}$& $5.39\pm0.03$& $95.71\pm0.23$& $1.2\pm0.4$& $(1.6\pm0.1)\times 10^3$& $1.1\pm0.1$\\
STRIP76& $105.9^{+18.1}_{-14.2}$& $6.56\pm0.54$& $97.47\pm0.11$& $7.7\pm7.5$& $26.5\pm19.5$& $0.9\pm0.3$\\
STRIP77& $299.2^{+4815.2}_{-648.1}$& $8.27\pm0.42$& $98.56\pm0.13$& $6.1\pm3.5$& $0.8\pm0.2$& $1.6\pm0.3$\\
STRIP78& $102.5^{+6795.4}_{-38.2}$& $7.41\pm0.74$& $98.67\pm0.01$& $1.0\pm0.2$& $67.5\pm1.5$& $3.6\pm0.2$\\
STRIP79& --& --& --& --& --& --\\
STRIP80& --& --& --& --& --& --\\
STRIP81& $108.4^{+61.1}_{-34.4}$& $8.64\pm0.57$& $99.76\pm0.40$& $1.7\pm1.1$& $15.5\pm4.5$& $1.2\pm0.2$\\
STRIP82& $183.2^{+252.1}_{-55.1}$& $6.10\pm0.40$& $97.33\pm0.10$& $3.3\pm1.6$& $19.5\pm3.5$& $1.2\pm0.1$\\
STRIP83& --& --& --& --& --& --\\
STRIP84& --& --& --& --& --& --\\

%% file: PartII.tex
\part*{Part II\vspace{0.5cm}\\The LSPE/STRIP scanning strategy}

%% file: chap5.tex
\chapter{Stripeline: the STRIP simulation pipeline}
\label{Chap:5}
\thispagestyle{plain}
In this chapter, I describe \textit{Stripeline}, the STRIP simulation pipeline, which allows end-to-end simulations of STRIP measurements. The code is written in the Julia programming language\footnote{\url{https://julialang.org/}.} and it is available online\footnote{\url{https://github.com/lspestrip/Stripeline.jl}.}. Julia enables us to execute high-performance for loops, while maintaining an user friendly interface as well as the Python portability. \par
Stripeline has been developed in the last three years by the STRIP data analysis team in Milan. In particular, I was mostly involved in developing the pointings generation codes, which I describe in the next sections. \par

\section{Software overview}
Stripeline allow us to simulate a STRIP observation of the sky through several modules that: collect all the information about the focal plane components (horn positions, horn/detector pairings, detector properties, etc.), simulate the scanning strategy, produce realizations of pseudo-instrumental noise and compute output maps from \textit{time-ordered data} (TOD). \par

\subsection{Instrument database}
The \textit{instrumentdb} module of Stripeline contains information about all the STRIP feedhorns and detectors. In particular, for each horn the database stores the optical characteristics (FWHM, spillover, cross-polarization level, directivity and ellipticity) as well as the orientation and the polarization angle\footnote{For each horn, the polarization angle is the reference angle starting from which the polarization of the incoming signal is measured.}. For each polarimeter, the database stores: central frequency, bandwidth, noise temperature, knee frequency, slope of the pink spectrum and white noise level. Moreover, it knows the appropriate pairing between each horn and the corresponding polarimeter. \par 
The instrument database retrieves information about horns and detectors from a YAML file\footnote{\url{https://yaml.org/}.}. There are a set of YAML files containing the default configuration for the STRIP instrument in the repository.\par

\subsection{Pointings generation}
The \textit{scanning} module of Stripeline implements several functions to simulate the motion of the telescope by computing the pointing of each horn as a function of time. The basic rotations around the three main axes of the telescope (see Fig. \ref{telesc}) are implemented through quaternions. Several methods to perform coordinate conversion, from local to absolute reference systems (Sect. \ref{pongen}), are implemented. The main output of this module is a sequence of sky coordinates and polarization angles for each horn. \par

\subsection{Map-making}
The Stripeline \textit{mapmaker} and \textit{noisegeneration} modules implement functions to convert TODs into maps and to produce realizations of pseudo-instrumental noise. We have realized two map-makers for STRIP: a \textit{binner} and a \textit{destriper}. We use the first one in the more simplistic case of TOD affected by white noise only. We use the second one when both white and 1/$f$ noises affect data.\par
\begin{figure}[H]
\centering
\includegraphics[scale=0.32]{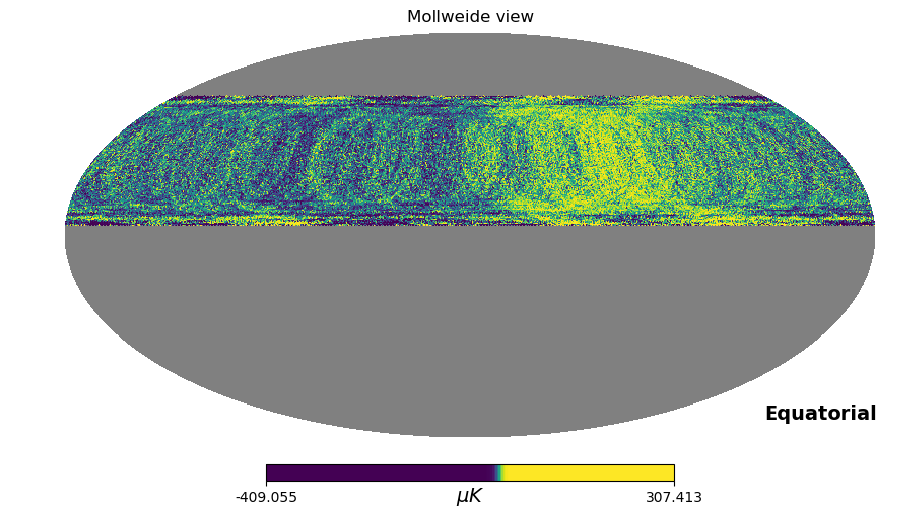}
\caption{Residual noise in a binned map obtained from a TOD affected by 1/$f$ noise. Stripes are due to correlated noise among close pixels.}\label{striper}
\end{figure}

The binning technique consists in computing the average value of the signal in each pixel of the sky. We perform a weighted average where each sample is weighted by the inverse of the white noise variance of the polarimeter that have observed a given pixel at the corresponding time. In this way, noisier polarimeters count less in the estimation of the map. This method is not effective when the TOD is affected by 1/$f$ noise because correlated noise influences measurements that are temporally close, leading to stripe-like structures in the sky map (Fig. \ref{striper}). In this case we use a destriper.\par
 \par
\begin{figure}[H]
\centering
\includegraphics[scale=0.4]{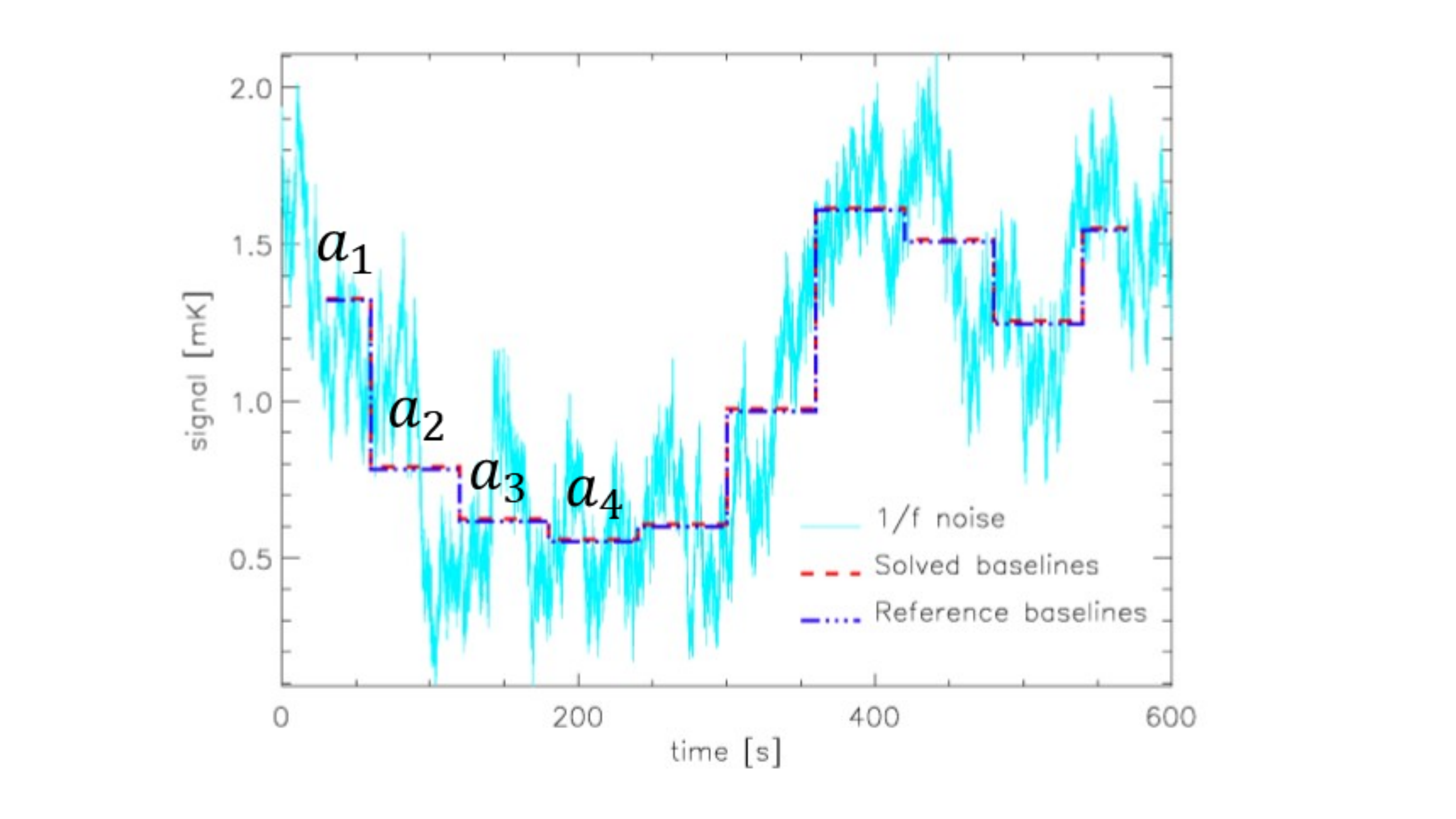}
\caption{A simulated TOD affected by 1/$f$ noise, modeled through baselines.}\label{baselines}
\end{figure}

The destriping tecnique is based on the assumptions that: ($\mathrm{i}$) the noise in TOD is the sum of white noise and 1/$f$ noise, and ($\mathrm{ii}$) the 1/$f$ noise can be modeled as a sequence of constant values, called \textit{baselines} (Fig. \ref{baselines}). In this way, the TOD, $y$, can be written as:
\be{TODmap}
y = P \cdot x + n \,,
\ee
where: $P$ is the pointing matrix, which is a $\mathrm{number \ of \ samples} \times \mathrm{number \ of \ sky \ pixels}$ sparse matrix filled with $1$ in correspondence of the observed pixels; $x$ is the unknown sky (I, Q or U); $n$ is the noise, which, in turn, can be modeled as:
\be{noiser}
n = F \cdot a + w \,,
\ee
where $F$ is a matrix that fixes the length of each baseline, $a$ is a vector containing the baseline values and $w$ is the white noise.\par 
 \par
It is possible to find $a$ and $x$ by solving Eqs. \ref{TODmap} and \ref{noiser} through a maximum-likelihood estimation. However, the matrices $P$ and $F$ cannot be explicitly computed because this would require too much memory\footnote{The pointing matrix, $P$, should contain $M \times N$ floating-points where $M$ is the number of pixels and $N$ is the number of samples. In the case of STRIP, the maps have $\simeq 800000$ pixels and there are: $49\;\mathrm{polarimeters} \times 2\;\mathrm{years} \times 365\;\mathrm{days} \times 86400\;\mathrm{seconds} \times 50\;\mathrm{Hz} \simeq 1.5^{11}\;\mathrm{samples}$, corresponding to $\simeq 1.2\;\mathrm{Terabyte}$ of data.}. To avoid this, we use an iterative method, the so-called \textit{conjugate gradient method}. This method is carried out by generating a succession of search directions, $p_k$, called \textit{conjugate gradients}, which are orthogonal among them. The solution at the $k$-th step is obtained from the solution at the previous step by adding $p_k$ and the iteration is repeated until convergence. For more detail about the destriper tecnique see, e.g., \citet{Suonio}.\par
\begin{figure}[H]
\centering
\includegraphics[scale=0.33]{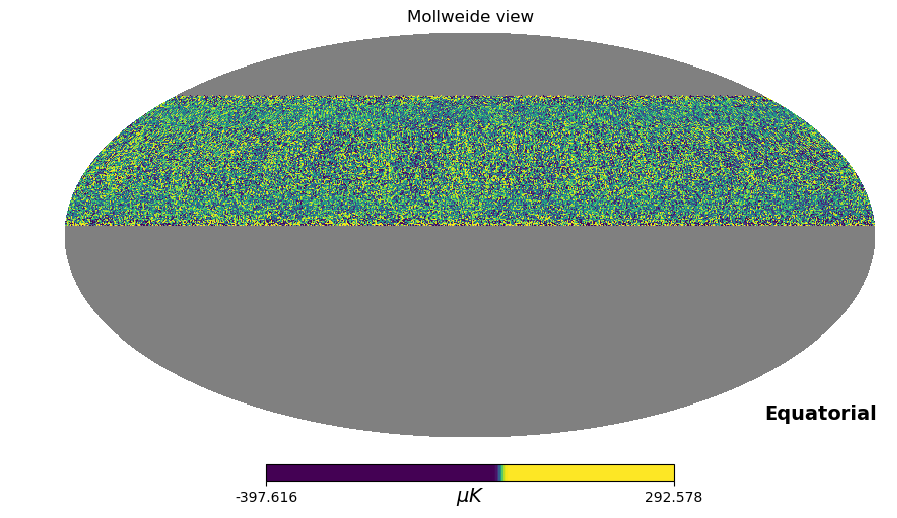}
\caption{Residual noise in a destriped map obtained from a TOD affected by 1/$f$ noise.}\label{striper2}
\end{figure}

The destriping technique allows us to remove the 1/$f$ noise from the maps, as shown in Fig. \ref{striper2}.
 \par
%% \begin{figure}[H]
%% \centering
%% \includegraphics[scale=0.45]{binned-destriped_map.png}
%% \caption{Difference between the residual noise in a binned and a destriped map, obtained from a TOD affected by 1/$f$ noise. The stripe-like structures are highlighted.}\label{striper3}
%% \end{figure} \par

\section{Coordinate reference systems and transformations}\label{pongen}
To simulate the pointing of each horn as a function of time, we must to define proper coordinate reference systems. \par
\begin{figure}[H]
\centering
\includegraphics[scale=0.3]{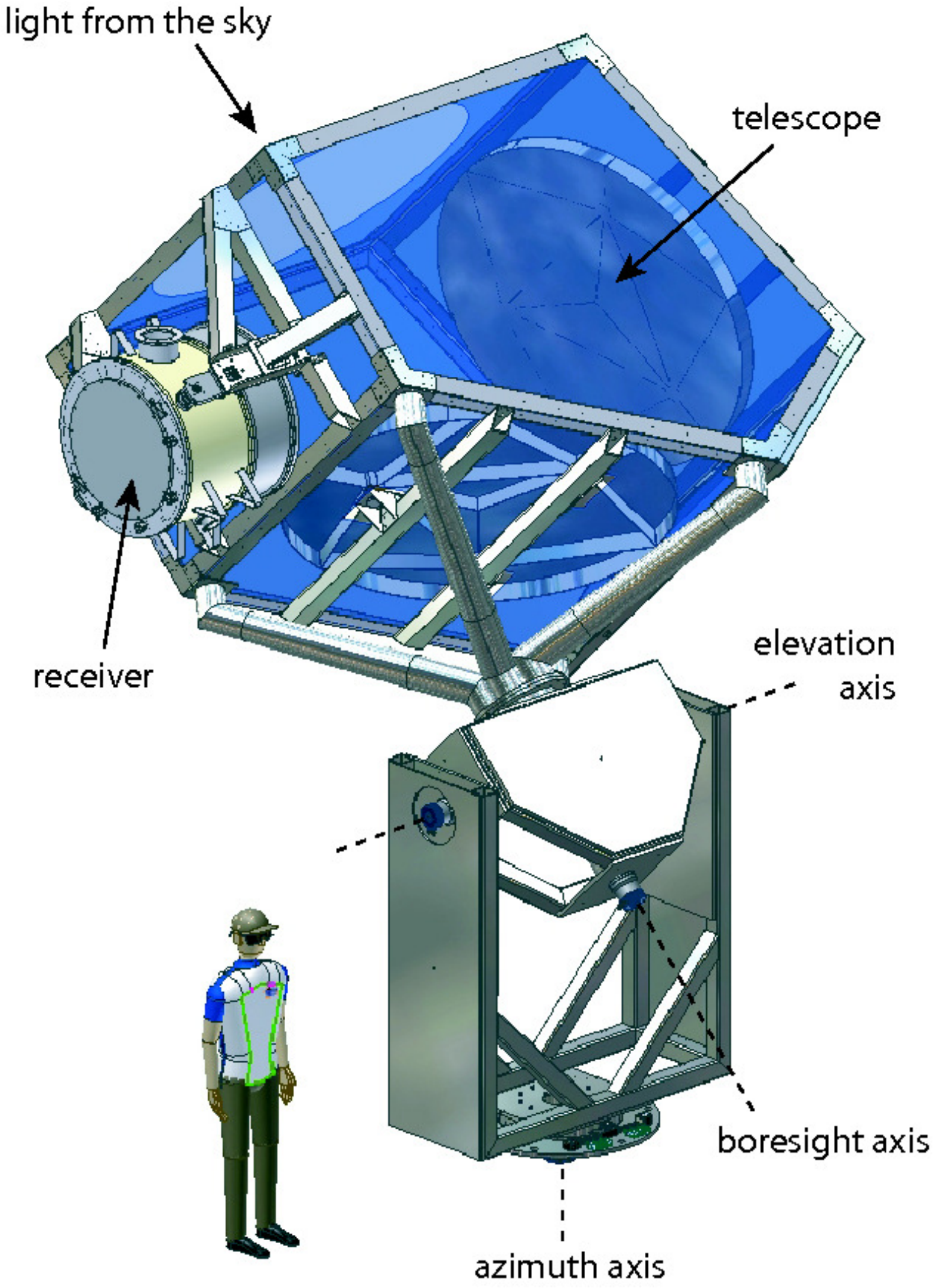}
\includegraphics[scale=0.17]{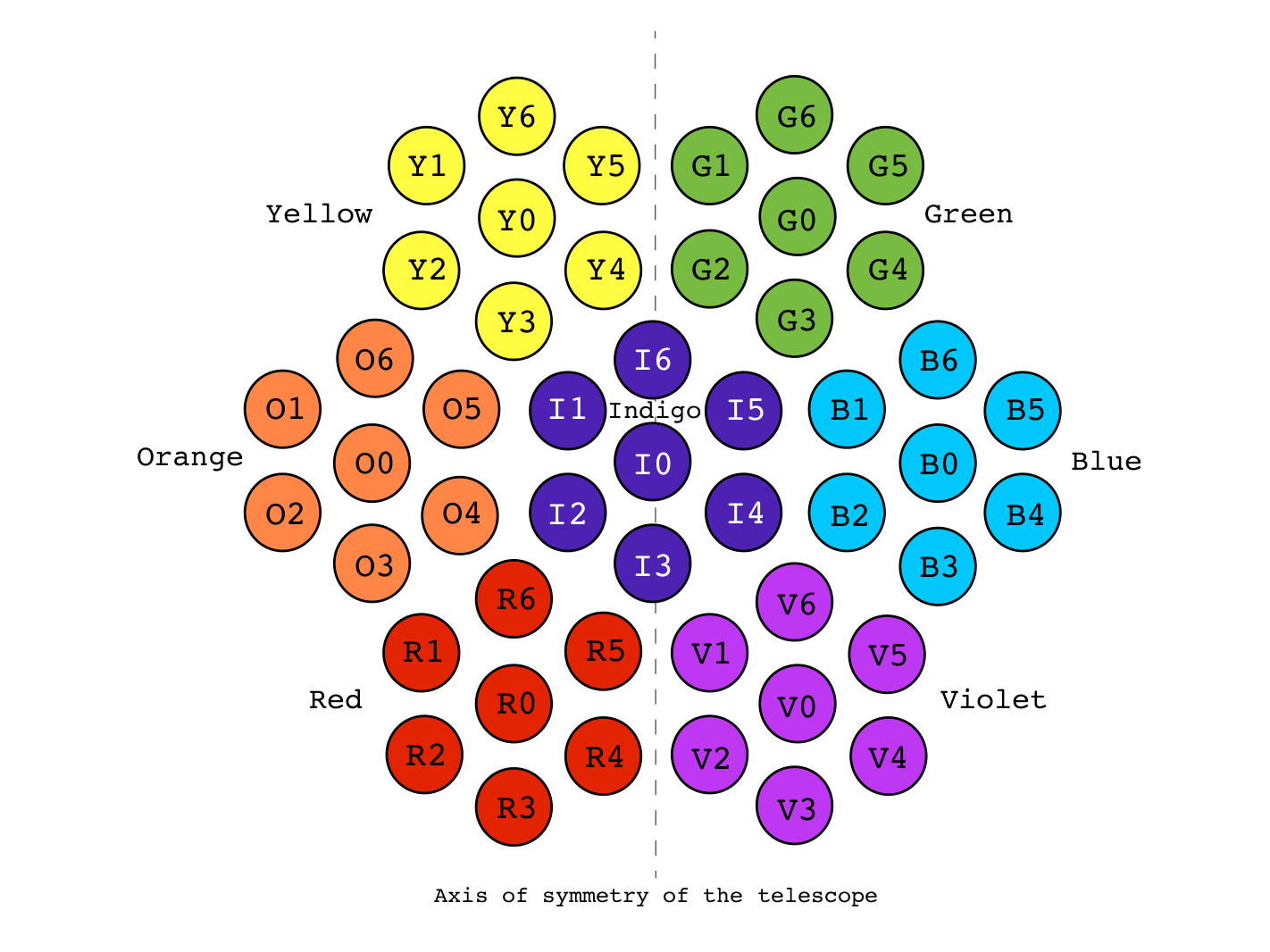}
\caption{Sketch of the STRIP telescope (left) and of the STRIP focal plane (right) with its naming convention.}\label{telesc}
\end{figure}

STRIP can rotate around the elevation and azimuth exes, shown in Fig.\ref{telesc}. \par
 \par
\begin{figure}[H]
\centering
\includegraphics[scale=0.6]{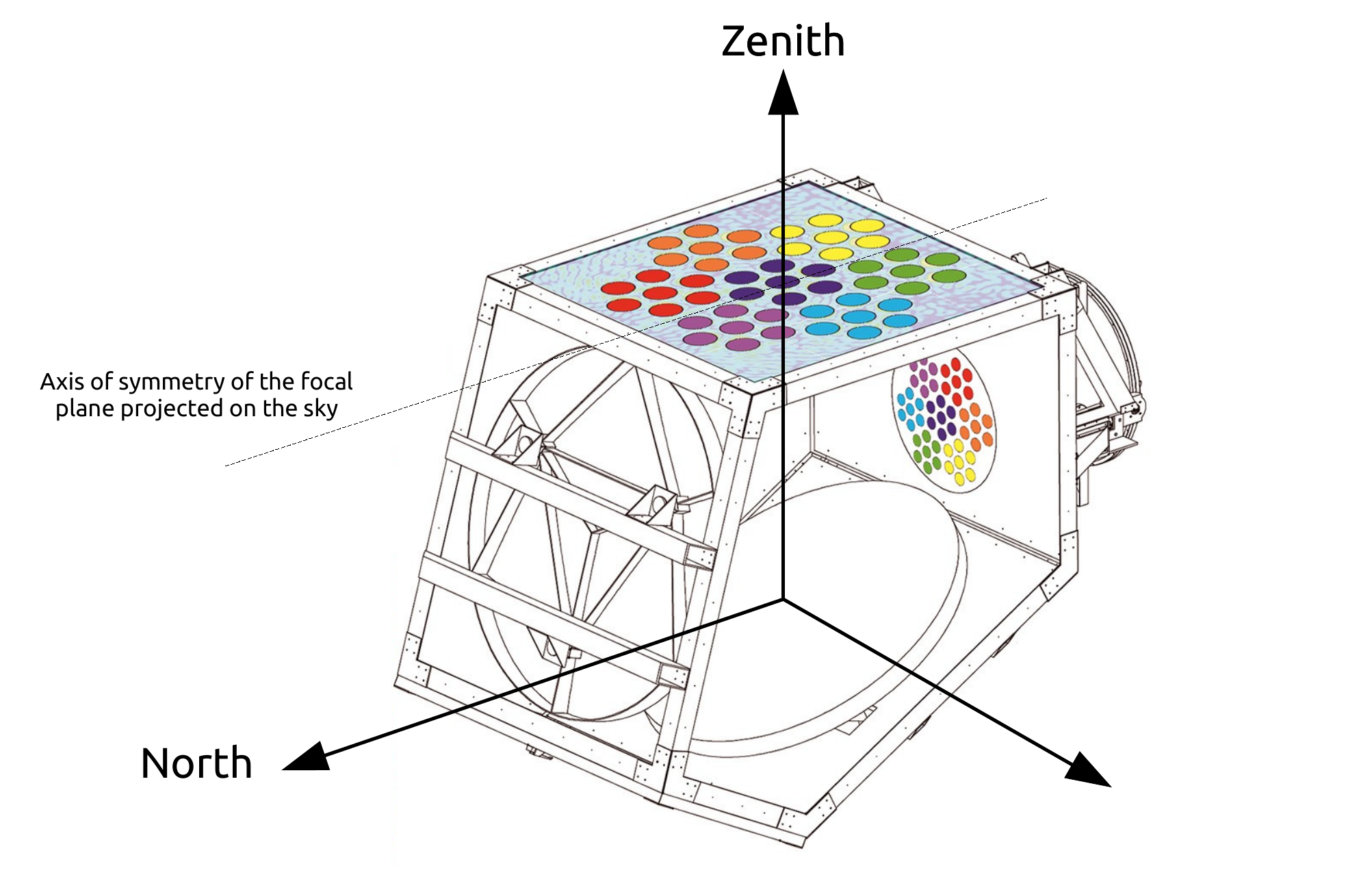}
\caption{The mount coordinate system (MCS).}\label{ref_telesc}
\end{figure} \par
\subsection{Local reference systems}
In the rest position, the line-of-sight (LOS) of the telescope looks at the local zenith of the site of observation. Furthermore, the projection on the sky of the axis of symmetry of the focal plane points towards the local north, so that the \textit{elevation angle} of telescope (the angular distance between the LOS and the Z-axis, also called \textit{zenith distance} or \textit{zenith angle}) is defined positive when the enclosure that contains the mirrors moves down. In this way, it is possible to define a local right-handed coordinate system where the Z-axis is directed towards the zenith and the X-axis points towards north: this is the ``mount coordinate system''(MCS) and is shown in Fig. \ref{ref_telesc}.\par

\begin{figure}[H]
\centering
\includegraphics[scale=0.1]{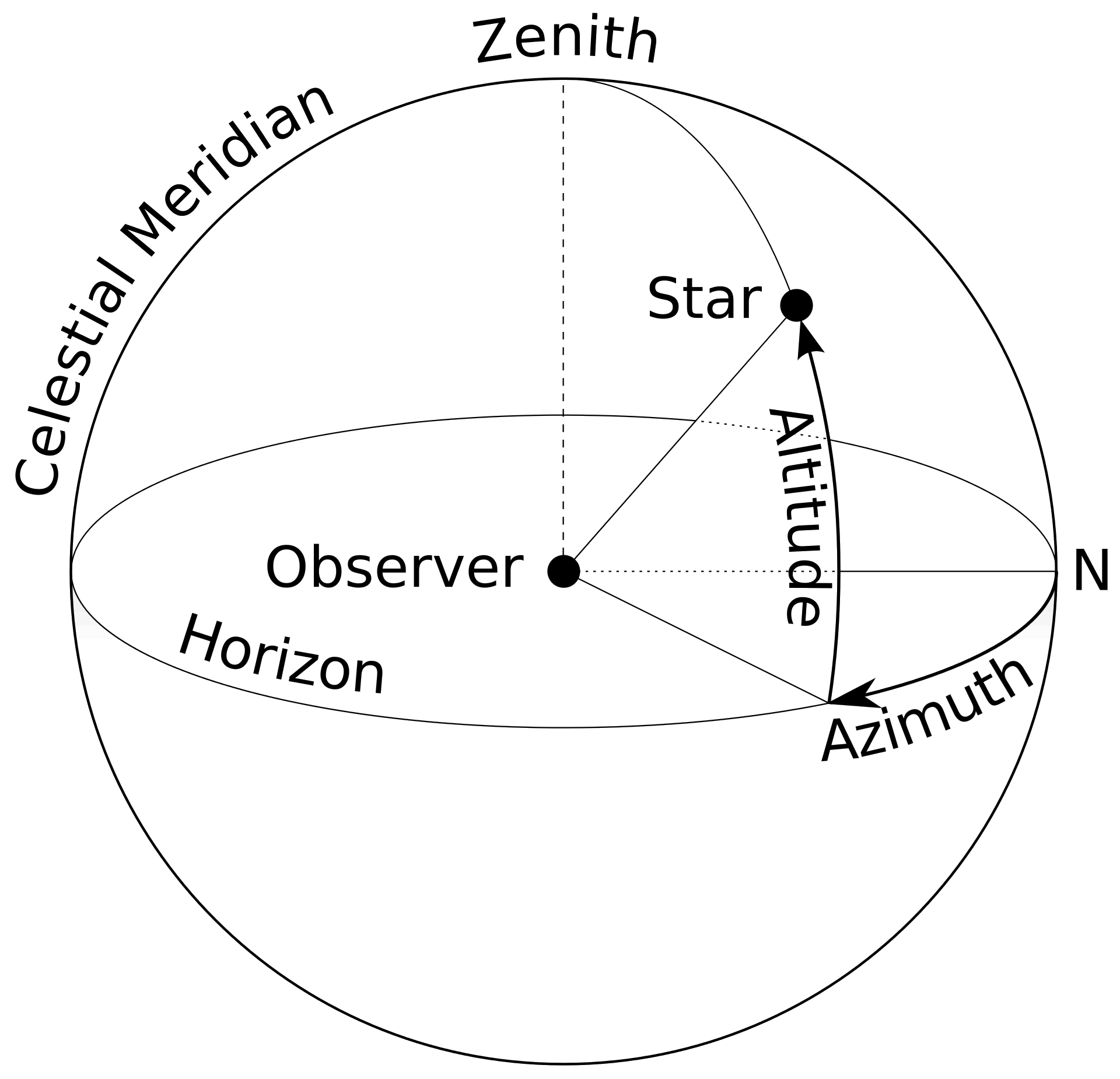}
\caption{Horizon coordinates (HC).}\label{HC}
\end{figure}

In the MCS, we can express spherical coordinates in terms of ``horizon coordinates'' (HC, Fig. \ref{HC}). In HC, the \textit{altitude} (Alt) is the angular distance of the object from the local horizon, along the local meridian. It is also the complementary angle of the zenith distance. The \textit{azimuth} (Az) is the angle around the horizon, looking from Earth's North Pole increasing clockwise.\par
 \par

\subsection{Absolute reference systems}
Pointings in the MCS must be converted into absolute reference system. In Stripeline, coordinate conversions are performed through the \textit{AstroLib} module of Julia\footnote{\url{https://github.com/JuliaAstro/AstroLib.jl}.}. The conversion precision is accurate to about $1\,\mathrm{arcsec}$ and secondary effects, such as Earth's precession and nutation, stellar aberration and atmospheric refraction, are taken into account\footnote{For more details about how this effects are computed see the \textit{AstroLib} documentation at: \url{https://juliaastro.github.io/AstroLib.jl/stable/ref.html}.}. \par
We use two kinds of absolute coordinates: the \textit{galactic} and \textit{equatorial} ones. \par
\begin{figure}[H]
\centering
\includegraphics[scale=4]{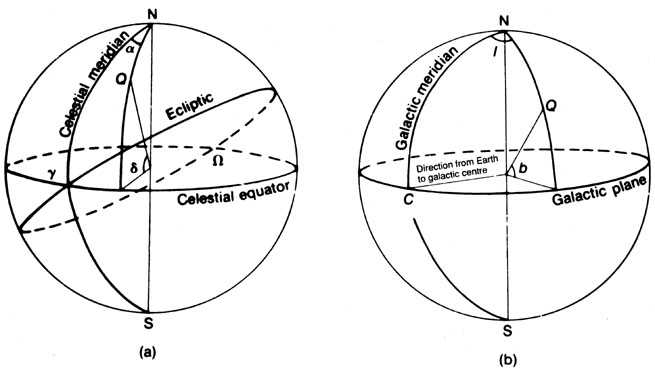}
\caption{Equatorial (a) and galactic (b) coordinate systems. In picture (a), $\gamma$, $\delta$ and $\alpha$ are, respectively, the vernal equinox, the declination and the right ascension.}\label{eqgal}
\end{figure}

The ``equatorial coordinate system'' (ECS, Fig. \ref{eqgal}) has its origin at the center of the Earth, the Z-axis is directed along the Earth's rotation axis, towards the Earth's Northern Emisphere, and the X-axis points towards the vernal equinox, which is the point where the ecliptic intersects the celestial equator in the Sun's ascending node. In this way, the fundamental plane consists on the projection of Earth's equator onto the celestial sphere. This defines a right-handed convention where coordinates increase counterclockwise on the fundamental plane, seen by above. Earth's precession and nutation make the orientation of this reference system not fixed on long periods. For this reason, it is necessary to specify the equinox of a particular date. We use the mean equinox of the standard epoch $\mathrm{J2000.0}$.\par
In the ECS, the spherical coordinates are called \textit{declination} and \textit{right ascension}. The declination (Dec) measures the angular distance of an object perpendicular to the celestial equator. It is positive northward and negative southward, and is conventionally measured in degrees. The right ascension (RA) measures the angular distance of an object eastward along the celestial equator from the vernal equinox to the hour circle passing through the object, the hour circle being orthogonal to the celestial equator. RA is usually measured in sidereal hours, minutes and seconds.\par
The ``galactic coordinate system'' (GCS, Fig. \ref{eqgal}) has its origin at the center of the Sun, the Z-axis is directed along the Galactic north pole, the X-axis points towards the center of the Milky Way, so that the fundamental plane is parallel to the Galactic plane. It is a right-handed system and is based on the ECS since the Galactic north pole and Galactic center are defined in equatorial coordinates. \textit{Galactic latitude} (l) and \textit{longitude} (b) measure respectively the angle (counterclockwise) of an object along the Galactic equator and the angle of the object above the Galactic plane (positive northward, negative southward). They are usually measured in degrees. \par
 \par
Conversions among several reference systems are implemented by means of some functions that take as input the exact UTC date and time at which the observation starts, the latitude and longitude of the location where the observation is made and a function that describe the motion of the telescope as a function of the time. \par

\section{Focal plane pointings}
\begin{figure}[H]
\centering
\includegraphics[scale=0.4]{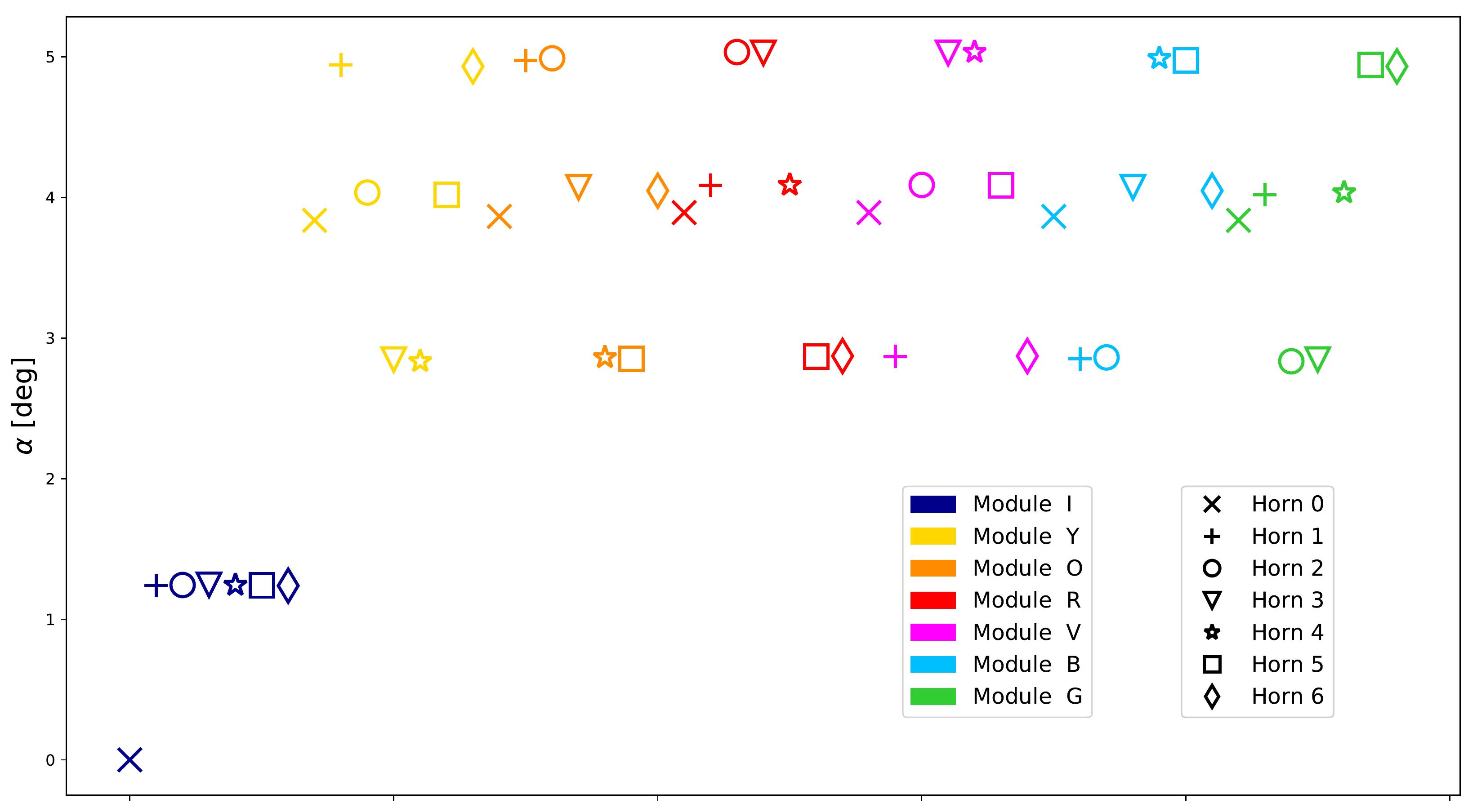}
\caption{Horns tilt angles when the telescope is in its rest position. Horns with the same distance from $\mathrm{I0}$ have the same tilt angle.}\label{thetas}
\end{figure}

In the rest position of the telescope, only the central horn (namely ``$\mathrm{I0}$'' in Fig. \ref{telesc}) looks at the zenith while the LOSs of the other horns are tilted with respect to it, with radial symmetry. Fig. \ref{thetas} reports the tilt angles, $\alpha_i$, of the horns in the rest position of the telescope. Horns at the same distance from $\mathrm{I0}$ have the same tilt angle. \par
 \par
\begin{figure}[H]
\centering
\includegraphics[scale=0.5]{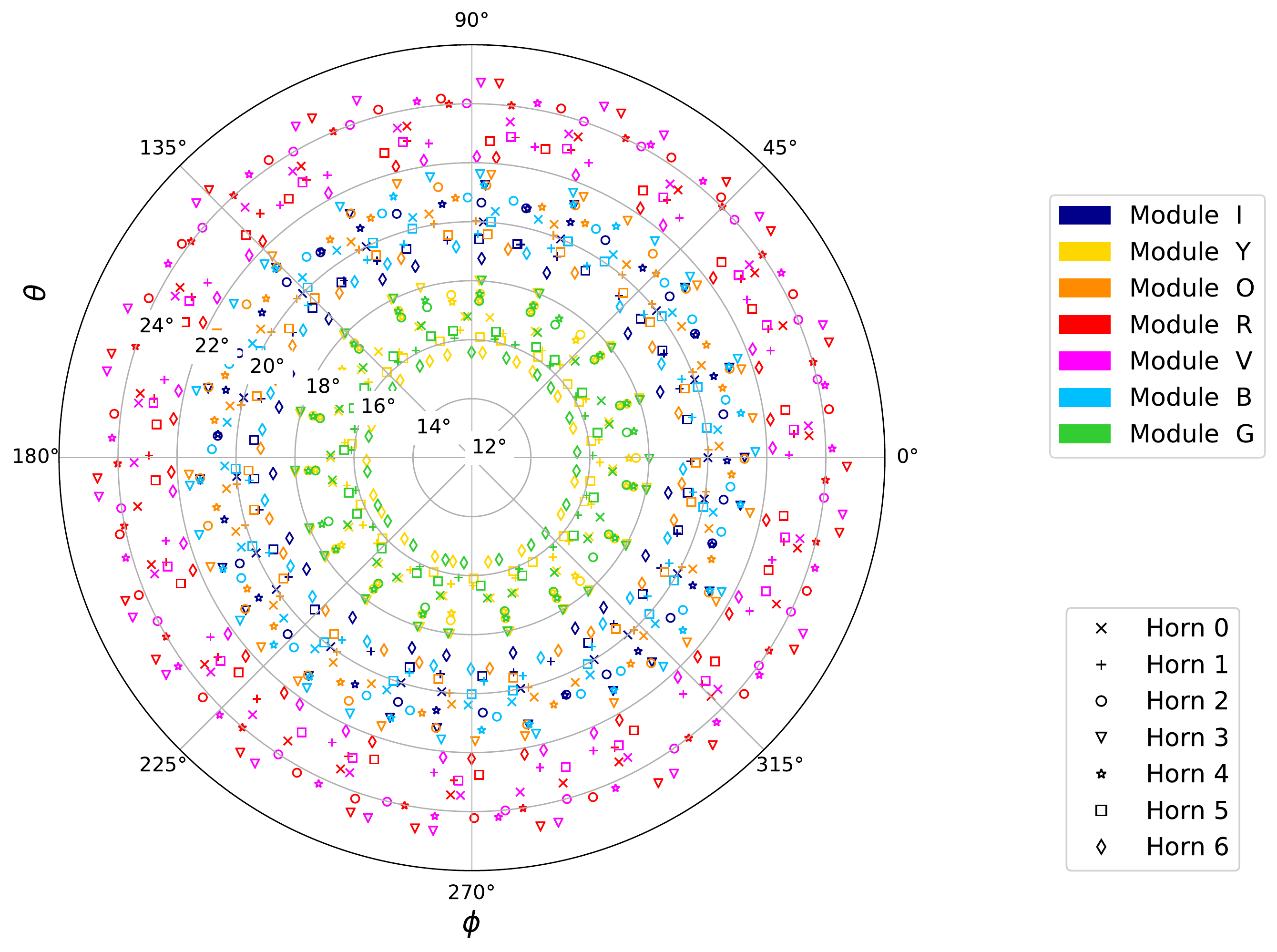}
\caption{Circles described by the horns on the sky (in MCS) after one revolution of the telescope, with elevation angle of $20\deg$. Only ``$\mathrm{I0}$'' has the same elevation angle of the telescope.}\label{sphertheta}
\end{figure}
\begin{figure}[H]
\centering
\includegraphics[scale=0.4]{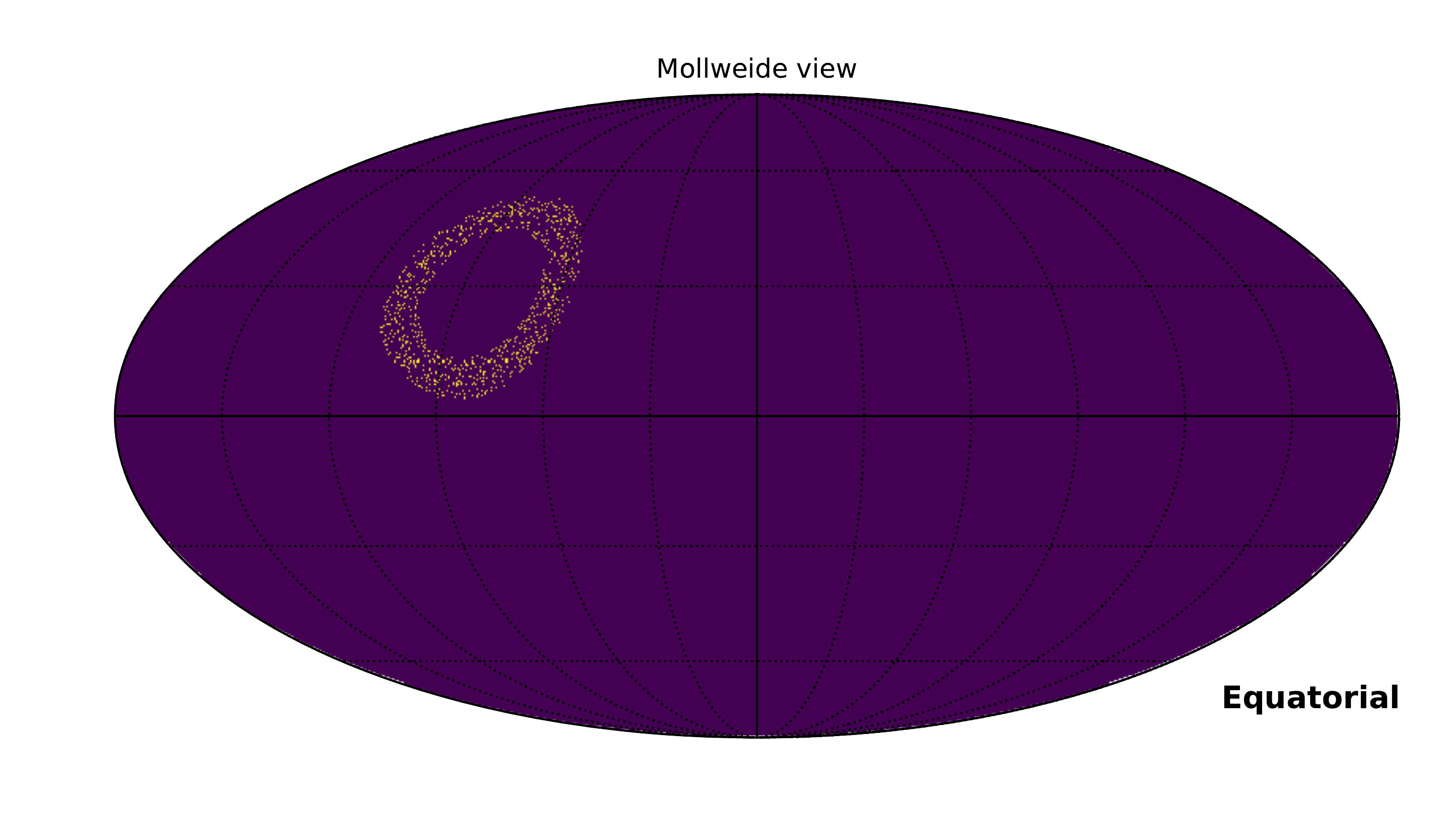}
\caption{Projection of the circles described by the horns on the sky (in ECS) after one revolution of the telescope, with elevation angle of $20\deg$.}\label{pointigns_proj}
\end{figure}

If the telescope changes its elevation, the elevation angles of the horns will depend on the tilt angle $\alpha_i$ and on the side where the telescope tilts. Besides, if the telescope rotates around the azimuth with a fixed elevation, each horn will describe a circle in the sky, whose amplitude will depend on its elevation angle. Assuming, for example, a positive elevation angle of the telescope of $20\deg$ the innermost circles are the ones described by the horns in the modules ``G'' and ``Y'' while the outermost ones are depicted by the modules ``R'' and ``V'' (Fig. \ref{sphertheta}). The same circles projected on a sky map (ECS), are shown in Fig. \ref{pointigns_proj}. \par
 \par

 \par

%% file: chap6.tex
\chapter{Scanning strategy analysis}
\label{Chap:6}
\thispagestyle{plain}
In this chapter, I illustrate the results of the scanning strategy analysis I performed for the LSPE/STRIP experiment exploiting Stripeline (Ch. \ref{Chap:4}). \par
I have simulated the motion of the telescope for two years from Tenerife and I have studied the coverage properties both in terms of sensitivity per pixel and of overlap with the LSPE/SWIPE coverage. I have studied the possibility to observe particular sky regions for calibration or scientific follow-ups. I have provided an estimation of the effective observation time and, finally, I have investigated a way to make the sensitivity per pixel distribution more uniform. Some of the results I present here have already been shown in \citet{Incardona}.

\section{Scanning strategy objectives}
There are three main goals that have driven the choice of the STRIP scanning strategy:
\begin{list}{\leftmargin 15pt \itemsep 0pt \topsep 3pt}
\item To trade-off the sky coverage with the sensitivity per pixel distribution. In particular, a large sky coverage ($\gtrsim 25\%$, which is desirable to access the cosmological $B$-modes angular scales) implies less redundancy and, therefore, larger noise per pixel.
\item To ensure the required sky coverage while maximizing the overlap with the sky region observed by SWIPE.
\item To include specific sources in our field of view both for calibration and astronomical follow-ups.
\end{list}\par

\subsection{Nominal scanning mode}
\begin{figure}[H]
  \centering
  \includegraphics[scale=0.4]{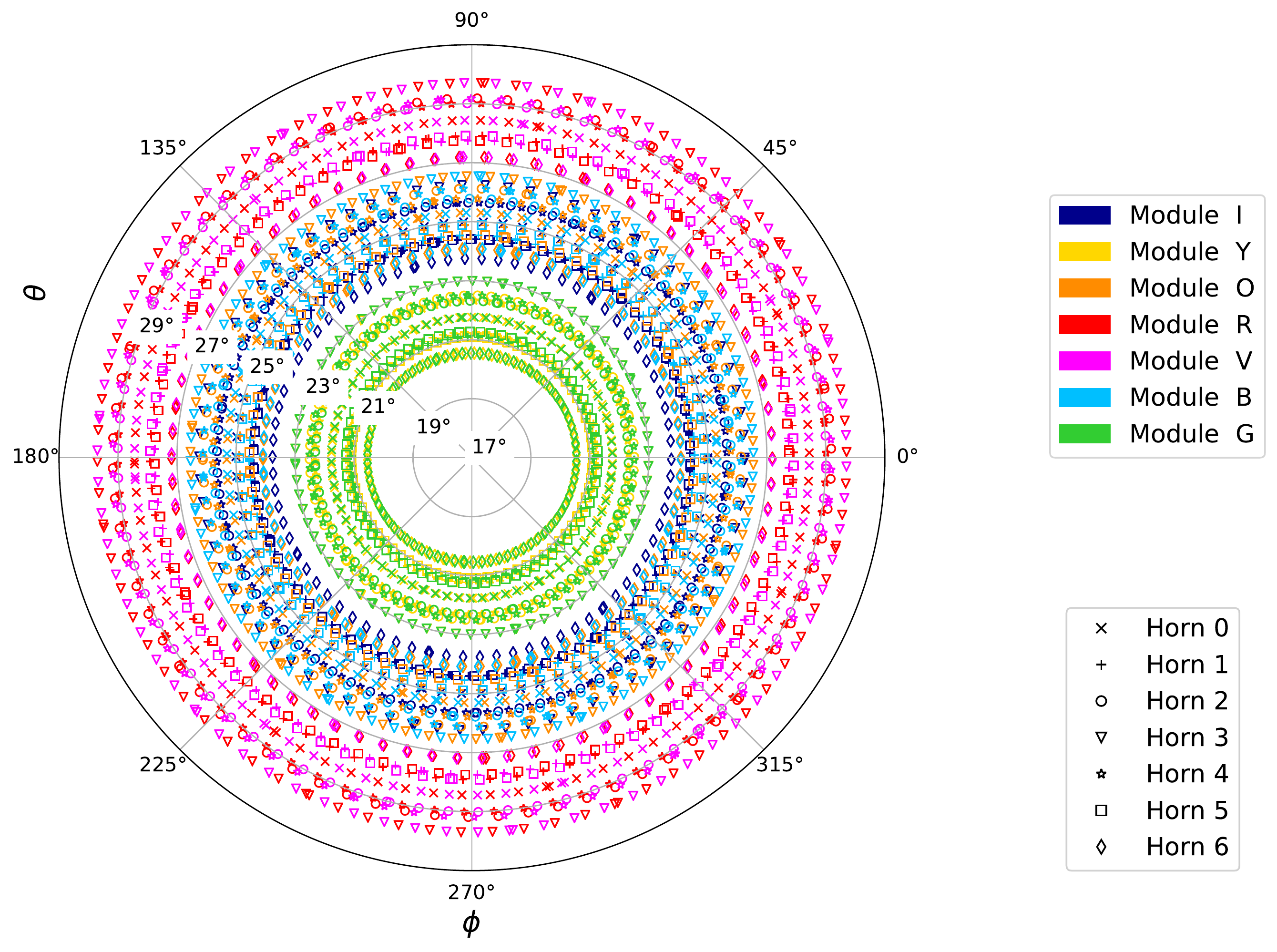}
  \caption[]{Pointings after one minute scan with $\omega_\mathrm{spin} = 1\,\mathrm{rpm}$, an elevation angle of $25\deg$ from the zenith and a reduced sampling rate of $4\,\mathrm{Hz}$.}\label{fig:focalplanepointings} 
\end{figure}

A good trade-off between the three conditions listed above is given by spinning the telescope around the azimuth axis with constant elevation and angular velocity. In this way, the beam pattern associated to each horn will describe a circle in the sky whose radius depends on its elevation angle (Fig. \ref{fig:focalplanepointings}). Furthermore, constant elevation spins will allow each horn to scan over atmospheric layers of constant airmass. This is useful to reduce atmosphere systematic effects. Combining the motion of the telescope with the rotation of the Earth, STRIP will observe a sky-band with an amplitude that depends on the elevation of the instrument (Fig. \ref{fig:STRIP}).

\section{Simulations}\label{sssim}
I set the Tenerife coordinates (Latitude = $28\deg \,18'\, 00''\, \mathrm{N}$, Longitude = $16\deg \,30'\, 35''\, \mathrm{W}$, height = $2390\,\mathrm{m\,a.s.l.}$) and let the telescope rotate around the azimuth axis with a rotational speed of $\omega_\mathrm{spin} = 1\,\mathrm{revolution \ per \ minute}$ ($\mathrm{rpm}$). I assumed two years of mission, starting on 1 April 2021, using the $49$ polarimeters operating at the central frequency of $43\,\mathrm{GHz}$, with a sampling rate ($\nu_\mathrm{sampling}$) of $50\,\mathrm{Hz}$ and variable duty cycle (see Sect. \ref{dc} for a precise definition) in order to take into account the presence of the Sun and the possibility of adverse meteorological conditions.\par
I performed simulations for several elevation angles\footnote{We do not take into account the growth of the cosmic variance due to limiting the sky coverage, since the STRIP data will be used principally to study the Synchrotron emission, so that we will not compute the CMB power spectra from the Q-band data only.} ranging from $5\deg$ to $50\deg$ (the maximum elevation angle that STRIP can reach), by steps of $5\deg$. For each simulation, I computed the \textit{hit count map}, a map reporting the number of hit per pixel, and the corresponding \textit{noise map}. Both these maps assess the sensitivity per pixel: the longer a pixel is observed, the greater will be the signal-to-noise ratio.\par
To represents the maps, I have used the \textit{HEALPix}\footnote{\url{https://healpix.sourceforge.io/}.} \citep{2005ApJ...622..759G} discretization of the sky, which divides the spherical surface in pixels of equal surface area. The resolution of the Healpix maps depends on the parameter $\mathrm{NSIDE}$ according to:
\be{npixels}
\mathrm{number \ of \ pixels} = 12 \times \mathrm{NSIDE}^2 \,.
\ee
The beam resolution of the STRIP telescope is $20\,\mathrm{arcmin}$ (Sect. \ref{STRIPlfi}), calling for a value of $\mathrm{NSIDE} = 256$ that ensure pixels smaller than the beam size projected on the sky. A higher resolution would be unnecessary and would slow the simulations.

\subsection{Hit count maps}
The elevation (zenith) angle impacts the sky coverage and the sensitivity.\par
\begin{figure}[H]
\centering
\includegraphics[scale=0.27]{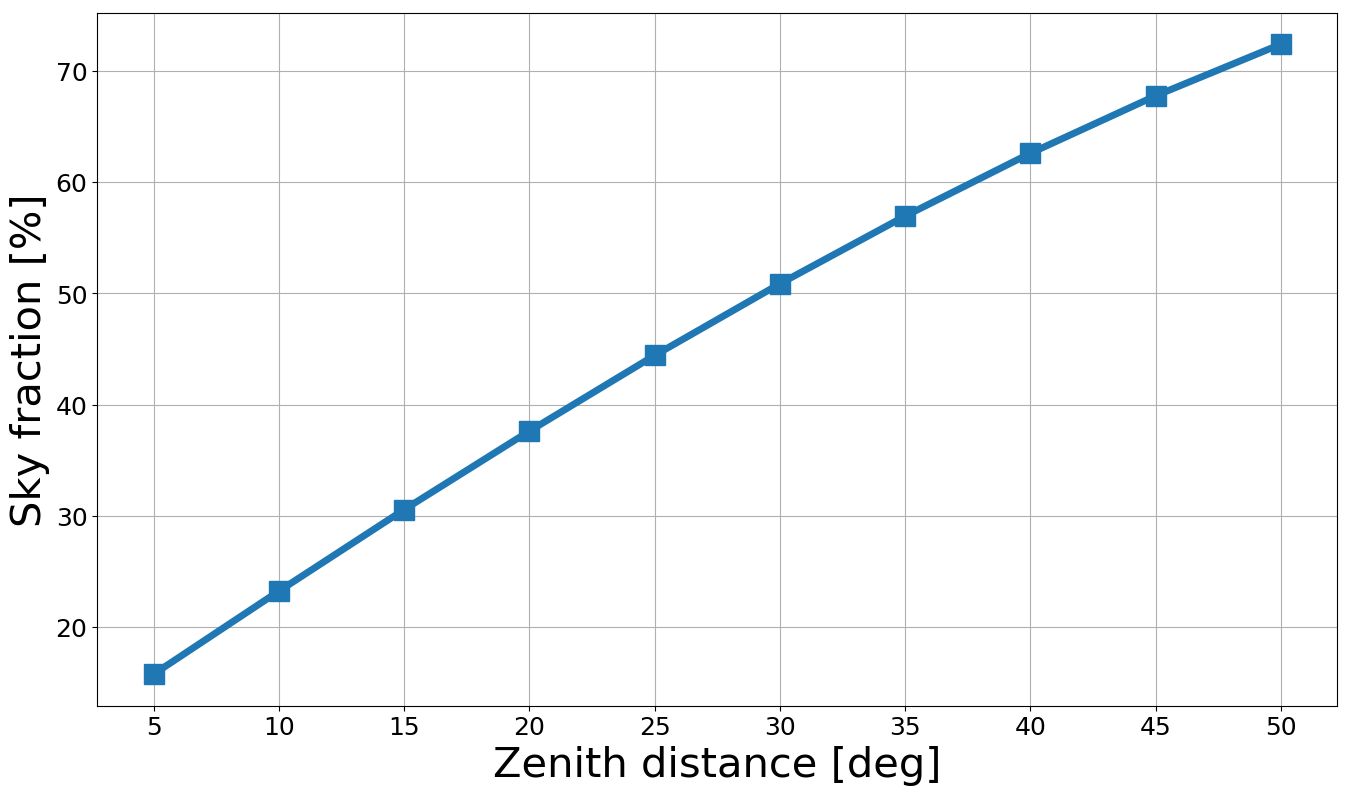}
\caption[]{Sky fraction observed by STRIP as a function of the elevation angle.}\label{stripf_sky} 
\end{figure}

In Fig. \ref{stripf_sky}, I report the sky fraction observed by STRIP as a function of the elevation angle. STRIP will observe a portion of the sky larger than $\sim 25\%$ for zenith angles larger than $\sim 10\deg$. \par
\begin{figure}[H]
\centering
   \hspace{6pt}\includegraphics[scale=0.085]{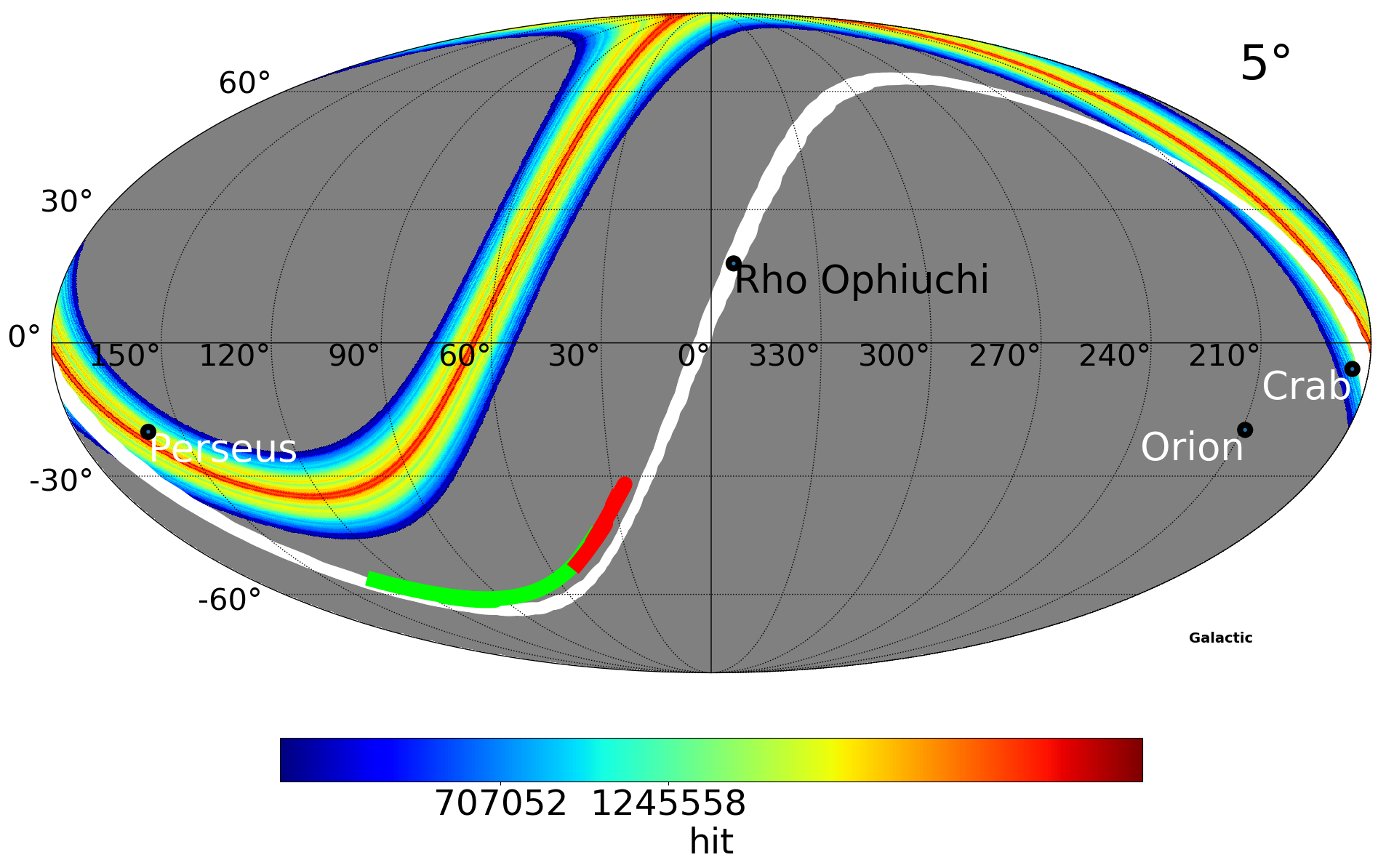}
   \includegraphics[scale=0.235]{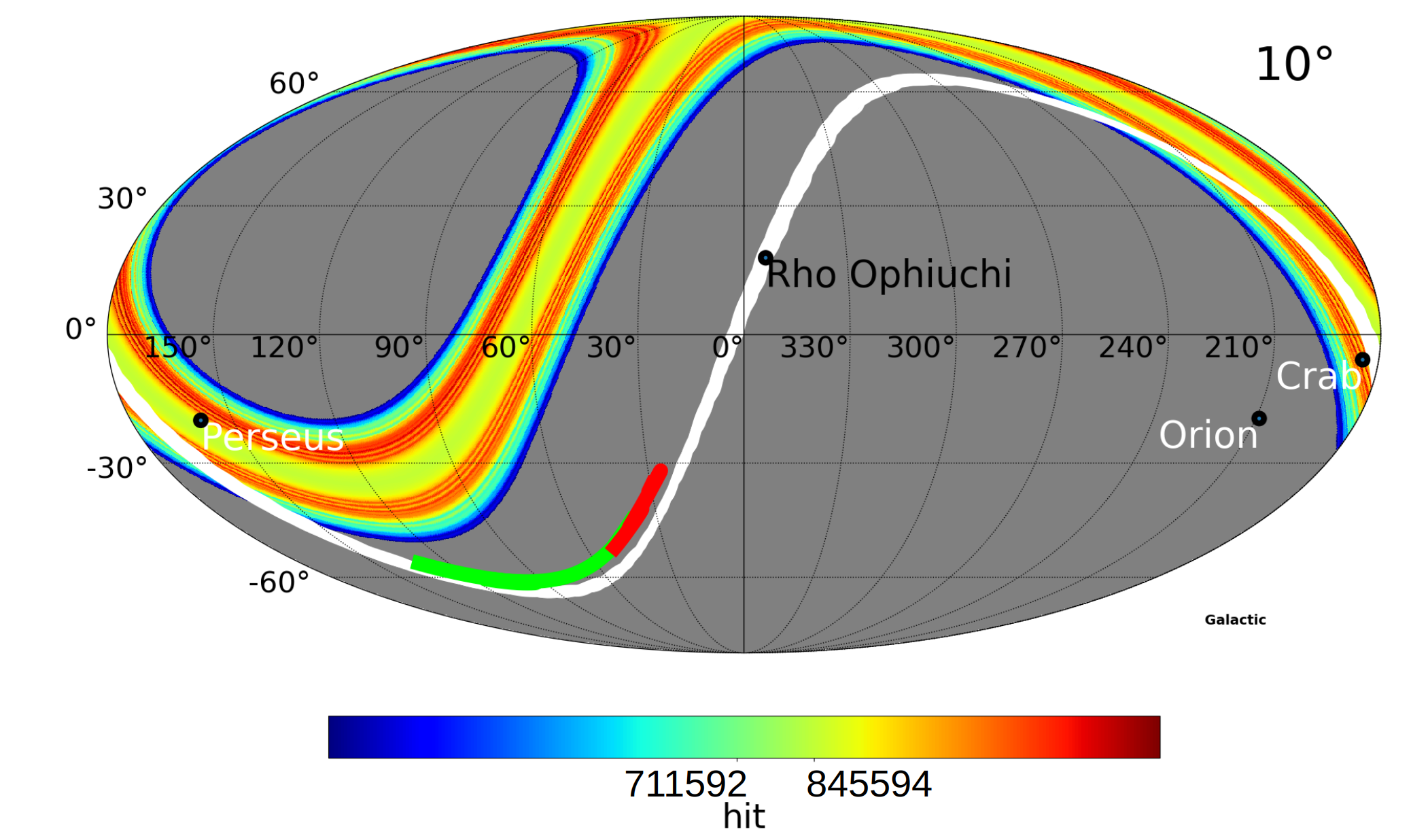}
   \includegraphics[scale=0.235]{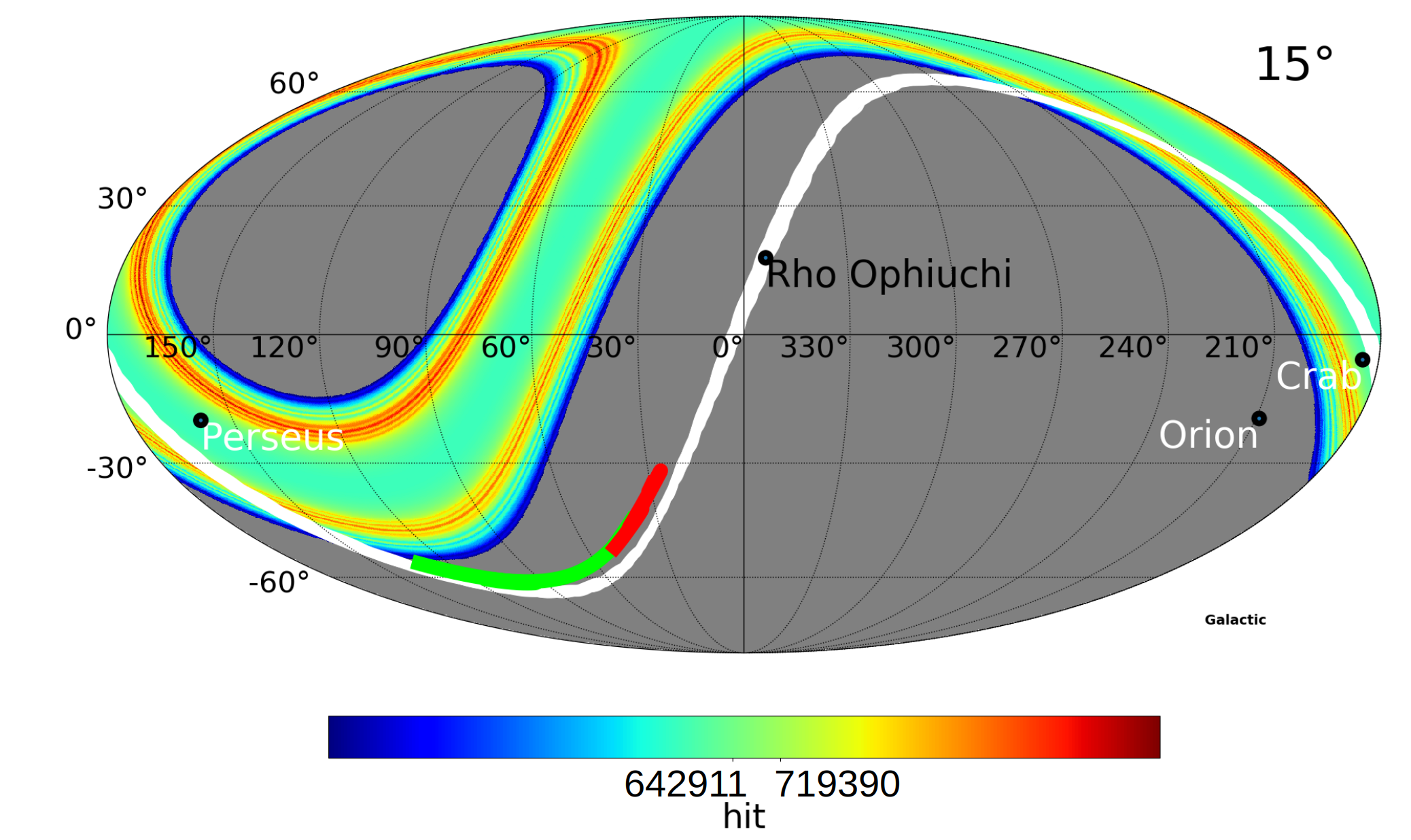}
   \includegraphics[scale=0.085]{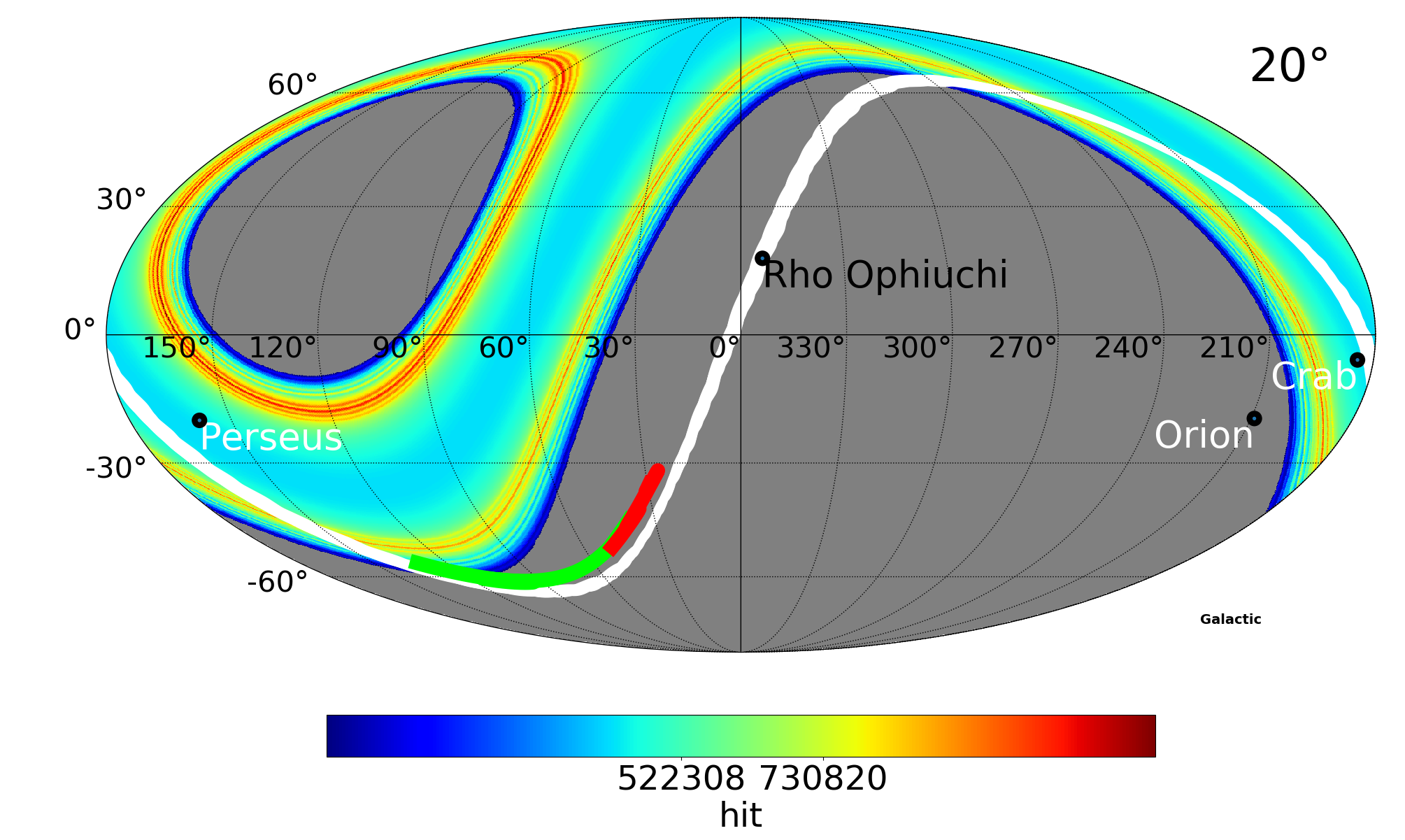}
   \includegraphics[scale=0.085]{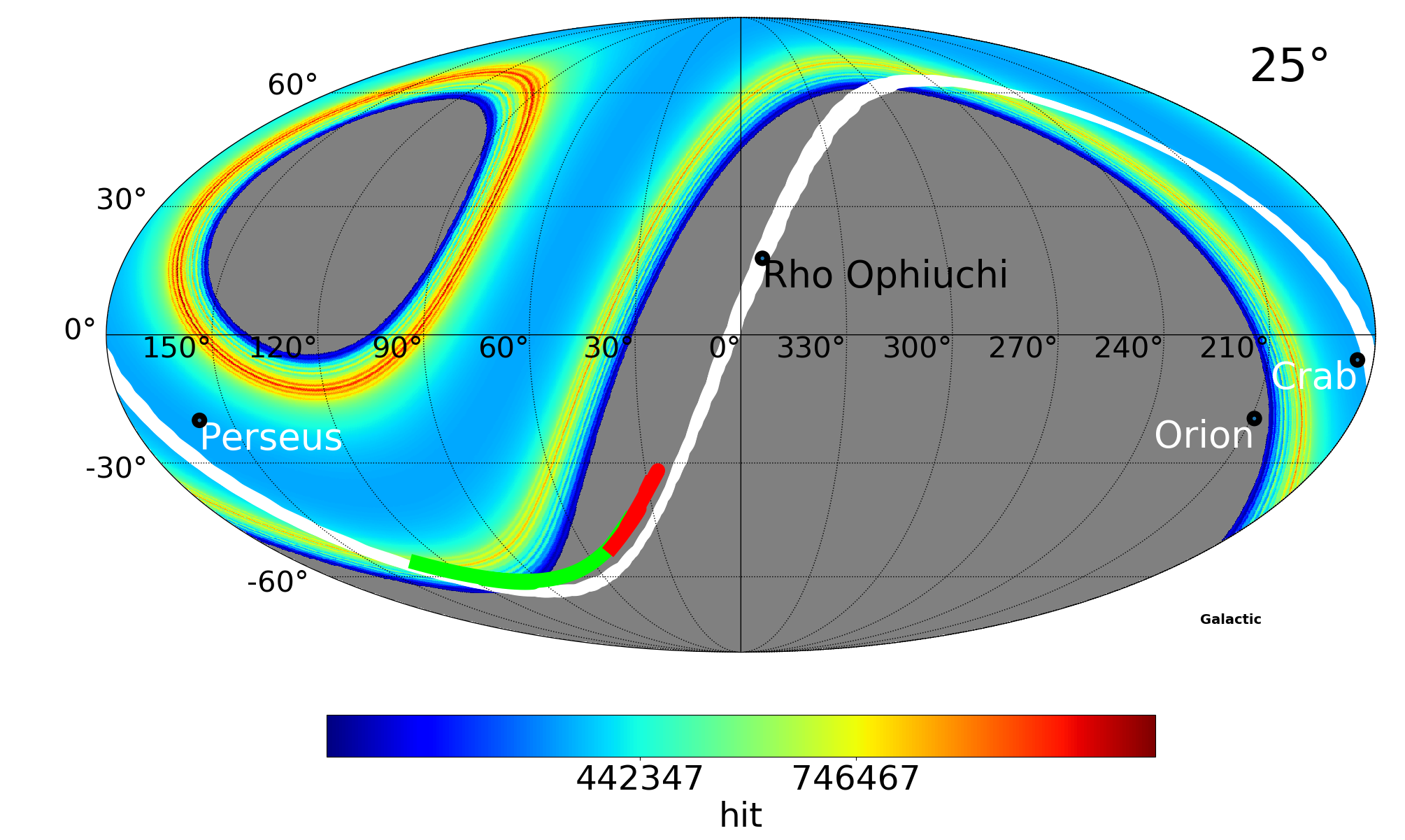}
   \includegraphics[scale=0.085]{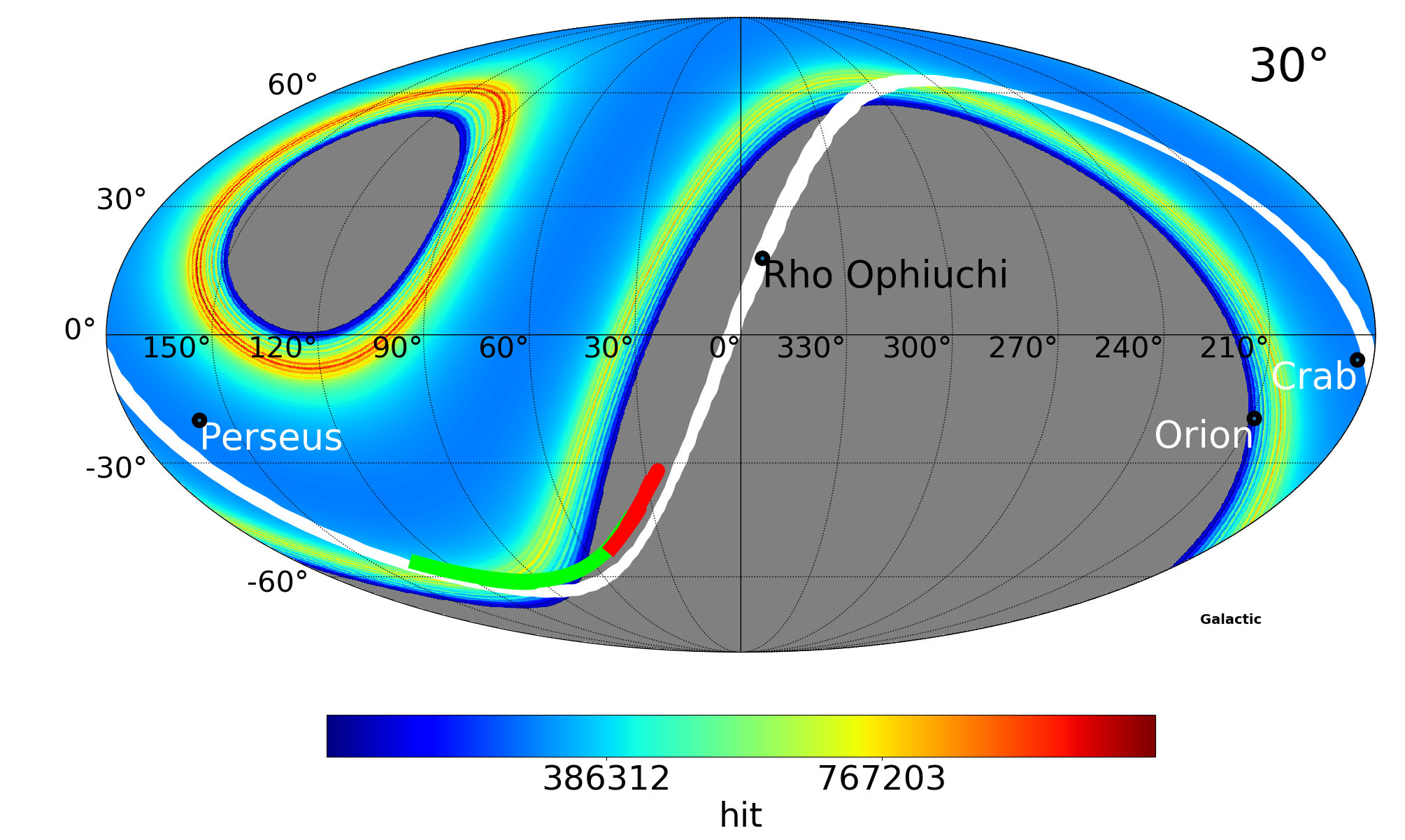}
   \includegraphics[scale=0.085]{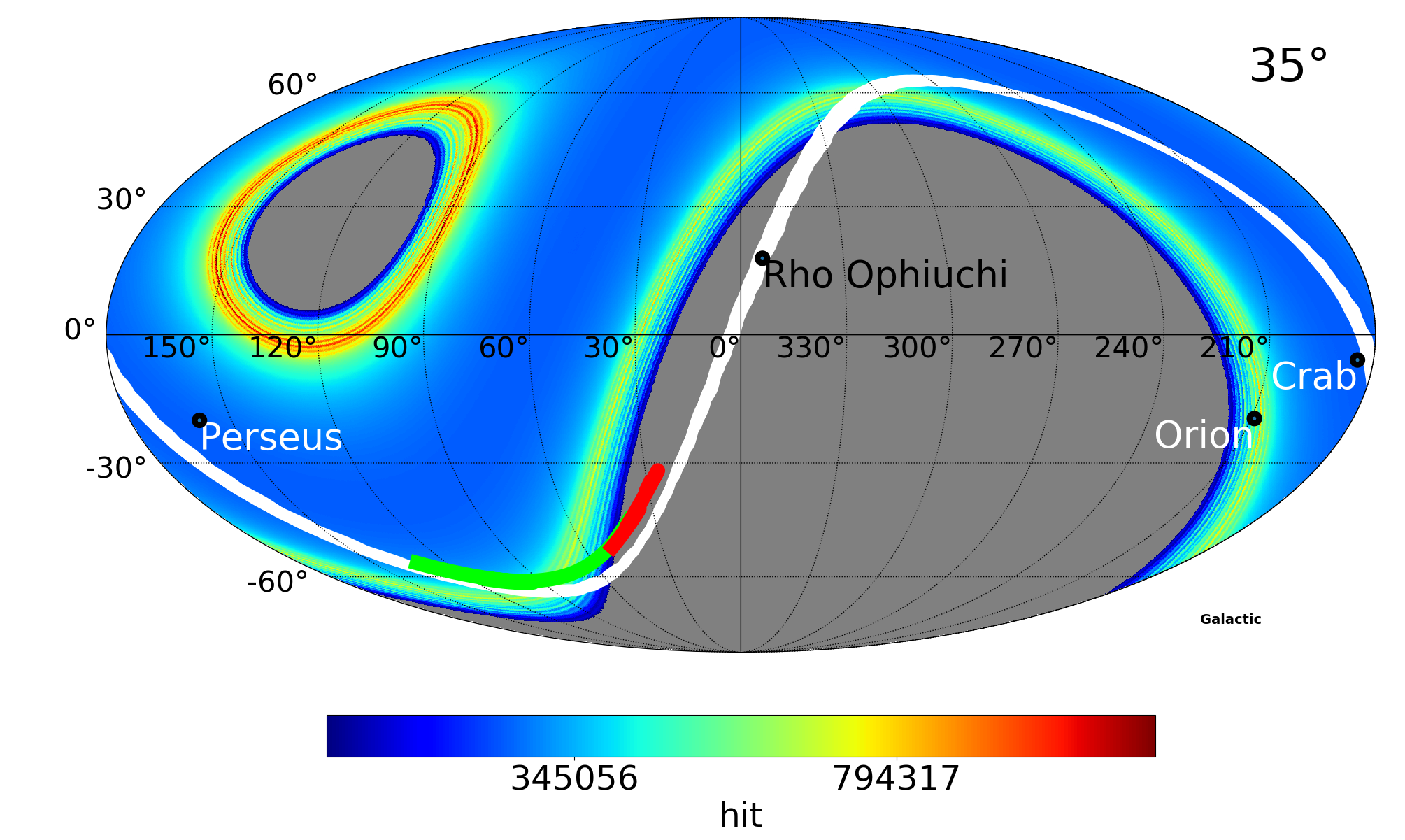}
   \includegraphics[scale=0.085]{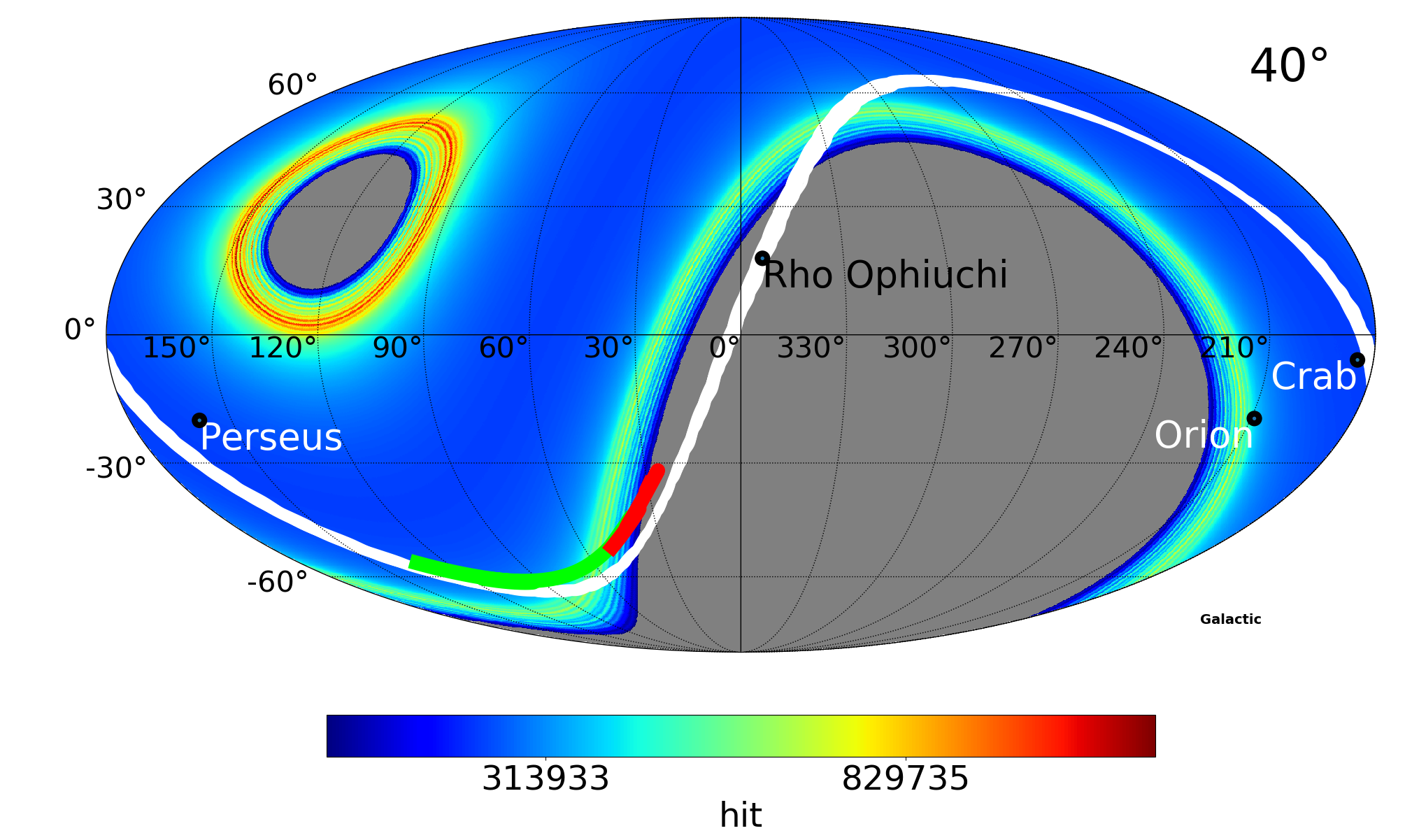}
   \includegraphics[scale=0.085]{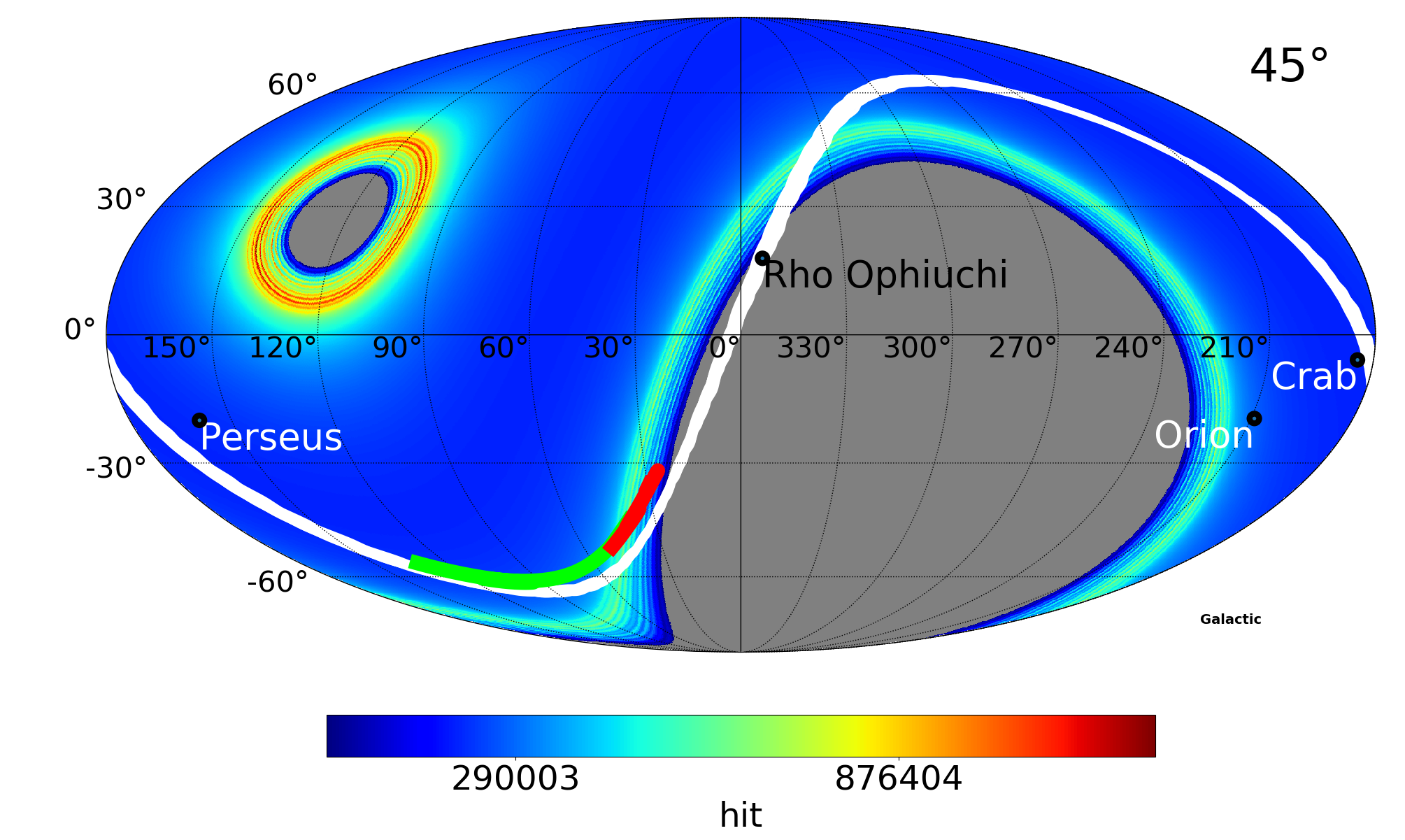}
   \includegraphics[scale=0.085]{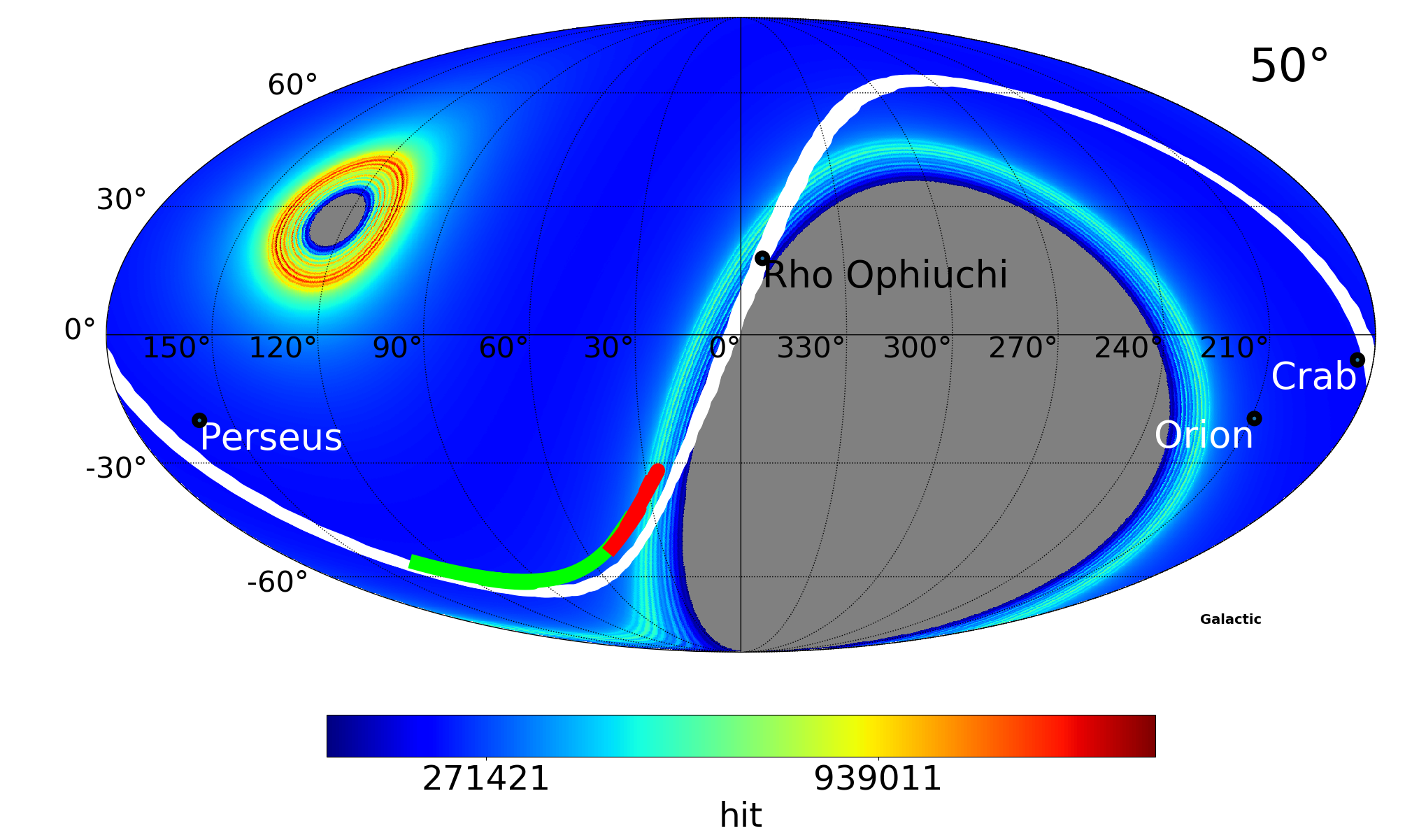}
   \caption[]{Hit count maps (in GCS) for several zenith angles (ranging from $5\deg$ to $50\deg$ by steps of $5\deg$, from the left to the right staring from the upper left). The linear color scale identifies the number of hits per pixel. In the color bars are reported the average value of each map and the value corresponding to the requirement. The maps show also the positions of the Crab and Orion nebulas, the Perseus molecular cloud, Rho Ophiuchi and the trajectories of Jupiter (green mark), Saturn (red mark) and the Moon (white curve).}\label{fig:STRIP} 
\end{figure}

In Fig. \ref{fig:STRIP}, I show the hit count maps for zenith angles ranging from $5\deg$ to $50\deg$, in the galactic coordinate system (GCS). An obvious consequence of increasing the sky coverage is the reduction of the average number of hits per pixel, with a consequent reduction of the sensitivity per pixel for a given integration time. It is possible to observe the sensitivity reduction by looking at the color scale in Fig. \ref{fig:STRIP}: red regions are more redundant than blue ones. We note also that, in the central regions, the number of hits corresponds approximately to the average value while at the edges of the map we find both the most and the least redundant regions.\par
In each map of Fig. \ref{fig:STRIP} the linear color scale identifies the number of hits per pixel. In the color bars are reported the average value of each map and the value corresponding to the STRIP sensitivity requirement (Table \ref{deltaT}). As the observed sky fraction increases, the average number of pixel decreases. Only in the first two cases, $5\deg$ and $10\deg$, the average number of hits is greater than the requirement.\par


\subsection{Noise maps}\label{noismaps}
The intensity of the astrophysical signal measured by STRIP is proportional to the antenna temperature\footnote{The antenna temperature is the convolution of the brightness temperature of a given source with the antenna radiation pattern.} $T_\mathrm{sys}$:
\be{Tsys}
T_\mathrm{sys} = T_\mathrm{CMB} + T_\mathrm{atm} + T_\mathrm{noise} + T_\mathrm{optics} + T_\mathrm{foregrounds} + T_\mathrm{ground} \,,
\ee
where the addends are the antenna temperatures at a given frequency of, respectively: CMB, atmosphere, radiometric components of the polarimeter (Sect. \ref{tnoise}), optical components of the telescope (mirrors, cryostat window, filters, feedhorn, OMT, etc.), astrophysical foregrounds (such as synchrotron or thermal dust) and ground. \par
The sensitivity of the polarimeters both in intensity and polarization is given by the \textit{radiometer equation} \citep{Kraus}:
\be{DeltaQrms}
\Delta T_\mathrm{rms} = \frac{1}{\sqrt{2}} \frac{T_\mathrm{sys}}{\sqrt{\beta \tau}} \,,
\ee
where $\beta$ is the bandwidth, $\tau$ is the integration time and the factor $1 / \sqrt{2}$ is typical of the architecture of the STRIP polarimeters.\par
The antenna temperature of the atmosphere depends on the airmass crossed by the line-of-sight (LOS) of the horn. The airmass depends by the zenith angle according to the secant law \citep{DanesePartridge}:
\be{airmass}
T_\mathrm{atm} = \frac{T^0_\mathrm{atm}}{\cos{\theta}} \,,
\ee
where $T^0_\mathrm{atm}$ is the antenna temperature at the zenith and $\theta$ is the elevation angle.\par
Furthermore, for a given sampling rate the integration time in each pixel depends on the number of hits: $\tau(\mathrm{pixel}) = \mathrm{hit}(\mathrm{pixel}) / \nu_\mathrm{sampling}$. In this was, I computed the noise maps that show, for a given zenith angle $\theta$, the noise level in each pixel:
\be{DeltaQrms}
\Delta T_\mathrm{rms}(\mathrm{pixel}) = \frac{1}{\sqrt{2}} \frac{T_\mathrm{CMB} + T^0_\mathrm{atm} / \cos{\theta} + T_\mathrm{noise}}{\sqrt{\beta \ \mathrm{hit}(\mathrm{pixel}) / \nu_\mathrm{sampling}}} \,.
\ee
I neglected the contribution of the foregrounds and of the ground and I made the approximation that all the LOSs of all the horns cross the same airmass as the central horn. The values I used to compute the noise maps for the $49$ polarimeters in the Q-band are listed in Table \ref{parausenoise}. $T_\mathrm{optics}$ has been estimated through specific electromagnetic simulations while the value of $T^0_\mathrm{atm}$ has been extrapolated at the STRIP frequencies through the AM \citep{paine_scott_2019_3406496} and ATM \citep{Pardo} models. %using a precipitable water vapour (PWV) of $0.4\,\mathrm{mm}
The values of $T_\mathrm{noise}$ and $\beta$ have been measured directly during the unit-level tests (Ch. \ref{Chap:4}).
\begin{table}[H]
\centering
\begin{tabular}{|c|c|c|c|c|c|}
  \hline
  $T_\mathrm{CMB}$ [$\mathrm{K}$] & $T^0_\mathrm{atm}$ [$\mathrm{K}$] & $T_\mathrm{noise}$ [$\mathrm{K}$] & $T_\mathrm{optics}$ [$\mathrm{K}$] & $\beta$ [$\mathrm{GHz}$] & $\nu_\mathrm{sampling}$ [$\mathrm{Hz}$] \\  \hline 
  $ 1.82 $ & $16.4$ & $33.6$ & $7.5$ & $7.3$ & $50$ \\ \hline
\end{tabular}\caption{List of parameters used to estimate the noise maps.}\label{parausenoise}
\end{table}\par
\begin{figure}[H]
\centering
   \includegraphics[scale=0.085]{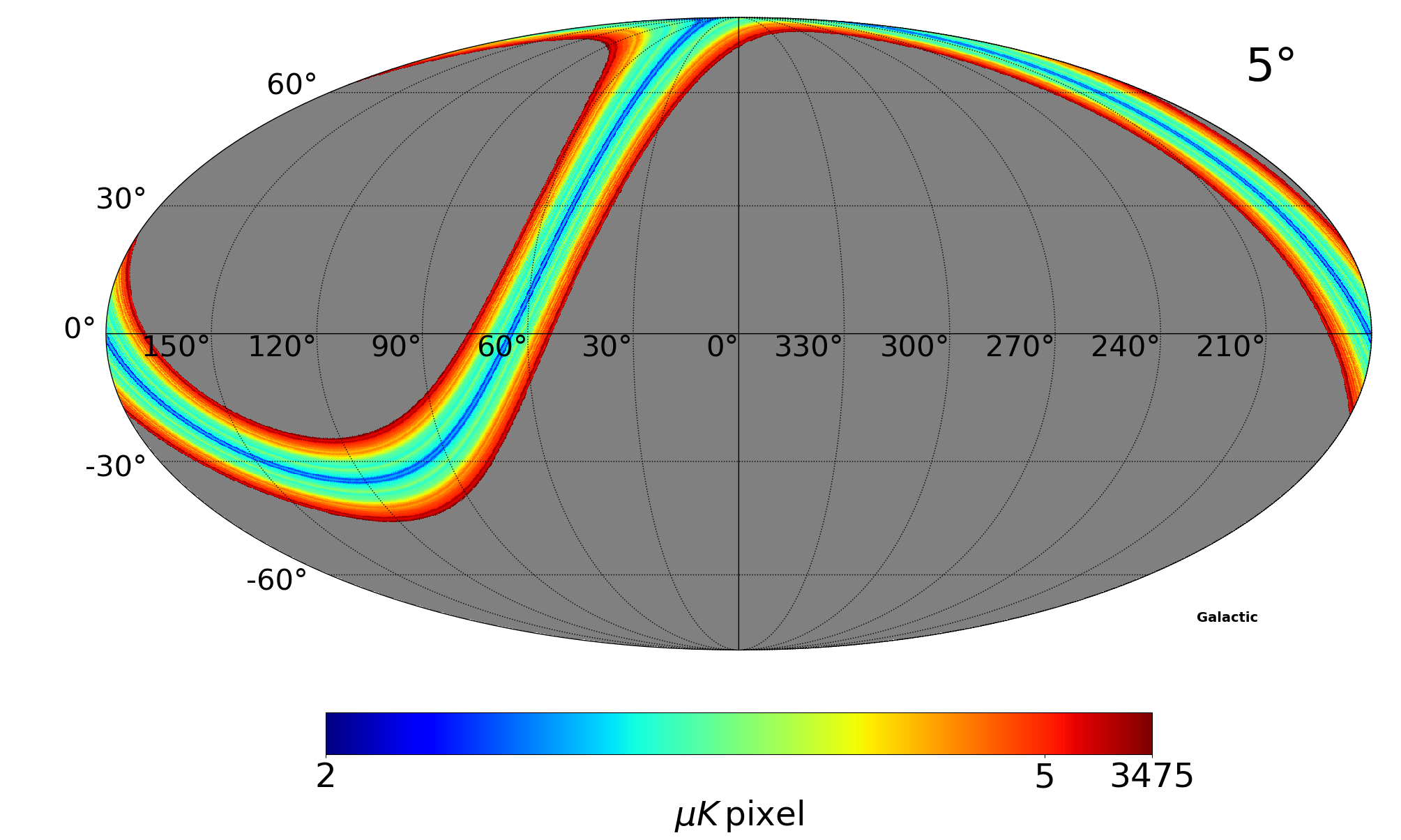}
   \includegraphics[scale=0.085]{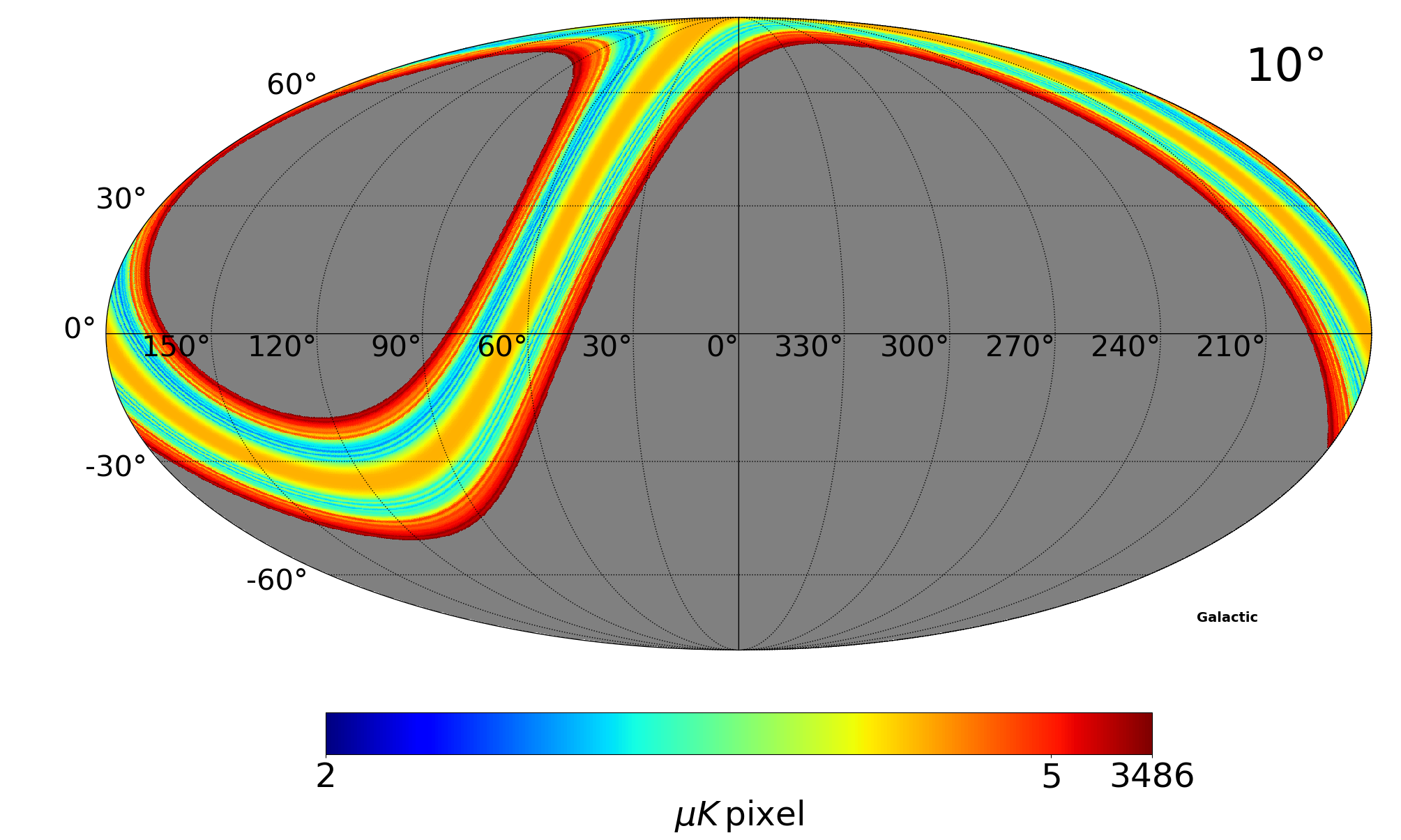}
   \includegraphics[scale=0.085]{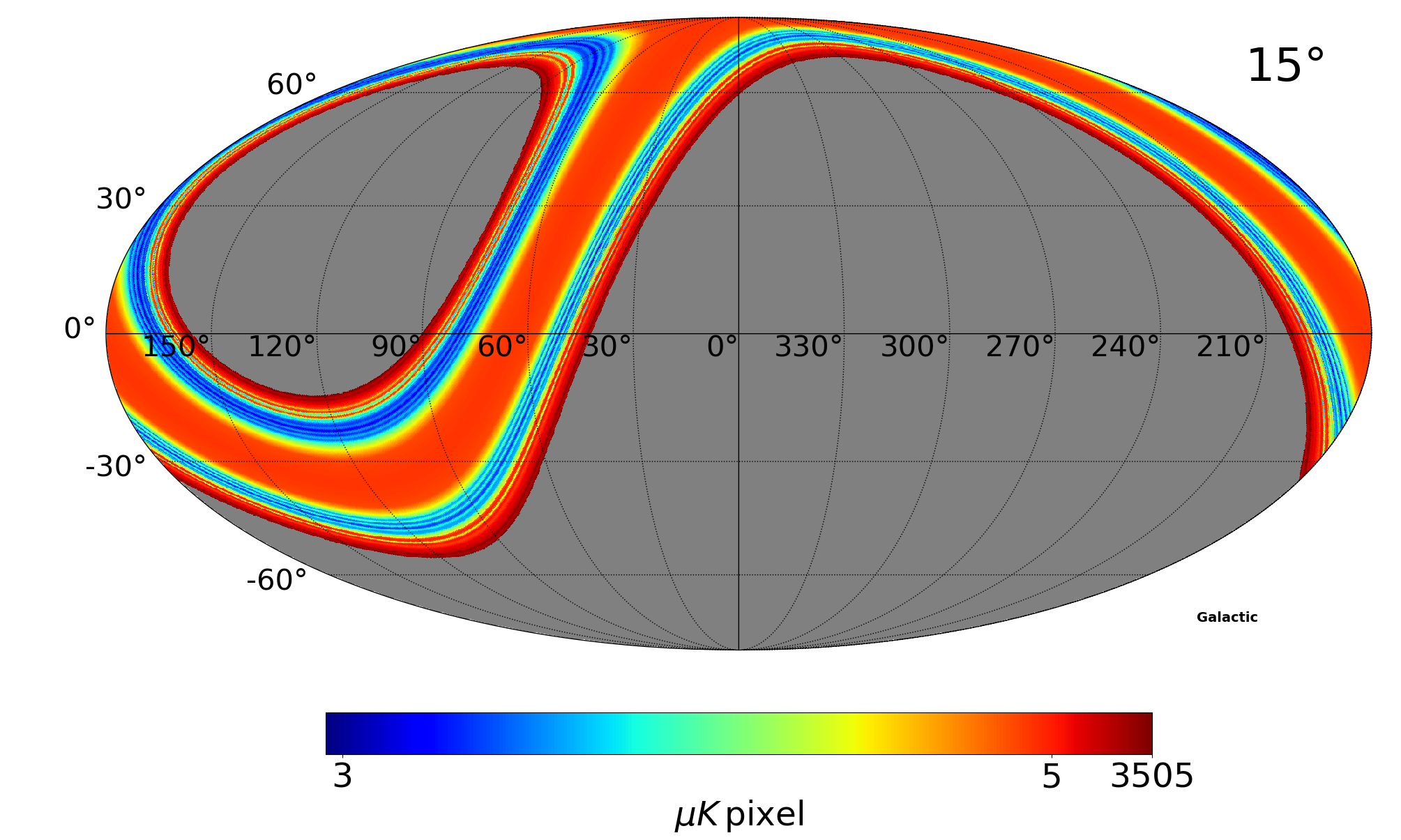}
   \includegraphics[scale=0.085]{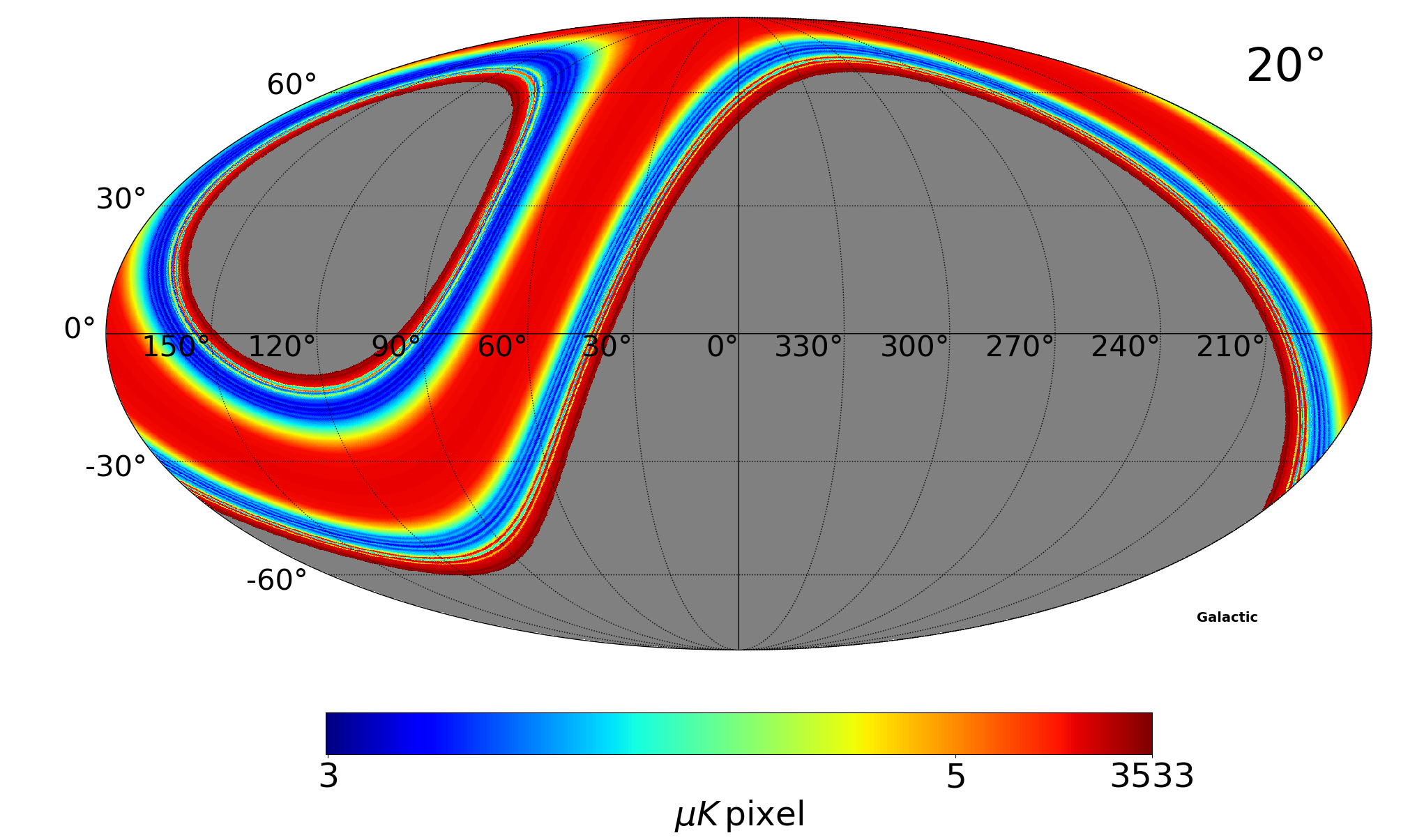}
   \includegraphics[scale=0.085]{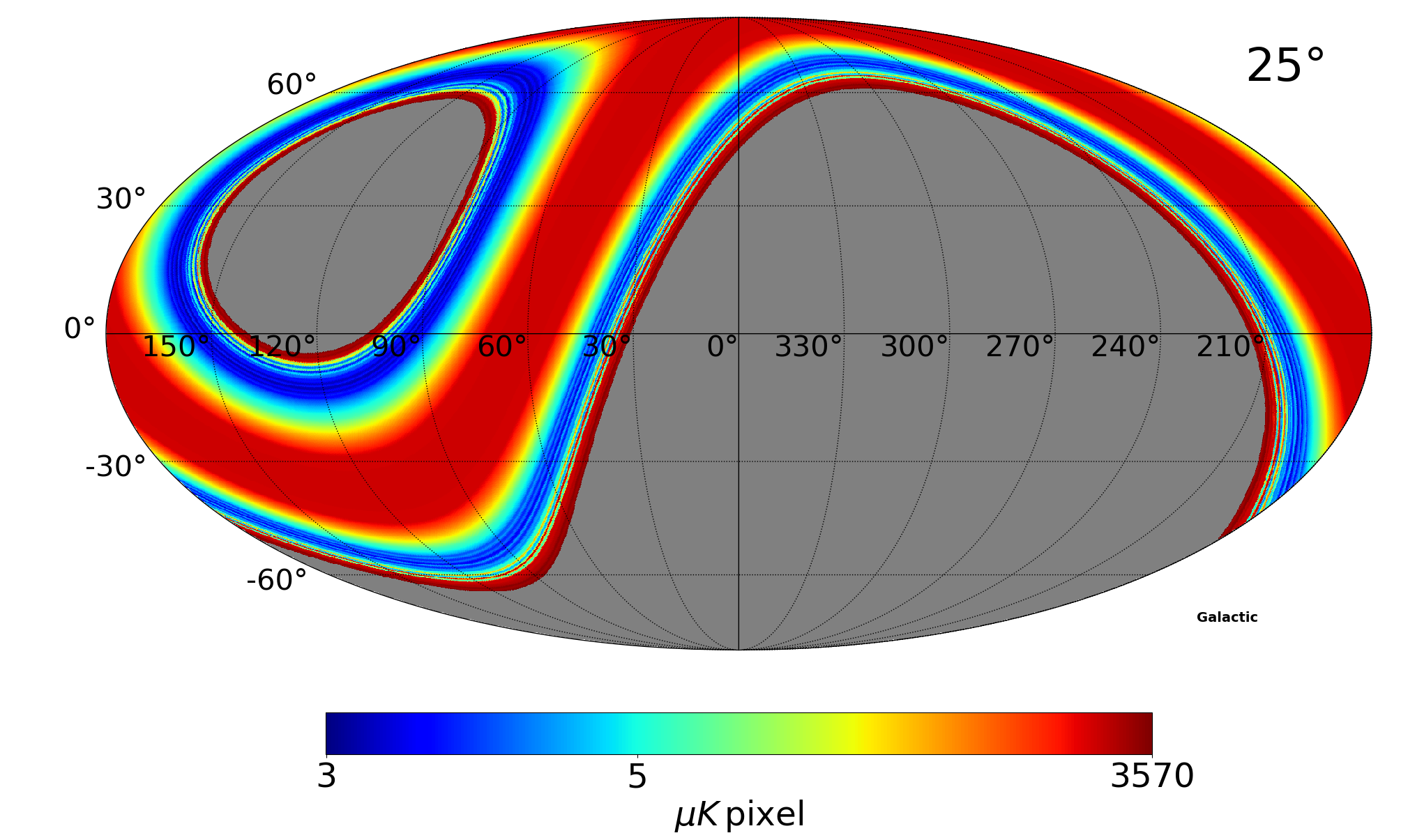}
   \includegraphics[scale=0.085]{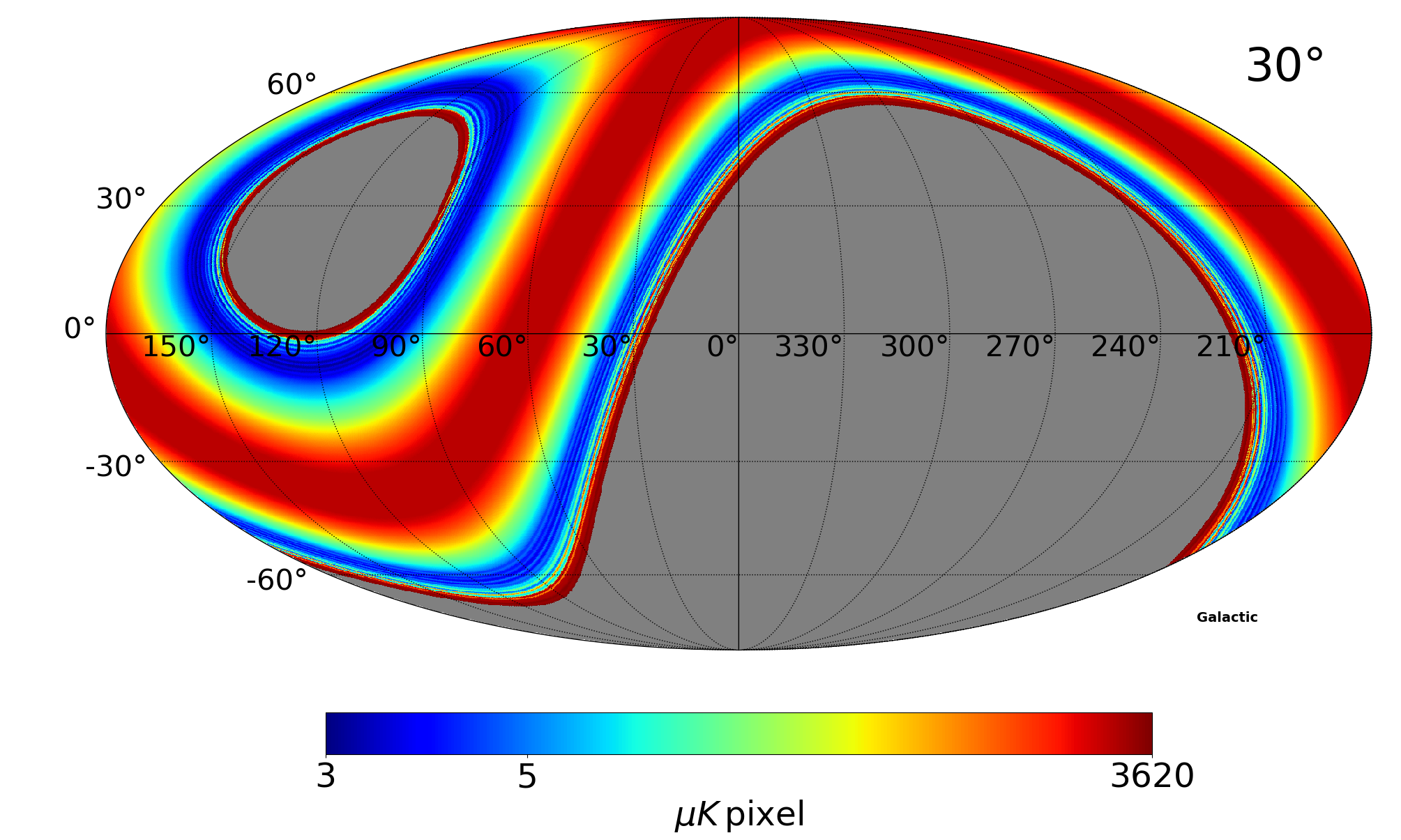}
   \includegraphics[scale=0.085]{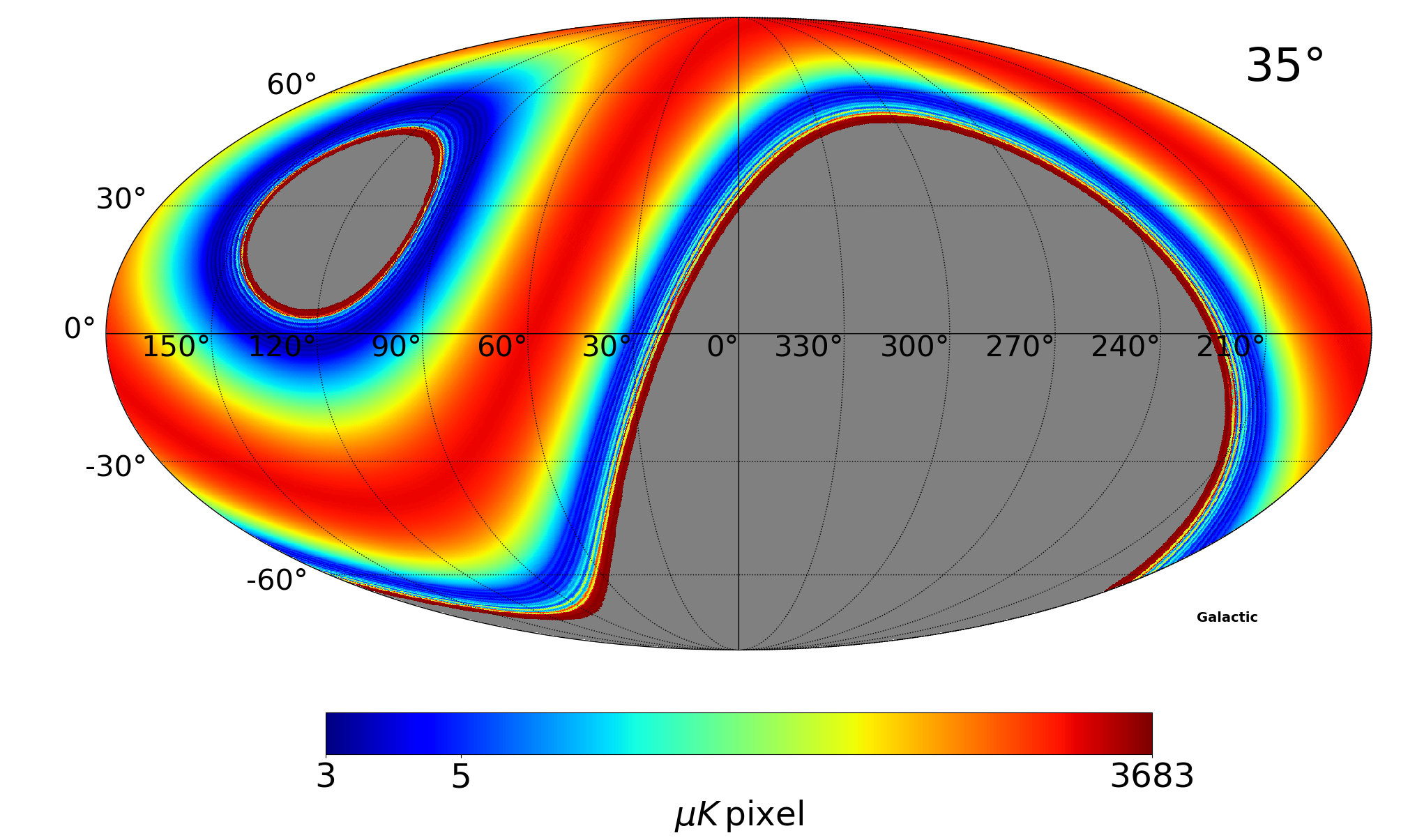}
   \includegraphics[scale=0.085]{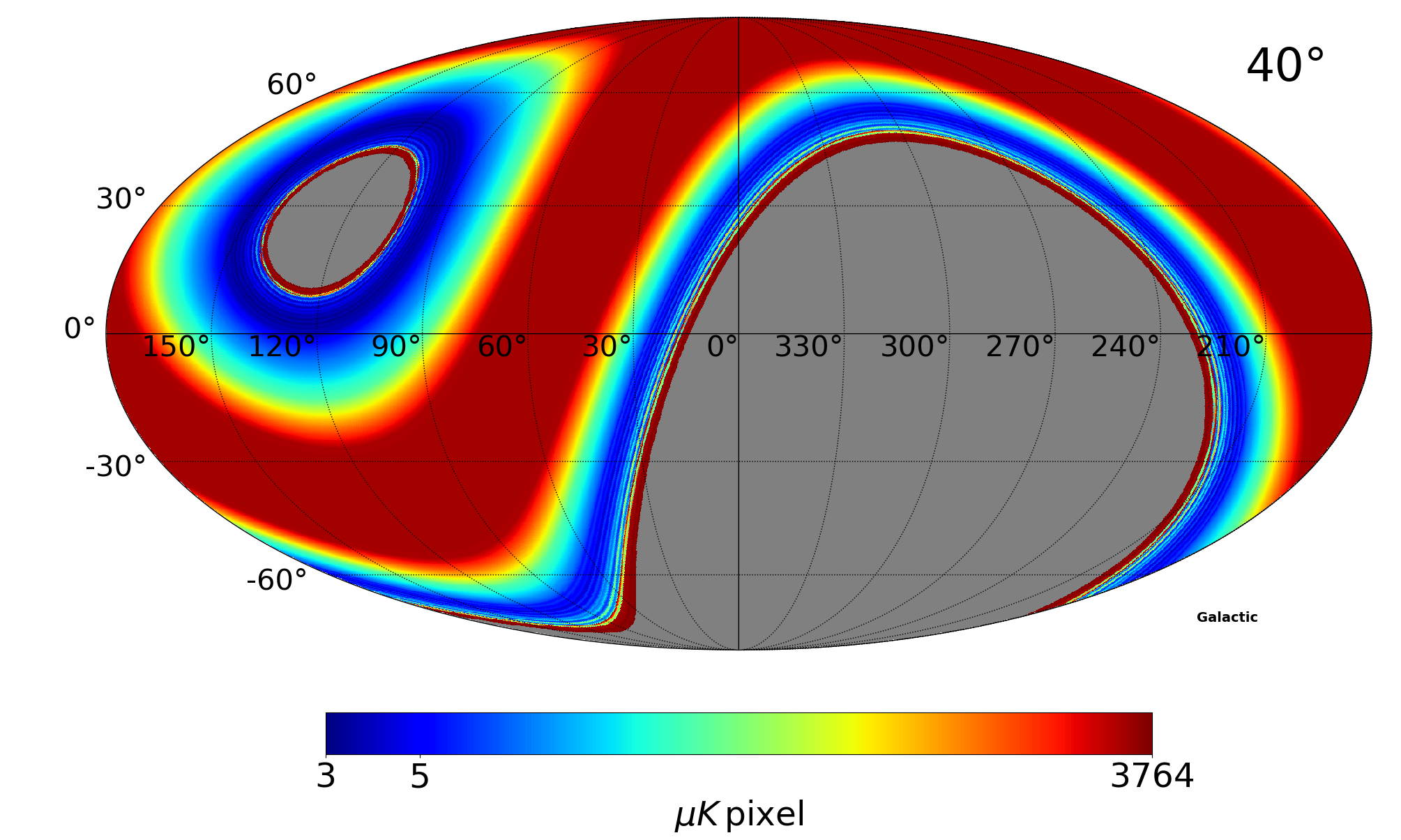}
   \includegraphics[scale=0.085]{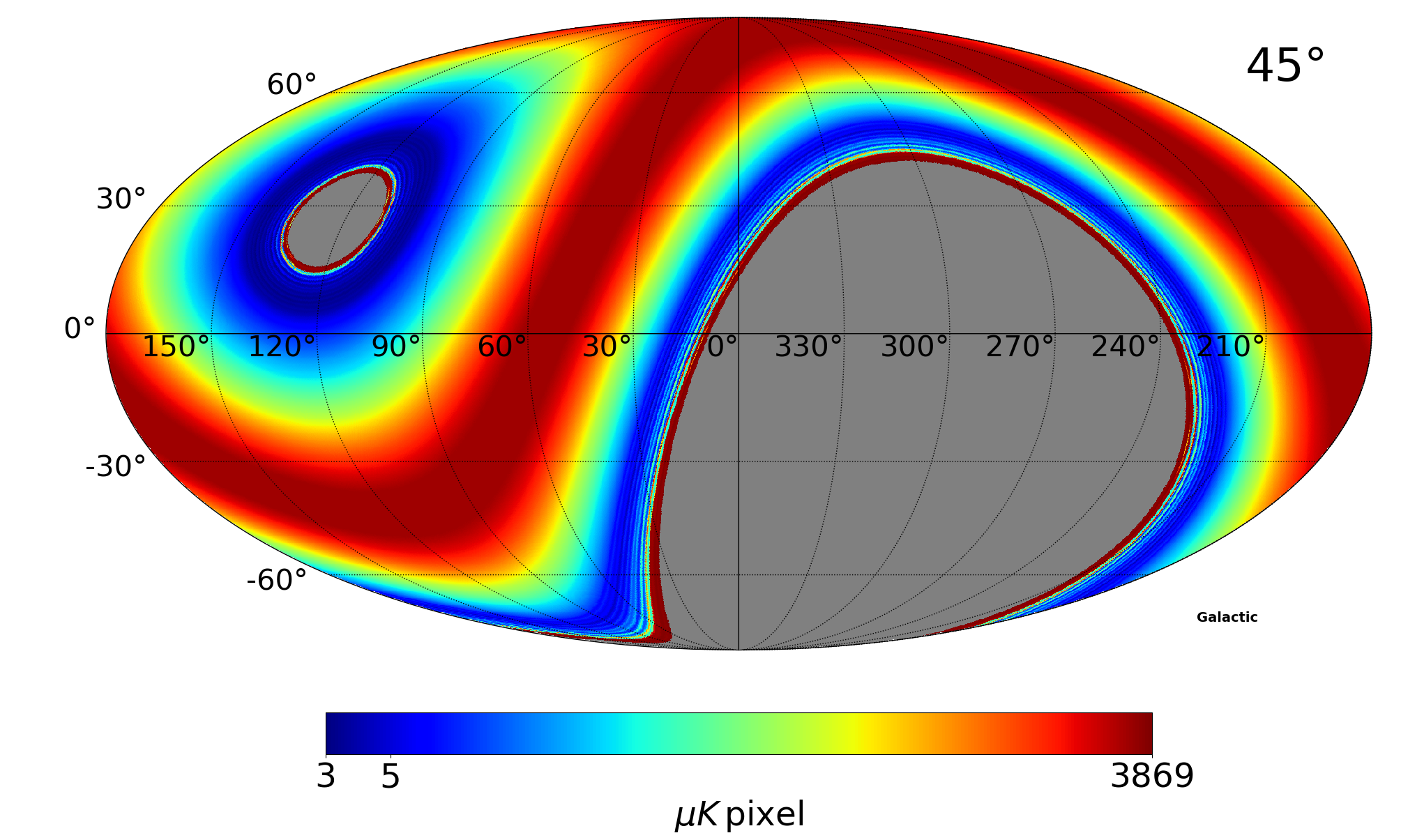}
   \includegraphics[scale=0.085]{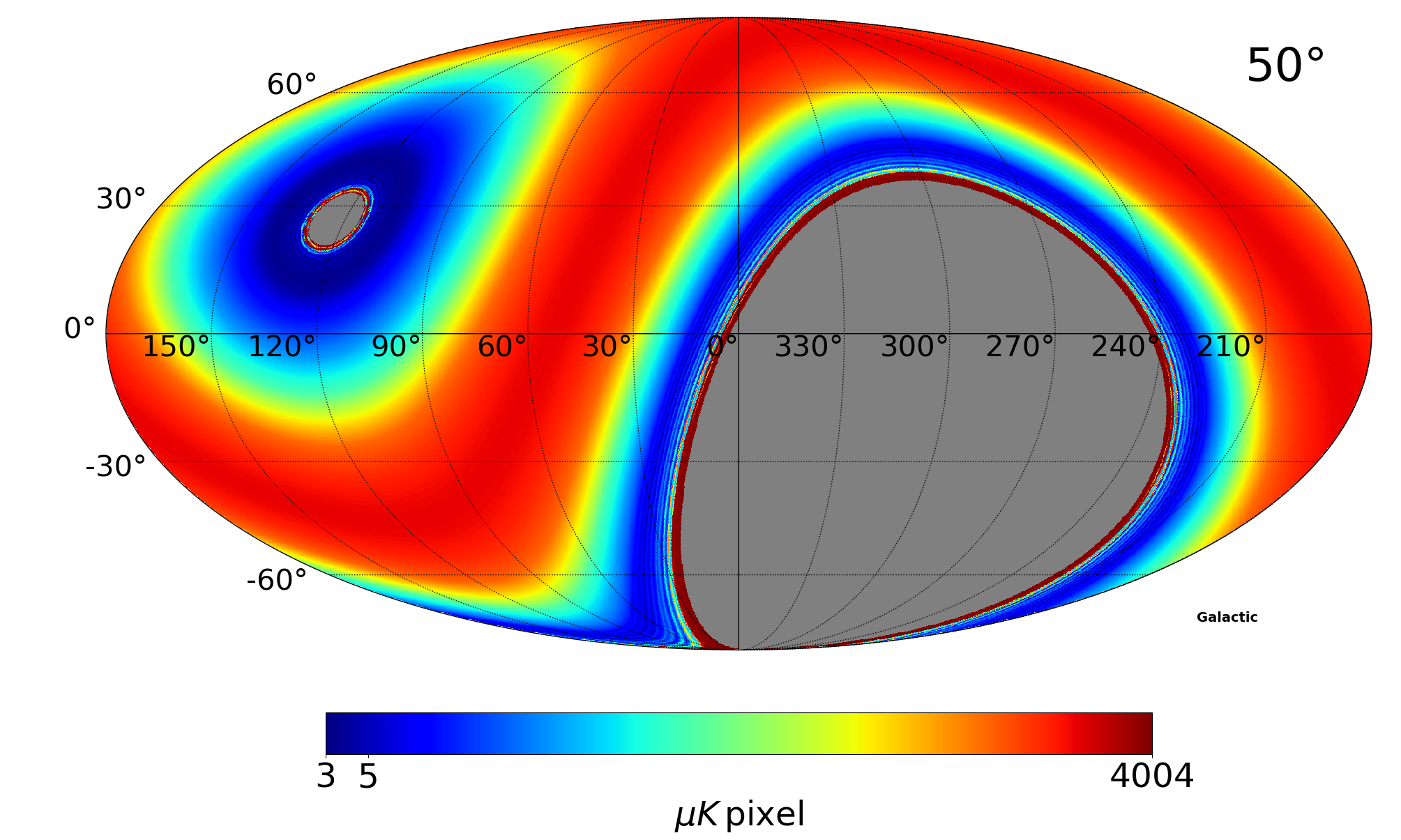}
   \caption[]{STRIP noise maps (in GCS) for several zenith angles (ranging from $5\deg$ to $50\deg$ by steps of $5\deg$, from the left to the right staring from the upper left).}\label{fig:STRIPnoise} 
\end{figure}

Noise maps are shown in Fig. \ref{fig:STRIPnoise} and reflect the same behavior observed for the hit count maps. 

\subsection{Matching the sky coverage between SWIPE and STRIP}
\begin{figure}[H]
\centering
   \includegraphics[scale=0.17]{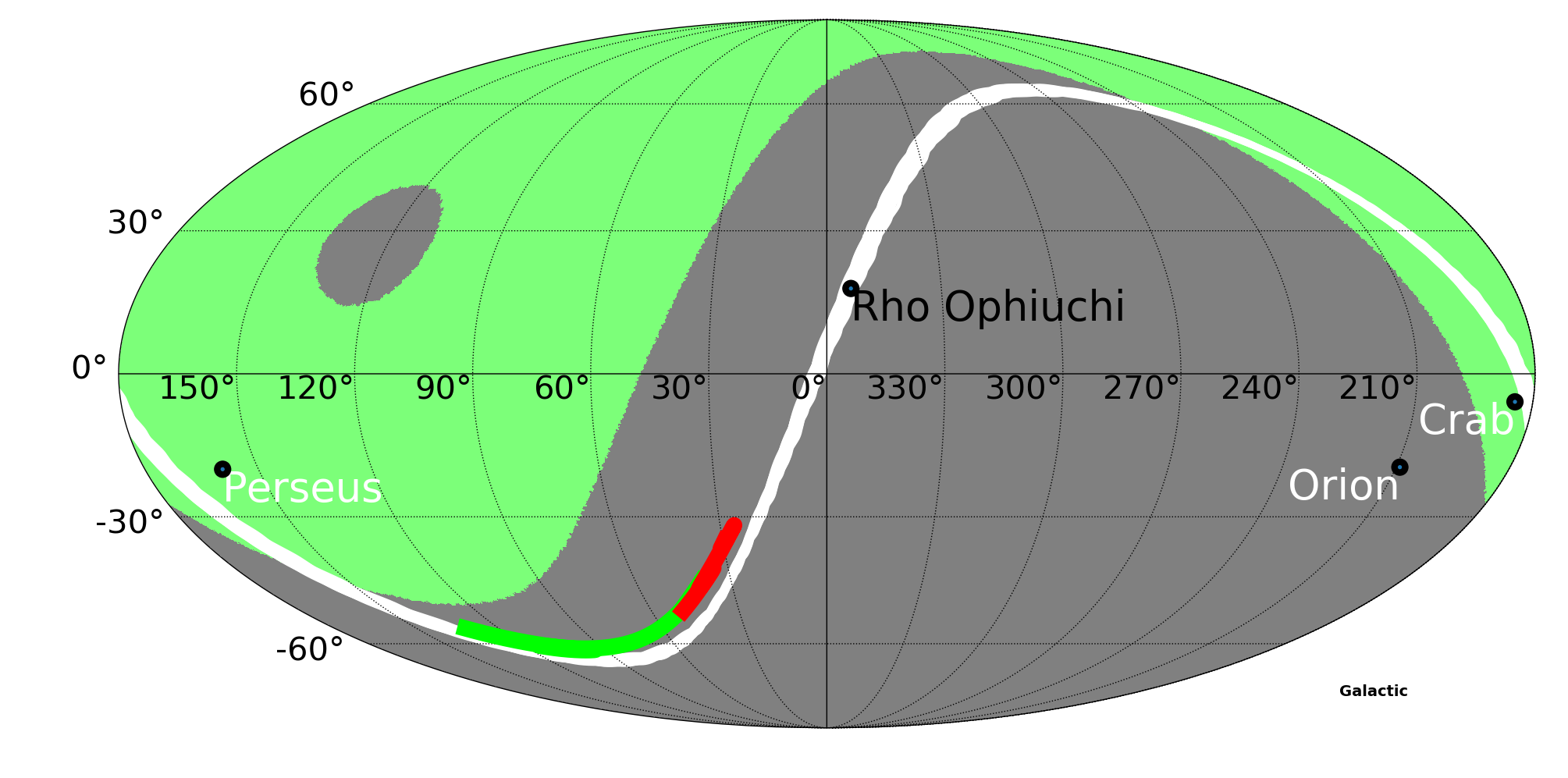}
\caption[]{The SWIPE sky coverage at $240\,\mathrm{GHz}$ (in GCS). The map shows also the positions of the Crab and Orion nebulas, the Perseus molecular cloud, Rho Ophiuchi and the trajectories of Jupiter (green mark), Saturn (red mark) and the Moon (white curve).}\label{fig:SWIPE}
\end{figure}

In this section, I show how we can overlap the STRIP sky coverage with that of SWIPE. SWIPE is expected to observe about $38\%$ of the northern sky covering the sky region shown in Fig. \ref{fig:SWIPE}.\par
\begin{figure}[H]
\centering
\includegraphics[scale=0.25]{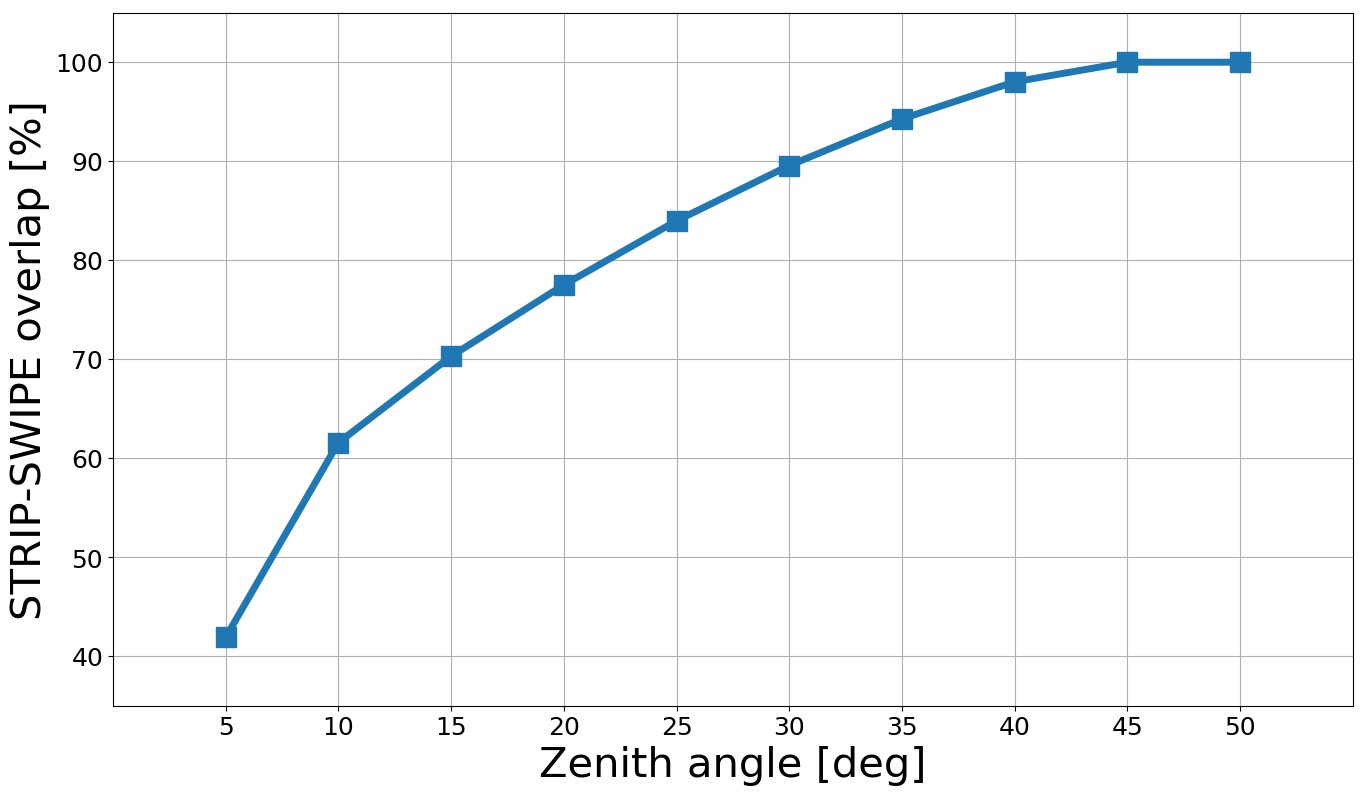}
\caption[]{STRIP-SWIPE overlap as a function of the zenith angle.}\label{fig:overlaperc} 
\end{figure}

\begin{figure}[H]
  \centering
  \includegraphics[scale=0.092]{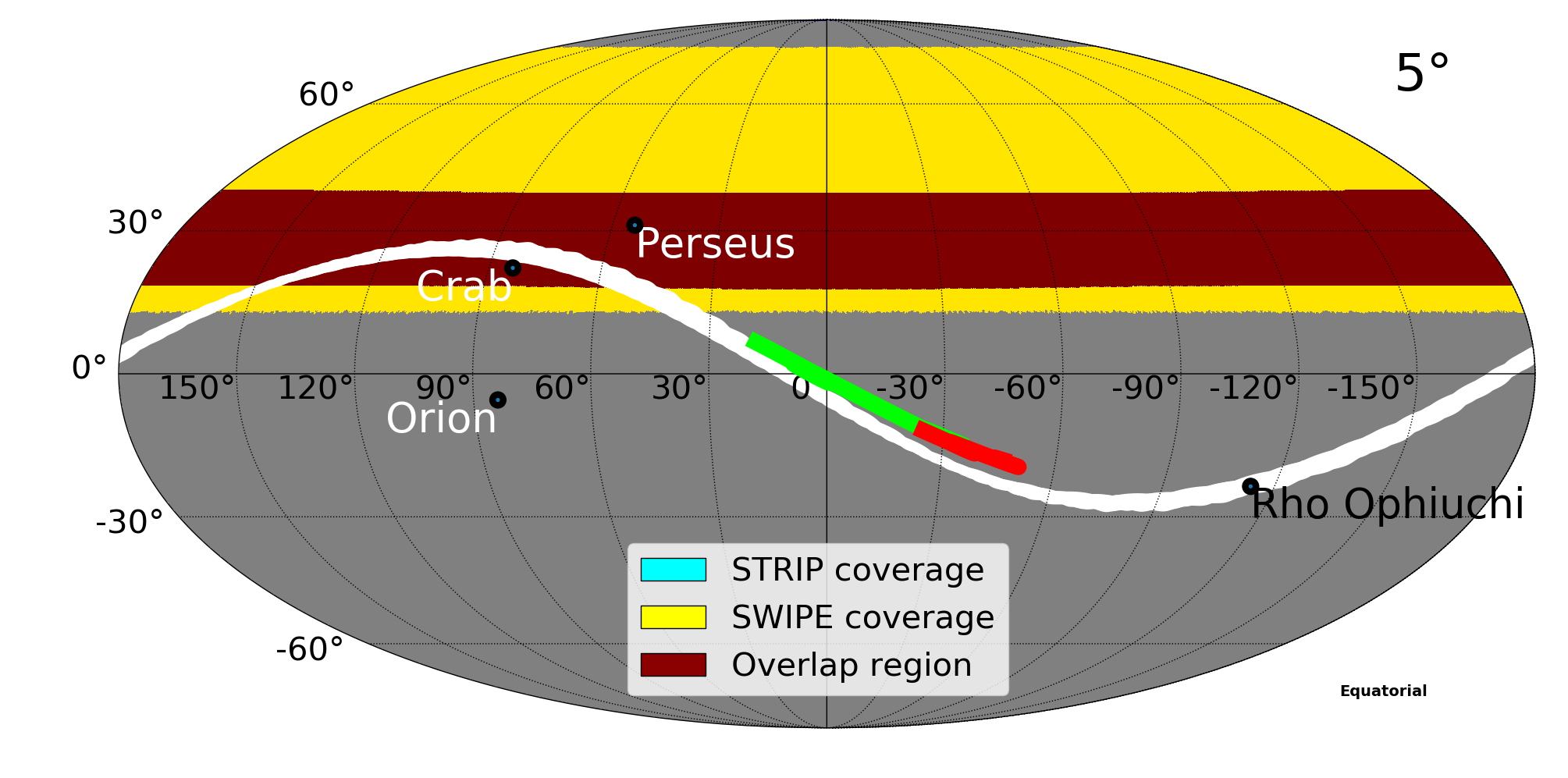}
  \includegraphics[scale=0.092]{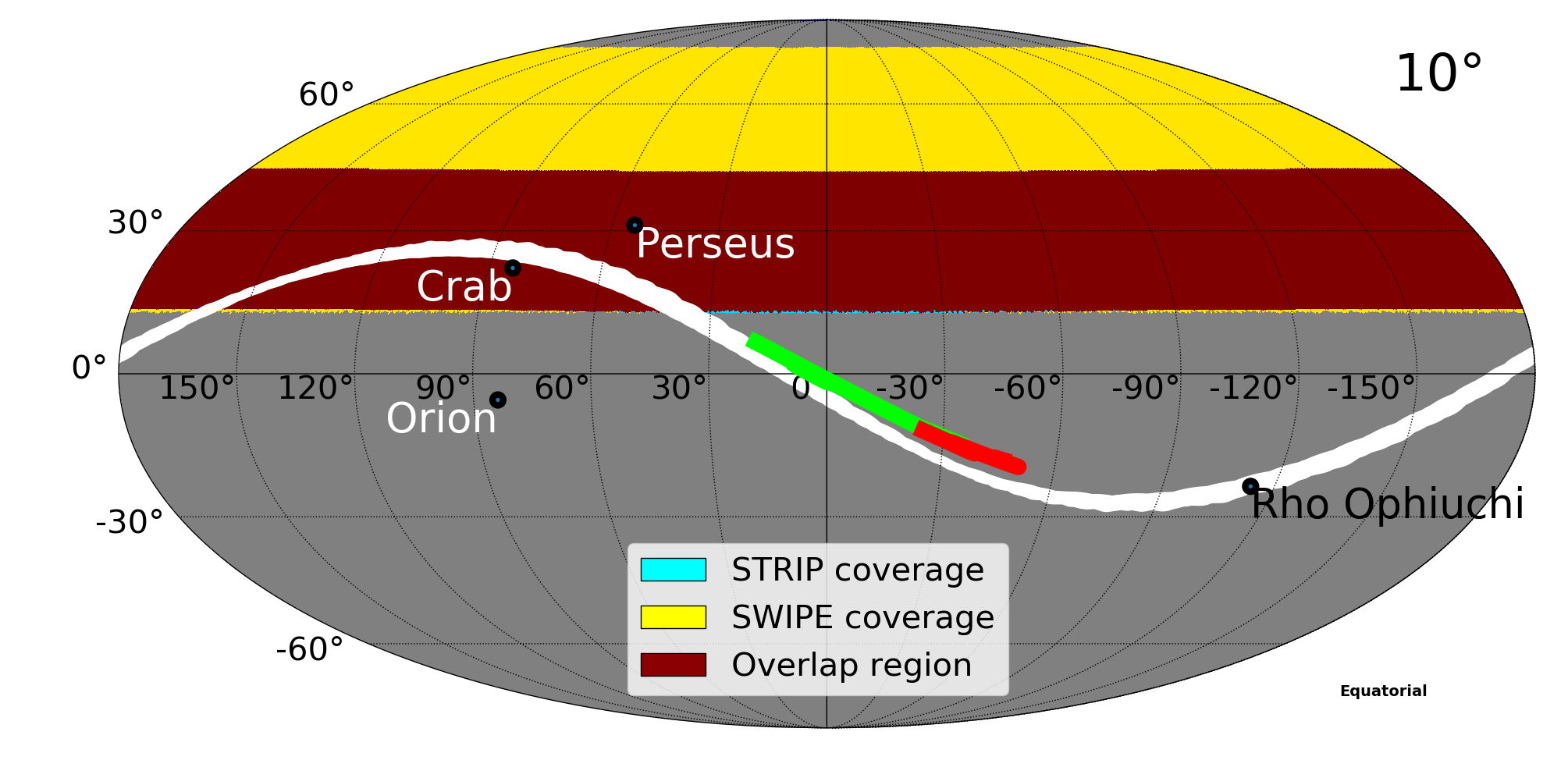}
  \includegraphics[scale=0.092]{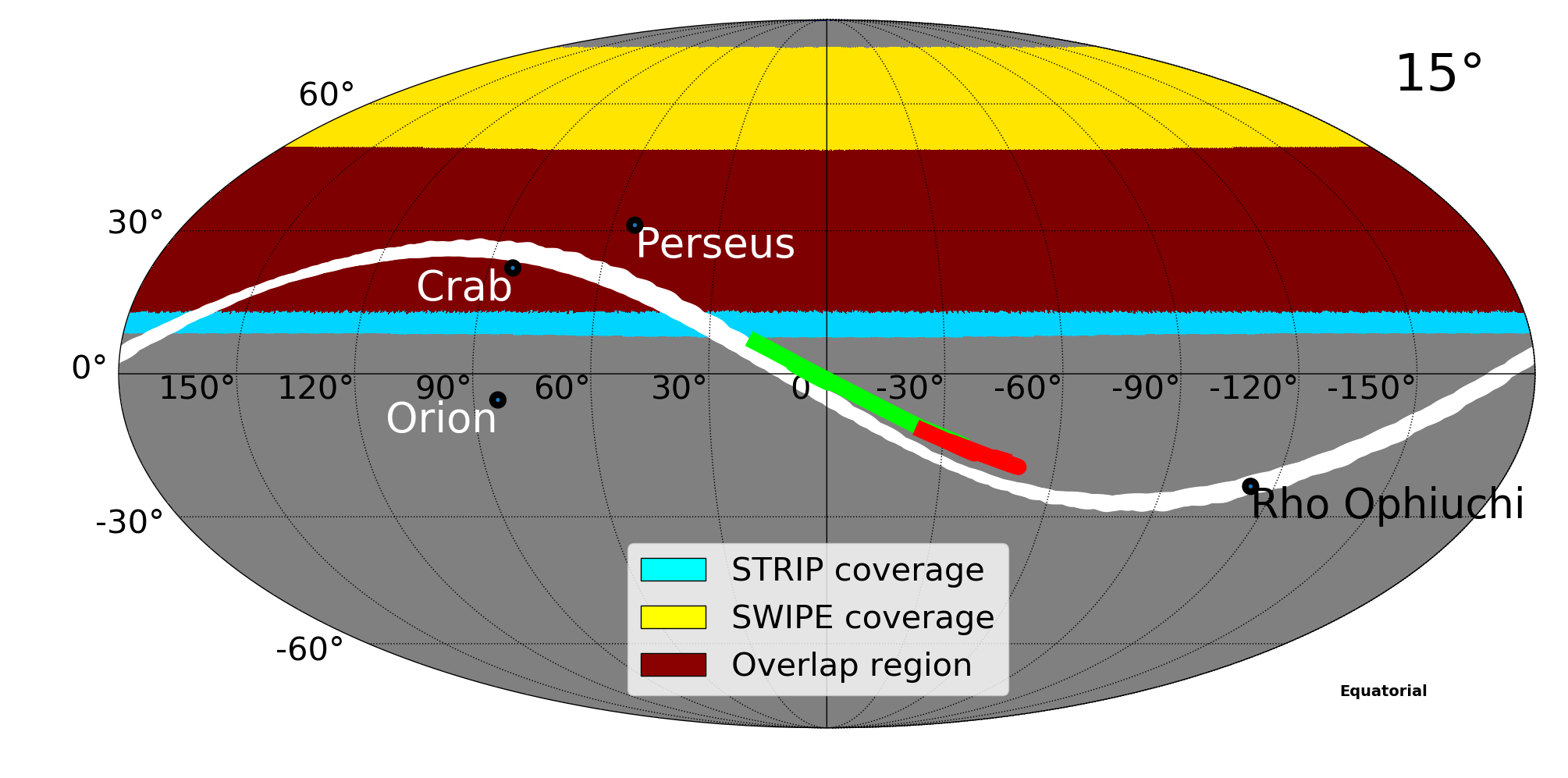}
  \includegraphics[scale=0.092]{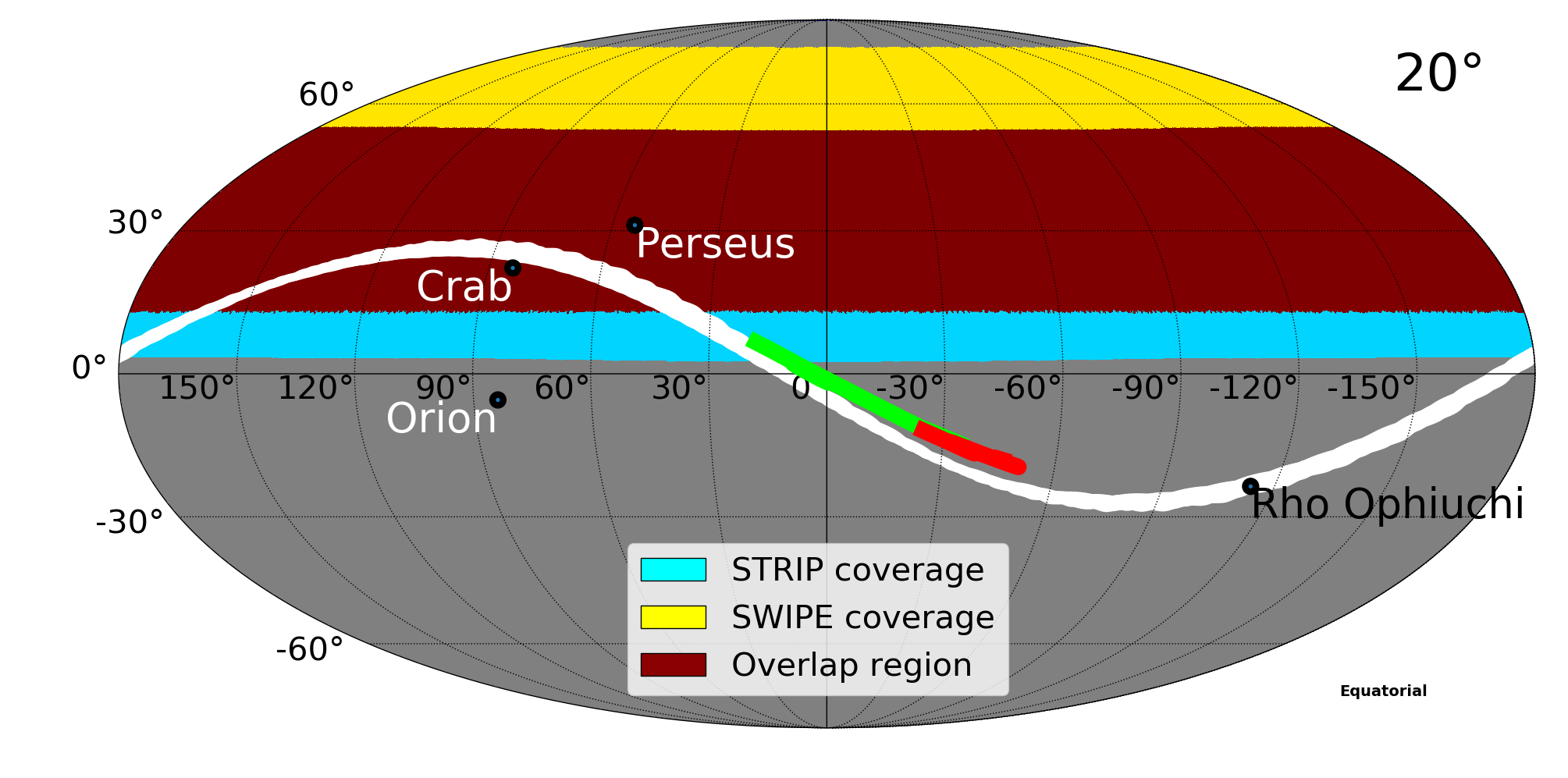}
  \includegraphics[scale=0.092]{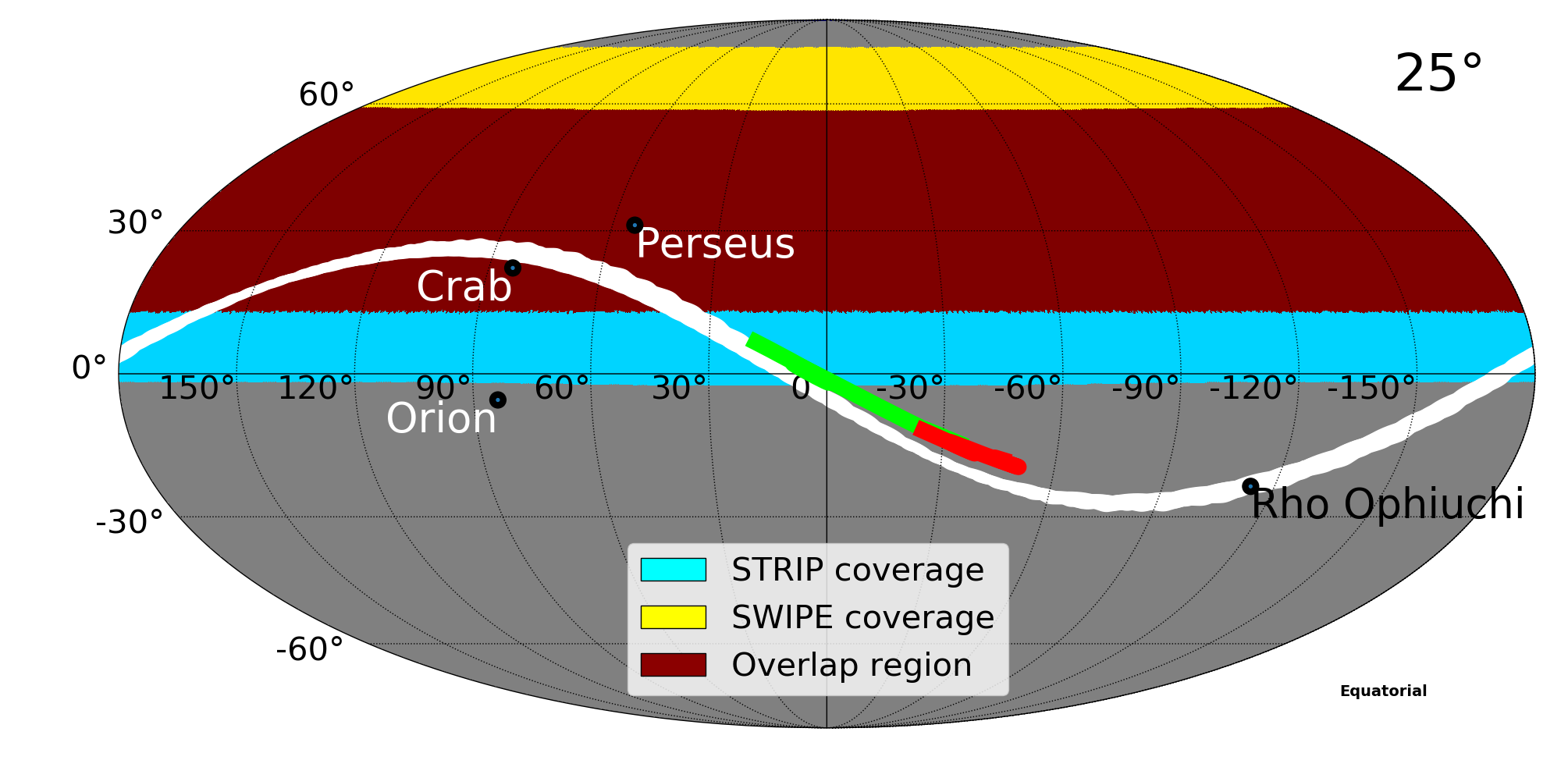}
  \includegraphics[scale=0.092]{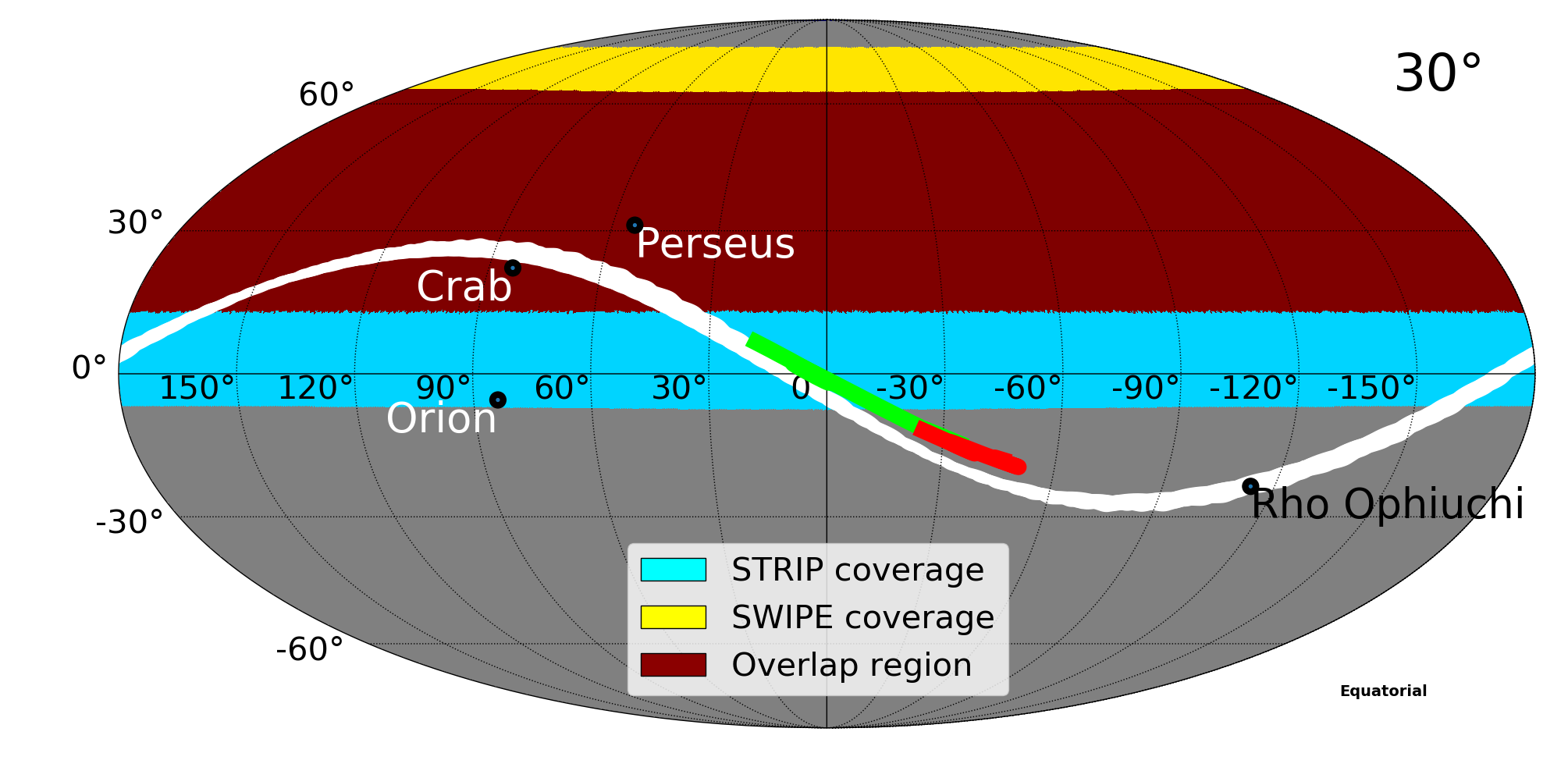}
  \includegraphics[scale=0.092]{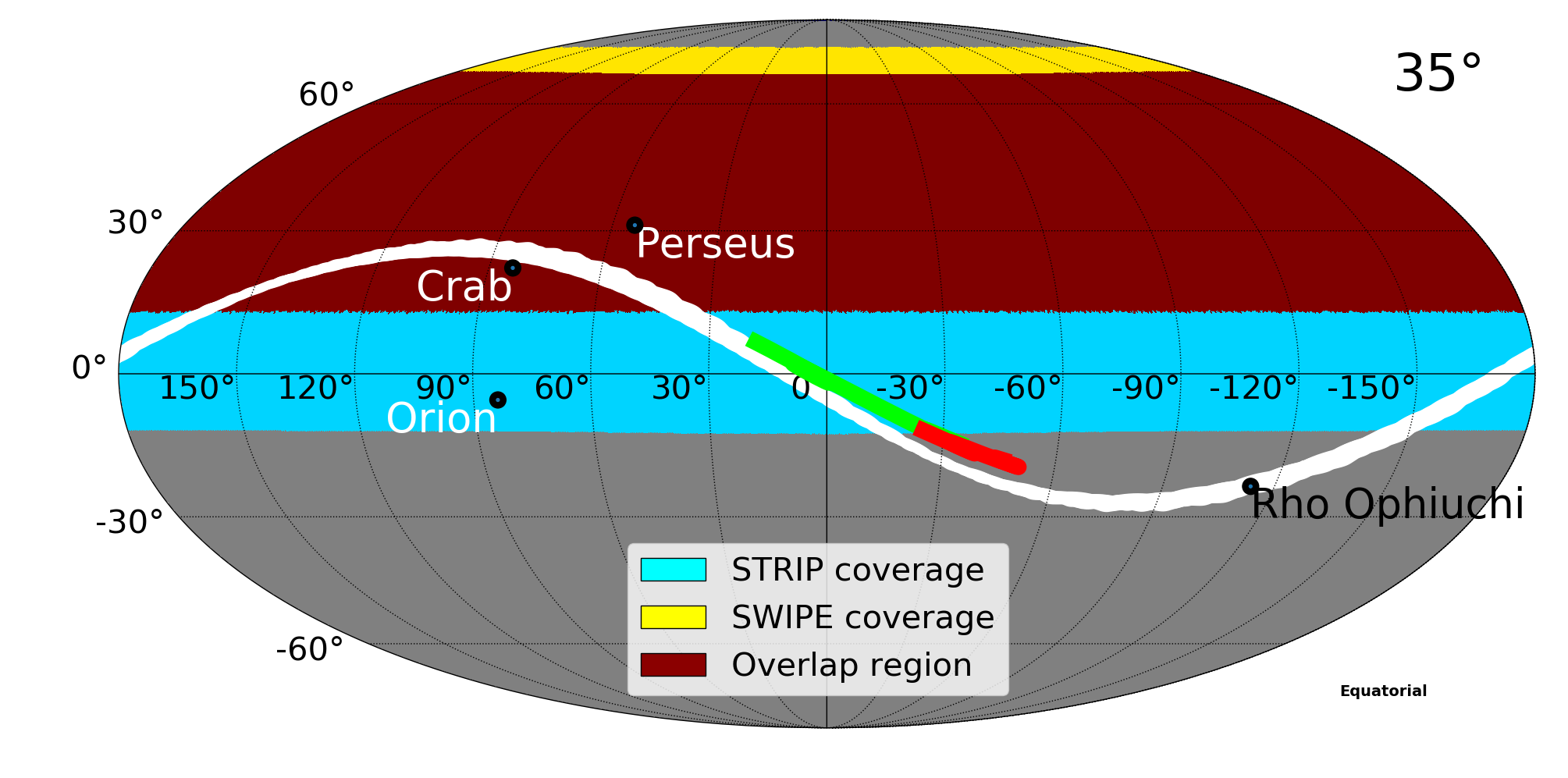}
  \includegraphics[scale=0.092]{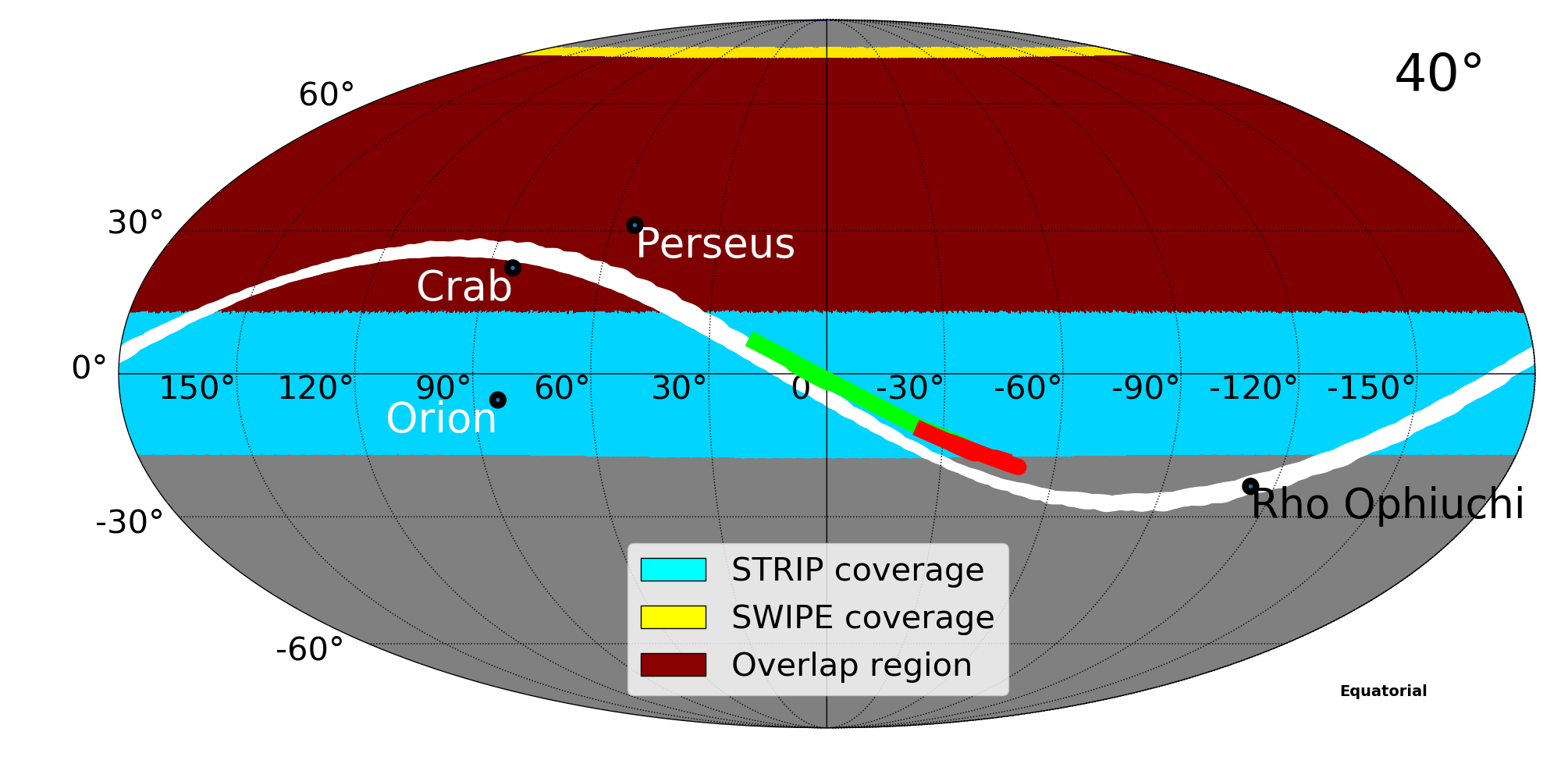}
  \includegraphics[scale=0.092]{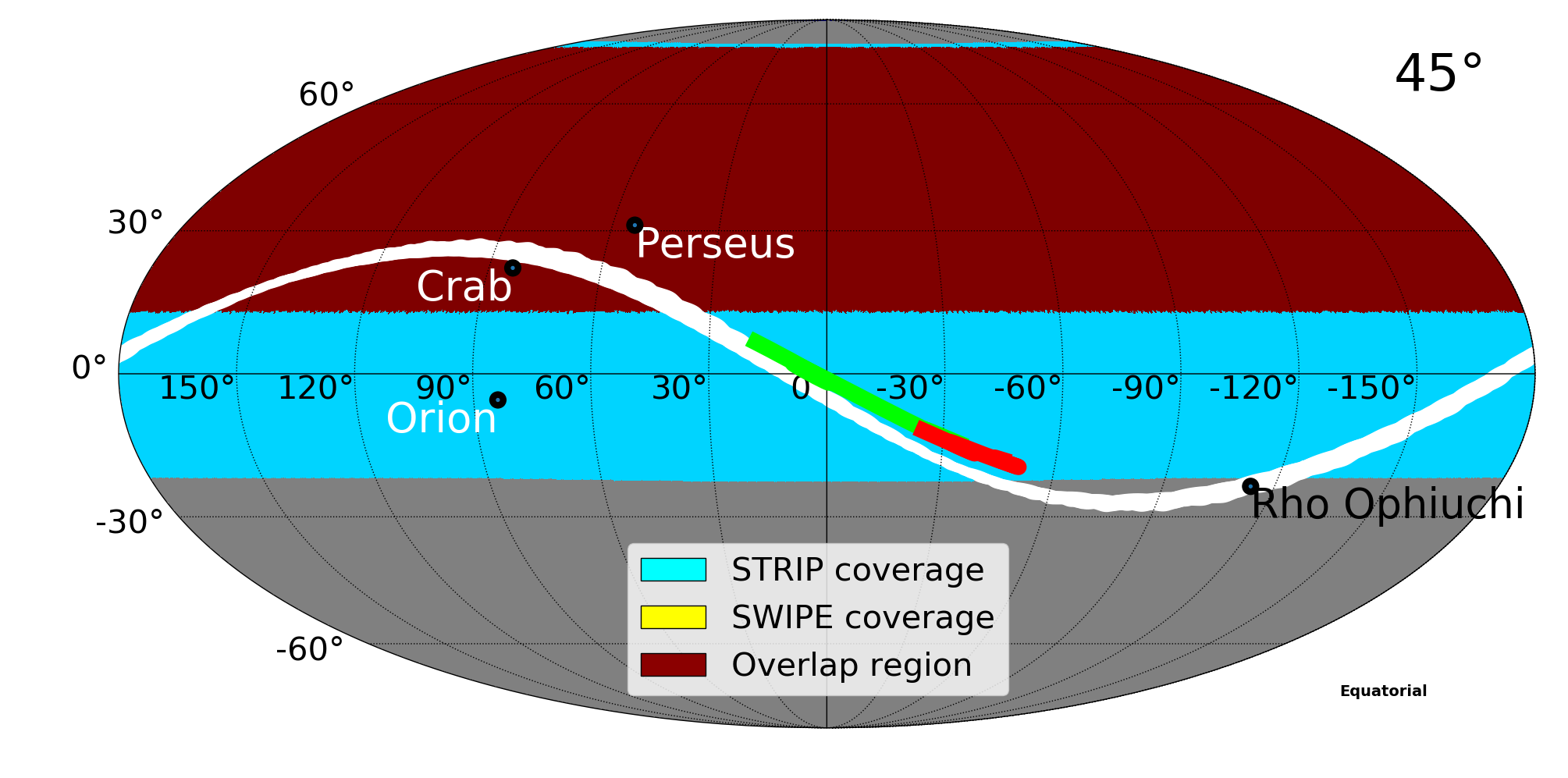}
  \includegraphics[scale=0.092]{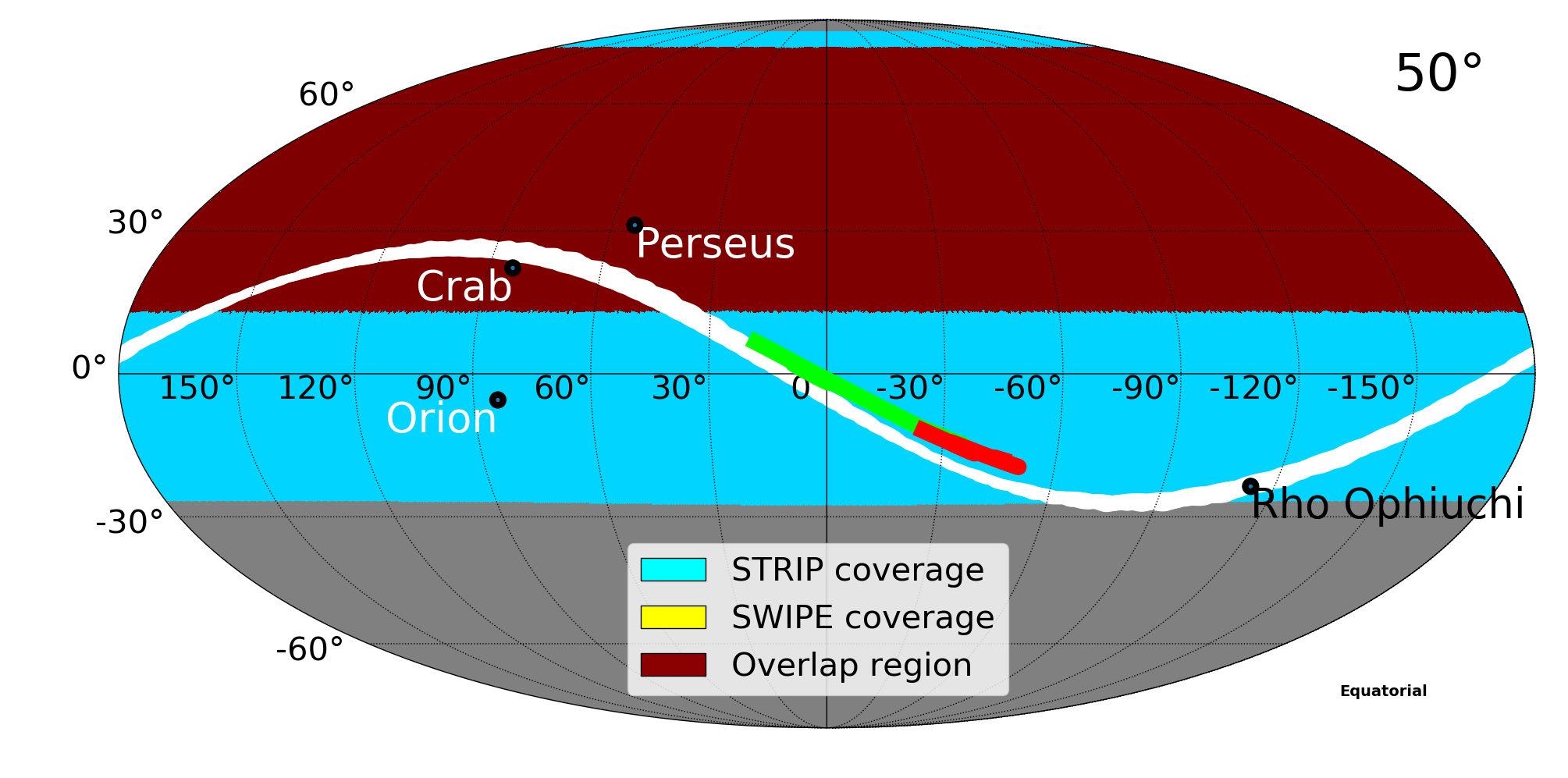}
  \caption[]{Maps of the STRIP-SWIPE overlap (in ECS) for various values of the zenith angle (ranging from $5\deg$ to $50\deg$ by steps of $5\deg$, from the left to the right staring from the upper left). The yellow area represents the SWIPE sky coverage, the cyan area represents the STRIP sky coverage, the dark-red area is the overlap. The maps also show the positions of the Crab and Orion nebulas, the Perseus molecular cloud, Rho Ophiuchi and the trajectories of Jupiter (green mark), Saturn (red mark) and the Moon (white curve).}\label{fig:overmaps} 
\end{figure}

In Fig. \ref{fig:overmaps} I represent the overlap, in the equatorial coordinate system (ECS), of the sky regions covered by SWIPE and STRIP as a function of zenith angle of the STRIP telescope. The yellow area represents the SWIPE coverage, the cyan area the STRIP coverage, the dark-red area the overlap region. The percentage of overlap as a function of the elevation angle is shown in Fig. \ref{fig:overlaperc}.\par

\subsection{Trade-off between sky coverage and sensitivity}\label{to}
\begin{figure} [H]
  \centering
  \includegraphics[scale=0.35]{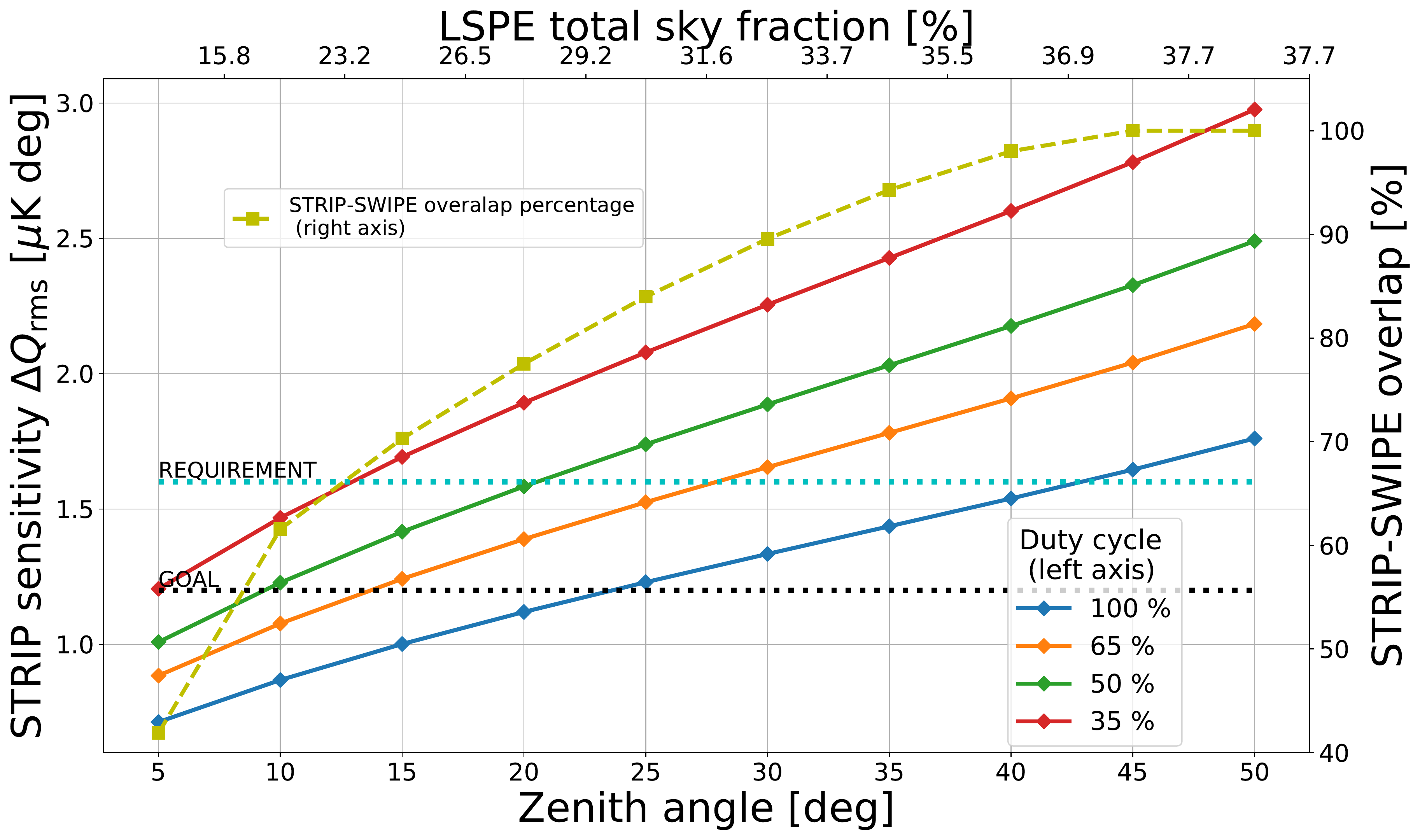}
  \caption[]{LSPE/STRIP sensitivity as a function of the zenith distance and of the fraction of usable duty cycle (left axis). The black and cyan dotted lines (left axis) represent, respectively, the sensitivity goal and requirement. The dashed gold line shows the STRIP/SWIPE overlap percentage (right axis). On the top X-axis are reported the corresponding cumulative sky fractions.}\label{fig:tradoffpicture}
\end{figure}

In Fig. \ref{fig:tradoffpicture}, I show the average STRIP sensitivity as a function of the zenith angle and of the fraction of usable duty cycle, compared with requirement ($1.6\,\mu\mathrm{K}\,\mathrm{deg}$, Sect. \ref{STRIPlfi}) and goal ($1.2\,\mu\mathrm{K}\,\mathrm{deg}$, Sect. \ref{STRIPasaw}) values. This is obtained by re-scaling the noise maps shown in Fig. \ref{fig:STRIPnoise} to pixel of $1\deg$ and to the proper duty cycle fraction and then by taking the average value of each map.\par
For zenith angles between $20\deg$ and $25\deg$ we can obtain an overlap of the order of $80\%$ but we also need to ensure a duty cycle greater than $50\%$ to satisfy the requirement. At the same angles a duty cycle of about $100\%$ is required to reach the goal. \par
A detailed discussion about the STRIP duty cycle is reported in Sect. \ref{dc}

\subsection{Observing particular sky regions}\label{pois}
The possibility to scan galactic sources is of particular interest for STRIP. In particular, the minimum requirement set for the scanning strategy is to properly cover the Crab Nebula, the Orion Nebula and the Perseus Molecular Cloud.\par
The Crab Nebula is the best known polarized calibrator in the sky. The Perseus Molecular Cloud instead, is interesting for the search of polarized \textit{Anomalous Microwave Emission} (AME). On the contrary, the Orion Nebula is an unpolarized source that will be useful to observe for checking any spurious polarizations of the instrument.\par
My simulations show that the Crab Nebula and Perseus are visible in all considered cases, while the Orion Nebula is visible for zenith angles $\ge\,30\deg$. Therefore, to observe Orion we will probably need deep scans, maintaining the elevation angle less than $30\deg$ in the nominal scanning mode, not to lose too much sensitivity.\par
Rho-Ophiuchi instead, which is the most widely known source of AME in the sky is not going to be visible from Tenerife in the nominal scanning mode. \par
We also require planets to fall within the main beam of each acquisition chain, as their angular size is much smaller than the beam size and can therefore be used to characterize the beam response itself, in particular Jupiter, Saturn and the Moon. \par
My simulations show that the Moon can be seen for all zenith angles in some times of the year (Fig. \ref{fig:STRIP}). It will be also possible to observe Jupiter during the continuous survey starting from an angle of $20\deg$ even if for a limited amount of time of the year, which will increase by increasing the elevation angle. As for Saturn, it will be quite low on the horizon, so it will not possible to observe it during the continuous survey, unless we increase the elevation between $35\deg$ and $50\deg$. Such high elevation angles are incompatible with the STRIP sensitivity requirements for the large scale survey. This means that, to observe Orion, Saturn and Rho-Ophiuchi dedicated observations are required.

\section{Duty cycle estimation}\label{dc}
In this section, I provide an estimation of the STRIP duty cycle by taking into account several causes preventing the telescope to observe for the whole duration of the mission. Precisely, the duty cycle is defined as the ratio between the effective time spent observing the CMB and the total time.  
\begin{list}{\leftmargin 15pt \itemsep 0pt \topsep 3pt}
\item{\bf Sun.}
  The astrophysical observations of STRIP can be performed both by night and day. During daylight hours, the concern is about avoiding the direct microwave emission of the Sun as well as the instrument overheating due to direct pointing. For this reason, I assume a safety margin of $\simeq 10\deg$ between the Sun position and the telescope pointing. Assuming also that the telescope spins with positive elevation, the horns pointing at the lowest altitudes are ``$\mathrm{R3}$'' and ``$\mathrm{V3}$'' (Fig. \ref{telesc}).\par
  In Fig. \ref{Altitudes}, I show, in horizontal coordinates (HC), the altitude of the Sun above the horizon during the two years of mission (from 1 April 2021 to 1 April 2023)  from ``Teide Observatory'' and the altitudes of $\mathrm{R3}$ and $\mathrm{V3}$ (which are equal), for two values of zenith angle of the STRIP telescope: $20\deg$ and $25\deg$. \par
  I have computed the time during which the LOSs of $\mathrm{R3}$ and $\mathrm{V3}$ are above the Sun by $10\deg$ as a function of the elevation angle of the STRIP telescope (Fig. \ref{dcVSzd}). For zenith distances between $20\deg$ and $25\deg$ the effective observation time is $\simeq 85\%$. \par
\begin{figure}[H]
  \centering
  \includegraphics[scale=0.3]{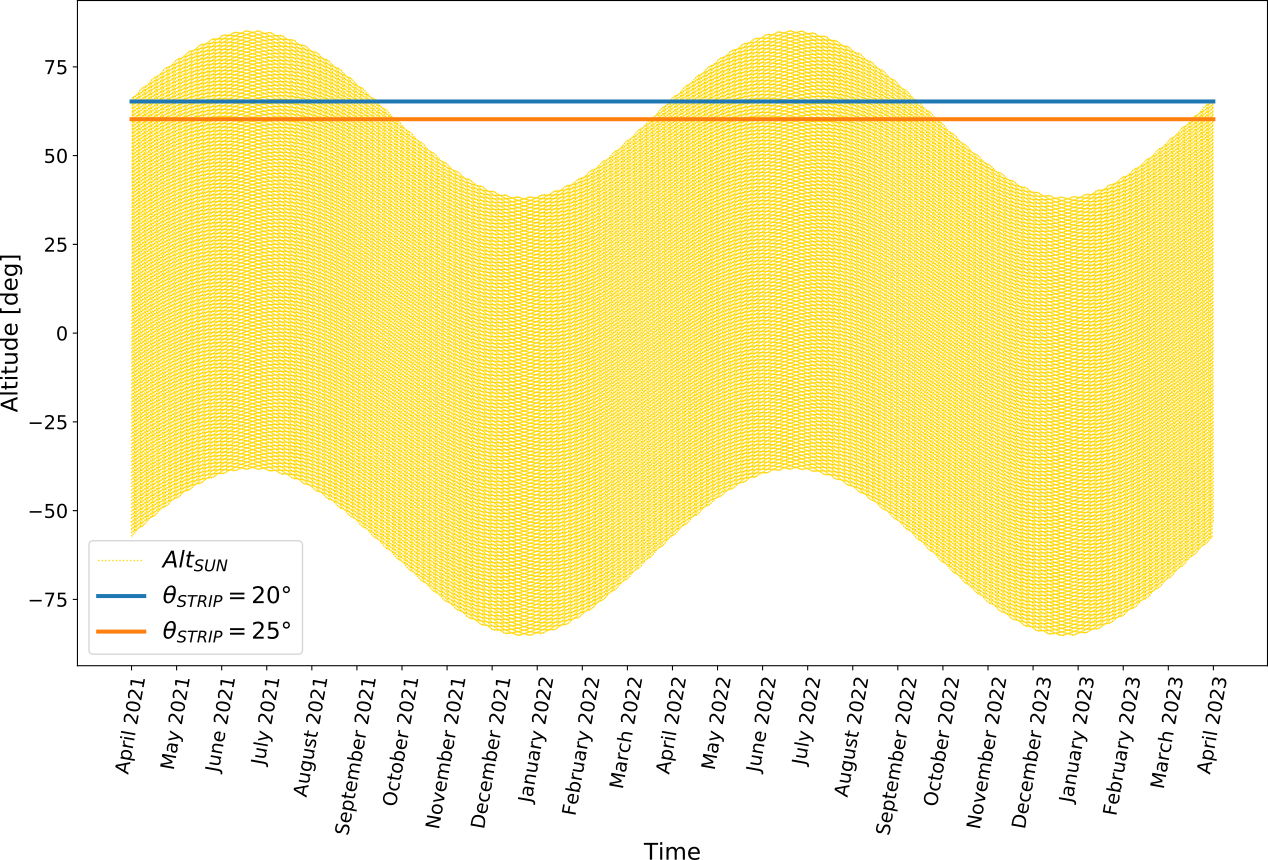}
  \caption[]{Altitudes (in HC) of the Sun (yellow curve) and of the horns $\mathrm{R3}$ and $\mathrm{V3}$ during the nominal scanning mode with telescope elevations of $20\deg$ (blue curve) and $25\deg$ (orange curve). Negative altitudes are below the Tenerife horizon.}\label{Altitudes}
\end{figure}\par
\begin{figure}[H]
  \centering
  \includegraphics[scale=0.4]{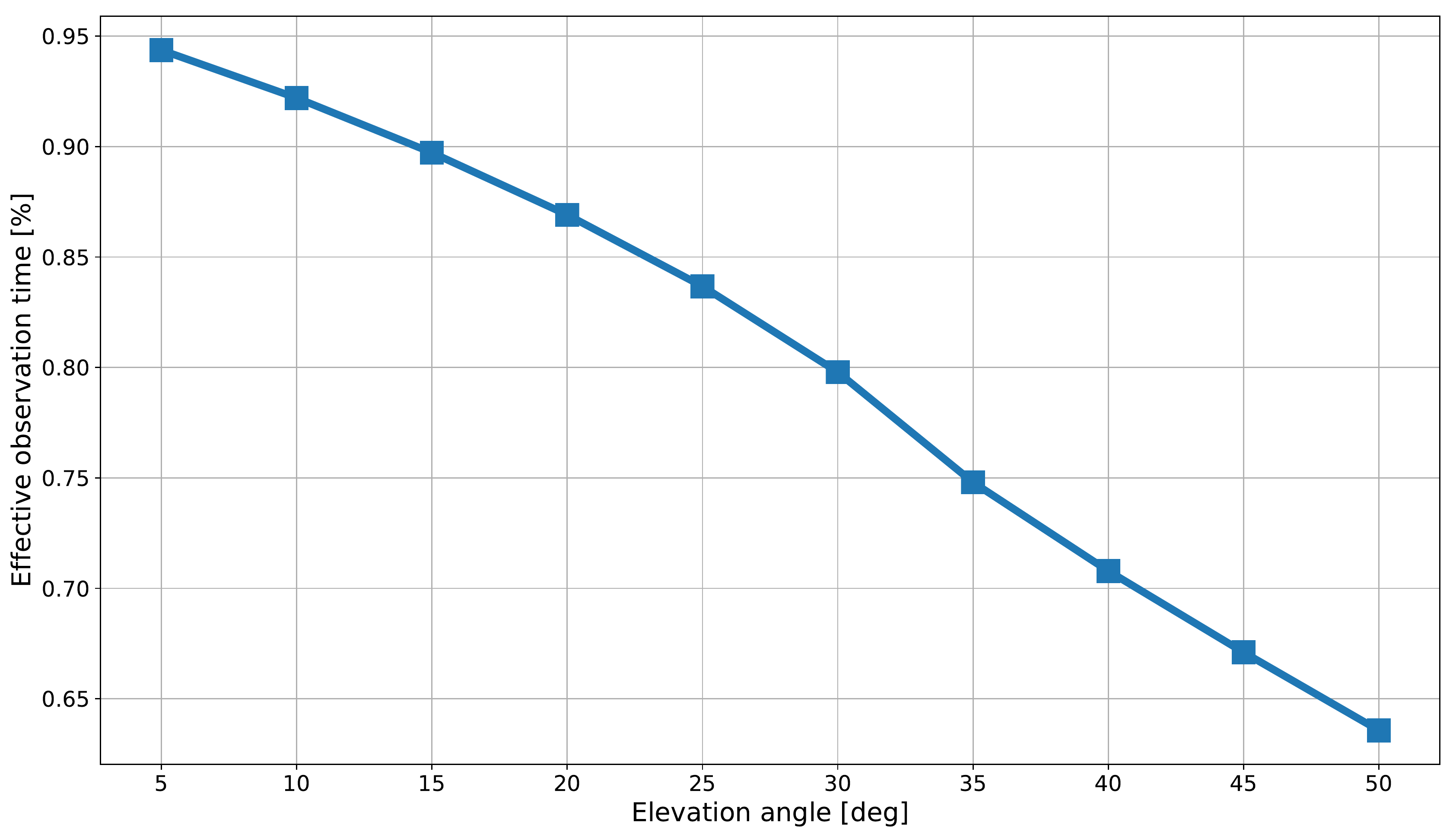}
  \caption[]{Effective observation time due to Sun as a function of the elevation angle.}\label{dcVSzd}
\end{figure}\par

\item{\bf ADC blanking time.}
  The acquisition cycles of the ADCs last about $125\,\mu\mathrm{sec}$ with a latency of about $10\,\mu\mathrm{sec}$. In this way, we lose $\sim 8\%$ of integration time.  
\item{\bf Calibrations.}
  A fraction of the STRIP observation time must be dedicated to calibrations. An absolute calibrator will be placed inside the telescope enclosure and it will be observed, periodically, for $\sim 300\,\mathrm{sec}$ every hour. This will impact the total duty cycle for $\sim 8\%$.\par
  Beam pattern calibration will be performed by means of a drone that will fly above the telescope carrying a microwave gun. This calibration will be performed only once, during the commissioning of the instrument, and it will not impact the duty cycle.\par
  Deep scans to observe target point sources that does not fall on the STRIP field of view during the nominal scanning mode, such as Saturn or Rho-Ophiuchi (Sect. \ref{pois}), will be performed occasionally with an negligible impact on the total duty cycle. 
\item{\bf Weather conditions.}
  Meteorological conditions could prevent STRIP to observe the sky. Of course, they are quite unpredictable events but it is possible to infer the fraction of usable time by the past years. By considering relative humidity, rain, wind speed and Calima\footnote{The Calima is a hot wind from the Sahara desert carrying dust and sand. It is typical of the Canary Islands region.} conditions in the last three years, it is possible to estimate the fraction of observing time to $\sim 82\%$.  
\item{\bf Atmosphere measurements.}
  A secondary goal of STRIP is to measure the atmospheric emission from Tenerife. These measurements require elevation scans of the atmosphere and can be performed when the Sun is high above the horizon by pointing the telescope in the opposite direction. In this way, atmosphere measurements will not impact the total duty cycle.   
\end{list} \par

By considering all the elements listed above, we can assert that a reasonable upper limit for the duty cycle of STRIP is about $50\%$, which is compatible with our sensitivity requirement at $20\deg \div 25\deg$ (Sect. \ref{to}).

\section{Optimizing the uniformity of the sky coverage}\label{optcov}
The step further to improve the scanning strategy is to make the STRIP coverage as uniform as possible to have homogeneous noise per pixel distributions over the whole sky. The nominal scanning mode, in fact, provides observations that are more redundant on the edge regions and less on the center, which is noisier. \par
A possible way to obtain a more uniform coverage is to modulate the elevation of the telescope while it is spinning. According to Eq. \ref{DeltaQrms}, this strategy introduces also a modulation of the atmospheric load. Then, to have each horn observing atmospheric layers of equal airmass, this modulation must be slow enough to make sure that after one revolution the telescope elevation is approximately constant. Otherwise, if the modulation is faster, we can measure, in principle, the atmospheric temperature with no need for dedicated scans.\par
This strategy hides two critical aspects. One is the choice of the modulation period that, besides, must be different from the period of the revolution of the telescope to avoid spin-synchronous systematic effects. The second is the instrumental I $\to$ Q/U leakage that affects the polarization measurements (Sect. \ref{miidbm}): depending on its level, the modulation could help us to characterize the leakage itself or could lead to a systematic effect.\par
It is possible to compute the minimum detectable signal due to change in elevation by imposing that the variation of the signal is smaller than the instrumental sensitivity at the given elevation angle. So that, by using Eqs. \ref{Tsys} and \ref{DeltaQrms}, and assuming continuous elevation scans at constant velocity from $5\deg$ to $35\deg$, I found that the period of the modulation must be $\sim 10 \,\mathrm{days}$ to ensure the validity of the approximation of constant airmass layers.\par 
I repeated the scanning strategy simulations of Sect. \ref{sssim} with the same parameters but, this time, I modulated the elevation of the telescope with ($\mathrm{i}$) a sinusoidal function and ($\mathrm{ii}$) the sine of an optimized \textit{bezier} curve\footnote{The bezier functions \citep{Riesenfeld:1975:AMC:1499949.1500072} are parametric curves defined by a set of $P_n$ control points, so that the first and last ones are always the end points of the curve while the intermediate ones generally do not lie on the curve. The number $n$ of control points is called the order of the curve ($n = 1$ for linear, $2$ for quadratic, etc.).}. I performed the simulations for three different modulation periods ($1\,\mathrm{hour}$, $1\,\mathrm{day}$ and $1\,\mathrm{month}$) and then I compared the results.\par
The idea at the base of using the sine of a bezier curve lies on the fact that when the telescope spins at lower zenith angles it observes, in equal time with respect to higher zenith angles, a smaller portion of the sky. The sine of the bezier function, with respect to the sine, allow the telescope to spend more observation time at higher zenith angles. I used a one dimensional bezier curve of the third-order, which has the following equation:
\be{bez}
y = a(1-x)^3 + 3b(1 - x)^2x + 3c(1 - x)x^2 + dx^3 \,,
\ee
where $a = 0$ and $d = 1$ are the end points of the curve, and $b$ and $c$ must be determined within the interval $0\div1$. I take the sine of Eq. \ref{bez} to obtain a periodic function.\par
To find the proper values of $b$ and $c$, I have created, for each couple of them within the range $0\div1$ by steps of $0.01$, a hit count map by using the sine of Eq. \ref{bez} to modulate the elevation. Then, I have associated to each map the following value of $\chi^2$:
\be{chi2}
\chi^2 = \frac{\sum_\mathrm{pixels}{\mathrm{hit}_\mathrm{pixel} - \mean{\mathrm{hit}}}}{\mean{\mathrm{hit}}} \,,
\ee
which quantifies the deviation from a uniform coverage.\par
\begin{figure}[H]
  \centering
  \includegraphics[scale=0.3]{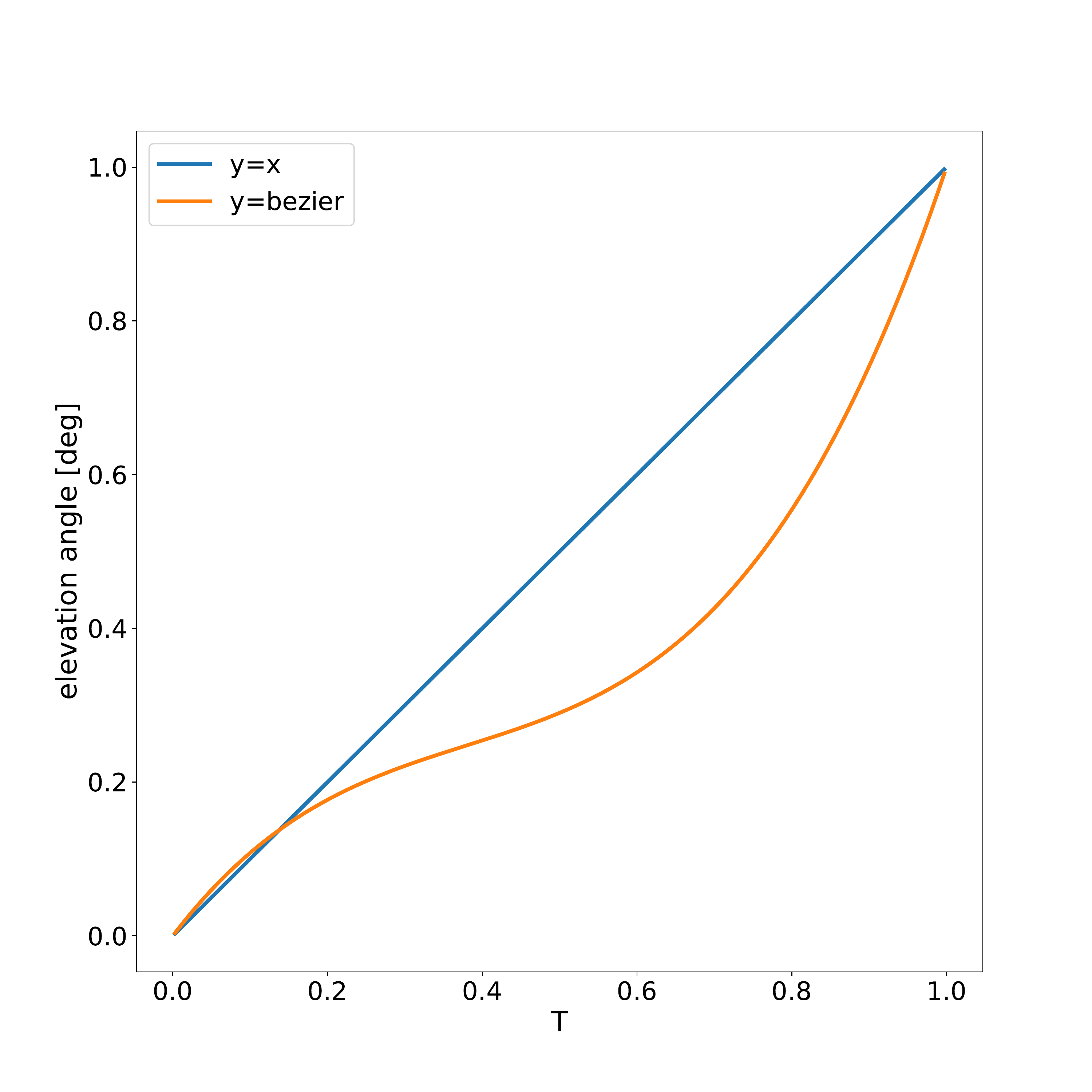}
  \caption[]{The bezier function of Eq. \ref{bez} with $a = 0$, $b = 0.44$, $c = 0$ and $d = 1$.}\label{beziplot}
\end{figure}

I have found that the minimum value of $\chi^2$ is given by:
\begin{align}\label{b,c}
b &= 0.44 \,,\\
c &= 0.00 \,,
\end{align}
corresponding to the bezier curve shown in Fig. \ref{beziplot}. \par
\begin{figure}[H]
  \centering
  \includegraphics[scale=0.3]{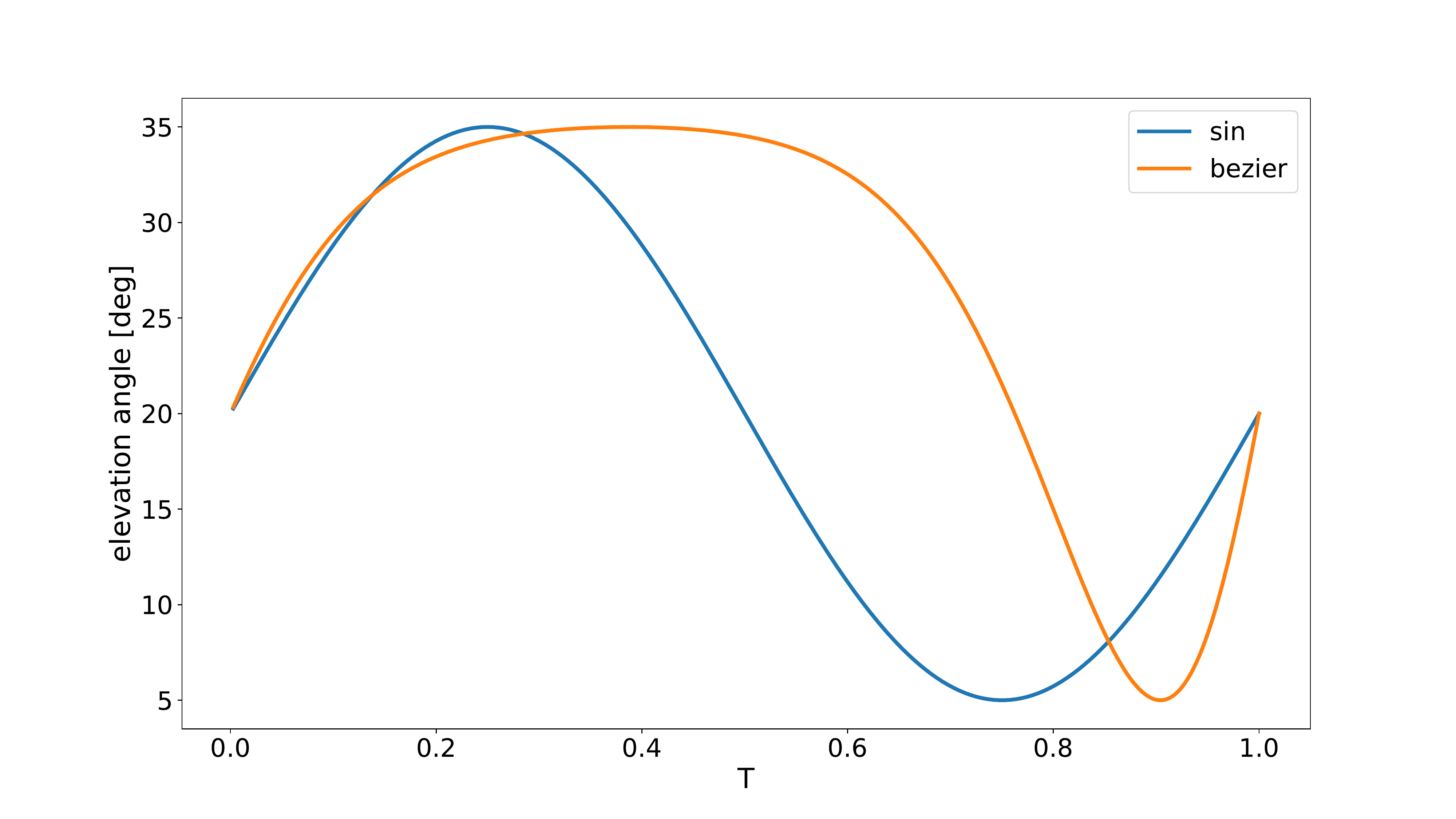}
  \caption[]{Comparison between the functions used to modulate the elevation. Within one period, the blue curve is Eq. \ref{funsinbez1} while the orange one is Eq. \ref{funsinbez2}.}\label{sinbeziplot}
\end{figure}

I want to modulate the elevation of the telescope between $5\deg$ and $35\deg$, so that, for a given period $T$ and as a function of the time $t$, the two function I used are given by (Fig. \ref{sinbeziplot}):
\begin{align}
f(t) &= 20 + 15 \sin{\Bigl(\frac{2 \pi t}{T}\Bigr)} \,, \label{funsinbez1}\\
g(t) &= 20 + 15 \sin{\Bigl\{\frac{2 \pi}{T}\bigl[1.32(1 - t)^2t + t^3\bigr]\Bigr\}} \,.\label{funsinbez2}
\end{align}\par
\begin{figure}[H]
  \centering
  \hspace{12pt}\includegraphics[scale=0.1]{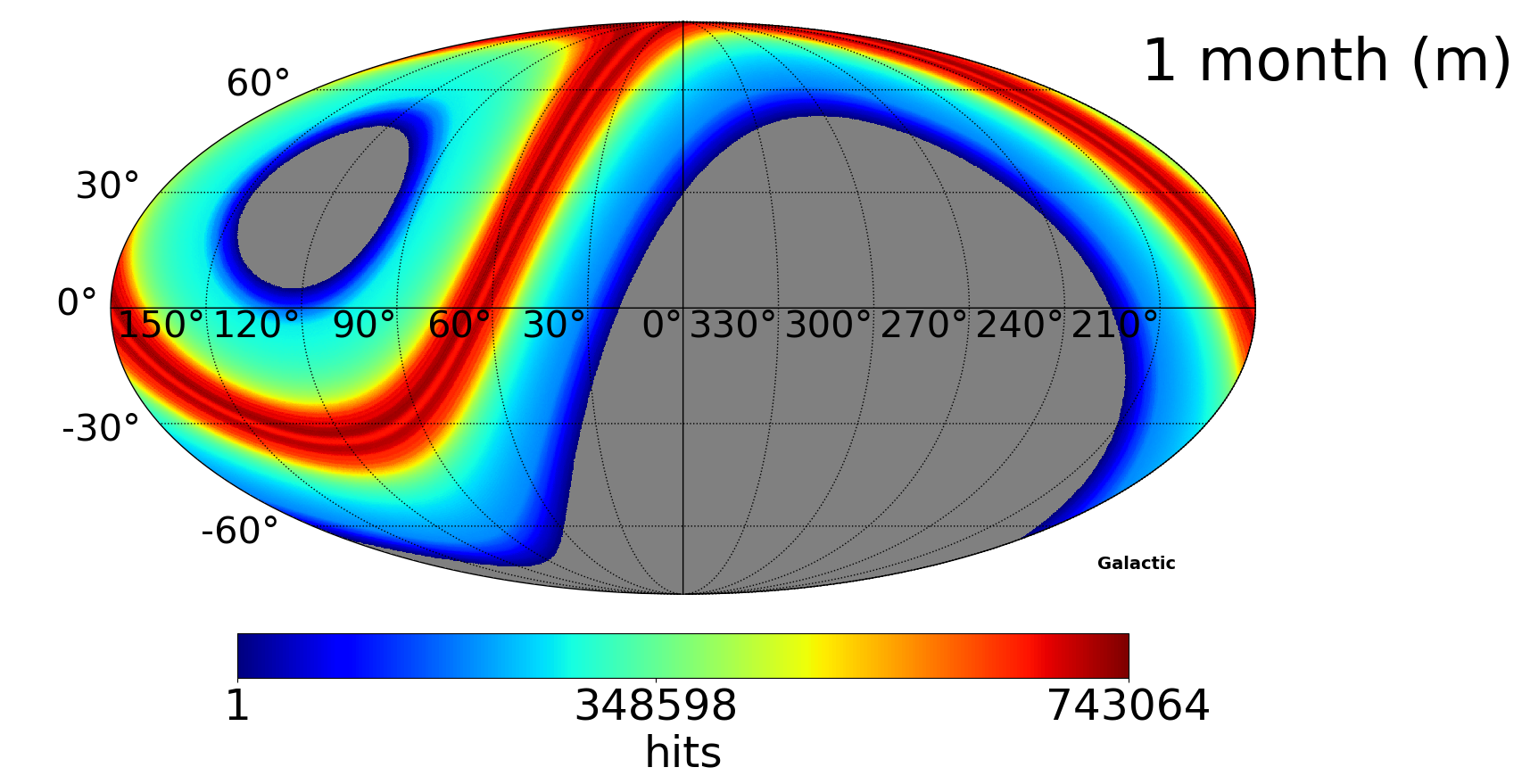}
  \hspace{-4pt}\includegraphics[scale=0.1]{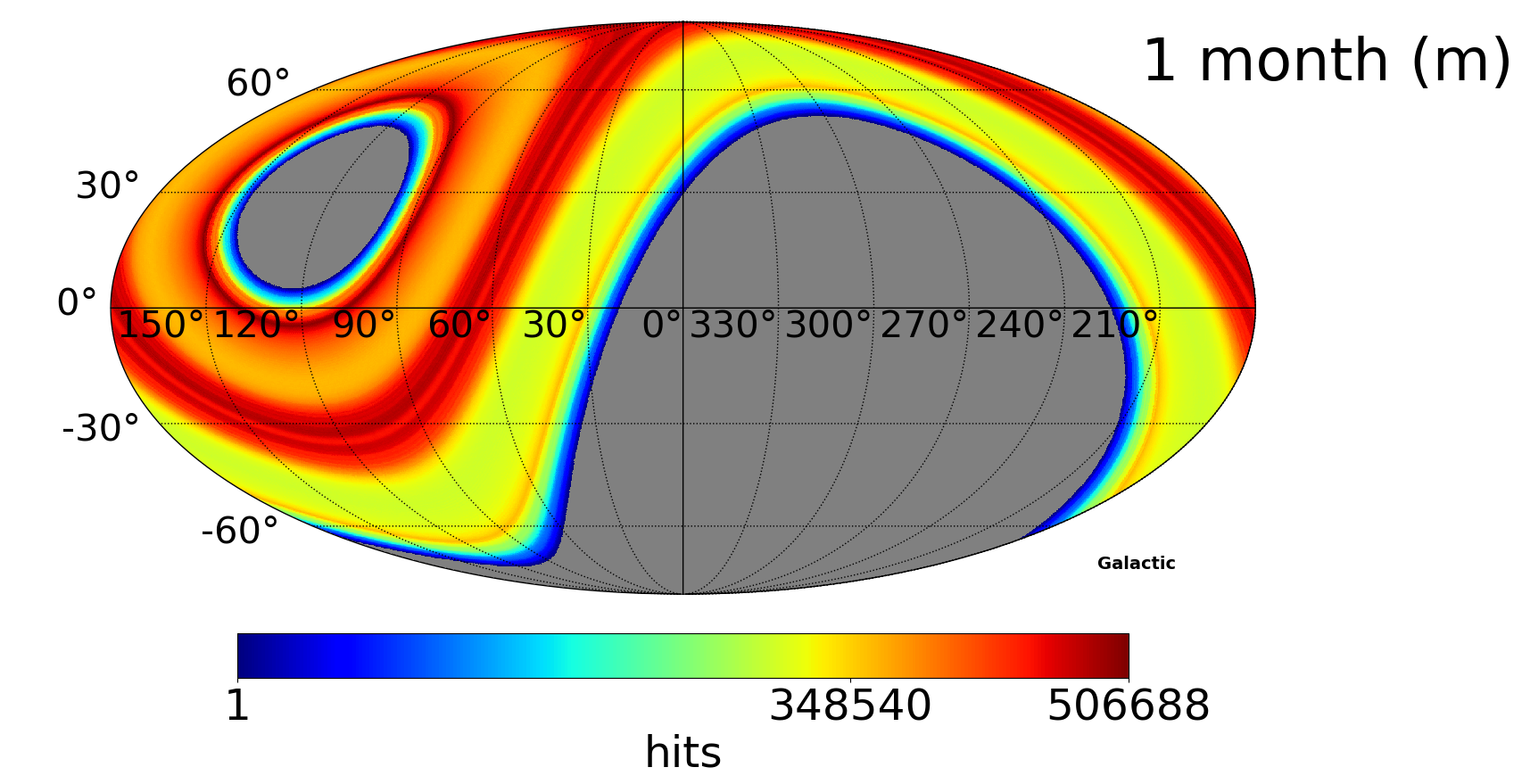}
  \includegraphics[scale=0.1]{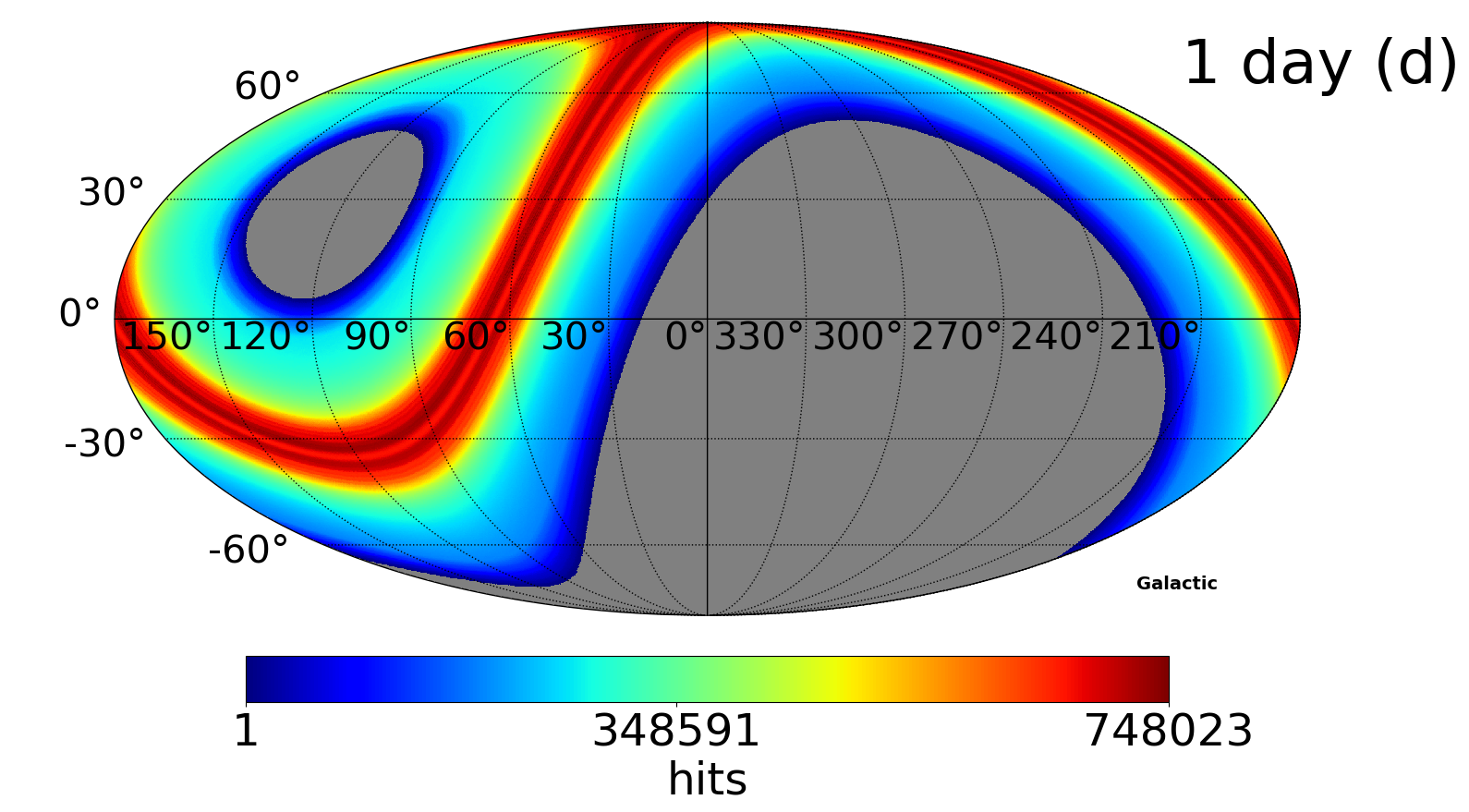}
  \includegraphics[scale=0.1]{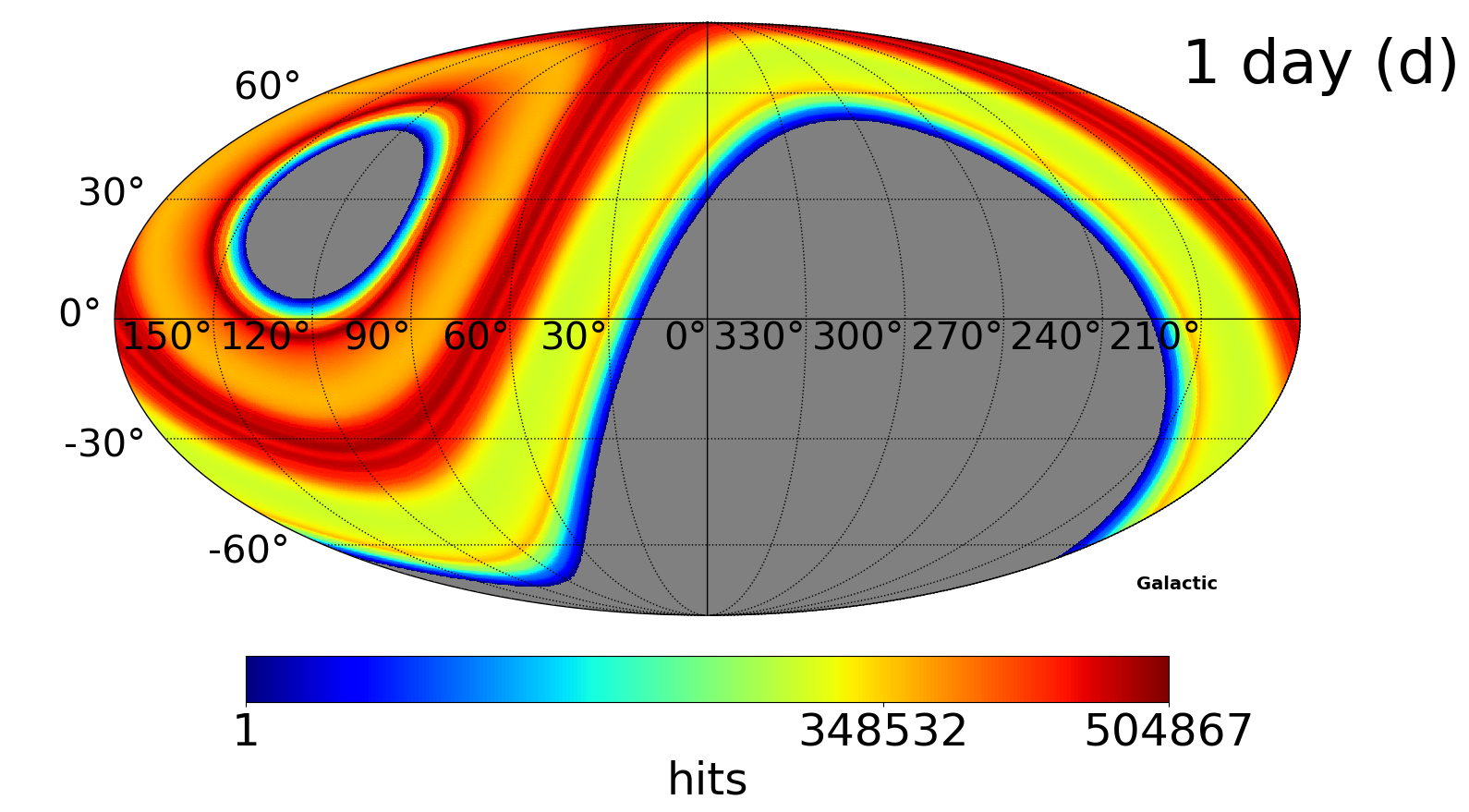}
  \includegraphics[scale=0.1]{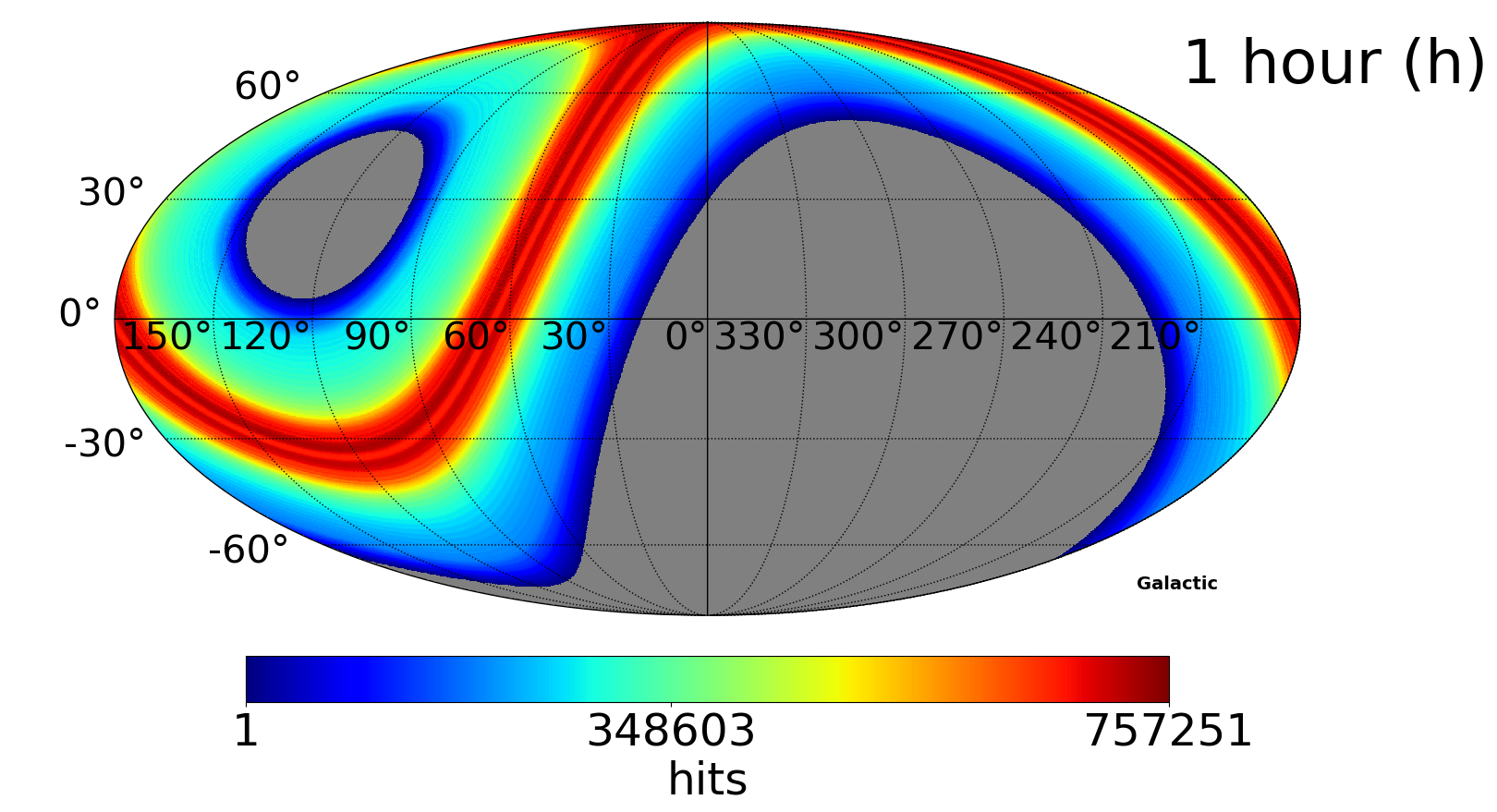}
  \includegraphics[scale=0.1]{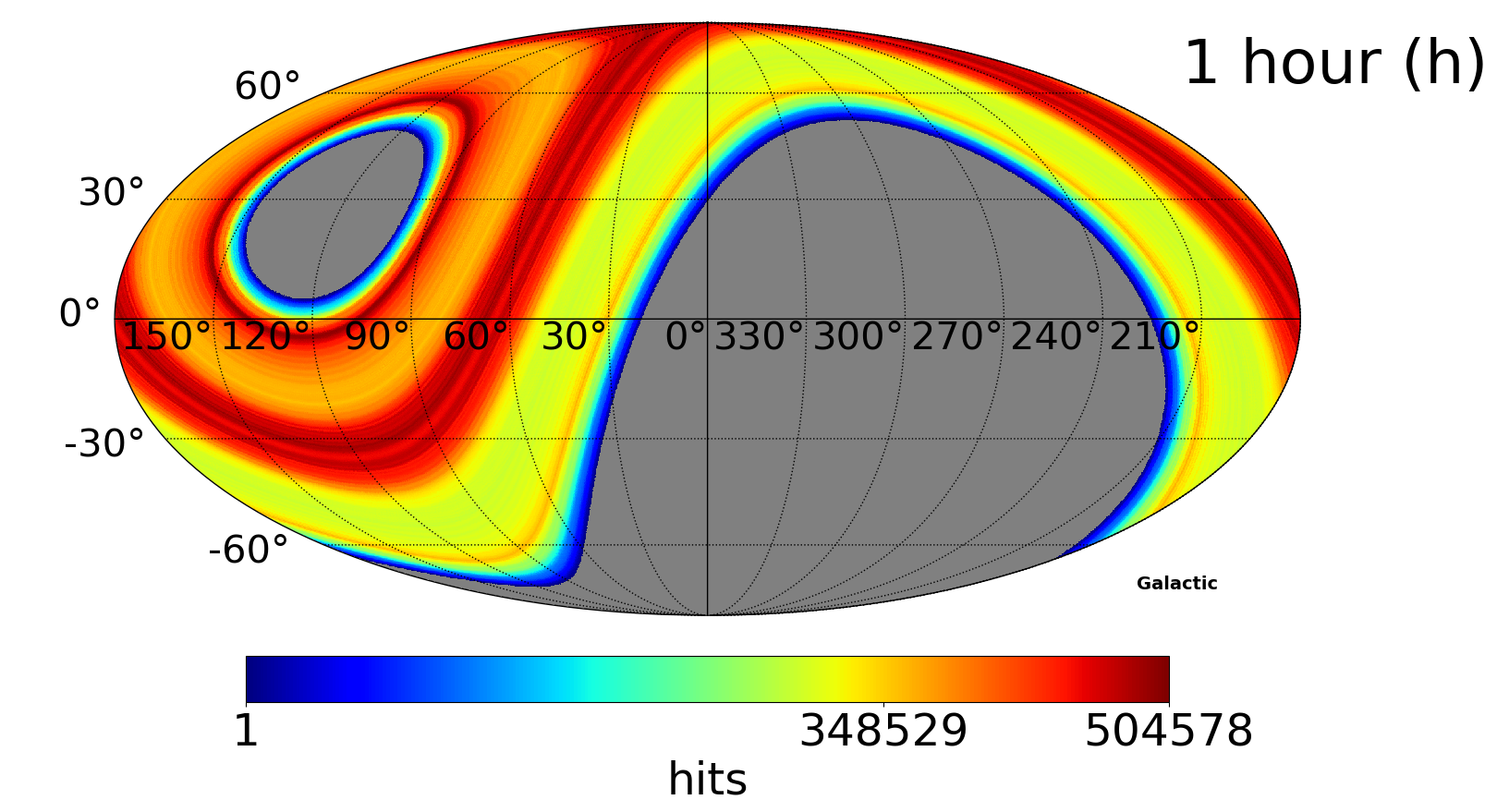}
  \caption[]{Hit count maps (in GCS) obtained by spinning the telescope with constant angular velocity and by modulating the elevation through Eq. \ref{funsinbez1} (left column) and Eq. \ref{funsinbez2} (right column), as a function of the modulation period. The minimum, the average  and the maximum number of hits are reported in the color bar.}\label{bezmaps}
\end{figure}

In Fig. \ref{bezmaps}, I show the hit count maps obtained by modulating the telescope elevation through Eqs. \ref{funsinbez1} and \ref{funsinbez2}. For each map, in the color bar, I report the minimum, the average and the maximum number of hits. We notice that the hit distribution depends strongly on the used function. In particular, the bezier function allow us to obtain a more uniform coverage. \par
\par
\par

\par

\begin{figure}[H]
  \centering
  \includegraphics[scale=0.1]{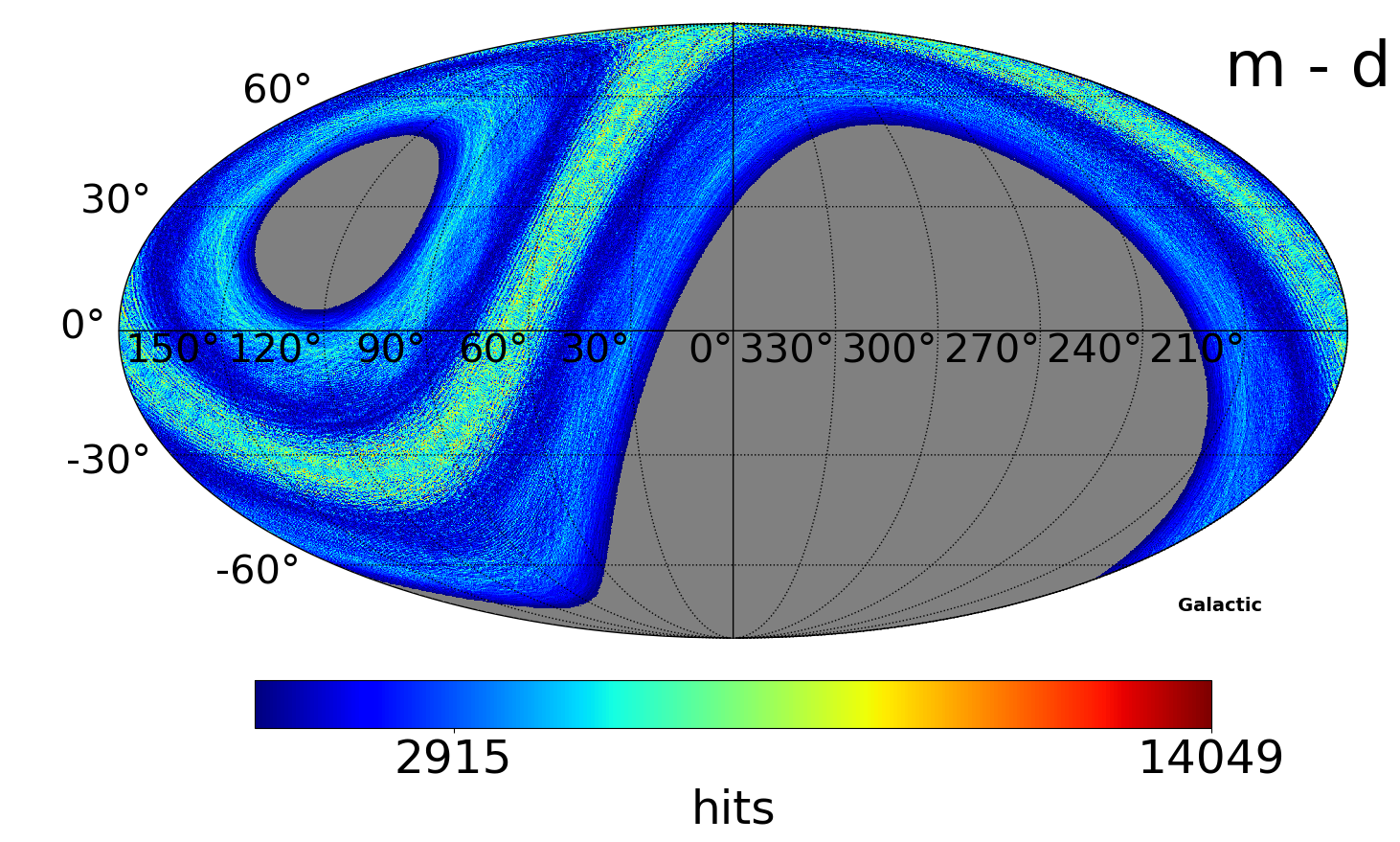}
  \includegraphics[scale=0.1]{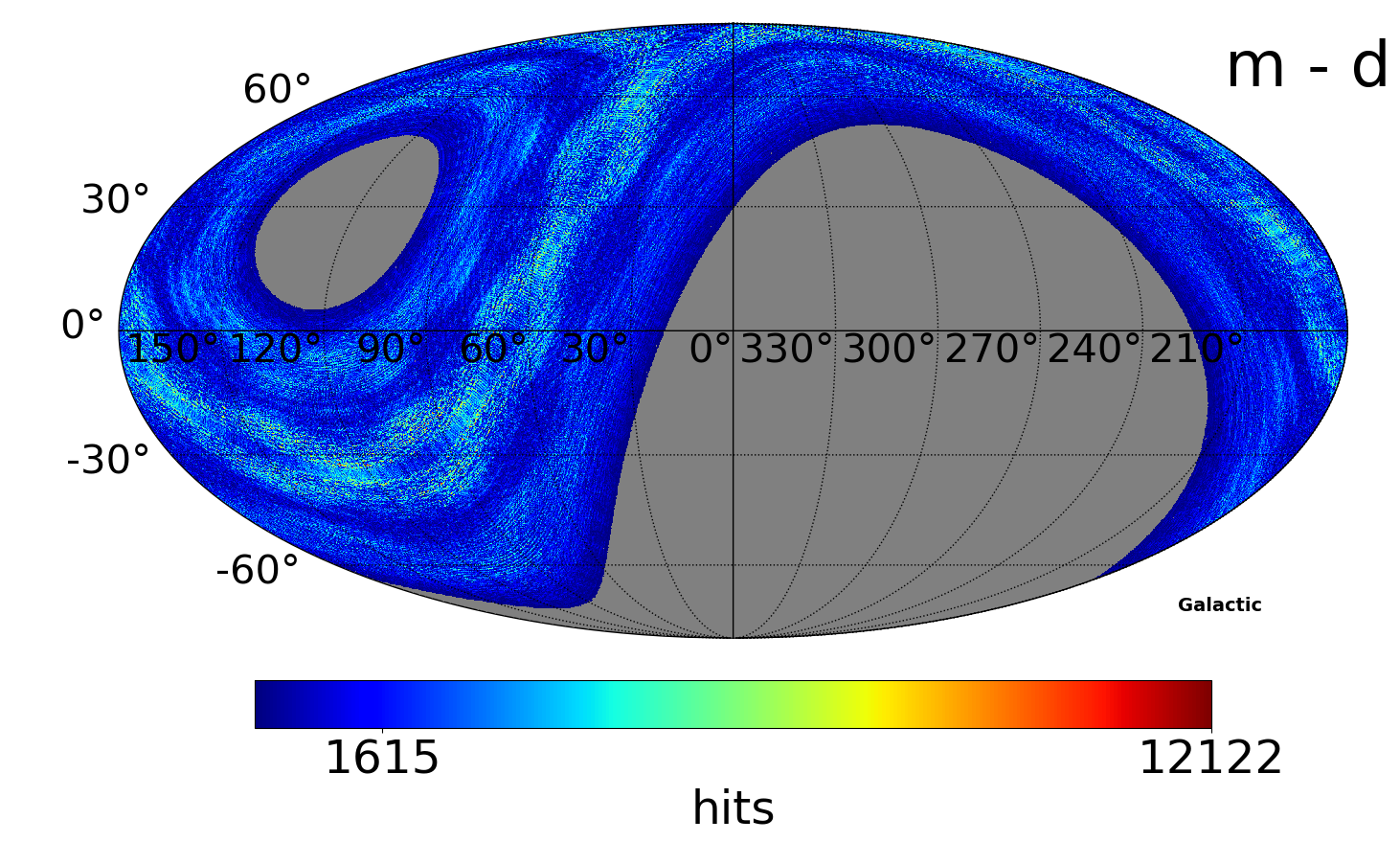}
  \includegraphics[scale=0.1]{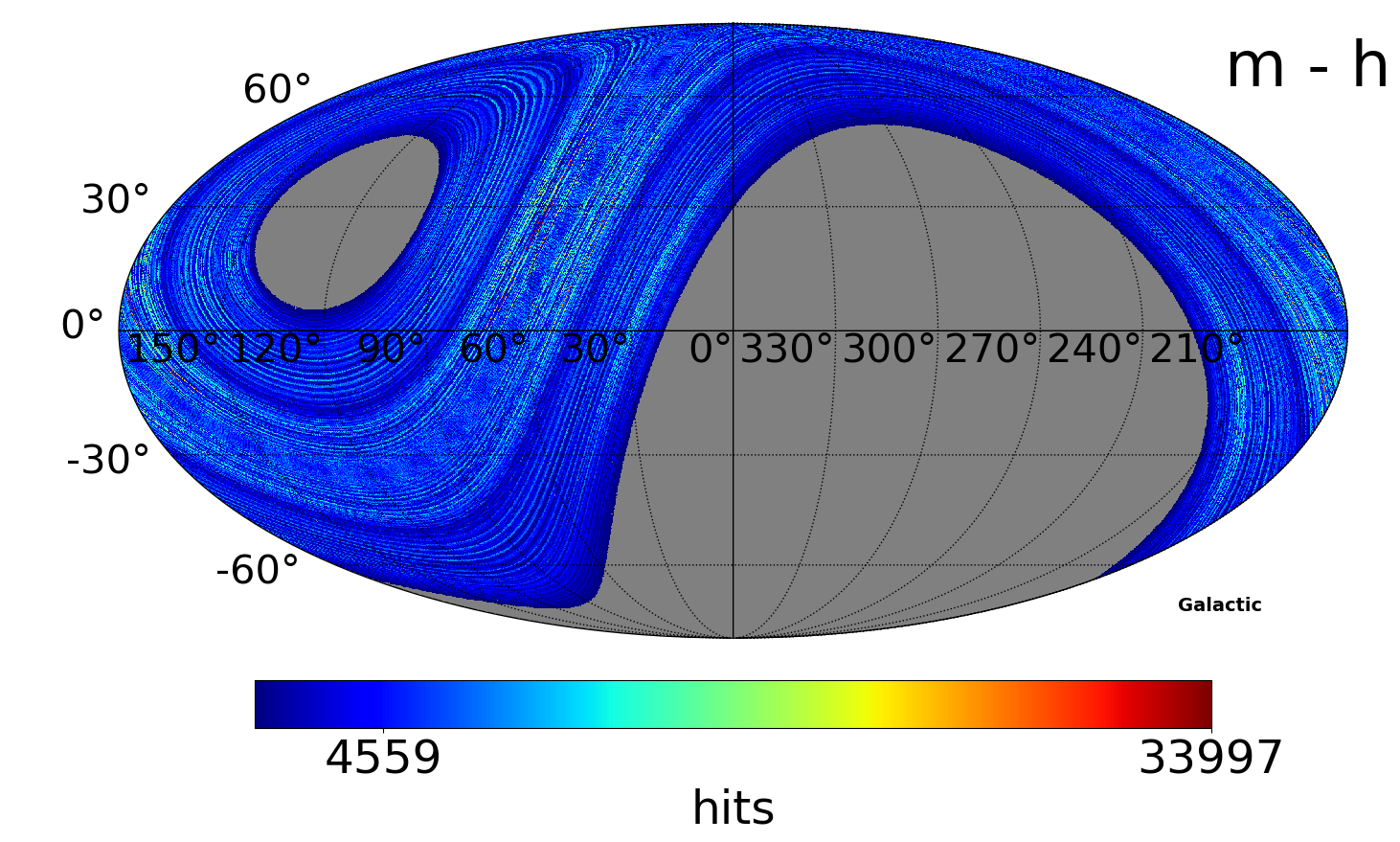}
  \includegraphics[scale=0.1]{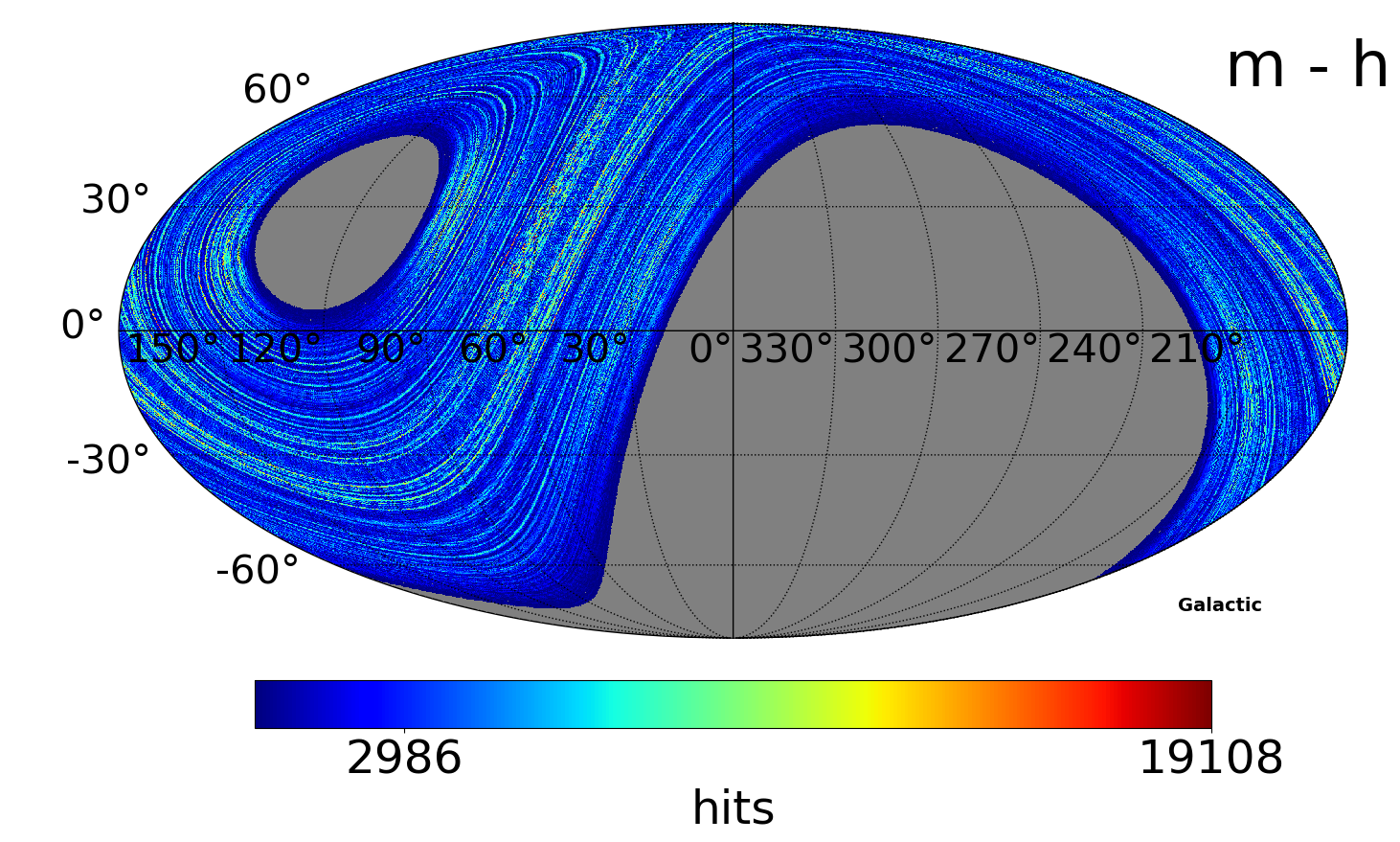}
  \includegraphics[scale=0.1]{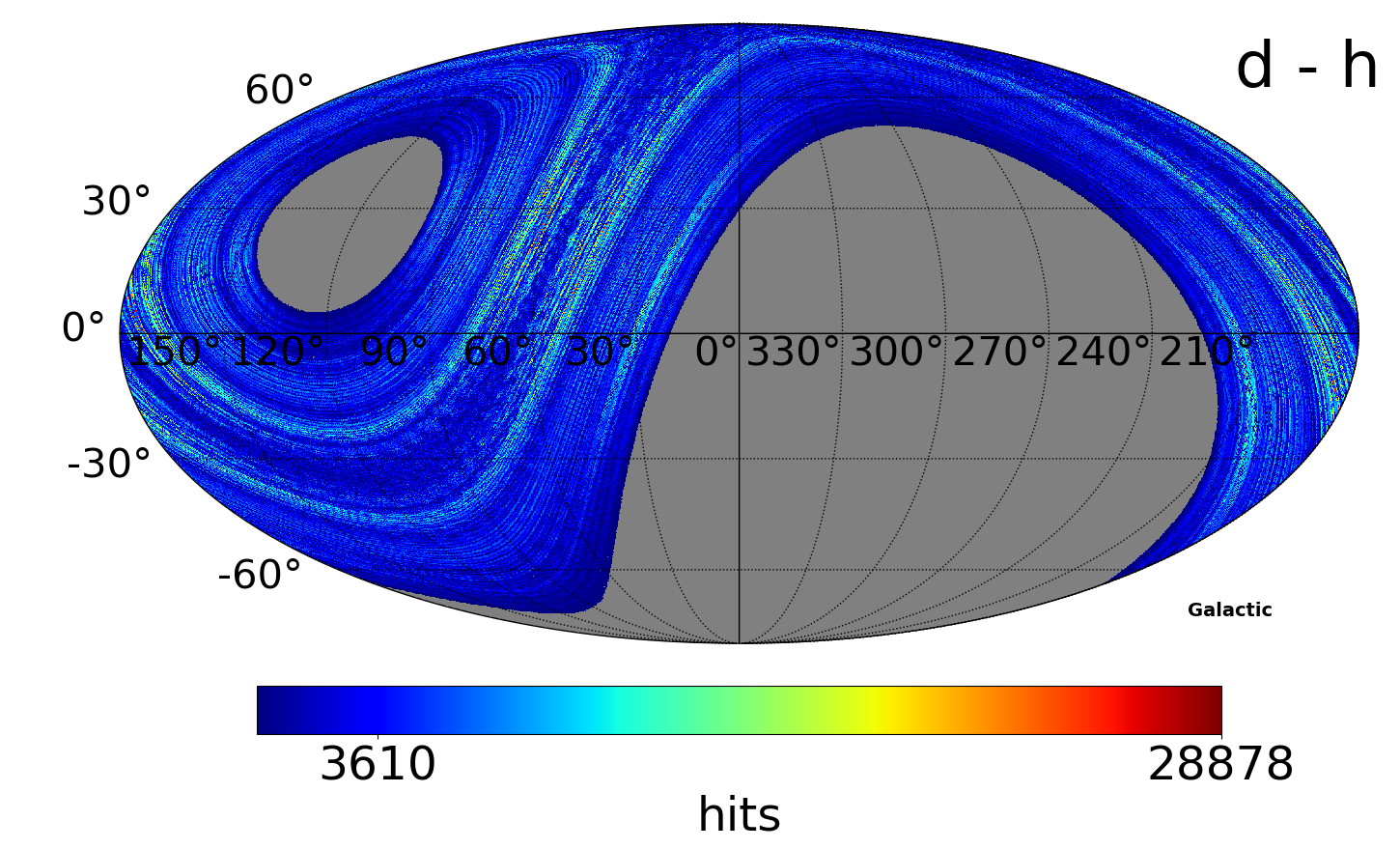}
  \includegraphics[scale=0.1]{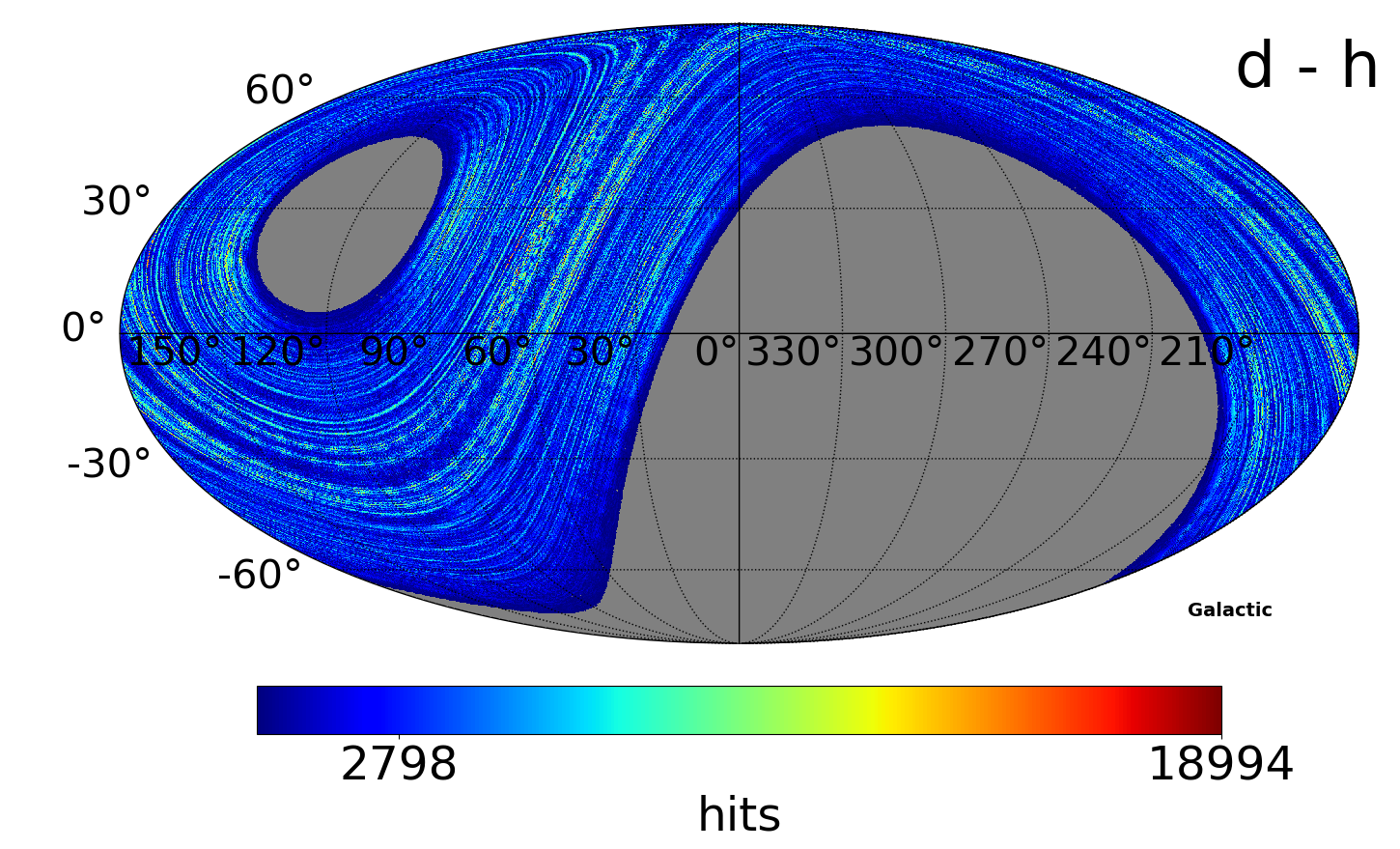}
  \caption[]{Residual maps (in GCS) of the hit count maps shown in Fig. \ref{bezmaps}. The rows refer to the absolute difference between the maps computed with, respectively, ($1\,\mathrm{month}$, $1\,\mathrm{day}$), ($1\,\mathrm{month}$, $1\,\mathrm{hour}$) and ($1\,\mathrm{day}$, $1\,\mathrm{hour}$) modulation periods. The left and right columns refer to the functions used to modulate the telescope elevation, i.e., respectively, Eqs. \ref{funsinbez1} and \ref{funsinbez2}. The average and the maximum number of hits are reported in the color bar.}\label{resbezmaps}
\end{figure}

In Fig. \ref{resbezmaps}, I report the residual differences among the hit maps of Fig. \ref{bezmaps}. Each row shows the absolute difference between the maps computed with, respectively, ($1\,\mathrm{month}$, $1\,\mathrm{day}$), ($1\,\mathrm{month}$, $1\,\mathrm{hour}$) and ($1\,\mathrm{day}$, $1\,\mathrm{hour}$) modulation periods. The column instead refers to the function used to modulate the elevation. I report, in the color bar, the average and the maximum number of hits of each map.\par
\begin{figure}[H]
  \centering
  \includegraphics[scale=0.25]{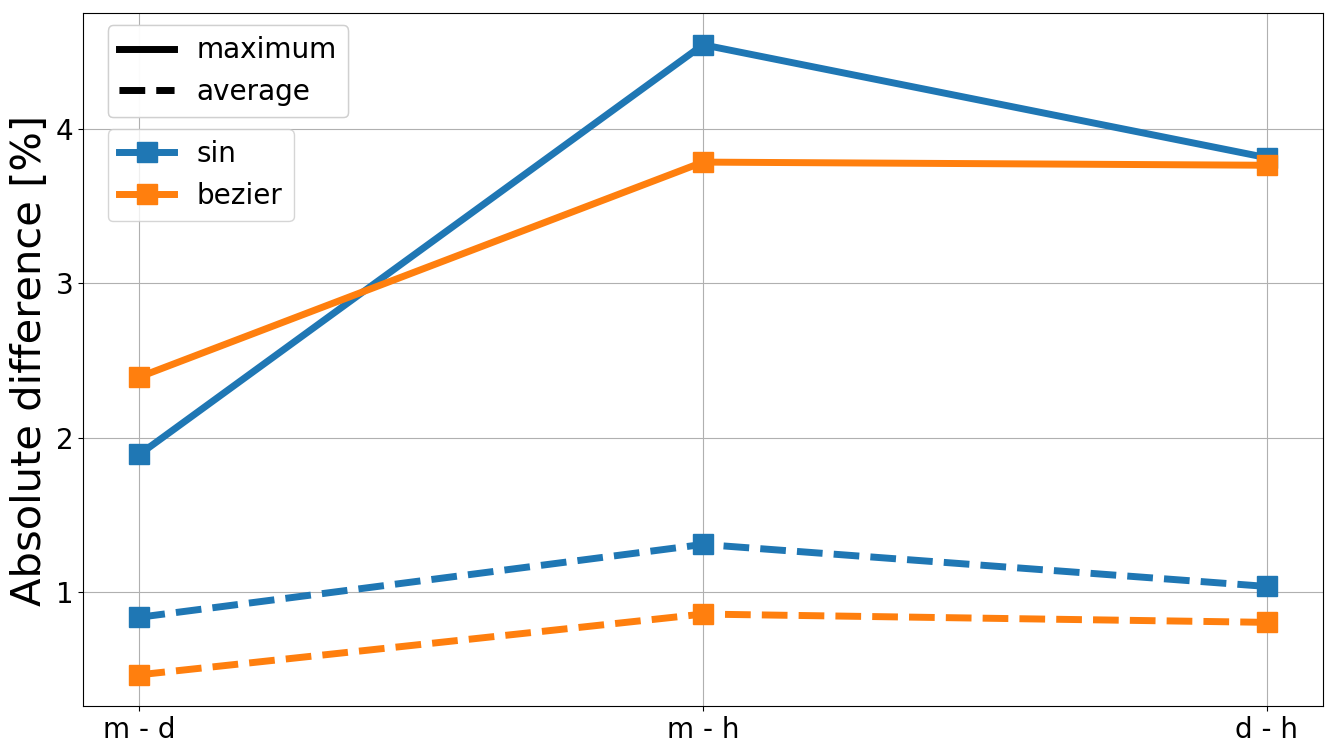}
  \caption[]{Percentage of absolute difference between the average (dashed line) and the maximum (continuous line) numbers of hits for the maps shown in Fig. \ref{resbezmaps}. Blue and orange curves refer to the functions used to modulate the telescope elevation, i.e., respectively, Eqs. \ref{funsinbez1} and \ref{funsinbez2}.}\label{absdiff}
\end{figure}

The differences among the maps are quantified in terms of absolute difference between the average and the maximum numbers of hits. From Fig. \ref{absdiff}, we notice that as for the average the differences are negligible ($\sim 1\%$) while for the maximum the differences are within a few percent.\par
\begin{figure}[H]
  \centering
  \includegraphics[scale=0.13]{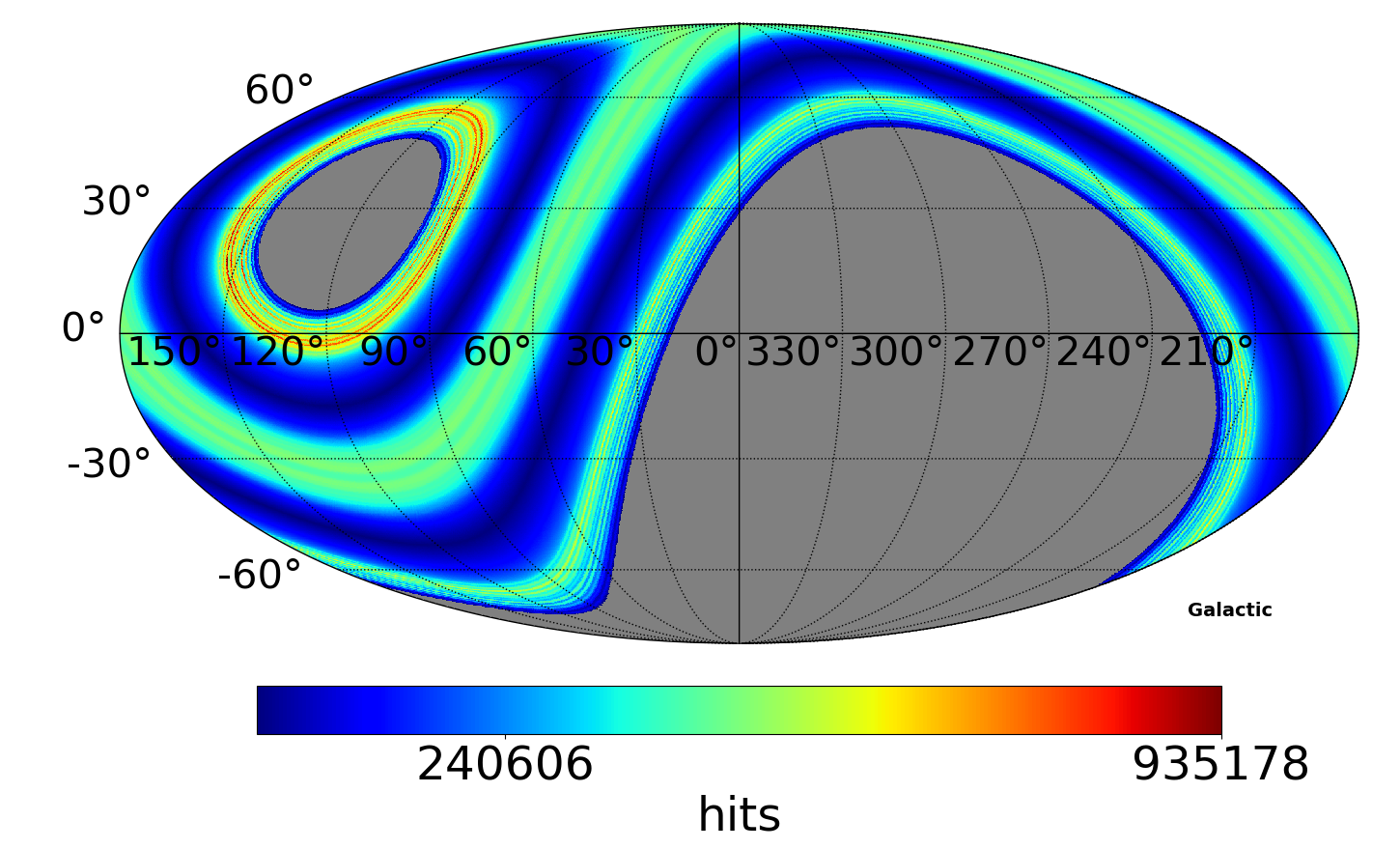}
  \includegraphics[scale=0.13]{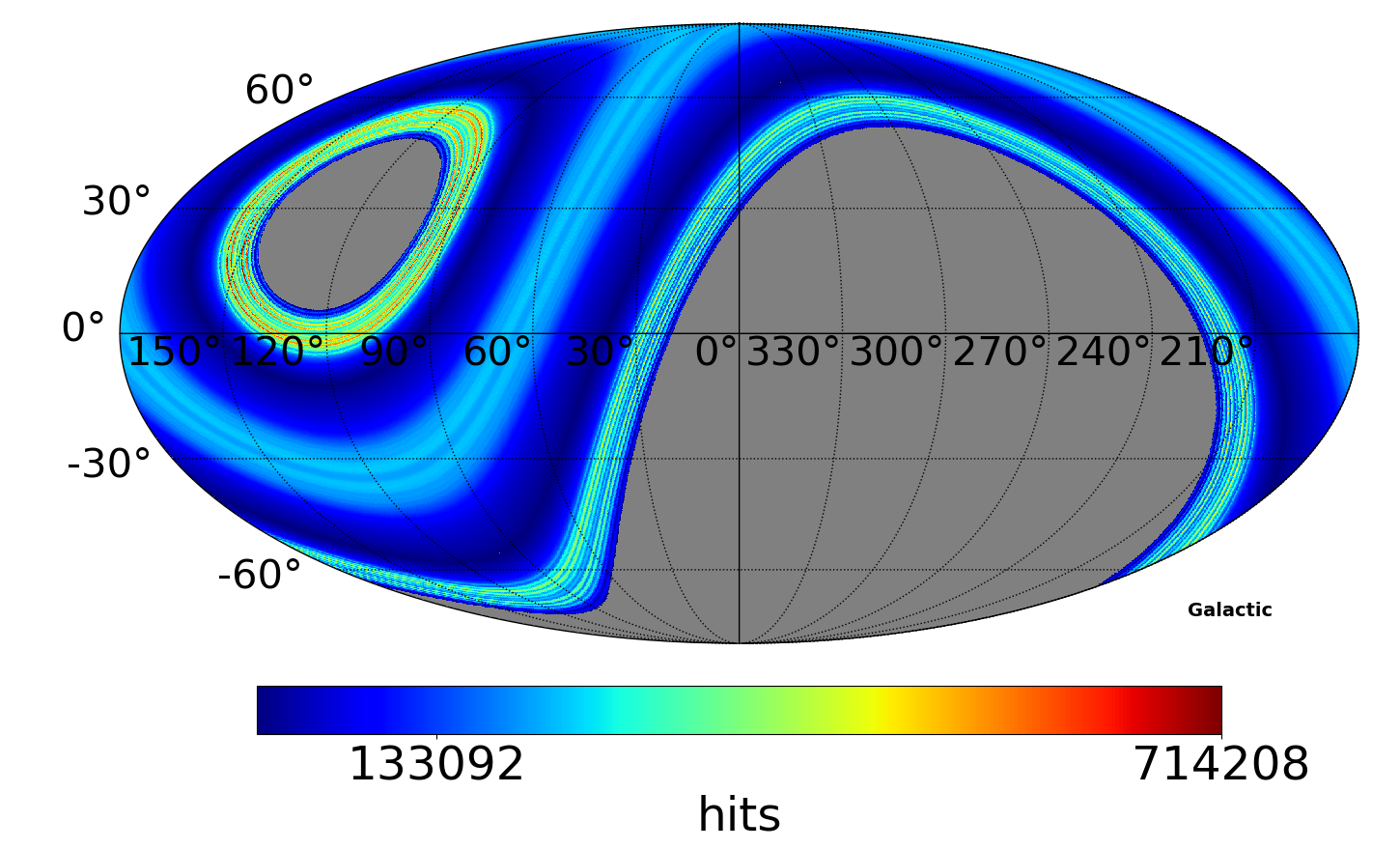}
  \caption[]{Residual maps (in GCS) between the hit count maps obtained by modulating the elevation with a period of $1\,\mathrm{day}$ and by the nominal scanning mode at $35\deg$. The left and right columns refer to the functions used to modulate the telescope elevation, i.e., respectively, Eqs. \ref{funsinbez1} and \ref{funsinbez2}. The average and the maximum number of hits are reported in the color bar.}\label{resmodnom}
\end{figure}

In Fig. \ref{resmodnom}, I show the residual difference between the hit maps obtained by modulating the elevation with a period of $1\,\mathrm{day}$ and the hit map obtained in the nominal scanning mode at $35\deg$, which is the case that guarantees the same observed sky fraction. \par
\begin{figure}[H]
  \centering
  \includegraphics[scale=0.25]{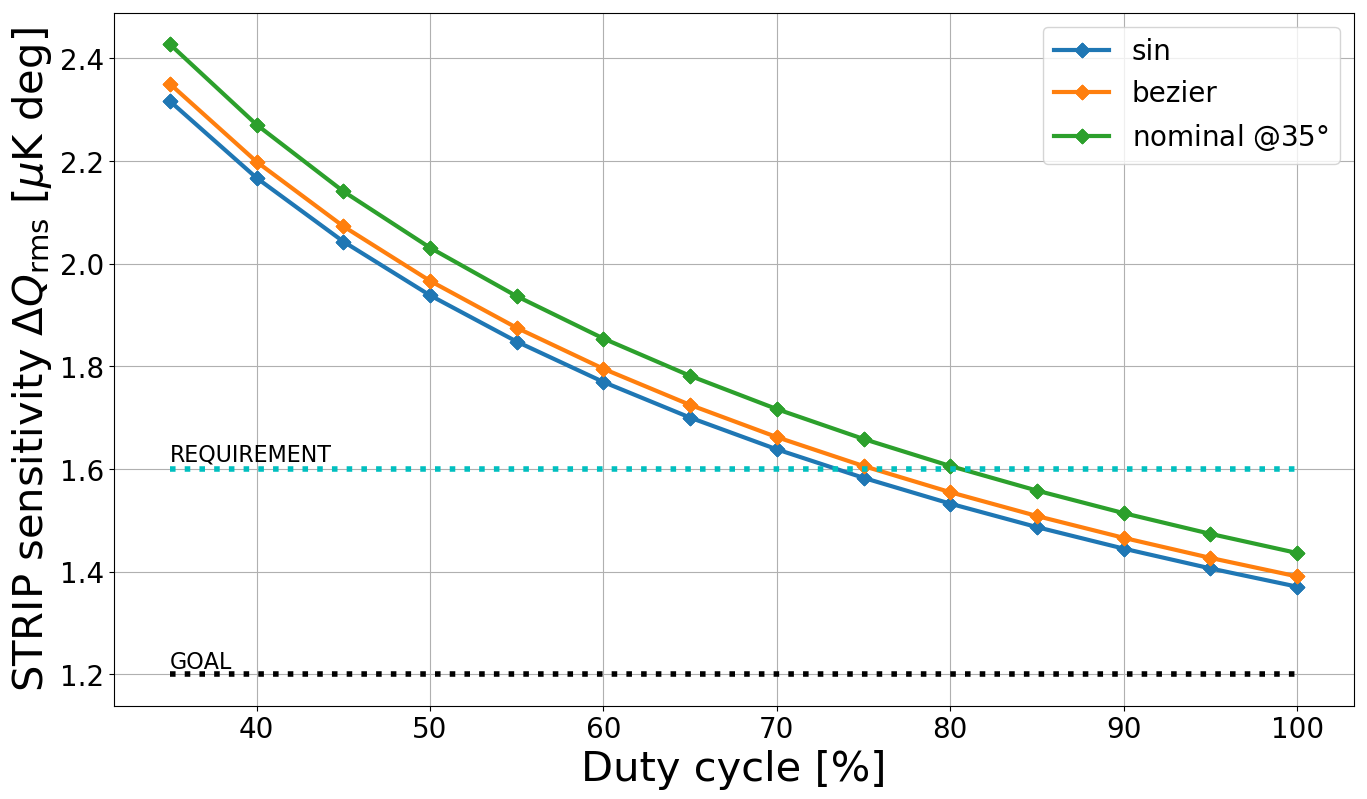}
  \caption[]{LSPE/STRIP sensitivity as a function of the fraction of usable duty cycle. The blue and orange curves refer to the functions used to modulate the telescope elevation, i.e., respectively, Eqs. \ref{funsinbez1} and \ref{funsinbez2} while the green curve corresponds to the nominal scanning mode at $35\deg$.The black and cyan dotted line represent, respectively, the sensitivity goal and requirement.}\label{resbezduty}
\end{figure}

   
Independently of the used function, the observed sky fraction is $\simeq 56\deg$ ensuring an overlap with the SWIPE coverage of $\simeq 94\%$. In Fig. \ref{resbezduty}, I plot, as a function of the fraction of usable duty cycle, the average sensitivity per pixels of $1\deg$ obtained by modulating the elevation and, for comparison, the average sensitivity in the nominal scanning mode at $35\deg$. We notice that, with respect to the nominal scanning mode, the modulations allow the instrument to increase the average sensitivity.\par 

\par

\par

\section{Final remarks}
Changing the elevation of the telescope while it is spinning at constant angular velocity can be an option to observe more uniformly the sky and, at the same time, to increase the observed sky fraction and the overlap with the SWIPE coverage. In both the analyzed cases, we have seen that the coverage does not depend significantly on the period of the modulation but we need to ensure an instrument duty cycle of about $75\%$ to reach the STRIP sensitivity requirement. \par
The method I investigate here requires parallel measurements of the brightness temperature of the Tenerife atmosphere. Such a signal, in fact, affects the STRIP measurements at each elevation angle and it changes over the seasons of the year. However, on short time-scales, the atmospheric signal can be considered stable. Thus, if the telescope elevation is fast enough, this method could be even used to measure directly the atmospheric temperature of Tenerife. Since the maximum angular velocity the elevation motor can reach is $\sim1\deg/\mathrm{sec}$, assuming to scan from $5\deg$ to $35\deg$ and back, these fast scans can last at least $\sim 60\,\mathrm{sec}$.\par
Further studies are required to assess the effectiveness of spinning the telescope with constant angular velocity and, simultaneously, modulating its elevation. In particular, a fine characterization of the instrumental I $\to$ Q, U leakage must be performed to evaluate the impact on the polarization measurements. Besides, cross-checks with the most recent atmospheric data from Teide Observatory are required to estimate the time-scales on which the atmospheric temperature varies and to define the modulation period. Finally, the possibility to use a different class of functions to modulate the elevation or to spin the telescope with a non-constant angular velocity (e.g., which changes as a function of the elevation) must be investigated. \par
The scanning strategy simulations I presented in this chapter should be repeated at the end of the system-level tests by using the estimation of the noise temperatures and of the 1/$f$ noise that these tests will provide. New simulations should be performed to assess the impact of the pointing errors, of the ground emission, and to fix the parameters of the ``deep fields'' required to observe particular sky regions (Sect. \ref{pois}). \par
Finally, we can obtain a thorough understanding of the observation scenario by studying the impact of the Earth atmosphere on the STRIP measurements both in intensity and polarization. The atmosphere, in fact, attenuates the incoming astrophysical signal, and emits radiation that could be polarized. We have seen that, in principle, it is possible to measure the atmospheric signal by varying the elevation of the instrument, since it implies a variation of the instrumental sensitivity (Sect. \ref{noismaps}). However, more simulations are required to assess the minimum number of elevation angles and the minimum time to spend at each angle, or, in the case of a slowly varying elevation, the scan period. At this purpose, a reliable model of the Tenerife atmosphere is needed, which takes into account its spatial and temporal fluctuations.  

%% file: Concl.tex
\chapter*{Conclusions}
\addcontentsline{toc}{chapter}{Conclusions}
\label{Chap:Conl}
\thispagestyle{plain}

My thesis has been carried out in the context of the LSPE project, an experiment that aims to constrain the ratio between the amplitudes of tensor and scalar modes to $\simeq 0.03$ at the $99.7\,\%$ confidence level and to study the polarized emission of the Milky Way. Within this collaboration, I was part of the simulation and data analysis group of STRIP, the low frequency instrument. \par
The first part of my thesis is dedicated to show the results of the unit-level tests on the STRIP detectors that have been held at ``Università degli Studi di Milano Bicocca'' from September 2017 to July 2018. In the second part, I presented the code I developed and the simulations I performed to study the STRIP scanning strategy.\par
At first, I illustrated the working principle of the STRIP detectors (Ch. \ref{Chap:3}) by going through the details of the mathematical model (Appendices \ref{App:1}, \ref{App:2}) and the electronic processes that drive the acquisition of the sky signal. I showed how, thanks to the combination of several RF components, the overall response of the detectors is proportional to the four Stokes parameters of the incident radiation field and, furthermore, how the electronics of the polarimeter is able to reduce the correlated noise, the 1/$f$ component, by many order of magnitudes thanks to the double demodulation process.\par
Then, I reported the main results of the unit-level tests campaign (Ch. \ref{Chap:4}) during which more than $800$ tests on $68$ polarimeters have been performed in order to select the $55$ ($49$ Q-band and $6$ W-band) with the best performance. For each polarimeter, we ran three tests in a cryogenic chamber cooled down to $20\,\mathrm{K}$: a bandpass characterization, a Y-factor test to estimate the noise temperature and a long acquisition to measure the noise characteristics. \par
As for the bandpass characterization (Sect. \ref{bp}), a clear dichotomy between the Q-band polarimeters inherited by QUIET and the other ones appeared in terms of bandwidth as the former showed a larger one ($\beta \sim 8\,\mathrm{GHz}$) with respect to the others ($\beta \sim 6.5\div 7.0\,\mathrm{GHz}$). All the polarimeters presented a remarkable agreement in the central frequency. I found that, the average values of bandwidth and central frequency for the $49$ Q-band and $6$ W-band polarimeters implemented in the STRIP focal plane are: $\beta^Q = 7.3 \pm 0.8 \,\mathrm{GHz}$, $\nu_c^Q = 43.4 \pm 0.2 \,\mathrm{GHz}$, $\beta^W = 7.9 \pm 1.9 \,\mathrm{GHz}$ and $\nu_c^W = 97.5 \pm 1.8 \,\mathrm{GHz}$. \par
We estimated the noise temperature of the detectors using the so-called Y-factor test (Sect. \ref{tnoise}). To run the test, we changed the temperature of one of the two thermal loads installed on the cryogenic chamber from $10\,\mathrm{K}$ to $50\,\mathrm{K}$ by steps of $10\,\mathrm{K}$ and we measured the response of the polarimeter. For each polarimeter, we computed the noise temperature for each temperature pairs ($10$ if there were $5$ steps) and then we evaluated the median value. Then, I found the median value of the noise temperature for the $49$ Q-band and $6$ W-band polarimeters that make up the STRIP focal plane: $T_\mathrm{noise}^Q = 33.6 \pm 5.8 \,\mathrm{K}$; $T_\mathrm{noise}^W = 105.0 \pm 17.0 \,\mathrm{K}$. The analysis showed high uncertainties. Possible sources of systematic errors could be non-linearity in detector (or ADC) response, uncertainty in ADC offset, imperfect balance between the two legs, non-idealities in the polarimeters or in the set-up, etc. More investigations could not be conducted since fundamental HK parameters were not recorded by the software. \par
We performed the noise analysis (Sect. \ref{noisechara}) over a sub-sample of the whole batch of polarimeters: only $42$ Q-band and $10$ W-band ones out of $70$ and $14$. The analysis consisted in measuring the polarimeter response to the signals emitted by the two thermal loads at the temperature of $\sim 20\,\mathrm{K}$, for a long time, ranging from $1\,\mathrm{hour}$ to $60\,\mathrm{hour}$, with the phase switches set on the switching mode. No signal was injected by the RF generator. Data have been acquired with a sampling rate of $25\,\mathrm{Hz}$. The results of this analysis showed non-Gaussian distributions with several outliers and very large error bars, which were probably due to instabilities of the power generator. The median values of knee frequency, slope of the 1/$f$ spectrum and WNL distributions in terms of the $Q$ and $U$ combinations of the detector outputs, for the Q-band polarimeters installed in the focal plane are: $\nu_\mathrm{knee}^{Q} = 75 \pm 59\,\mathrm{mHz}$, $\nu_\mathrm{knee}^{U} = 45 \pm 34\,\mathrm{mHz}$, $\alpha^{Q} = 1.3 \pm 0.4$, $\alpha^{U} = 0.9 \pm 0.3$, $\mathrm{WNL}^{Q} = 1.8 \pm 0.5\,\mathrm{mK^2 / Hz}$, $\mathrm{WNL}^{U} = 1.5 \pm 0.4\,\mathrm{mK^2 / Hz}$; while for the W-band ones are: $\nu_\mathrm{knee}^{Q} = 63 \pm 5\,\mathrm{mHz}$, $\nu_\mathrm{knee}^{U} = 78 \pm 76\,\mathrm{mHz}$, $\alpha^{Q} = 1.2 \pm 0.2$, $\alpha^{U} = 1.0 \pm 0.2$, $\mathrm{WNL}^{Q} = 6.1 \pm 2.4\,\mathrm{mK^2 / Hz}$, $\mathrm{WNL}^{U} = 4.0 \pm 1.7\,\mathrm{mK^2 / Hz}$. \par
In the second part of my thesis, I presented Stripeline, the STRIP simulation pipeline, which I contributed to develop and that is written in the Julia programming language (Ch. \ref{Chap:5}). I gave an overlook on the several modules of Stripeline that: collect all the information about the focal plane components (horn positions, horn/detector pairings, detector properties, etc.), simulate the scanning strategy, produce realizations of pseudo-instrumental noise and compute output maps from TODs. Then, I illustrated the coordinate systems that are used in Stripeline. Coordinate conversions are accurate to about $1\,\mathrm{arcsec}$ and take into account secondary effects, such as Earth's precession and nutation, stellar aberration and atmospheric refraction. \par
I illustrated the results of the scanning strategy analysis I performed for the LSPE/STRIP experiment (Ch. \ref{Chap:6}). The analysis was driven by three main goals: to observe the same sky region of SWIPE, to obtain at the same time a good sensitivity per sky pixel and a wide sky coverage, to include specific sources in the field of observation. A good trade-off between these three conditions was obtained by spinning the telescope around the azimuth axis with constant elevation and angular velocity. In this way, each receiver observes also the same air column collecting the same signal due to the atmosphere.\par
In my simulations, I set the Tenerife coordinates (Latitude = $28\deg \,18'\, 00''\, \mathrm{N}$, Longitude = $16\deg \,30'\, 35''\, \mathrm{W}$, height = $2390\,\mathrm{m\,a.s.l.}$) and let the telescope rotate around the azimuth axis with a rotational speed of $\omega_\mathrm{spin} = 1\,\mathrm{rpm}$. I assumed two years of mission, starting on 1 April 2021, using the $49$ polarimeters operating at the central frequency of $43\,\mathrm{GHz}$, with a sampling rate of $50\,\mathrm{Hz}$ and variable duty cycle. I performed simulations for several zenith angles, ranging from $5\deg$ to $50\deg$ (the maximum the STRIP telescope can reach), by steps of $5\deg$. For each simulation, I computed the hit count map and the noise map by using the detector properties that have been measured during the unit-level tests. I have studied the coverage properties of these maps both in terms of sensitivity per pixel and of overlap with the LSPE/SWIPE coverage. I found that (Sect. \ref{to}), for elevation angles between $20\deg$ and $25\deg$, we can obtain an overlap of the order of $80\%$ with SWIPE but we also need to ensure a duty cycle greater than $50\%$ to satisfy the STRIP sensitivity requirement, which is set to $1.6\,\mu\mathrm{K}\,\mathrm{deg}$ (on average). At these angles a duty cycle of about $100\%$ is required to reach the goal ($1.2\,\mu\mathrm{K}\,\mathrm{deg}$, on average). \par
My simulations showed that STRIP can observe periodically the Crab Nebula as well as the Perseus molecular complex (Sect. \ref{pois}). The Crab is one of the best known polarized sources in the sky and it must be observed for calibration purposes. The second one is source of AME that could be characterized both in intensity and polarization. I showed that the Moon and Jupiter are going to be visible from Tenerife for some periods of the year: they could be observed to characterize the beam response. To observe Orion, Saturn and Rho-Ophiuchi we probably need deep scans, maintaining the elevation angle less than $30\deg$ in the nominal scanning mode, not to lose too much sensitivity. Orion is an unpolarized source that would be useful to observe for checking any spurious polarizations of the instrument while Rho-Ophiuchi is the most widely known source of AME. \par
I have provided an estimation of the effective observation time of STRIP (Sect. \ref{dc}) by considering several causes preventing the telescope to observe for the whole duration of the mission (Sun presence, ADC blanking time, calibration time, adverse meteorological conditions, elevation scans for atmosphere measurements). I found that a reasonable upper limit for the duty cycle of STRIP can be assessed to be about $50\%$, which is compatible with the sensitivity requirement at $20\deg \div 25\deg$.\par
I presented a way to make the STRIP sky coverage more uniform (Sect. \ref{optcov}) since, in the nominal scanning mode, the hit maps are more redundant on the edge regions and less on the center ones. I proposed to introduce a modulation of the elevation of the telescope while it is spinning, which must be slow enough to make sure that after one revolution the telescope elevation is approximately constant, in order to have each horn observing atmospheric layers of equal airmass. I computed that, assuming continuous elevation scans at constant velocity from $5\deg$ to $35\deg$, the minimum period of the modulation to ensure the validity of the approximation of constant airmass layers is $\sim 10 \,\mathrm{days}$. \par
I repeated the scanning strategy simulation by modulating the elevation of the telescope with a sinusoidal function and with the sine of an optimized bezier curve. I performed the simulations for three different modulation periods ($1\,\mathrm{hour}$, $1\,\mathrm{day}$ and $1\,\mathrm{month}$) and then I compared the results. I found that the hit distribution depends strongly on the used function and that the bezier function allow us to obtain a more uniform coverage. Furthermore, I discovered that the coverage does not depend significantly on the period of the modulation ($\sim 1\%$ average difference, $\sim 2\% \div 5\%$ maximum difference) and that, with respect to the nominal scanning mode at $35\deg$, the modulations allow the instrument to increase the average sensitivity. At the same time, the instrument duty cycle must be ensured to be about $75\%$ to reach the STRIP sensitivity requirement.\par
Further studies are required to assess the effectiveness of spinning the telescope with constant angular velocity and, simultaneously, modulating its elevation. In particular, a fine characterization of the instrumental I $\to$ Q, U leakage must be performed to evaluate the impact on the polarization measurements. Besides, cross-checks with the most recent atmospheric data from Teide Observatory are required to estimate the time-scales on which the atmospheric temperature varies and to define the modulation period. Finally, the possibility to use a different class of functions to modulate the elevation or to spin the telescope with a non-constant angular velocity (e.g., which changes as a function of the elevation) must be investigated. \par
The scanning strategy simulations I presented in Ch. \ref{Chap:6} should be repeated at the end of the system-level tests by using the estimation of the noise temperatures and of the 1/$f$ noise that these tests will provide. New simulations should be performed to assess the impact of the pointing errors, of the ground emission, and to fix the parameters of the ``deep fields'' required to observe particular sky regions (Sect. \ref{pois}). \par
Finally, we can obtain a thorough understanding of the observation scenario by studying the impact of the Earth atmosphere on the STRIP measurements both in intensity and polarization. The atmosphere, in fact, attenuates the incoming astrophysical signal, and emits radiation that could be polarized. We have seen that, in principle, it is possible to measure the atmospheric signal by varying the elevation of the instrument, since it implies a variation of the instrumental sensitivity (Sect. \ref{noismaps}). However, more simulations are required to assess the minimum number of elevation angles and the minimum time to spend at each angle, or, in the case of a slowly varying elevation, the scan period. At this purpose, a reliable model of the Tenerife atmosphere is needed, which takes into account its spatial and temporal fluctuations.

%% file: App.tex
\part*{Appendices}

%% file: App1.tex
\chapter{Math of the polarimeter model}
\label{App:1}
\thispagestyle{plain}
\numberwithin{equation}{section}

In this appendix, I want to discuss the mathematical details of the polarimeter model illustrated in Ch. \ref{Chap:3}. \par
Following the description of Sect. \ref{mamo}, I define:
\be{LR}
\begin{pmatrix} L \\ R \end{pmatrix} = O \cdot I_F \cdot \mathbf{E} = \frac{1}{\sqrt2} \begin{pmatrix} 1 & i \\ 1 & -i \end{pmatrix} \cdot \begin{pmatrix} E_x \\ E_y \end{pmatrix} = \begin{pmatrix} \frac{E_x + i E_y}{\sqrt2} \\ \frac{E_x - i E_y}{\sqrt2} \end{pmatrix} \,,
\ee
and:
\be{EaEb}
\begin{pmatrix} E_A \\ E_B \end{pmatrix} = \begin{pmatrix} g_A e^{i\phi_A} (L + N_A) \\ g_B e^{i\phi_B} (R + N_B) \end{pmatrix} \,,
\ee
where $\phi_{A/B} = 0,\pi$ and $N_A,\,N_B$ are the noise introduced at the amplifiers level. According to Eqs.\ref{signalQ} and \ref{signalU}, $E_A$ and $E_B$ are the signals at the entrance of the $180\deg$ hybrid.\par
By applying the operator $H_{180}$ (Eq.\ref{H180}) to $E_A$ and $E_B$, I can define also:
\be{DaDb}
\begin{pmatrix} D_A \\ D_B \end{pmatrix} = \begin{pmatrix} \frac{E_A + E_B}{\sqrt2} \\ \frac{E_A - E_B}{\sqrt2} \end{pmatrix} \,,
\ee
which are the values of the signals after the $180\deg$ coupler. \par
With this notation, and applying the operator of Eq. \ref{ps} to Eq.\ref{DaDb}, the amplitudes of the fields at detectors $\mathrm{Q1}$ and $\mathrm{Q2}$ are given by:
\be{q1q2}
\begin{pmatrix} E_{Q1} \\ E_{Q2} \end{pmatrix} = \begin{pmatrix} \frac{D_A}{\sqrt2} \\ \frac{D_B}{\sqrt2} \end{pmatrix} \,.
\ee\par
By applying the operator $H_{90}$ (Eq.\ref{H90}) to Eq.\ref{q1q2}, it is possible to obtain the amplitudes of the fields at detectors $\mathrm{U1}$ and $\mathrm{U2}$: 
\be{u1u2}
\begin{pmatrix} E_{U1} \\ E_{U2} \end{pmatrix} = \begin{pmatrix} \frac{iE_{Q1} + E_{Q2}}{\sqrt2} \\ \frac{E_{Q1} + iE_{Q2}}{\sqrt2} \end{pmatrix} \,.
\ee \par

In this context all the signals are assumed Gaussians with zero mean:
\be{gaussign}
S = (S_x + i S_y) e^{i\omega t}\, ,
\ee
where $S_x$ and $S_y$ are random numbers gathered by a normal distribution with standard deviation $\sigma_S$. So that:
\be{Econd}
  \begin{aligned}
    \mean{\abs S} &= 0 \,, \\
    \mean{\abs S ^2} &= \mean{\abs {S_x }^2} + \mean{\abs {S_y} ^2} = 2 \sigma_S \,.
  \end{aligned}
\ee \par
 
\section{Signal expected at detector Q1}
I want to make explicit Eq.\ref{q1q2}, in the case of detector $\mathrm{Q1}$:
\begin{align}\label{eq1}
E_{Q1} &= \frac{D_A}{\sqrt2} = \frac{E_A + E_B}{2} = \frac{g_A e^{i\phi_A} (L + N_A) + g_B e^{i\phi_B} (R + N_B)}{2} = \notag \\
      &\phantom{}= \frac{g_A e^{i\phi_A}}{2}\Bigl[ \frac{E_x + i E_y}{\sqrt2} + N_A\Bigr] + \frac{g_B e^{i\phi_B}}{2}\Bigl[ \frac{E_x - i E_y}{\sqrt2} + N_B \Bigr] = \notag \\
      &\phantom{}= \frac{g_A e^{i\phi_A} + g_B e^{i\phi_B}}{2\sqrt2} E_x + \frac{g_A e^{i\phi_A} - g_B e^{i\phi_B}}{2\sqrt2} i E_y + \frac{g_A e^{i\phi_A}}{2} N_A + \frac{g_B e^{i\phi_B}}{2} N_B \,.
\end{align}\par

The diodes convert the RF signal into an electric signal proportionally to the power of the incident field (Eq.\ref{diode}). To get the power, the square module of Eq.\ref{eq1} must be computed:
\begin{align}\label{eq1_1}
  \abs{ E_{Q1} }^2 = A^2 \Bigl[ &(g_A e^{i\phi_A} + g_B e^{i\phi_B})^* E_x^* - (g_A e^{i\phi_A} - g_B e^{i\phi_B})^*i E_y^* + \sqrt2 g_A e^{-i\phi_A} N_A^* + \notag \\
    &\phantom{}  \sqrt2 g_B e^{-i\phi_B} N_B^* \Bigr] \cdot \Bigl[ (g_A e^{i\phi_A} + g_B e^{i\phi_B}) E_x + (g_A e^{i\phi_A} - g_B e^{i\phi_B})i E_y + \notag \\
    &\phantom{}  \sqrt2 g_A e^{i\phi_A} N_A + \sqrt2 g_B e^{i\phi_B} N_B \Bigr] \,,
\end{align}
where $A = \frac{1}{2\sqrt2}$.\par
By doing the products, it is possible to obtain:
\begin{align}\label{eq1_2}
  \abs{ E_{Q1} }^2 = A^2 \Bigl\{ &\Bigl[ g_A^2 + g_Ag_B e^{-i\Delta\phi} + g_Ag_B e^{i\Delta\phi} + g_B^2 \Bigr] E_x^2 + \notag \\
  &\phantom{} \Bigl[ g_A^2 - g_Ag_B e^{-i\Delta\phi} + g_Ag_B e^{i\Delta\phi} - g_B^2 \Bigr] i E_x^* E_y - \notag \\
  &\phantom{} \Bigl[ g_A^2 + g_Ag_B e^{-i\Delta\phi} - g_Ag_B e^{i\Delta\phi} - g_B^2 \Bigr] i E_x E_y^* + \notag \\
  &\phantom{} \Bigl[ g_A^2 - g_Ag_B e^{-i\Delta\phi} - g_Ag_B e^{i\Delta\phi} + g_B^2 \Bigr] E_y^2 + \notag \\
  &\phantom{} 2 g_A^2 N_A^2 + 2 g_B^2 N_B^2 + \mathrm{ products \ between \ uncorrelated \ signals} \Bigr\} \,,
\end{align}
with $\Delta\phi = \phi_A - \phi_B$. \par

Eq. \ref{eq1_2} can be written also as:
\begin{align}\label{eq1_3}
  \abs{ E_{Q1} }^2 = A^2 \Bigl[ &(g_A^2 + g_B^2) (E_x^2 + E_y^2) +  g_Ag_B (e^{i\Delta\phi} + e^{-i\Delta\phi}) (E_x^2 - E_y^2) + \notag \\
  &\phantom{}  (g_A^2 - g_B^2) i (E_x^* E_y - E_x E_y^*) + g_Ag_B (e^{i\Delta\phi} - e^{-i\Delta\phi}) i (E_x^* E_y + E_x E_y^*) + \notag \\
  &\phantom{} 2 g_A^2 N_A^2 + 2 g_B^2 N_B^2 + \mathrm{ products \ between \ uncorrelated \ signals} \Bigr] \,.
\end{align}\par

The diode also performs the temporal average over a sampling period, so that:
\begin{align}\label{esq1}
  \mean{\abs{E_{Q1}} ^2} = \frac14 \Bigl[ &\frac{g_A^2 + g_B^2}{2}\Bigl(\mean{E_x^2} + \mean{E_y^2}\Bigr) + g_Ag_B \cos(\Delta\phi) \Bigl(\mean{E_x^2} - \mean{E_y^2}\Bigr) - \notag \\
  &\phantom{} \frac{g_A^2 - g_B^2}{2} i \Bigl(\mean{E_x E_y^*} - \mean{E_x^* E_y}\Bigr) - g_Ag_B \sin(\Delta\phi) \Bigl(\mean{E_x^* E_y} + \mean{E_x E_y^*}\Bigr) + \notag \\
  &\phantom{} g_A^2\mean{N_A^2} + g_B^2\mean{N_B^2} \Bigr] \,,
\end{align}
where the gains of the amplifiers have been assumed stable in the sampling interval and the Euler's formulae have been used. Furthermore, has been exploited the fact that the average products between uncorrelated signals is zero. \par

Eventually, it is possible to write Eq.\ref{esq1} as a function of the Stokes parameters, defined by Eq.\ref{stokes}:
\begin{align}\label{esq1stokes}
  \mean{\abs{E_{Q1}} ^2} = \frac14 \Bigl[ &\frac{g_A^2 + g_B^2}{2} I + g_Ag_B \cos(\Delta\phi) Q - g_Ag_B \sin(\Delta\phi) U - \frac{g_A^2 - g_B^2}{2} V + \notag \\
    &\phantom{} g_A^2\mean{N_A^2} + g_B^2\mean{N_B^2} \Bigr] \,.
\end{align}

\section{Signal expected at detector Q2} 
I want to make explicit Eq.\ref{q1q2}, in the case of detector $\mathrm{Q2}$:
\begin{align}\label{eq12}
E_{Q2} &= \frac{D_A}{\sqrt2} = \frac{E_A - E_B}{2} = \frac{g_A e^{i\phi_A} (L + N_A) - g_B e^{i\phi_B} (R + N_B)}{2} = \notag \\
      &\phantom{}= \frac{g_A e^{i\phi_A}}{2}\Bigl[ \frac{E_x + i E_y}{\sqrt2} + N_A\Bigr] - \frac{g_B e^{i\phi_B}}{2}\Bigl[ \frac{E_x - i E_y}{\sqrt2} + N_B \Bigr] = \notag \\
      &\phantom{}= \frac{g_A e^{i\phi_A} - g_B e^{i\phi_B}}{2\sqrt2} E_x + \frac{g_A e^{i\phi_A} + g_B e^{i\phi_B}}{2\sqrt2} i E_y + \frac{g_A e^{i\phi_A}}{2} N_A - \frac{g_B e^{i\phi_B}}{2} N_B \,.
\end{align}\par

The diodes convert the RF signal into an electric signal proportionally to the power of the incident field (Eq.\ref{diode}). To get the power, the square module of Eq.\ref{eq12} must be computed:
\begin{align}\label{eq12_1}
  \abs{ E_{Q2} }^2 = A^2 \Bigl[ &(g_A e^{i\phi_A} - g_B e^{i\phi_B})^* E_x^* - (g_A e^{i\phi_A} + g_B e^{i\phi_B})^*i E_y^* + \sqrt2 g_A e^{-i\phi_A} N_A^* - \notag \\
    &\phantom{} \sqrt2 g_B e^{-i\phi_B} N_B^* \Bigr] \cdot \Bigl[ (g_A e^{i\phi_A} - g_B e^{i\phi_B}) E_x + (g_A e^{i\phi_A} + g_B e^{i\phi_B})i E_y + \notag \\
    &\phantom{} \sqrt2 g_A e^{i\phi_A} N_A - \sqrt2 g_B e^{i\phi_B} N_B \Bigr] \,,
\end{align}
where $A = \frac{1}{2\sqrt2}$.\par
By doing the products, it is possible to obtain:
\begin{align}\label{eq12_2}
  \abs{ E_{Q2} }^2 = A^2 \Bigl\{ &\Bigl[ g_A^2 - g_Ag_B e^{-i\Delta\phi} - g_Ag_B e^{i\Delta\phi} + g_B^2 \Bigr] E_x^2 + \notag \\
  &\phantom{} \Bigl[ g_A^2 + g_Ag_B e^{-i\Delta\phi} - g_Ag_B e^{i\Delta\phi} - g_B^2 \Bigr] i E_x^* E_y - \notag \\
  &\phantom{} \Bigl[ g_A^2 - g_Ag_B e^{-i\Delta\phi} + g_Ag_B e^{i\Delta\phi} - g_B^2 \Bigr] i E_x E_y^* + \notag \\
  &\phantom{} \Bigl[ g_A^2 + g_Ag_B e^{-i\Delta\phi} + g_Ag_B e^{i\Delta\phi} + g_B^2 \Bigr] E_y^2 + \notag \\
  &\phantom{} 2 g_A^2 N_A^2 + 2 g_B^2 N_B^2 + \mathrm{ products \ between \ uncorrelated \ signals} \Bigr\} \,,
\end{align}
with $\Delta\phi = \phi_A - \phi_B$. \par

Eq. \ref{eq12_2} can be written also as:
\begin{align}\label{eq12_3}
  \abs{ E_{Q2} }^2 = A^2 \Bigl[ &(g_A^2 + g_B^2) (E_x^2 + E_y^2) -  g_Ag_B (e^{i\Delta\phi} + e^{-i\Delta\phi}) (E_x^2 - E_y^2) + \notag \\
  &\phantom{}  (g_A^2 - g_B^2) i (E_x^* E_y - E_x E_y^*) - g_Ag_B (e^{i\Delta\phi} - e^{-i\Delta\phi}) i (E_x^* E_y + E_x E_y^*) + \notag \\
  &\phantom{} 2 g_A^2 N_A^2 + 2 g_B^2 N_B^2 + \mathrm{ products \ between \ uncorrelated \ signals} \Bigr] \,.
\end{align}\par

The diode also performs the temporal average over a sampling period, so that:
\begin{align}\label{esq12}
  \mean{\abs{E_{Q2}} ^2} = \frac14 \Bigl[ &\frac{g_A^2 + g_B^2}{2}\Bigl(\mean{E_x^2} + \mean{E_y^2}\Bigr) - g_Ag_B \cos(\Delta\phi) \Bigl(\mean{E_x^2} - \mean{E_y^2}\Bigr) - \notag \\
  &\phantom{} \frac{g_A^2 - g_B^2}{2} i \Bigl(\mean{E_x E_y^*} - \mean{E_x^* E_y}\Bigr) + g_Ag_B \sin(\Delta\phi) \Bigl(\mean{E_x^* E_y} + \mean{E_x E_y^*}\Bigr) + \notag \\
  &\phantom{} g_A^2\mean{N_A^2} + g_B^2\mean{N_B^2} \Bigr] \,,
\end{align}
where the gains of the amplifiers have been assumed stable in the sampling interval and the Euler's formulae have been used. Furthermore, has been exploited the fact that the average products between uncorrelated signals is zero. \par

Eventually, it is possible to write Eq.\ref{esq12} as a function of the Stokes parameters, defined by Eq.\ref{stokes}:
\begin{align}\label{esq12stokes}
  \mean{\abs{E_{Q2}} ^2} = \frac14 \Bigl[ &\frac{g_A^2 + g_B^2}{2} I - g_Ag_B \cos(\Delta\phi) Q + g_Ag_B \sin(\Delta\phi) U - \frac{g_A^2 - g_B^2}{2} V + \notag \\
    &\phantom{} g_A^2\mean{N_A^2} + g_B^2\mean{N_B^2} \Bigr] \,.
\end{align}

\section{Signal expected at detector U1}
I want to make explicit Eq.\ref{u1u2}, in the case of detector $\mathrm{U1}$:
\begin{align}\label{eu1}
E_{U1} = &\frac{i E_{Q1} + E_{Q2}}{\sqrt2} = \frac{g_A e^{i\phi_A} + g_B e^{i\phi_B}}{4} i E_x - \frac{g_A e^{i\phi_A} - g_B e^{i\phi_B}}{4} E_y +  \frac{g_A e^{i\phi_A}}{4} \sqrt2 i N_A - \notag \\
      &\phantom{} \frac{g_B e^{i\phi_B}}{4} \sqrt2 i N_B + \frac{g_A e^{i\phi_A} - g_B e^{i\phi_B}}{4} E_x + \frac{g_A e^{i\phi_A} + g_B e^{i\phi_B}}{4} i E_y + \notag \\
      &\phantom{} \frac{g_A e^{i\phi_A}}{4} \sqrt2 N_A - \frac{g_B e^{i\phi_B}}{4} \sqrt2 N_B = \notag \\
      &\phantom{} \hspace{-13pt}= \frac{g_A e^{i\phi_A} (i+1) + g_B e^{i\phi_B} (i-1)}{4} E_x + \frac{g_A e^{i\phi_A} (i-1) + g_B e^{i\phi_B} (i+1)}{4} E_y + \notag \\
      &\phantom{} g_A e^{i\phi_A} \sqrt2 N_A \frac{i+1}{4} + g_B e^{i\phi_B} \sqrt2 N_B \frac{i-1}{4}  \,,
\end{align}\par
where Eqs. \ref{eq1} and \ref{eq12} have been used. 

The diodes convert the RF signal into an electric signal proportionally to the power of the incident field (Eq.\ref{diode}). To get the power, the square module of Eq.\ref{eu1} must be computed:
\begin{align}\label{eu1_1}
  \abs{ E_{U1} }^2 = B^2 \Bigl\{ \Bigl[ &g_A e^{i\phi_A} (i+1) + g_B e^{i\phi_B} (i-1)\Bigr] E_x + \Bigl[g_A e^{i\phi_A}(i-1) + g_B e^{i\phi_B}(i+1) \Bigr] E_y + \notag \\
    &\phantom{} g_A e^{i\phi_A} \sqrt2 N_A (i+1) + g_B e^{i\phi_B} \sqrt2 N_B (i-1) \Bigr\}  \cdot \notag \\
    &\phantom{} \hspace{-10pt}\Bigl\{ \Bigl[ g_A e^{-i\phi_A} (1-i) - g_B e^{-i\phi_B} (1+i)\Bigr] E_x^* - \notag \\
    &\phantom{} \hspace{-3pt}\Bigl[g_A e^{-i\phi_A}(1+i) - g_B e^{-i\phi_B}(1-i) \Bigr] E_y^* + g_A e^{-i\phi_A} \sqrt2 N_A^*(1-i) - \notag \\
    &\phantom{} g_B e^{-i\phi_B} \sqrt2 N_B^*(1+i) \Bigr\}\,,
\end{align}
where $B = \frac14$.\par
By doing the products, it is possible to obtain:
\begin{align}\label{eu1_2}
  \abs{ E_{U1} }^2 = B^2 \Bigl\{ &\Bigl[2g_A^2 - 2 i g_Ag_B e^{i\Delta\phi} + 2 i g_Ag_B e^{-i\Delta\phi} + 2 g_B^2 \Bigr] E_x^2 + \notag \\
  &\phantom{} \Bigl[ 2g_A^2 + 2 i g_Ag_B e^{i\Delta\phi} - 2 i g_Ag_B e^{-i\Delta\phi} + 2g_B^2 \Bigr] E_y^2 - \notag \\
  &\phantom{} \Bigl[ -2 i g_A^2 + 2g_Ag_B e^{i\Delta\phi} + 2g_Ag_B e^{-i\Delta\phi} + 2 i g_B^2 \Bigr] E_x E_y^* + \notag \\
  &\phantom{} \Bigl[ 2 i g_A^2 + 2g_Ag_B e^{i\Delta\phi} + 2g_Ag_B e^{-i\Delta\phi} - 2 i g_B^2 \Bigr] E_x^* E_y + \notag \\
  &\phantom{} 4 g_A^2 N_A^2 + 4 g_B^2 N_B^2 + \mathrm{ products \ between \ uncorrelated \ signals} \Bigr\} \,,
\end{align}
with $\Delta\phi = \phi_A - \phi_B$. \par

Eq. \ref{eu1_2} can be written also as:
\begin{align}\label{eu1_3}
  \abs{ E_{U1} }^2 = \frac18 \Bigl\{ &\Bigl[g_A^2 + g_B^2 + 2g_Ag_B \sin(\Delta\phi) \Bigr] E_x^2 + \Bigl[g_A^2 + g_B^2 - 2g_Ag_B \sin(\Delta\phi) \Bigr] E_y^2 + \notag \\
  &\phantom{} \Bigl[i (g_A^2 - g_B^2) + 2g_Ag_B \cos(\Delta\phi) \Bigr] E_x^*E_y -  \notag \\
  &\phantom{} \Bigl[i (g_A^2 - g_B^2) - 2g_Ag_B \cos(\Delta\phi) \Bigr] E_xE_y^* + 2 g_A^2 N_A^2 + 2 g_B^2 N_B^2 + \notag \\
  &\phantom{} \mathrm{ products \ between \ uncorrelated \ signals} \Bigr\} \,.
\end{align}\par

The diode also performs the temporal average over a sampling period, so that:
\begin{align}\label{esu1}
  \mean{\abs{E_{U1}} ^2} = \frac14 \Bigl[ &\frac{g_A^2 + g_B^2}{2}\Bigl(\mean{E_x^2} + \mean{E_y^2}\Bigr) + g_Ag_B \sin(\Delta\phi) \Bigl(\mean{E_x^2} - \mean{E_y^2}\Bigr) - \notag \\
  &\phantom{} \frac{g_A^2 - g_B^2}{2} i \Bigl(\mean{E_x E_y^*} - \mean{E_x^* E_y}\Bigr) + g_Ag_B \cos(\Delta\phi) \Bigl(\mean{E_x^* E_y} + \mean{E_x E_y^*}\Bigr) + \notag \\
  &\phantom{} g_A^2\mean{N_A^2} + g_B^2\mean{N_B^2} \Bigr] \,,
\end{align}
where the gains of the amplifiers have been assumed stable in the sampling interval and the Euler's formulae have been used. Furthermore, has been exploited the fact that the average products between uncorrelated signals is zero. \par

Eventually, it is possible to write Eq.\ref{esu1} as a function of the Stokes parameters, defined by Eq.\ref{stokes}:
\begin{align}\label{esu1stokes}
  \mean{\abs{E_{U1}} ^2} = \frac14 \Bigl[ &\frac{g_A^2 + g_B^2}{2} I + g_Ag_B \sin(\Delta\phi) Q + g_Ag_B \cos(\Delta\phi) U - \frac{g_A^2 - g_B^2}{2} V + \notag \\
    &\phantom{} g_A^2\mean{N_A^2} + g_B^2\mean{N_B^2} \Bigr] \,.
\end{align}

\section{Signal expected at detector U2} 
I want to make explicit Eq.\ref{u1u2}, in the case of detector $\mathrm{U2}$:
\begin{align}\label{eu12}
E_{U2} = &\frac{E_{Q1} + i E_{Q2}}{\sqrt2} = \frac{g_A e^{i\phi_A} + g_B e^{i\phi_B}}{4} E_x + \frac{g_A e^{i\phi_A} - g_B e^{i\phi_B}}{4} i E_y + \frac{g_A e^{i\phi_A}}{4} \sqrt2 N_A + \notag \\
      &\phantom{} \frac{g_B e^{i\phi_B}}{4} \sqrt2 N_B + \frac{g_A e^{i\phi_A} - g_B e^{i\phi_B}}{4} i E_x - \frac{g_A e^{i\phi_A} + g_B e^{i\phi_B}}{4} E_y + \notag \\
      &\phantom{} \frac{g_A e^{i\phi_A}}{4} \sqrt2 i N_A - \frac{g_B e^{i\phi_B}}{4} \sqrt2 i N_B = \notag \\
      &\phantom{} \hspace{-13pt}= \frac{g_A e^{i\phi_A} (i+1) - g_B e^{i\phi_B} (i-1)}{4} E_x + \frac{g_A e^{i\phi_A} (i-1) - g_B e^{i\phi_B} (i+1)}{4} E_y + \notag \\
      &\phantom{} g_A e^{i\phi_A} \sqrt2 N_A \frac{i+1}{4} + g_B e^{i\phi_B} \sqrt2 N_B \frac{1-i}{4}  \,,
\end{align}\par
where Eqs. \ref{eq1} and \ref{eq12} have been used. 

The diodes convert the RF signal into an electric signal proportionally to the power of the incident field (Eq.\ref{diode}). To get the power, the square module of Eq.\ref{eu12} must be computed:
\begin{align}\label{eu12_1}
  \abs{ E_{U2} }^2 = B^2 \Bigl\{ \Bigl[ &g_A e^{i\phi_A} (i+1) - g_B e^{i\phi_B} (i-1)\Bigr] E_x + \Bigl[g_A e^{i\phi_A}(i-1) - g_B e^{i\phi_B}(i+1) \Bigr] E_y + \notag \\
    &\phantom{} g_A e^{i\phi_A} \sqrt2 N_A (i+1) + g_B e^{i\phi_B} \sqrt2 N_B (1-i) \Bigr\}  \cdot \notag \\
    &\phantom{} \hspace{-10pt}\Bigl\{ \Bigl[ g_A e^{-i\phi_A} (1-i) + g_B e^{-i\phi_B} (1+i)\Bigr] E_x^* - \notag \\
    &\phantom{} \hspace{-3pt}\Bigl[g_A e^{-i\phi_A}(1+i) + g_B e^{-i\phi_B}(1-i) \Bigr] E_y^* + g_A e^{-i\phi_A} \sqrt2 N_A^*(1-i) + \notag \\
    &\phantom{} g_B e^{-i\phi_B} \sqrt2 N_B^*(1+i) \Bigr\}\,,
\end{align}
where $B = \frac14$.\par
By doing the products, it is possible to obtain:
\begin{align}\label{eu12_2}
  \abs{ E_{U2} }^2 = B^2 \Bigl\{ &\Bigl[2g_A^2 + 2 i g_Ag_B e^{i\Delta\phi} - 2 i g_Ag_B e^{-i\Delta\phi} + 2 g_B^2 \Bigr] E_x^2 + \notag \\
  &\phantom{} \Bigl[ 2g_A^2 - 2 i g_Ag_B e^{i\Delta\phi} + 2 i g_Ag_B e^{-i\Delta\phi} + 2g_B^2 \Bigr] E_y^2 - \notag \\
  &\phantom{} \Bigl[ 2 i g_A^2 + 2g_Ag_B e^{i\Delta\phi} + 2g_Ag_B e^{-i\Delta\phi} - 2 i g_B^2 \Bigr] E_x E_y^* + \notag \\
  &\phantom{} \Bigl[ 2 i g_A^2 - 2g_Ag_B e^{i\Delta\phi} - 2g_Ag_B e^{-i\Delta\phi} - 2 i g_B^2 \Bigr] E_x^* E_y + \notag \\
  &\phantom{} 4 g_A^2 N_A^2 + 4 g_B^2 N_B^2 + \mathrm{ products \ between \ uncorrelated \ signals} \Bigr\} \,,
\end{align}
with $\Delta\phi = \phi_A - \phi_B$. \par

Eq. \ref{eu12_2} can be written also as:
\begin{align}\label{eu12_3}
  \abs{ E_{U2} }^2 = \frac18 \Bigl\{ &\Bigl[g_A^2 + g_B^2 - 2g_Ag_B \sin(\Delta\phi) \Bigr] E_x^2 + \Bigl[g_A^2 + g_B^2 + 2g_Ag_B \sin(\Delta\phi) \Bigr] E_y^2 + \notag \\
  &\phantom{} \Bigl[i (g_A^2 - g_B^2) - 2g_Ag_B \cos(\Delta\phi) \Bigr] E_x^*E_y - \notag \\
  &\phantom{} \Bigl[i (g_A^2 - g_B^2) + 2g_Ag_B \cos(\Delta\phi) \Bigr] E_xE_y^* + 2 g_A^2 N_A^2 + 2 g_B^2 N_B^2 + \notag \\
  &\phantom{} \mathrm{ products \ between \ uncorrelated \ signals} \Bigr\} \,.
\end{align}\par

The diode also performs the temporal average over a sampling period, so that:
\begin{align}\label{esu12}
  \mean{\abs{E_{U2}} ^2} = \frac14 \Bigl[ &\frac{g_A^2 + g_B^2}{2}\Bigl(\mean{E_x^2} + \mean{E_y^2}\Bigr) - g_Ag_B \sin(\Delta\phi) \Bigl(\mean{E_x^2} - \mean{E_y^2}\Bigr) - \notag \\
  &\phantom{} \frac{g_A^2 - g_B^2}{2} i \Bigl(\mean{E_x E_y^*} - \mean{E_x^* E_y}\Bigr) - g_Ag_B \cos(\Delta\phi) \Bigl(\mean{E_x^* E_y} + \mean{E_x E_y^*}\Bigr) + \notag \\
  &\phantom{} g_A^2\mean{N_A^2} + g_B^2\mean{N_B^2} \Bigr] \,,
\end{align}
where the gains of the amplifiers have been assumed stable in the sampling interval and the Euler's formulae have been used. Furthermore, has been exploited the fact that the average products between uncorrelated signals is zero. \par

Eventually, it is possible to write Eq.\ref{esu12} as a function of the Stokes parameters, defined by Eq.\ref{stokes}:
\begin{align}\label{esu12stokes}
  \mean{\abs{E_{U2}} ^2} = \frac14 \Bigl[ &\frac{g_A^2 + g_B^2}{2} I - g_Ag_B \sin(\Delta\phi) Q - g_Ag_B \cos(\Delta\phi) U - \frac{g_A^2 - g_B^2}{2} V + \notag \\
    &\phantom{} g_A^2\mean{N_A^2} + g_B^2\mean{N_B^2} \Bigr] \,.
\end{align}

%% file: App2.tex
\chapter{Math of the polarimeter model in the unit-level tests configuration}
\label{App:2}
\thispagestyle{plain}
\numberwithin{equation}{section}

In this appendix, I want to discuss the mathematical details of the polarimeter model in the unit level tests configuration illustrated in Ch. \ref{Chap:4}. \par
Following the description of Sect. \ref{setup}, I define:
\be{e1e2}
\begin{pmatrix} E_1 \\ E_2 \end{pmatrix} = T \cdot \mathbf{E} = \frac{1}{\sqrt2} \begin{pmatrix} 1 & 1 \\ 1 & -1 \end{pmatrix} \cdot\begin{pmatrix} E_A' \\ E_B \end{pmatrix} = \begin{pmatrix} \frac{E_A' + E_B}{\sqrt2} \\ \frac{E_A' - E_B}{\sqrt2} \end{pmatrix} \,,
\ee
and:
\be{HaHb}
\begin{pmatrix} H_A \\ H_B \end{pmatrix} = \begin{pmatrix} g_A e^{i\phi_A} (E_1 + N_A) \\ g_B e^{i\phi_B} (E_2 + N_B) \end{pmatrix} \,,
\ee
where $\phi_{A/B} = 0,\pi$ and $N_A,\,N_B$ are the noise introduced at the amplifiers level. According to Eqs.\ref{signalQ} and \ref{signalU}, $H_A$ and $H_B$ are the signals at the entrance of the $180\deg$ hybrid.\par
By applying the operator $H_{180}$ (Eq.\ref{H180}) to $H_A$ and $H_B$, I can define also:
\be{CaCb}
\begin{pmatrix} C_A \\ C_B \end{pmatrix} = \begin{pmatrix} \frac{H_A + H_B}{\sqrt2} \\ \frac{H_A - H_B}{\sqrt2} \end{pmatrix} \,,
\ee
which are the values of the signals after the $180\deg$ coupler. \par
With this notation, and applying the operator of Eq. \ref{ps} to Eq.\ref{CaCb}, the amplitudes of the fields at detectors $\mathrm{Q1}$ and $\mathrm{Q2}$ are given by:
\be{q1q2B}
\begin{pmatrix} E_{Q1} \\ E_{Q2} \end{pmatrix} = \begin{pmatrix} \frac{C_A}{\sqrt2} \\ \frac{C_B}{\sqrt2} \end{pmatrix} \,.
\ee\par
By applying the operator $H_{90}$ (Eq.\ref{H90}) to Eq.\ref{q1q2B}, it is possible to obtain the amplitudes of the fields at detectors $\mathrm{U1}$ and $\mathrm{U2}$: 
\be{u1u2B}
\begin{pmatrix} E_{U1} \\ E_{U2} \end{pmatrix} = \begin{pmatrix} \frac{iE_{Q1} + E_{Q2}}{\sqrt2} \\ \frac{E_{Q1} + iE_{Q2}}{\sqrt2} \end{pmatrix} \,.
\ee \par

In this context all the signals are assumed Gaussians with zero mean:
\be{gaussign}
S = (S_x + i S_y) e^{i\omega t}\, ,
\ee
where $S_x$ and $S_y$ are random numbers gathered by a normal distribution with standard deviation $\sigma_S$. So that:
\be{Econd}
  \begin{aligned}
    \mean{\abs S} &= 0 \,, \\
    \mean{\abs S ^2} &= \mean{\abs {S_x }^2} + \mean{\abs {S_y} ^2} = 2 \sigma_S \,.
  \end{aligned}
\ee \par

\section{Signal expected at detector Q1}
I want to make explicit Eq.\ref{q1q2B}, in the case of detector $\mathrm{Q1}$:
\begin{align}\label{eq1B}
  E_{Q1} = &\frac{C_A}{\sqrt2} = \frac{H_A + H_B}{2} = \frac{g_A e^{i\phi_A}}{2} \Bigl(\frac{E_A' + E_B}{\sqrt2} + N_A\Bigl) + \frac{g_B e^{i\phi_B}}{2} \Bigl(\frac{E_A' - E_B}{\sqrt2} + N_B\Bigl) = \notag \\
   &\phantom{} \hspace{-10pt}= \frac{g_A e^{i\phi_A} + g_B e^{i\phi_B}}{2\sqrt2} E_A' + \frac{g_A e^{i\phi_A} - g_B e^{i\phi_B}}{2\sqrt2} E_B + \frac{g_A e^{i\phi_A}}{2} N_A + \frac{g_B e^{i\phi_B}}{2} N_B\,.
\end{align}\par

The diodes convert the RF signal into an electric signal proportionally to the power of the incident field (Eq.\ref{diode}). To get the power, the square module of Eq.\ref{eq1B} must be computed:
\begin{align}\label{eq1B_1}
  \abs{ E_{Q1} }^2 = A^2 \Bigl[ &(g_A e^{i\phi_A} + g_B e^{i\phi_B}) E_A' + (g_A e^{i\phi_A} - g_B e^{i\phi_B}) E_B + \sqrt2 g_A e^{i\phi_A} N_A + \notag \\
    &\phantom{}  \sqrt2 g_B e^{i\phi_B} N_B \Bigr] \cdot \Bigl[ (g_A e^{-i\phi_A} + g_B e^{-i\phi_B}) E_A'^* + (g_A e^{-i\phi_A} - g_B e^{-i\phi_B}) E_B^* + \notag \\
    &\phantom{}  \sqrt2 g_A e^{-i\phi_A} N_A^* + \sqrt2 g_B e^{-i\phi_B} N_B^* \Bigr] \,,
\end{align}
where $A = \frac{1}{2\sqrt2}$.\par
By doing the products, it is possible to obtain:
\begin{align}\label{eq1B_2}
  \abs{ E_{Q1} }^2 = A^2 \Bigl\{ &\Bigl[ g_A^2 + g_Ag_B e^{i\Delta\phi} + g_Ag_B e^{-i\Delta\phi} + g_B^2 \Bigr] E_A'^2 + \notag \\
  &\phantom{} \Bigl[ g_A^2 - g_Ag_B e^{i\Delta\phi} - g_Ag_B e^{-i\Delta\phi} + g_B^2 \Bigr] E_B^2 + \notag \\
  &\phantom{} 2 g_A^2 N_A^2 + 2 g_B^2 N_B^2 + \mathrm{ products \ between \ uncorrelated \ signals} \Bigr\} \,,
\end{align}
with $\Delta\phi = \phi_A - \phi_B$. \par

Eq. \ref{eq1B_2} can be written also as:
\begin{align}\label{eq1B_3}
  \abs{ E_{Q1} }^2 = A^2 \Bigl[ &(g_A^2 + g_B^2) (E_A'^2 + E_B^2) + g_Ag_B (e^{i\Delta\phi} + e^{-i\Delta\phi}) (E_A'^2 - E_B) + \notag \\
  &\phantom{} 2 g_A^2 N_A^2 + 2 g_B^2 N_B^2 + \mathrm{ products \ between \ uncorrelated \ signals} \Bigr] \,.
\end{align}\par

The diode also performs the temporal average over a sampling period, so that:
\begin{align}\label{esq1B}
  \mean{\abs{E_{Q1}} ^2} = \frac14 \Bigl[ &\frac{g_A^2 + g_B^2}{2}\Bigl(\mean{E_A'^2} + \mean{E_B^2}\Bigr) + g_Ag_B \cos(\Delta\phi) \Bigl(\mean{E_A'^2} - \mean{E_B^2}\Bigr) + \notag \\
  &\phantom{} g_A^2\mean{N_A^2} + g_B^2\mean{N_B^2} \Bigr] \,,
\end{align}
where the gains of the amplifiers have been assumed stable in the sampling interval and the Euler's formulae have been used. Furthermore, has been exploited the fact that the average products between uncorrelated signals is zero. \par

Eventually, since $E_A$ and $E_{RF}$ are also uncorrelated:
\be{meaneaearf}
\mean{E_A'^2} = \mean{E_A^2} + \mean{E_{RF}^2} \,,
\ee
and it is possible to write Eq.\ref{esq1B} as:
\begin{align}\label{esq1Bfinal}
  \mean{\abs{E_{Q1}} ^2} = \frac14 \Bigl[ &\frac{g_A^2 + g_B^2}{2}\Bigl(\mean{E_A^2} + \mean{E_{RF}^2} + \mean{E_B^2}\Bigr) + \notag \\
  &\phantom{} g_Ag_B \cos(\Delta\phi) \Bigl(\mean{E_A^2} + \mean{E_{RF}^2} - \mean{E_B^2}\Bigr) + \notag \\
  &\phantom{} g_A^2\mean{N_A^2} + g_B^2\mean{N_B^2} \Bigr] \,.
\end{align}

\section{Signal expected at detector Q2}
I want to make explicit Eq.\ref{q1q2B}, in the case of detector $\mathrm{Q2}$:
\begin{align}\label{eq2B}
  E_{Q2} = &\frac{C_B}{\sqrt2} = \frac{H_A - H_B}{2} = \frac{g_A e^{i\phi_A}}{2} \Bigl(\frac{E_A' + E_B}{\sqrt2} + N_A\Bigl) - \frac{g_B e^{i\phi_B}}{2} \Bigl(\frac{E_A' - E_B}{\sqrt2} + N_B\Bigl) = \notag \\
   &\phantom{} \hspace{-10pt}= \frac{g_A e^{i\phi_A} - g_B e^{i\phi_B}}{2\sqrt2} E_A' + \frac{g_A e^{i\phi_A} + g_B e^{i\phi_B}}{2\sqrt2} E_B + \frac{g_A e^{i\phi_A}}{2} N_A - \frac{g_B e^{i\phi_B}}{2} N_B\,.
\end{align}\par

The diodes convert the RF signal into an electric signal proportionally to the power of the incident field (Eq.\ref{diode}). To get the power, the square module of Eq.\ref{eq2B} must be computed:
\begin{align}\label{eq2B_1}
  \abs{ E_{Q2} }^2 = A^2 \Bigl[ &(g_A e^{i\phi_A} - g_B e^{i\phi_B}) E_A' + (g_A e^{i\phi_A} + g_B e^{i\phi_B}) E_B + \sqrt2 g_A e^{i\phi_A} N_A - \notag \\
    &\phantom{}  \sqrt2 g_B e^{i\phi_B} N_B \Bigr] \cdot \Bigl[ (g_A e^{-i\phi_A} - g_B e^{-i\phi_B}) E_A'^* + (g_A e^{-i\phi_A} + g_B e^{-i\phi_B}) E_B^* + \notag \\
    &\phantom{}  \sqrt2 g_A e^{-i\phi_A} N_A^* - \sqrt2 g_B e^{-i\phi_B} N_B^* \Bigr] \,,
\end{align}
where $A = \frac{1}{2\sqrt2}$.\par
By doing the products, it is possible to obtain:
\begin{align}\label{eq2B_2}
  \abs{ E_{Q2} }^2 = A^2 \Bigl\{ &\Bigl[ g_A^2 - g_Ag_B e^{i\Delta\phi} - g_Ag_B e^{-i\Delta\phi} + g_B^2 \Bigr] E_A'^2 + \notag \\
  &\phantom{} \Bigl[ g_A^2 + g_Ag_B e^{i\Delta\phi} + g_Ag_B e^{-i\Delta\phi} + g_B^2 \Bigr] E_B^2 + \notag \\
  &\phantom{} 2 g_A^2 N_A^2 + 2 g_B^2 N_B^2 + \mathrm{ products \ between \ uncorrelated \ signals} \Bigr\} \,,
\end{align}
with $\Delta\phi = \phi_A - \phi_B$. \par

Eq. \ref{eq2B_2} can be written also as:
\begin{align}\label{eq2B_3}
  \abs{ E_{Q2} }^2 = A^2 \Bigl[ &(g_A^2 + g_B^2) (E_A'^2 + E_B^2) - g_Ag_B (e^{i\Delta\phi} + e^{-i\Delta\phi}) (E_A'^2 - E_B) + \notag \\
  &\phantom{} 2 g_A^2 N_A^2 + 2 g_B^2 N_B^2 + \mathrm{ products \ between \ uncorrelated \ signals} \Bigr] \,.
\end{align}\par

The diode also performs the temporal average over a sampling period, so that:
\begin{align}\label{esq2B}
  \mean{\abs{E_{Q2}} ^2} = \frac14 \Bigl[ &\frac{g_A^2 + g_B^2}{2}\Bigl(\mean{E_A'^2} + \mean{E_B^2}\Bigr) - g_Ag_B \cos(\Delta\phi) \Bigl(\mean{E_A'^2} - \mean{E_B^2}\Bigr) + \notag \\
  &\phantom{} g_A^2\mean{N_A^2} + g_B^2\mean{N_B^2} \Bigr] \,,
\end{align}
where the gains of the amplifiers have been assumed stable in the sampling interval and the Euler's formulae have been used. Furthermore, has been exploited the fact that the average products between uncorrelated signals is zero. \par

Eventually, since $E_A$ and $E_{RF}$ are also uncorrelated:
\be{meaneaearf}
\mean{E_A'^2} = \mean{E_A^2} + \mean{E_{RF}^2} \,,
\ee
and it is possible to write Eq.\ref{esq2B} as:
\begin{align}\label{esq2Bfinal}
  \mean{\abs{E_{Q2}} ^2} = \frac14 \Bigl[ &\frac{g_A^2 + g_B^2}{2}\Bigl(\mean{E_A^2} + \mean{E_{RF}^2} + \mean{E_B^2}\Bigr) - \notag \\
  &\phantom{} g_Ag_B \cos(\Delta\phi) \Bigl(\mean{E_A^2} + \mean{E_{RF}^2} - \mean{E_B^2}\Bigr) + \notag \\
  &\phantom{} g_A^2\mean{N_A^2} + g_B^2\mean{N_B^2} \Bigr] \,.
\end{align}

\section{Signal expected at detector U1}
I want to make explicit Eq.\ref{u1u2B}, in the case of detector $\mathrm{U1}$:
\begin{align}\label{eu1B}
E_{U1} = &\frac{i E_{Q1} + E_{Q2}}{\sqrt2} = \frac{g_A e^{i\phi_A} + g_B e^{i\phi_B}}{4} i E_A' + \frac{g_A e^{i\phi_A} - g_B e^{i\phi_B}}{4} i E_B +  \frac{g_A e^{i\phi_A}}{4} \sqrt2 i N_A + \notag \\
      &\phantom{} \frac{g_B e^{i\phi_B}}{4} \sqrt2 i N_B + \frac{g_A e^{i\phi_A} - g_B e^{i\phi_B}}{4} E_A' + \frac{g_A e^{i\phi_A} + g_B e^{i\phi_B}}{4} E_B + \notag \\
      &\phantom{} \frac{g_A e^{i\phi_A}}{4} \sqrt2 N_A - \frac{g_B e^{i\phi_B}}{4} \sqrt2 N_B = \notag \\
      &\phantom{} \hspace{-13pt}= \frac14 \Bigl\{ \Bigl[ g_A e^{i\phi_A} (i+1) + g_B e^{i\phi_B} (i-1) \Bigr] E_A' + \Bigl[g_A e^{i\phi_A} (i+1) - g_B e^{i\phi_B} (i-1) \Bigr] E_B + \notag \\
      &\phantom{} g_A e^{i\phi_A} \sqrt2 N_A (i+1) + g_B e^{i\phi_B} \sqrt2 N_B (i-1)  \Bigr\} \,.
\end{align}\par

The diodes convert the RF signal into an electric signal proportionally to the power of the incident field (Eq.\ref{diode}). To get the power, the square module of Eq.\ref{eu1B} must be computed:
\begin{align}\label{eu1B_1}
  \abs{ E_{U1} }^2 = B^2 \Bigl\{ \Bigl[ &g_A e^{i\phi_A} (i+1) + g_B e^{i\phi_B} (i-1)\Bigr] E_A' + \Bigl[g_A e^{i\phi_A}(i+1) - g_B e^{i\phi_B}(i-1) \Bigr] E_B + \notag \\
    &\phantom{} g_A e^{i\phi_A} \sqrt2 N_A (i+1) + g_B e^{i\phi_B} \sqrt2 N_B (i-1) \Bigr\}  \cdot \notag \\
    &\phantom{} \hspace{-10pt}\Bigl\{ \Bigl[ g_A e^{-i\phi_A} (1-i) - g_B e^{-i\phi_B} (1+i) \Bigr] E_A'^* + \notag \\
    &\phantom{} \hspace{-3pt}\Bigl[g_A e^{-i\phi_A}(1-i) + g_B e^{-i\phi_B}(1+i) \Bigr] E_B^* + g_A e^{-i\phi_A} \sqrt2 N_A^*(1-i) - \notag \\
    &\phantom{} g_B e^{-i\phi_B} \sqrt2 N_B^*(1+i) \Bigr\}\,,
\end{align}
where $B = \frac14$.\par
By doing the products, it is possible to obtain:
\begin{align}\label{eu1B_2}
  \abs{ E_{U1} }^2 = B^2 \Bigl\{ &\Bigl[ 2 g_A^2 - 2 i g_Ag_B e^{i\Delta\phi} + 2 i g_Ag_B e^{-i\Delta\phi} + 2 g_B^2 \Bigr] E_A'^2 + \notag \\
  &\phantom{} \Bigl[ 2 g_A^2 + 2 i g_Ag_B e^{i\Delta\phi} - 2 i g_Ag_B e^{-i\Delta\phi} + 2 g_B^2 \Bigr] E_B^2 + \notag \\
  &\phantom{} 4 g_A^2 N_A^2 + 4 g_B^2 N_B^2 + \mathrm{ products \ between \ uncorrelated \ signals} \Bigr\} \,,
\end{align}
with $\Delta\phi = \phi_A - \phi_B$. \par

Eq. \ref{eu1B_2} can be written also as:
\begin{align}\label{eu1B_3}
  \abs{ E_{U1} }^2 = B^2 \Bigl\{ &\Bigl[ 2g_A^2 + 4 g_Ag_B \frac{e^{i\Delta\phi} - e^{-i\Delta\phi}}{2i} + 2g_B^2 \Bigr] E_A'^2 + \notag \\
  &\phantom{} \Bigl[ 2g_A^2 - 4 g_Ag_B \frac{e^{i\Delta\phi} - e^{-i\Delta\phi}}{2i} + 2g_B^2 \Bigr] E_B^2 + \notag \\
  &\phantom{} 4 g_A^2 N_A^2 + 4 g_B^2 N_B^2 + \mathrm{ products \ between \ uncorrelated \ signals} \Bigr\} \,.
\end{align}\par

The diode also performs the temporal average over a sampling period, so that:
\begin{align}\label{esu1B}
  \mean{\abs{E_{U1}} ^2} = \frac14 \Bigl[ &\frac{g_A^2 + g_B^2}{2}\Bigl(\mean{E_A'^2} + \mean{E_B^2}\Bigr) + g_Ag_B \sin(\Delta\phi) \Bigl(\mean{E_A'^2} - \mean{E_B^2}\Bigr) + \notag \\
  &\phantom{} g_A^2\mean{N_A^2} + g_B^2\mean{N_B^2} \Bigr] \,,
\end{align}
where the gains of the amplifiers have been assumed stable in the sampling interval and the Euler's formulae have been used. Furthermore, has been exploited the fact that the average products between uncorrelated signals is zero. \par

Eventually, since $E_A$ and $E_{RF}$ are also uncorrelated:
\be{meaneaearf}
\mean{E_A'^2} = \mean{E_A^2} + \mean{E_{RF}^2} \,,
\ee
and it is possible to write Eq.\ref{esu1B} as:
\begin{align}\label{esu1Bfinal}
  \mean{\abs{E_{U1}} ^2} = \frac14 \Bigl[ &\frac{g_A^2 + g_B^2}{2}\Bigl(\mean{E_A^2} + \mean{E_{RF}^2} + \mean{E_B^2}\Bigr) + \notag \\
  &\phantom{} g_Ag_B \sin(\Delta\phi) \Bigl(\mean{E_A^2} + \mean{E_{RF}^2} - \mean{E_B^2}\Bigr) + \notag \\
  &\phantom{} g_A^2\mean{N_A^2} + g_B^2\mean{N_B^2} \Bigr] \,.
\end{align}

\section{Signal expected at detector U2}
I want to make explicit Eq.\ref{u1u2B}, in the case of detector $\mathrm{U2}$:
\begin{align}\label{eu2B}
E_{U2} = &\frac{E_{Q1} + i E_{Q2}}{\sqrt2} = \frac{g_A e^{i\phi_A} + g_B e^{i\phi_B}}{4} E_A' + \frac{g_A e^{i\phi_A} - g_B e^{i\phi_B}}{4} E_B + \frac{g_A e^{i\phi_A}}{4} \sqrt2 N_A + \notag \\
      &\phantom{} \frac{g_B e^{i\phi_B}}{4} \sqrt2 N_B + \frac{g_A e^{i\phi_A} - g_B e^{i\phi_B}}{4} i E_A' + \frac{g_A e^{i\phi_A} + g_B e^{i\phi_B}}{4} i E_B + \notag \\
      &\phantom{} \frac{g_A e^{i\phi_A}}{4} \sqrt2 i N_A - \frac{g_B e^{i\phi_B}}{4} \sqrt2 i N_B = \notag \\
      &\phantom{} \hspace{-13pt}= \frac14 \Bigl\{ \Bigl[ g_A e^{i\phi_A} (i+1) - g_B e^{i\phi_B} (i-1) \Bigr] E_A' + \Bigl[g_A e^{i\phi_A} (i+1) + g_B e^{i\phi_B} (i-1) \Bigr] E_B + \notag \\
      &\phantom{} g_A e^{i\phi_A} \sqrt2 N_A (i+1) - g_B e^{i\phi_B} \sqrt2 N_B (i-1)  \Bigr\} \,.
\end{align}\par

The diodes convert the RF signal into an electric signal proportionally to the power of the incident field (Eq.\ref{diode}). To get the power, the square module of Eq.\ref{eu2B} must be computed:
\begin{align}\label{eu2B_1}
  \abs{ E_{U2} }^2 = B^2 \Bigl\{ \Bigl[ &g_A e^{i\phi_A} (i+1) - g_B e^{i\phi_B} (i-1)\Bigr] E_A' + \Bigl[g_A e^{i\phi_A}(i+1) + g_B e^{i\phi_B}(i-1) \Bigr] E_B + \notag \\
    &\phantom{} g_A e^{i\phi_A} \sqrt2 N_A (i+1) - g_B e^{i\phi_B} \sqrt2 N_B (i-1) \Bigr\}  \cdot \notag \\
    &\phantom{} \hspace{-10pt}\Bigl\{ \Bigl[ g_A e^{-i\phi_A} (1-i) + g_B e^{-i\phi_B} (1+i) \Bigr] E_A'^* + \notag \\
    &\phantom{} \hspace{-3pt}\Bigl[g_A e^{-i\phi_A}(1-i) - g_B e^{-i\phi_B}(1+i) \Bigr] E_B^* + g_A e^{-i\phi_A} \sqrt2 N_A^*(1-i) + \notag \\
    &\phantom{} g_B e^{-i\phi_B} \sqrt2 N_B^*(1+i) \Bigr\}\,,
\end{align}
where $B = \frac14$.\par
By doing the products, it is possible to obtain:
\begin{align}\label{eu2B_2}
  \abs{ E_{U2} }^2 = B^2 \Bigl\{ &\Bigl[ 2 g_A^2 + 2 i g_Ag_B e^{i\Delta\phi} - 2 i g_Ag_B e^{-i\Delta\phi} + 2 g_B^2 \Bigr] E_A'^2 + \notag \\
  &\phantom{} \Bigl[ 2 g_A^2 - 2 i g_Ag_B e^{i\Delta\phi} + 2 i g_Ag_B e^{-i\Delta\phi} + 2 g_B^2 \Bigr] E_B^2 + \notag \\
  &\phantom{} 4 g_A^2 N_A^2 + 4 g_B^2 N_B^2 + \mathrm{ products \ between \ uncorrelated \ signals} \Bigr\} \,,
\end{align}
with $\Delta\phi = \phi_A - \phi_B$. \par

Eq. \ref{eu2B_2} can be written also as:
\begin{align}\label{eu2B_3}
  \abs{ E_{U2} }^2 = B^2 \Bigl\{ &\Bigl[ 2g_A^2 - 4 g_Ag_B \frac{e^{i\Delta\phi} - e^{-i\Delta\phi}}{2i} + 2g_B^2 \Bigr] E_A'^2 + \notag \\
  &\phantom{} \Bigl[ 2g_A^2 + 4 g_Ag_B \frac{e^{i\Delta\phi} - e^{-i\Delta\phi}}{2i} + 2g_B^2 \Bigr] E_B^2 + \notag \\
  &\phantom{} 4 g_A^2 N_A^2 + 4 g_B^2 N_B^2 + \mathrm{ products \ between \ uncorrelated \ signals} \Bigr\} \,.
\end{align}\par

The diode also performs the temporal average over a sampling period, so that:
\begin{align}\label{esu2B}
  \mean{\abs{E_{U2}} ^2} = \frac14 \Bigl[ &\frac{g_A^2 + g_B^2}{2}\Bigl(\mean{E_A'^2} + \mean{E_B^2}\Bigr) - g_Ag_B \sin(\Delta\phi) \Bigl(\mean{E_A'^2} - \mean{E_B^2}\Bigr) + \notag \\
  &\phantom{} g_A^2\mean{N_A^2} + g_B^2\mean{N_B^2} \Bigr] \,,
\end{align}
where the gains of the amplifiers have been assumed stable in the sampling interval and the Euler's formulae have been used. Furthermore, has been exploited the fact that the average products between uncorrelated signals is zero. \par

Eventually, since $E_A$ and $E_{RF}$ are also uncorrelated:
\be{meaneaearf}
\mean{E_A'^2} = \mean{E_A^2} + \mean{E_{RF}^2} \,,
\ee
and it is possible to write Eq.\ref{esu2B} as:
\begin{align}\label{esu2Bfinal}
  \mean{\abs{E_{U2}} ^2} = \frac14 \Bigl[ &\frac{g_A^2 + g_B^2}{2}\Bigl(\mean{E_A^2} + \mean{E_{RF}^2} + \mean{E_B^2}\Bigr) - \notag \\
  &\phantom{} g_Ag_B \sin(\Delta\phi) \Bigl(\mean{E_A^2} + \mean{E_{RF}^2} - \mean{E_B^2}\Bigr) + \notag \\
  &\phantom{} g_A^2\mean{N_A^2} + g_B^2\mean{N_B^2} \Bigr] \,.
\end{align}

%% file: pubs.tex
\chapter*{List of Publications}
\thispagestyle{plain}
\hfill {\it As of \today}
\vspace{0.2cm}\\
{\bf Refereed publications}
\\
\cite{2019Univ....5...42M} \\
\cite{2018BAAA...60..107B} \\
\vspace{2cm}

\noindent {\bf Publications under review}
\\
\cite{2018arXiv181200785M} \\
\cite{2018arXiv180103730M} \\
\vspace{2cm}

\noindent {\bf Publications in preparation}
\\
\cite{piacentini:inpress} \\
\vspace{2cm}

\noindent {\bf Publications in conference proceedings}
\\
\cite{Incardona} \\
\cite{2018SPIE10708E..45S} \\
\cite{2018SPIE10708E..3VM} \\
\cite{2018SPIE10708E..2IO} \\
\cite{2018SPIE10708E..2BO} \\
\cite{2018SPIE10708E..1GF} \\
\vspace{2cm}